\documentclass[a4paper,11pt]{article}

\usepackage{jheppub} 
                     
\usepackage[T1]{fontenc} 
\usepackage{amsmath,amsthm}
\usepackage{amssymb,amsfonts}
\usepackage{graphicx}
\usepackage{dsfont}
\usepackage{relsize}
\usepackage{stmaryrd}
\usepackage{lmodern}
\usepackage{slantsc}
\usepackage{scalefnt}
\usepackage{subfigure}
\usepackage{xspace}
\usepackage{boxedminipage}
\usepackage{ifpdf}
\usepackage{multirow}
\usepackage{xifthen}
\usepackage{tensor}
\usepackage{array}
\usepackage{tabu}
\usepackage{tikz}
\usepackage{color}
\usepackage{diagbox}
\usetikzlibrary{arrows,matrix,calc,scopes,decorations.markings}
\allowdisplaybreaks[1]
\usepackage[mathcal]{euscript}
\usepackage{bbold}

\title{Fourier-transformed  gauge theory models of three-dimensional topological orders  with gapped boundaries}
\date{\today}
\author[a,b]{Siyuan Wang}
 \author[a,b]{Yanyan Chen}
 \author[a,b]{Hongyu Wang}
 \author[c,1]{Yuting Hu}
 \author[a,b,1]{Yidun Wan\note{Corresponding author}}
 \affiliation[a]{State Key Laboratory of Surface Physics, Department of Physics, Center for Field Theory and Particle Physics, and Institute for Nanoelectronic devices and Quantum computing, Fudan University, Shanghai 200433, China}
 \affiliation[b]{Shanghai Qi Zhi Institute, Shanghai 200030, China}
 \affiliation[c]{School of Physics, Hangzhou Normal University, Hangzhou 311121, China}
 \emailAdd{siyuanwang18@fudan.edu.cn, yanyanchen235@gmail.com, \\ wanghy17@fudan.edu.cn, yuting.phys@gmail.com, ydwan@fudan.edu.cn}

\abstract{In this paper, we apply the method of Fourier transform and basis rewriting developed in \cite{Wang2020b} for the two-dimensional quantum double model of topological orders to the three-dimensional gauge theory model (with a gauge group $G$) of three-dimensional topological orders. We find that the gapped boundary condition of the gauge theory model is characterized by a Frobenius algebra in the representation category $\mathcal Rep(G)$ of $G$, which also describes the charge splitting and condensation on the boundary. We also show that our Fourier transform maps the three-dimensional gauge theory model with input data $G$ to the Walker-Wang model with input data $\mathcal Rep(G)$ on a trivalent lattice with dangling edges, after truncating the Hilbert space by projecting all dangling edges to the trivial representation of $G$. This Fourier transform also provides a systematic construction of the gapped boundary theory of the Walker-Wang model. This establishes a correspondence between two types of topological field theories: the extended Dijkgraaf-Witten and extended Crane-Yetter theories.}

\begin{document}

\maketitle 

\flushbottom
\section{Introduction}\label{sec:intro}
Exactly solvable lattice models are a useful tool for studying topologically ordered matter phases (topological orders for short). For instances, the Kitaev quantum double (QD) model\cite{Kitaev2003a}, the Levin-Wen (LW) model\cite{Levin2004}, and the twisted quantum double (TQD) model\cite{Hu2012a} describe two-dimensional topological orders, while the Walker-Wang (WW) model\cite{Walker2011} and the twisted gauge theory (TGT) model\cite{Wan2014} describe three-dimensional topological orders. 

Despite the apparent differences between the QD double model and LW model, the two are intimately related: They are dual to each other in the sense that the QD model with a finite gauge group $G$ as its input data can be mapped to the LW model with input data $\mathcal{R}ep(G)$ (the representation category of $G$) via Fourier transform\cite{Buerschaper2009,Buerschaper2013,Wang2020b}. Both the QD and LW models have been extended to the case with gapped boundaries, called the extended QD\cite{Bravyi1998,Beigi2011,Cong2017,Bullivant2017} and LW models\cite{Kitaev2012,Hu2017,Hu2017a}. While the gapped boundary of the extended QD model is specified by a subgroup $K\subseteq G$ of the bulk gauge group $G$, that of the extended LW model is specified by a Frobenius algebra object of the input fusion category of the bulk. The extended QD and LW models can also be mapped to each other via Fourier transform and basis rewriting\cite{Wang2020b}. The boundary input data $K\subseteq G$ of the extended QD model is mapped to the boundary input Frobenius algebra $A_{G,K}\in \mathcal{O}bj(\mathcal{R}ep(G))$. The gapped boundaries of 2-dimensional topological orders are well understood through anyon condensation\cite{Bais2009,Kitaev2012,Levin2013,Kong2013,HungWan2014,Hung2015,Wan2017,Hu2022}, which has been extensively studied from both category\cite{Kong2013} and lattice model perspectives\cite{Cong2016a,Cong2017}, where the charge condensation at the boundary is observed from the input data of the boundary via the Fourier transform\cite{Wang2020b}. This Fourier transform and basis rewriting also manifest the full electromagnetic (EM) duality in these models\cite{Wang2020b}. It is believed that the understanding of the relationship between the QD model and the LW model is now complete. The EM duality in the TQD model is also studied\cite{Hu2022}. 

In contrast to two dimensions, the models of 3-dimensional topological orders and their exact relations are less understood. It is believed that the TGT model describes all possible $3$-dimensional topological orders with bosonic point-like excitations\cite{Lan2018b}. The TGT model is specified by an input finite gauge group $G$ and a 4-cocycle $\omega\in H^4[G,U(1)]$, and the gapped boundary of the TGT model is specified by a subgroup $K$ and a 3-cochain $\alpha\in Z^3[G,U(1)]$\cite{Wang2018}. Furthermore, we will explore the charge condensation of the GT model through concrete model construction, since it reveals the relationship between the bulk and boundary of the GT model. In two dimensions, the bulk charges may split and partially or fully condense at the boundary, depending on the gapped boundary condition\cite{Wang2020b}. In three dimensions, however, the phenomenon of boundary charge condensation has not been fully understood\cite{Luo2022}. While layer construction offers another approach to understanding charge condensation at the boundary of certain $3$-dimensional topological orders\cite{jian2014,Gaiotto2019}, most current understandings of this phenomenon are either categorical and abstract or limited to specific cases\cite{Zhao2022a,Luo2022}. Therefore, investigating the splitting and partial condensation phenomena in concrete lattice models of three-dimensional topological orders with general input data would be worthwhile. In this paper, we generalize the method of Fourier transform and basis rewriting developed in \cite{Wang2020b} from two to three dimensions to investigate the gapped boundaries of a $3$-dimensional GT model. We prove that the boundary theory of the Fourier-transformed model is also characterized by the Frobenius algebra. Then similar to the 2-dimensional extended LW model with input data $\mathcal{R}ep(G)$, the charge of the condensates is described by Frobenius algebra, and the splitting process origins from the multiplicity of the objects of $\mathcal{R}ep(G)$. Thus we explain these phenomena using only the input data of the 3-dimensional model. 

On the other hand, the $3$-dimensional WW model is specified by an input unitary braided fusion category (UBFC) and describes a family of three-dimensional topological orders\cite{Cho2011,Wang2017a}. In this paper, we show that our Fourier transform, extending the mapping process from QD models to LW models into three dimensions, indeed maps a $3$-dimensional GT model with input data $G$ to a WW model with  UBFC $\mathcal{R}ep(G)$ as its input data on the level of Hilbert spaces and Hamiltonians. Furthermore, the Fourier transform of the gapped boundaries of the $3$-dimensional GT model also gives a systematic construction of the gapped boundaries of the WW model. Prior to our work, aside from a few special cases like the three-dimensional toric code and double-semion\cite{VonKeyserlingk2013}, no systematic construction of the gapped-boundary Hamiltonian of the WW model has been established. Our understanding of the relationship between the TGT model and the WW model is also limited to some special cases, such as the example of $\mathbb{Z}_2 \times \mathbb{Z}_2$ twisted gauge theory model, which is shown to share the same modular matrices with a corresponding WW model\cite{Wang2017a}. 

In this paper, we focus on three-dimensional GT models defined on cubic lattices with boundaries, with input data being a finite group $G$ in the bulk and a subgroup $K\subseteq G$ within the boundary. The more complicated cases, i.e., the twisted gauge theory models and twisted boundaries, are our ongoing work and will be reported elsewhere. The Fourier-transformed basis of the Hilbert space of the GT model leads us to rewrite the model on a slightly different trivalent lattice $\tilde{\Gamma}$ with a tail (dangling edge) attached to each vertex. This lattice is precisely where the WW model lives on with an enlarged Hilbert space. This enlargement is necessary since the original WW model has a Hilbert space insufficient for accommodating the full spectrum of charge excitation. After the Fourier transform, the bulk input data becomes a UBFC $\mathcal{R}ep(G)$, while the boundary degrees of freedom are projected into Frobenius algebra $A_{G,K}$, as in the case of Fourier-transformed QD model in two dimensions. We also show that the Fourier-transformed GT model with input data $G$ on the revised lattice $\tilde{\Gamma}$ can be mapped to a WW model with input data $\mathcal{R}ep(G)$ on the same lattice after truncating the Hilbert space by projecting all dangling edges to trivial representation. Since both GT and WW models with gapped boundaries serve as Hamiltonian extensions of the extended Dijkgraaf-Witten and extended Crane-Yetter topological field theories, our results also establish a correspondence between these two types of topological field theories.
 
Our paper is organized as follows. Section \ref{sec:2} reviews the three-dimensional GT model with gapped boundaries. Section \ref{sec:3} Fourier transforms and rewrites the extended GT model. Section \ref{sec:4} verifies the emergent Frobenius algebra structure on the boundary. Section \ref{sec:5} proves that our Fourier transform indeed maps the three-dimensional GT model to the WW model. Finally, the appendices collect a review of the WW model and certain details to avoid clutter in the main text.  

\section{Three-dimensional GT model with gapped boundaries}\label{sec:2}
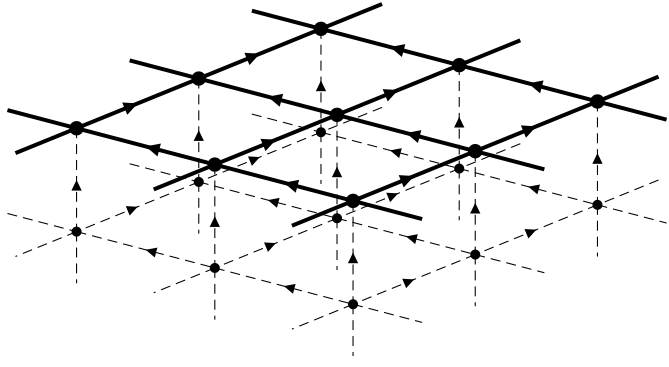
\begin{figure}[ht]
    \centering
\begin{tikzpicture}[x=0.75pt,y=0.75pt,yscale=-1,xscale=1]

\draw [line width=1.5]    (226.5,31.13) -- (261,40.26) ;
\draw [line width=1.5]    (261,40.26) -- (330.01,58.53) ;
\draw [shift={(330.01,58.53)}, rotate = 14.83] [color={rgb, 255:red, 0; green, 0; blue, 0 }  ][fill={rgb, 255:red, 0; green, 0; blue, 0 }  ][line width=1.5]      (0, 0) circle [x radius= 2.61, y radius= 2.61]   ;
\draw [shift={(295.51,49.4)}, rotate = 14.83] [fill={rgb, 255:red, 0; green, 0; blue, 0 }  ][line width=0.08]  [draw opacity=0] (6.97,-3.35) -- (0,0) -- (6.97,3.35) -- cycle    ;
\draw [shift={(261,40.26)}, rotate = 14.83] [color={rgb, 255:red, 0; green, 0; blue, 0 }  ][fill={rgb, 255:red, 0; green, 0; blue, 0 }  ][line width=1.5]      (0, 0) circle [x radius= 2.61, y radius= 2.61]   ;
\draw [line width=1.5]    (330.01,58.53) -- (399.01,76.79) ;
\draw [shift={(364.51,67.66)}, rotate = 14.83] [fill={rgb, 255:red, 0; green, 0; blue, 0 }  ][line width=0.08]  [draw opacity=0] (6.97,-3.35) -- (0,0) -- (6.97,3.35) -- cycle    ;
\draw [line width=1.5]    (399.01,76.79) -- (433.51,85.93) ;
\draw [line width=1.5]    (261,40.26) -- (200,65.26) ;
\draw [shift={(230.5,52.76)}, rotate = 157.71] [fill={rgb, 255:red, 0; green, 0; blue, 0 }  ][line width=0.08]  [draw opacity=0] (6.97,-3.35) -- (0,0) -- (6.97,3.35) -- cycle    ;
\draw [line width=1.5]    (330.01,58.53) -- (269.01,83.53) ;
\draw [shift={(299.51,71.03)}, rotate = 157.71] [fill={rgb, 255:red, 0; green, 0; blue, 0 }  ][line width=0.08]  [draw opacity=0] (6.97,-3.35) -- (0,0) -- (6.97,3.35) -- cycle    ;
\draw [line width=1.5]    (399.01,76.79) -- (338.01,101.79) ;
\draw [shift={(368.51,89.29)}, rotate = 157.71] [fill={rgb, 255:red, 0; green, 0; blue, 0 }  ][line width=0.08]  [draw opacity=0] (6.97,-3.35) -- (0,0) -- (6.97,3.35) -- cycle    ;
\draw [line width=1.5]    (200,65.26) -- (269.01,83.53) ;
\draw [shift={(269.01,83.53)}, rotate = 14.83] [color={rgb, 255:red, 0; green, 0; blue, 0 }  ][fill={rgb, 255:red, 0; green, 0; blue, 0 }  ][line width=1.5]      (0, 0) circle [x radius= 2.61, y radius= 2.61]   ;
\draw [shift={(234.51,74.4)}, rotate = 14.83] [fill={rgb, 255:red, 0; green, 0; blue, 0 }  ][line width=0.08]  [draw opacity=0] (6.97,-3.35) -- (0,0) -- (6.97,3.35) -- cycle    ;
\draw [shift={(200,65.26)}, rotate = 14.83] [color={rgb, 255:red, 0; green, 0; blue, 0 }  ][fill={rgb, 255:red, 0; green, 0; blue, 0 }  ][line width=1.5]      (0, 0) circle [x radius= 2.61, y radius= 2.61]   ;
\draw [line width=1.5]    (269.01,83.53) -- (338.01,101.79) ;
\draw [shift={(338.01,101.79)}, rotate = 14.83] [color={rgb, 255:red, 0; green, 0; blue, 0 }  ][fill={rgb, 255:red, 0; green, 0; blue, 0 }  ][line width=1.5]      (0, 0) circle [x radius= 2.61, y radius= 2.61]   ;
\draw [shift={(303.51,92.66)}, rotate = 14.83] [fill={rgb, 255:red, 0; green, 0; blue, 0 }  ][line width=0.08]  [draw opacity=0] (6.97,-3.35) -- (0,0) -- (6.97,3.35) -- cycle    ;
\draw  [dash pattern={on 3.75pt off 2.25pt}]  (261,40.26) -- (261,92.26) ;
\draw [shift={(261,66.26)}, rotate = 90] [fill={rgb, 255:red, 0; green, 0; blue, 0 }  ][line width=0.08]  [draw opacity=0] (5.36,-2.57) -- (0,0) -- (5.36,2.57) -- cycle    ;
\draw [line width=1.5]    (322,15.26) -- (261,40.26) ;
\draw [shift={(291.5,27.76)}, rotate = 157.71] [fill={rgb, 255:red, 0; green, 0; blue, 0 }  ][line width=0.08]  [draw opacity=0] (6.97,-3.35) -- (0,0) -- (6.97,3.35) -- cycle    ;
\draw [line width=1.5]    (200,65.26) -- (169.5,77.76) ;
\draw  [dash pattern={on 3.75pt off 2.25pt}]  (261,92.26) -- (330.01,110.53) ;
\draw [shift={(295.51,101.4)}, rotate = 14.83] [fill={rgb, 255:red, 0; green, 0; blue, 0 }  ][line width=0.08]  [draw opacity=0] (5.36,-2.57) -- (0,0) -- (5.36,2.57) -- cycle    ;
\draw  [dash pattern={on 3.75pt off 2.25pt}]  (330.01,110.53) -- (399.01,128.79) ;
\draw [shift={(364.51,119.66)}, rotate = 14.83] [fill={rgb, 255:red, 0; green, 0; blue, 0 }  ][line width=0.08]  [draw opacity=0] (5.36,-2.57) -- (0,0) -- (5.36,2.57) -- cycle    ;
\draw  [dash pattern={on 3.75pt off 2.25pt}]  (330.01,58.53) -- (330.01,110.53) ;
\draw [shift={(330.01,84.53)}, rotate = 90] [fill={rgb, 255:red, 0; green, 0; blue, 0 }  ][line width=0.08]  [draw opacity=0] (5.36,-2.57) -- (0,0) -- (5.36,2.57) -- cycle    ;
\draw  [dash pattern={on 3.75pt off 2.25pt}]  (261,92.26) -- (200,117.26) ;
\draw [shift={(200,117.26)}, rotate = 157.71] [color={rgb, 255:red, 0; green, 0; blue, 0 }  ][fill={rgb, 255:red, 0; green, 0; blue, 0 }  ][line width=0.75]      (0, 0) circle [x radius= 2.01, y radius= 2.01]   ;
\draw [shift={(230.5,104.76)}, rotate = 157.71] [fill={rgb, 255:red, 0; green, 0; blue, 0 }  ][line width=0.08]  [draw opacity=0] (5.36,-2.57) -- (0,0) -- (5.36,2.57) -- cycle    ;
\draw [shift={(261,92.26)}, rotate = 157.71] [color={rgb, 255:red, 0; green, 0; blue, 0 }  ][fill={rgb, 255:red, 0; green, 0; blue, 0 }  ][line width=0.75]      (0, 0) circle [x radius= 2.01, y radius= 2.01]   ;
\draw  [dash pattern={on 3.75pt off 2.25pt}]  (399.01,76.79) -- (399.01,128.79) ;
\draw [shift={(399.01,102.79)}, rotate = 90] [fill={rgb, 255:red, 0; green, 0; blue, 0 }  ][line width=0.08]  [draw opacity=0] (5.36,-2.57) -- (0,0) -- (5.36,2.57) -- cycle    ;
\draw  [dash pattern={on 3.75pt off 2.25pt}]  (200,65.26) -- (200,117.26) ;
\draw [shift={(200,91.26)}, rotate = 90] [fill={rgb, 255:red, 0; green, 0; blue, 0 }  ][line width=0.08]  [draw opacity=0] (5.36,-2.57) -- (0,0) -- (5.36,2.57) -- cycle    ;
\draw [line width=1.5]    (165.5,56.13) -- (200,65.26) ;
\draw  [dash pattern={on 3.75pt off 2.25pt}]  (200,117.26) -- (169.5,129.76) ;
\draw  [dash pattern={on 3.75pt off 2.25pt}]  (200,117.26) -- (269.01,135.53) ;
\draw [shift={(234.51,126.4)}, rotate = 14.83] [fill={rgb, 255:red, 0; green, 0; blue, 0 }  ][line width=0.08]  [draw opacity=0] (5.36,-2.57) -- (0,0) -- (5.36,2.57) -- cycle    ;
\draw  [dash pattern={on 3.75pt off 2.25pt}]  (269.01,135.53) -- (338.01,153.79) ;
\draw [shift={(338.01,153.79)}, rotate = 14.83] [color={rgb, 255:red, 0; green, 0; blue, 0 }  ][fill={rgb, 255:red, 0; green, 0; blue, 0 }  ][line width=0.75]      (0, 0) circle [x radius= 2.01, y radius= 2.01]   ;
\draw [shift={(303.51,144.66)}, rotate = 14.83] [fill={rgb, 255:red, 0; green, 0; blue, 0 }  ][line width=0.08]  [draw opacity=0] (5.36,-2.57) -- (0,0) -- (5.36,2.57) -- cycle    ;
\draw  [dash pattern={on 3.75pt off 2.25pt}]  (330.01,110.53) -- (269.01,135.53) ;
\draw [shift={(269.01,135.53)}, rotate = 157.71] [color={rgb, 255:red, 0; green, 0; blue, 0 }  ][fill={rgb, 255:red, 0; green, 0; blue, 0 }  ][line width=0.75]      (0, 0) circle [x radius= 2.01, y radius= 2.01]   ;
\draw [shift={(299.51,123.03)}, rotate = 157.71] [fill={rgb, 255:red, 0; green, 0; blue, 0 }  ][line width=0.08]  [draw opacity=0] (5.36,-2.57) -- (0,0) -- (5.36,2.57) -- cycle    ;
\draw [shift={(330.01,110.53)}, rotate = 157.71] [color={rgb, 255:red, 0; green, 0; blue, 0 }  ][fill={rgb, 255:red, 0; green, 0; blue, 0 }  ][line width=0.75]      (0, 0) circle [x radius= 2.01, y radius= 2.01]   ;
\draw  [dash pattern={on 3.75pt off 2.25pt}]  (399.01,128.79) -- (338.01,153.79) ;
\draw [shift={(368.51,141.29)}, rotate = 157.71] [fill={rgb, 255:red, 0; green, 0; blue, 0 }  ][line width=0.08]  [draw opacity=0] (5.36,-2.57) -- (0,0) -- (5.36,2.57) -- cycle    ;
\draw [line width=1.5]    (391.01,33.53) -- (330.01,58.53) ;
\draw [shift={(360.51,46.03)}, rotate = 157.71] [fill={rgb, 255:red, 0; green, 0; blue, 0 }  ][line width=0.08]  [draw opacity=0] (6.97,-3.35) -- (0,0) -- (6.97,3.35) -- cycle    ;
\draw [line width=1.5]    (460.01,51.79) -- (399.01,76.79) ;
\draw [shift={(399.01,76.79)}, rotate = 157.71] [color={rgb, 255:red, 0; green, 0; blue, 0 }  ][fill={rgb, 255:red, 0; green, 0; blue, 0 }  ][line width=1.5]      (0, 0) circle [x radius= 2.61, y radius= 2.61]   ;
\draw [shift={(429.51,64.29)}, rotate = 157.71] [fill={rgb, 255:red, 0; green, 0; blue, 0 }  ][line width=0.08]  [draw opacity=0] (6.97,-3.35) -- (0,0) -- (6.97,3.35) -- cycle    ;
\draw [shift={(460.01,51.79)}, rotate = 157.71] [color={rgb, 255:red, 0; green, 0; blue, 0 }  ][fill={rgb, 255:red, 0; green, 0; blue, 0 }  ][line width=1.5]      (0, 0) circle [x radius= 2.61, y radius= 2.61]   ;
\draw [line width=1.5]    (322,15.26) -- (391.01,33.53) ;
\draw [shift={(391.01,33.53)}, rotate = 14.83] [color={rgb, 255:red, 0; green, 0; blue, 0 }  ][fill={rgb, 255:red, 0; green, 0; blue, 0 }  ][line width=1.5]      (0, 0) circle [x radius= 2.61, y radius= 2.61]   ;
\draw [shift={(356.51,24.4)}, rotate = 14.83] [fill={rgb, 255:red, 0; green, 0; blue, 0 }  ][line width=0.08]  [draw opacity=0] (6.97,-3.35) -- (0,0) -- (6.97,3.35) -- cycle    ;
\draw [shift={(322,15.26)}, rotate = 14.83] [color={rgb, 255:red, 0; green, 0; blue, 0 }  ][fill={rgb, 255:red, 0; green, 0; blue, 0 }  ][line width=1.5]      (0, 0) circle [x radius= 2.61, y radius= 2.61]   ;
\draw [line width=1.5]    (391.01,33.53) -- (460.01,51.79) ;
\draw [shift={(425.51,42.66)}, rotate = 14.83] [fill={rgb, 255:red, 0; green, 0; blue, 0 }  ][line width=0.08]  [draw opacity=0] (6.97,-3.35) -- (0,0) -- (6.97,3.35) -- cycle    ;
\draw [line width=1.5]    (460.01,51.79) -- (494.51,60.93) ;
\draw [line width=1.5]    (352.5,2.76) -- (322,15.26) ;
\draw [line width=1.5]    (421.51,21.03) -- (391.01,33.53) ;
\draw [line width=1.5]    (490.51,39.29) -- (460.01,51.79) ;
\draw [line width=1.5]    (287.5,6.13) -- (322,15.26) ;
\draw  [dash pattern={on 3.75pt off 2.25pt}]  (322,15.26) -- (322,67.26) ;
\draw [shift={(322,41.26)}, rotate = 90] [fill={rgb, 255:red, 0; green, 0; blue, 0 }  ][line width=0.08]  [draw opacity=0] (5.36,-2.57) -- (0,0) -- (5.36,2.57) -- cycle    ;
\draw  [dash pattern={on 3.75pt off 2.25pt}]  (391.01,33.53) -- (391.01,85.53) ;
\draw [shift={(391.01,59.53)}, rotate = 90] [fill={rgb, 255:red, 0; green, 0; blue, 0 }  ][line width=0.08]  [draw opacity=0] (5.36,-2.57) -- (0,0) -- (5.36,2.57) -- cycle    ;
\draw  [dash pattern={on 3.75pt off 2.25pt}]  (460.01,51.79) -- (460.01,103.79) ;
\draw [shift={(460.01,77.79)}, rotate = 90] [fill={rgb, 255:red, 0; green, 0; blue, 0 }  ][line width=0.08]  [draw opacity=0] (5.36,-2.57) -- (0,0) -- (5.36,2.57) -- cycle    ;
\draw  [dash pattern={on 3.75pt off 2.25pt}]  (322,67.26) -- (261,92.26) ;
\draw [shift={(291.5,79.76)}, rotate = 157.71] [fill={rgb, 255:red, 0; green, 0; blue, 0 }  ][line width=0.08]  [draw opacity=0] (5.36,-2.57) -- (0,0) -- (5.36,2.57) -- cycle    ;
\draw  [dash pattern={on 3.75pt off 2.25pt}]  (391.01,85.53) -- (330.01,110.53) ;
\draw [shift={(360.51,98.03)}, rotate = 157.71] [fill={rgb, 255:red, 0; green, 0; blue, 0 }  ][line width=0.08]  [draw opacity=0] (5.36,-2.57) -- (0,0) -- (5.36,2.57) -- cycle    ;
\draw  [dash pattern={on 3.75pt off 2.25pt}]  (460.01,103.79) -- (399.01,128.79) ;
\draw [shift={(399.01,128.79)}, rotate = 157.71] [color={rgb, 255:red, 0; green, 0; blue, 0 }  ][fill={rgb, 255:red, 0; green, 0; blue, 0 }  ][line width=0.75]      (0, 0) circle [x radius= 2.01, y radius= 2.01]   ;
\draw [shift={(429.51,116.29)}, rotate = 157.71] [fill={rgb, 255:red, 0; green, 0; blue, 0 }  ][line width=0.08]  [draw opacity=0] (5.36,-2.57) -- (0,0) -- (5.36,2.57) -- cycle    ;
\draw [shift={(460.01,103.79)}, rotate = 157.71] [color={rgb, 255:red, 0; green, 0; blue, 0 }  ][fill={rgb, 255:red, 0; green, 0; blue, 0 }  ][line width=0.75]      (0, 0) circle [x radius= 2.01, y radius= 2.01]   ;
\draw  [dash pattern={on 3.75pt off 2.25pt}]  (322,67.26) -- (391.01,85.53) ;
\draw [shift={(391.01,85.53)}, rotate = 14.83] [color={rgb, 255:red, 0; green, 0; blue, 0 }  ][fill={rgb, 255:red, 0; green, 0; blue, 0 }  ][line width=0.75]      (0, 0) circle [x radius= 2.01, y radius= 2.01]   ;
\draw [shift={(356.51,76.4)}, rotate = 14.83] [fill={rgb, 255:red, 0; green, 0; blue, 0 }  ][line width=0.08]  [draw opacity=0] (5.36,-2.57) -- (0,0) -- (5.36,2.57) -- cycle    ;
\draw [shift={(322,67.26)}, rotate = 14.83] [color={rgb, 255:red, 0; green, 0; blue, 0 }  ][fill={rgb, 255:red, 0; green, 0; blue, 0 }  ][line width=0.75]      (0, 0) circle [x radius= 2.01, y radius= 2.01]   ;
\draw  [dash pattern={on 3.75pt off 2.25pt}]  (391.01,85.53) -- (460.01,103.79) ;
\draw [shift={(425.51,94.66)}, rotate = 14.83] [fill={rgb, 255:red, 0; green, 0; blue, 0 }  ][line width=0.08]  [draw opacity=0] (5.36,-2.57) -- (0,0) -- (5.36,2.57) -- cycle    ;
\draw  [dash pattern={on 3.75pt off 2.25pt}]  (269.01,83.53) -- (269.01,135.53) ;
\draw [shift={(269.01,109.53)}, rotate = 90] [fill={rgb, 255:red, 0; green, 0; blue, 0 }  ][line width=0.08]  [draw opacity=0] (5.36,-2.57) -- (0,0) -- (5.36,2.57) -- cycle    ;
\draw  [dash pattern={on 3.75pt off 2.25pt}]  (338.01,101.79) -- (338.01,153.79) ;
\draw [shift={(338.01,127.79)}, rotate = 90] [fill={rgb, 255:red, 0; green, 0; blue, 0 }  ][line width=0.08]  [draw opacity=0] (5.36,-2.57) -- (0,0) -- (5.36,2.57) -- cycle    ;
\draw  [dash pattern={on 3.75pt off 2.25pt}]  (269.01,135.53) -- (238.51,148.03) ;
\draw  [dash pattern={on 3.75pt off 2.25pt}]  (338.01,153.79) -- (307.51,166.29) ;
\draw  [dash pattern={on 3.75pt off 2.25pt}]  (460.01,103.79) -- (494.51,112.93) ;
\draw  [dash pattern={on 3.75pt off 2.25pt}]  (399.01,128.79) -- (433.51,137.93) ;
\draw  [dash pattern={on 3.75pt off 2.25pt}]  (338.01,153.79) -- (372.51,162.93) ;
\draw  [dash pattern={on 3.75pt off 2.25pt}]  (165.5,108.13) -- (200,117.26) ;
\draw  [dash pattern={on 3.75pt off 2.25pt}]  (226.5,83.13) -- (261,92.26) ;
\draw  [dash pattern={on 3.75pt off 2.25pt}]  (287.5,58.13) -- (322,67.26) ;
\draw  [dash pattern={on 3.75pt off 2.25pt}]  (490.51,91.29) -- (460.01,103.79) ;
\draw  [dash pattern={on 3.75pt off 2.25pt}]  (421.51,73.03) -- (391.01,85.53) ;
\draw  [dash pattern={on 3.75pt off 2.25pt}]  (352.5,54.76) -- (322,67.26) ;
\draw  [dash pattern={on 3.75pt off 2.25pt}]  (269.01,135.53) -- (269.01,161.53) ;
\draw  [dash pattern={on 3.75pt off 2.25pt}]  (338.01,153.79) -- (338.01,179.79) ;
\draw  [dash pattern={on 3.75pt off 2.25pt}]  (330.01,110.53) -- (330.01,136.53) ;
\draw  [dash pattern={on 3.75pt off 2.25pt}]  (399.01,128.79) -- (399.01,154.79) ;
\draw  [dash pattern={on 3.75pt off 2.25pt}]  (460.01,103.79) -- (460.01,129.79) ;
\draw  [dash pattern={on 3.75pt off 2.25pt}]  (391.01,85.53) -- (391.01,111.53) ;
\draw  [dash pattern={on 3.75pt off 2.25pt}]  (322,67.26) -- (322,93.26) ;
\draw  [dash pattern={on 3.75pt off 2.25pt}]  (261,92.26) -- (261,118.26) ;
\draw  [dash pattern={on 3.75pt off 2.25pt}]  (200,117.26) -- (200,143.26) ;
\draw [line width=1.5]    (269.01,83.53) -- (238.51,96.03) ;
\draw [line width=1.5]    (338.01,101.79) -- (307.51,114.29) ;
\draw [line width=1.5]    (338.01,101.79) -- (372.51,110.93) ;
\end{tikzpicture}
    \caption{A portion of an oriented trivalent lattice, on which the three-dimensional GT model with gapped boundaries is defined. Each edge of the lattice is graced with a group element of a finite gauge group $G$. The thick lines comprise the boundary while the dashed lines comprise the bulk.}
    \label{fig:1}
\end{figure}

The GT model with gapped boundaries is a Hamiltonian extension of the Dijkgraaf-Witten topological gauge theory with a finite gauge group. Without taking twist into consideration, the model can be defined on an arbitrary lattice with one or multiple boundaries. Topological invariance allows the model to be defined on a fixed lattice for computational convenience. In this paper, we consider an oriented cubic lattice $\Gamma$ which boundaries, part of which is shown in \autoref{fig:1}. 

The input data of the model is a finite gauge group $G$, whose elements are assigned to the edges of $\Gamma$. The total Hilbert space is spanned by all possible configurations of the group elements of $G$ on the edges of $\Gamma$ and is the tensor product of the local Hilbert spaces respectively on the edges. Namely 
\begin{equation}\label{eq:2.1}
    \mathcal{H}^\text{GT}_G=\bigotimes_{e\in\Gamma}\mathcal{H}_e=\bigotimes_{e\in\Gamma}\operatorname{span}_{g_e\in G}\{|g_e\rangle\},
\end{equation}
where $e$ is an edge in $\Gamma$. Reversing the orientation of an edge graces with a group element $g$ changes the group element to its inverse $\bar g:=g^{-1}$. The inner product of the Hilbert space is simply $\langle g'_e|g_e\rangle=\delta_{g'g}$.  The Hamiltonian of the model is the sum of a bulk Hamiltonian and a boundary Hamiltonian: 
\begin{equation}\label{eq:2.2}
    H^\text{GT}_{G,K}=H^\text{GT}_{G}+\overline{H^\text{GT}_K}.
\end{equation}
Here and hereafter, an operator with an overline is a boundary operator. The bulk Hamiltonian consists of the sum of vertex operators and that of plaquette operators: 
\begin{equation}\label{eq:2.3}
    H^\text{GT}_{G}=-\sum_{v\in\Gamma\setminus\partial\Gamma}A^\text{GT}_{v}-\sum_{p\in\Gamma\setminus\partial\Gamma}B^\text{GT}_p,
\end{equation}
where the sums run over all vertices and plaquettes in the bulk of $\Gamma$. The vertex operator $A^\text{GT}_{v}$ acts locally on the six edges incident at the vertex $v$: 
\begin{equation}\label{eq:2.4}
    A_{v}^{\text{GT}} \Bigg|\begin{tikzpicture}[x=0.75pt,y=0.75pt,yscale=-1,xscale=1, baseline=(XXXX.south) ]
\path (0,91);\path (95.99015045166016,0);\draw    ($(current bounding box.center)+(0,0.3em)$) node [anchor=south] (XXXX) {};
\draw    (0.99,28.99) -- (46.41,44.44) ;
\draw [shift={(23.7,36.72)}, rotate = 18.79] [fill={rgb, 255:red, 0; green, 0; blue, 0 }  ][line width=0.08]  [draw opacity=0] (5.36,-2.57) -- (0,0) -- (5.36,2.57) -- cycle    ;
\draw    (46.41,44.44) -- (91.83,59.89) ;
\draw [shift={(69.12,52.16)}, rotate = 18.79] [fill={rgb, 255:red, 0; green, 0; blue, 0 }  ][line width=0.08]  [draw opacity=0] (5.36,-2.57) -- (0,0) -- (5.36,2.57) -- cycle    ;
\draw    (46.41,44.44) -- (85.83,22.89) ;
\draw [shift={(66.12,33.66)}, rotate = 151.33] [fill={rgb, 255:red, 0; green, 0; blue, 0 }  ][line width=0.08]  [draw opacity=0] (5.36,-2.57) -- (0,0) -- (5.36,2.57) -- cycle    ;
\draw    (6.99,65.99) -- (46.41,44.44) ;
\draw [shift={(46.41,44.44)}, rotate = 331.33] [color={rgb, 255:red, 0; green, 0; blue, 0 }  ][fill={rgb, 255:red, 0; green, 0; blue, 0 }  ][line width=0.75]      (0, 0) circle [x radius= 2.01, y radius= 2.01]   ;
\draw [shift={(26.7,55.22)}, rotate = 151.33] [fill={rgb, 255:red, 0; green, 0; blue, 0 }  ][line width=0.08]  [draw opacity=0] (5.36,-2.57) -- (0,0) -- (5.36,2.57) -- cycle    ;
\draw    (46.41,44.44) -- (46.41,1.44) ;
\draw [shift={(46.41,22.94)}, rotate = 90] [fill={rgb, 255:red, 0; green, 0; blue, 0 }  ][line width=0.08]  [draw opacity=0] (5.36,-2.57) -- (0,0) -- (5.36,2.57) -- cycle    ;
\draw    (46.41,87.44) -- (46.41,44.44) ;
\draw [shift={(46.41,65.94)}, rotate = 90] [fill={rgb, 255:red, 0; green, 0; blue, 0 }  ][line width=0.08]  [draw opacity=0] (5.36,-2.57) -- (0,0) -- (5.36,2.57) -- cycle    ;
\draw (20,59.4) node [anchor=north west][inner sep=0.75pt]    {$g$};
\draw (59,20) node [anchor=north west][inner sep=0.75pt]    {\small $h$};
\draw (14,37) node [anchor=north west][inner sep=0.75pt]    {\small $i$};
\draw (66,36) node [anchor=north west][inner sep=0.75pt]    {\small $j$};
\draw (48,62) node [anchor=north west][inner sep=0.75pt]    {\small $k$};
\draw (36,17) node [anchor=north west][inner sep=0.75pt]    {\small $l$};
\draw (35,49.4) node [anchor=north west][inner sep=0.75pt]    {\small $v$};
\end{tikzpicture}
\Bigg\rangle =\frac{1}{|G|}\sum _{x\in G} \Bigg|\begin{tikzpicture}[x=0.75pt,y=0.75pt,yscale=-1,xscale=1, baseline=(XXXX.south) ]
\path (0,91);\path (95.99015045166016,0);\draw    ($(current bounding box.center)+(0,0.3em)$) node [anchor=south] (XXXX) {};
\draw    (0.99,28.99) -- (46.41,44.44) ;
\draw [shift={(23.7,36.72)}, rotate = 18.79] [fill={rgb, 255:red, 0; green, 0; blue, 0 }  ][line width=0.08]  [draw opacity=0] (5.36,-2.57) -- (0,0) -- (5.36,2.57) -- cycle    ;
\draw    (46.41,44.44) -- (91.83,59.89) ;
\draw [shift={(69.12,52.16)}, rotate = 18.79] [fill={rgb, 255:red, 0; green, 0; blue, 0 }  ][line width=0.08]  [draw opacity=0] (5.36,-2.57) -- (0,0) -- (5.36,2.57) -- cycle    ;
\draw    (46.41,44.44) -- (85.83,22.89) ;
\draw [shift={(66.12,33.66)}, rotate = 151.33] [fill={rgb, 255:red, 0; green, 0; blue, 0 }  ][line width=0.08]  [draw opacity=0] (5.36,-2.57) -- (0,0) -- (5.36,2.57) -- cycle    ;
\draw    (6.99,65.99) -- (46.41,44.44) ;
\draw [shift={(46.41,44.44)}, rotate = 331.33] [color={rgb, 255:red, 0; green, 0; blue, 0 }  ][fill={rgb, 255:red, 0; green, 0; blue, 0 }  ][line width=0.75]      (0, 0) circle [x radius= 2.01, y radius= 2.01]   ;
\draw [shift={(26.7,55.22)}, rotate = 151.33] [fill={rgb, 255:red, 0; green, 0; blue, 0 }  ][line width=0.08]  [draw opacity=0] (5.36,-2.57) -- (0,0) -- (5.36,2.57) -- cycle    ;
\draw    (46.41,44.44) -- (46.41,1.44) ;
\draw [shift={(46.41,22.94)}, rotate = 90] [fill={rgb, 255:red, 0; green, 0; blue, 0 }  ][line width=0.08]  [draw opacity=0] (5.36,-2.57) -- (0,0) -- (5.36,2.57) -- cycle    ;
\draw    (46.41,87.44) -- (46.41,44.44) ;
\draw [shift={(46.41,65.94)}, rotate = 90] [fill={rgb, 255:red, 0; green, 0; blue, 0 }  ][line width=0.08]  [draw opacity=0] (5.36,-2.57) -- (0,0) -- (5.36,2.57) -- cycle    ;
\draw (17,61.4) node [anchor=north west][inner sep=0.75pt]    {$xg$};
\draw (54,20) node [anchor=north west][inner sep=0.75pt]    {$h\bar{x}$};
\draw (12,37) node [anchor=north west][inner sep=0.75pt]    {$i\bar{x}$};
\draw (66,39) node [anchor=north west][inner sep=0.75pt]    {$xj$};
\draw (48,62) node [anchor=north west][inner sep=0.75pt]    {$xk$};
\draw (29,17) node [anchor=north west][inner sep=0.75pt]    {$l\bar{x}$};
\draw (35,49.4) node [anchor=north west][inner sep=0.75pt]    {$v$};
\end{tikzpicture}
\Bigg\rangle,
\end{equation}
which is understood as a discrete gauge transformation averaged over $G$. Clearly, $A^\text{GT}_{v}$ is a projector because $(A^\text{GT}_{v})^2=A^\text{GT}_{v}$, and it projects out any such local states that are not invariant under the gauge transformation. In gauge theory terminologies, $A^\text{GT}_{v}$ imposes a local Gauss constraint. The plaquette operator acts locally on the four edges bounding the plaquette $p$: 
\begin{equation}\label{eq:2.5}
    B_{p}^{\text{GT}} \Bigg|\begin{tikzpicture}[x=0.75pt,y=0.75pt,yscale=-1,xscale=1, baseline=(XXXX.south) ]
\path (0,67);\path (133.9581298828125,0);\draw    ($(current bounding box.center)+(0,0.3em)$) node [anchor=south] (XXXX) {};
\draw  [draw opacity=0][fill={rgb, 255:red, 155; green, 155; blue, 155 }  ,fill opacity=0.5 ] (63.83,14.89) -- (109.25,30.34) -- (69.83,51.89) -- (24.41,36.44) -- cycle ;
\draw    (24.41,36.44) -- (69.83,51.89) ;
\draw [shift={(47.12,44.16)}, rotate = 18.79] [fill={rgb, 255:red, 0; green, 0; blue, 0 }  ][line width=0.08]  [draw opacity=0] (5.36,-2.57) -- (0,0) -- (5.36,2.57) -- cycle    ;
\draw    (24.41,36.44) -- (63.83,14.89) ;
\draw [shift={(44.12,25.66)}, rotate = 151.33] [fill={rgb, 255:red, 0; green, 0; blue, 0 }  ][line width=0.08]  [draw opacity=0] (5.36,-2.57) -- (0,0) -- (5.36,2.57) -- cycle    ;
\draw    (63.83,14.89) -- (109.25,30.34) ;
\draw [shift={(86.54,22.61)}, rotate = 18.79] [fill={rgb, 255:red, 0; green, 0; blue, 0 }  ][line width=0.08]  [draw opacity=0] (5.36,-2.57) -- (0,0) -- (5.36,2.57) -- cycle    ;
\draw    (69.83,51.89) -- (109.25,30.34) ;
\draw [shift={(89.54,41.11)}, rotate = 151.33] [fill={rgb, 255:red, 0; green, 0; blue, 0 }  ][line width=0.08]  [draw opacity=0] (5.36,-2.57) -- (0,0) -- (5.36,2.57) -- cycle    ;
\draw    (1.7,28.72) -- (24.41,36.44) ;
\draw    (69.83,51.89) -- (92.54,59.61) ;
\draw    (109.25,30.34) -- (131.96,38.06) ;
\draw    (41.12,7.16) -- (63.83,14.89) ;
\draw    (63.83,14.89) -- (83.54,4.11) ;
\draw    (109.25,30.34) -- (128.96,19.56) ;
\draw    (50.12,62.66) -- (69.83,51.89) ;
\draw    (4.7,47.22) -- (24.41,36.44) ;
\draw    (69.83,38.48) -- (69.83,51.89) ;
\draw    (69.83,51.89) -- (69.83,65.29) ;
\draw    (109.25,30.34) -- (109.25,43.74) ;
\draw    (109.25,16.93) -- (109.25,30.34) ;
\draw  [dash pattern={on 3.75pt off 2.25pt}]  (63.83,14.89) -- (63.83,28.29) ;
\draw    (63.83,1.48) -- (63.83,14.89) ;
\draw    (24.41,36.44) -- (24.41,49.85) ;
\draw    (24.41,23.03) -- (24.41,36.44) ;
\draw (35,14) node [anchor=north west][inner sep=0.75pt]    {\small $g$};
\draw (39,46) node [anchor=north west][inner sep=0.75pt]    {\small $h$};
\draw (86.54,42) node [anchor=north west][inner sep=0.75pt]    {\small $i$};
\draw (84,8) node [anchor=north west][inner sep=0.75pt]    {\small $j$};
\draw (58.95,31.37) node [anchor=north west][inner sep=0.75pt]    {\small $p$};
\end{tikzpicture}
\Bigg\rangle =\delta _{g\bar{j}\bar{i} h,e} \Bigg|\begin{tikzpicture}[x=0.75pt,y=0.75pt,yscale=-1,xscale=1, baseline=(XXXX.south) ]
\path (0,67);\path (133.9581298828125,0);\draw    ($(current bounding box.center)+(0,0.3em)$) node [anchor=south] (XXXX) {};
\draw  [draw opacity=0][fill={rgb, 255:red, 155; green, 155; blue, 155 }  ,fill opacity=0.5 ] (63.83,14.89) -- (109.25,30.34) -- (69.83,51.89) -- (24.41,36.44) -- cycle ;
\draw    (24.41,36.44) -- (69.83,51.89) ;
\draw [shift={(47.12,44.16)}, rotate = 18.79] [fill={rgb, 255:red, 0; green, 0; blue, 0 }  ][line width=0.08]  [draw opacity=0] (5.36,-2.57) -- (0,0) -- (5.36,2.57) -- cycle    ;
\draw    (24.41,36.44) -- (63.83,14.89) ;
\draw [shift={(44.12,25.66)}, rotate = 151.33] [fill={rgb, 255:red, 0; green, 0; blue, 0 }  ][line width=0.08]  [draw opacity=0] (5.36,-2.57) -- (0,0) -- (5.36,2.57) -- cycle    ;
\draw    (63.83,14.89) -- (109.25,30.34) ;
\draw [shift={(86.54,22.61)}, rotate = 18.79] [fill={rgb, 255:red, 0; green, 0; blue, 0 }  ][line width=0.08]  [draw opacity=0] (5.36,-2.57) -- (0,0) -- (5.36,2.57) -- cycle    ;
\draw    (69.83,51.89) -- (109.25,30.34) ;
\draw [shift={(89.54,41.11)}, rotate = 151.33] [fill={rgb, 255:red, 0; green, 0; blue, 0 }  ][line width=0.08]  [draw opacity=0] (5.36,-2.57) -- (0,0) -- (5.36,2.57) -- cycle    ;
\draw    (1.7,28.72) -- (24.41,36.44) ;
\draw    (69.83,51.89) -- (92.54,59.61) ;
\draw    (109.25,30.34) -- (131.96,38.06) ;
\draw    (41.12,7.16) -- (63.83,14.89) ;
\draw    (63.83,14.89) -- (83.54,4.11) ;
\draw    (109.25,30.34) -- (128.96,19.56) ;
\draw    (50.12,62.66) -- (69.83,51.89) ;
\draw    (4.7,47.22) -- (24.41,36.44) ;
\draw    (69.83,38.48) -- (69.83,51.89) ;
\draw    (69.83,51.89) -- (69.83,65.29) ;
\draw    (109.25,30.34) -- (109.25,43.74) ;
\draw    (109.25,16.93) -- (109.25,30.34) ;
\draw  [dash pattern={on 3.75pt off 2.25pt}]  (63.83,14.89) -- (63.83,28.29) ;
\draw    (63.83,1.48) -- (63.83,14.89) ;
\draw    (24.41,36.44) -- (24.41,49.85) ;
\draw    (24.41,23.03) -- (24.41,36.44) ;
\draw (35,14) node [anchor=north west][inner sep=0.75pt]    {\small $g$};
\draw (39,46) node [anchor=north west][inner sep=0.75pt]    {\small $h$};
\draw (86.54,42) node [anchor=north west][inner sep=0.75pt]    {\small $i$};
\draw (84,8) node [anchor=north west][inner sep=0.75pt]    {\small $j$};
\draw (58.95,31.37) node [anchor=north west][inner sep=0.75pt]    {\small $p$};
\end{tikzpicture}
\Bigg\rangle,
\end{equation}
which is also a projector. In gauge theory terminologies, $B^\text{GT}_p$ imposes a local flatness condition in $p$. All plaquette operators and vertex operators commute. 

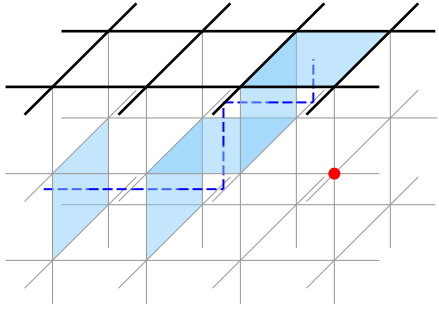
\begin{figure}[ht]
    \centering 
    \begin{tikzpicture}[x=0.75pt,y=0.75pt,yscale=-1,xscale=1]
    
    \draw [color={rgb, 255:red, 155; green, 155; blue, 155 }  ,draw opacity=1 ]   (273.88,60.12) -- (273.88,16.49) ;
    \draw [color={rgb, 255:red, 155; green, 155; blue, 155 }  ,draw opacity=1 ]   (320.78,60.12) -- (320.78,16.49) ;
    \draw [color={rgb, 255:red, 155; green, 155; blue, 155 }  ,draw opacity=1 ]   (381.64,46.11) -- (367.67,60.12) ;
    \draw [color={rgb, 255:red, 0; green, 0; blue, 255 }  ,draw opacity=1, line width=0.75pt ] [dash pattern={on 3.75pt off 1.5pt}]  (376.14,52.31) -- (376.14,30.5) ;
    \draw [color={rgb, 255:red, 0; green, 0; blue, 255 }  ,draw opacity=1, line width=0.75pt ] [dash pattern={on 3.75pt off 1.5pt}]  (331.27,52.31) -- (353.71,52.31) ;
    \draw [color={rgb, 255:red, 155; green, 155; blue, 155 }  ,draw opacity=1 ]   (334.74,89.74) -- (320.78,103.76) ;
    \draw [color={rgb, 255:red, 0; green, 0; blue, 255 }  ,draw opacity=1, line width=0.75pt ] [dash pattern={on 3.75pt off 1.5pt}]  (331.27,95.95) -- (331.27,74.13) ;
    \draw [color={rgb, 255:red, 0; green, 0; blue, 255 }  ,draw opacity=1, line width=0.75pt ] [dash pattern={on 3.75pt off 1.5pt}]  (286.4,95.95) -- (308.83,95.95) ;
    \draw [color={rgb, 255:red, 0; green, 0; blue, 255 }  ,draw opacity=1 , line width=0.75pt] [dash pattern={on 3.75pt off 1.5pt}]  (241.53,95.95) -- (263.96,95.95) ;
    \draw [color={rgb, 255:red, 0; green, 0; blue, 0 }  ,draw opacity=1, line width=1pt ]   (273.88,16.49) -- (320.78,16.49) ;
    \draw [color={rgb, 255:red, 0; green, 0; blue, 0 }  ,draw opacity=1, line width=1pt ]   (273.88,16.49) -- (245.95,44.51) ;
    \draw [color={rgb, 255:red, 155; green, 155; blue, 155 }  ,draw opacity=1 ]   (245.95,88.14) -- (245.95,44.51) ;
    \draw [color={rgb, 255:red, 155; green, 155; blue, 155 }  ,draw opacity=1 ]   (273.88,60.12) -- (320.78,60.12) ;
    \draw [color={rgb, 255:red, 0; green, 0; blue, 0 }  ,draw opacity=1, line width=1pt ]   (320.78,16.49) -- (367.67,16.49) ;
    \draw [color={rgb, 255:red, 0; green, 0; blue, 0 }  ,draw opacity=1, line width=1pt ]   (320.78,16.49) -- (292.84,44.51) ;
    \draw [color={rgb, 255:red, 155; green, 155; blue, 155 }  ,draw opacity=1 ]   (292.84,88.14) -- (292.84,44.51) ;
    \draw [color={rgb, 255:red, 155; green, 155; blue, 155 }  ,draw opacity=1 ]   (386.63,88.14) -- (339.74,88.14) ;
    \draw [color={rgb, 255:red, 155; green, 155; blue, 155 }  ,draw opacity=1 ]   (367.67,60.12) -- (414.57,60.12) ;
    \draw [color={rgb, 255:red, 0; green, 0; blue, 0 }  ,draw opacity=1, line width=1pt ]   (413.56,16.49) -- (437,16.49) ;
    \draw [color={rgb, 255:red, 0; green, 0; blue, 0 }  ,draw opacity=1, line width=1pt ]   (409.08,44.51) -- (385.63,44.51) ;
    \draw [color={rgb, 255:red, 155; green, 155; blue, 155 }  ,draw opacity=1 ]   (386.63,88.14) -- (386.63,44.51) ;
    \draw [color={rgb, 255:red, 155; green, 155; blue, 155 }  ,draw opacity=1 ]   (414.57,60.12) -- (414.57,16.49) ;
    \draw [color={rgb, 255:red, 155; green, 155; blue, 155 }  ,draw opacity=1 ]   (414.57,60.12) -- (386.63,88.14) ;
    \draw [color={rgb, 255:red, 155; green, 155; blue, 155 }  ,draw opacity=1 ]   (409.08,88.14) -- (385.63,88.14) ;
    \draw [color={rgb, 255:red, 155; green, 155; blue, 155 }  ,draw opacity=1 ]   (414.57,60.12) -- (438.01,60.12) ;
    \draw [color={rgb, 255:red, 155; green, 155; blue, 155 }  ,draw opacity=1 ]   (292.84,131.78) -- (245.95,131.78) ;
    \draw [color={rgb, 255:red, 155; green, 155; blue, 155 }  ,draw opacity=1 ]   (273.88,103.76) -- (320.78,103.76) ;
    \draw [color={rgb, 255:red, 155; green, 155; blue, 155 }  ,draw opacity=1 ]   (339.74,131.78) -- (292.84,131.78) ;
    \draw [color={rgb, 255:red, 155; green, 155; blue, 155 }  ,draw opacity=1 ]   (320.78,103.76) -- (367.67,103.76) ;
    \draw [color={rgb, 255:red, 155; green, 155; blue, 155 }  ,draw opacity=1 ]   (339.74,131.78) -- (339.74,88.14) ;
    \draw [color={rgb, 255:red, 155; green, 155; blue, 155 }  ,draw opacity=1 ]   (367.67,103.76) -- (367.67,60.12) ;
    \draw [color={rgb, 255:red, 155; green, 155; blue, 155 }  ,draw opacity=1 ]   (367.67,103.76) -- (339.74,131.78) ;
    \draw [color={rgb, 255:red, 155; green, 155; blue, 155 }  ,draw opacity=1 ]   (386.63,131.78) -- (339.74,131.78) ;
    \draw [color={rgb, 255:red, 155; green, 155; blue, 155 }  ,draw opacity=1 ]   (366.66,103.76) -- (413.56,103.76) ;
    \draw [color={rgb, 255:red, 155; green, 155; blue, 155 }  ,draw opacity=1 ]   (386.63,131.78) -- (386.63,88.14) ;
    \draw [color={rgb, 255:red, 155; green, 155; blue, 155 }  ,draw opacity=1 ]   (414.57,103.76) -- (414.57,60.12) ;
    \draw [color={rgb, 255:red, 155; green, 155; blue, 155 }  ,draw opacity=1 ]   (414.57,103.76) -- (386.63,131.78) ;
    \draw [color={rgb, 255:red, 155; green, 155; blue, 155 }  ,draw opacity=1 ]   (413.56,103.76) -- (437,103.76) ;
    \draw [color={rgb, 255:red, 155; green, 155; blue, 155 }  ,draw opacity=1 ]   (385.63,131.78) -- (409.08,131.78) ;
    \draw [color={rgb, 255:red, 155; green, 155; blue, 155 }  ,draw opacity=1 ]   (222.5,131.78) -- (245.95,131.78) ;
    \draw [color={rgb, 255:red, 155; green, 155; blue, 155 }  ,draw opacity=1 ]   (250.43,103.76) -- (273.88,103.76) ;
    \draw [color={rgb, 255:red, 0; green, 0; blue, 0 }  ,draw opacity=1, line width=1pt ]   (222.5,44.51) -- (245.95,44.51) ;
    \draw [color={rgb, 255:red, 0; green, 0; blue, 0 }  ,draw opacity=1, line width=1pt ]   (250.43,16.49) -- (273.88,16.49) ;
    \draw [color={rgb, 255:red, 155; green, 155; blue, 155 }  ,draw opacity=1 ]   (222.5,88.14) -- (245.95,88.14) ;
    \draw [color={rgb, 255:red, 155; green, 155; blue, 155 }  ,draw opacity=1 ]   (250.43,60.12) -- (273.88,60.12) ;
    \draw [color={rgb, 255:red, 0; green, 0; blue, 0 }  ,draw opacity=1, line width=1pt ]   (245.95,44.51) -- (231.98,58.52) ;
    \draw [color={rgb, 255:red, 0; green, 0; blue, 0 }  ,draw opacity=1, line width=1pt ]   (292.84,44.51) -- (278.88,58.52) ;
    \draw [color={rgb, 255:red, 0; green, 0; blue, 0 }  ,draw opacity=1, line width=1pt ]   (386.64,44.51) -- (372.67,58.52) ;
    \draw [color={rgb, 255:red, 0; green, 0; blue, 0 }  ,draw opacity=1, line width=1pt ]   (287.85,2.48) -- (273.88,16.49) ;
    \draw [color={rgb, 255:red, 0; green, 0; blue, 0 }  ,draw opacity=1, line width=1pt ]   (334.74,2.48) -- (320.78,16.49) ;
    \draw [color={rgb, 255:red, 0; green, 0; blue, 0 }  ,draw opacity=1, line width=1pt ]   (381.64,2.48) -- (367.67,16.49) ;
    \draw [color={rgb, 255:red, 0; green, 0; blue, 0 }  ,draw opacity=1 , line width=1pt]   (428.53,2.48) -- (414.57,16.49) ;
    \draw [color={rgb, 255:red, 155; green, 155; blue, 155 }  ,draw opacity=1 ]   (245.95,88.14) -- (231.98,102.15) ;
    \draw [color={rgb, 255:red, 155; green, 155; blue, 155 }  ,draw opacity=1 ]   (287.85,46.11) -- (273.88,60.12) ;
    \draw [color={rgb, 255:red, 155; green, 155; blue, 155 }  ,draw opacity=1 ]   (245.95,131.78) -- (231.98,145.79) ;
    \draw [color={rgb, 255:red, 155; green, 155; blue, 155 }  ,draw opacity=1 ]   (287.85,89.74) -- (273.88,103.76) ;
    \draw [color={rgb, 255:red, 155; green, 155; blue, 155 }  ,draw opacity=1 ]   (292.84,131.78) -- (278.88,145.79) ;
    \draw [color={rgb, 255:red, 155; green, 155; blue, 155 }  ,draw opacity=1 ]   (339.74,131.78) -- (325.77,145.79) ;
    \draw [color={rgb, 255:red, 155; green, 155; blue, 155 }  ,draw opacity=1 ]   (381.64,89.74) -- (367.67,103.76) ;
    \draw [color={rgb, 255:red, 155; green, 155; blue, 155 }  ,draw opacity=1 ]   (386.63,131.78) -- (372.67,145.79) ;
    \draw [color={rgb, 255:red, 155; green, 155; blue, 155 }  ,draw opacity=1 ]   (428.53,89.74) -- (414.57,103.76) ;
    \draw [color={rgb, 255:red, 155; green, 155; blue, 155 }  ,draw opacity=1 ]   (334.74,46.11) -- (320.78,60.12) ;
    \draw [color={rgb, 255:red, 155; green, 155; blue, 155 }  ,draw opacity=1 ]   (428.53,46.11) -- (414.57,60.12) ;
    \draw [color={rgb, 255:red, 0; green, 0; blue, 0 }  ,draw opacity=1, line width=1pt ]   (292.84,44.51) -- (245.95,44.51) ;
    \draw [color={rgb, 255:red, 155; green, 155; blue, 155 }  ,draw opacity=1 ]   (339.74,88.14) -- (325.78,102.15) ;
    \draw [color={rgb, 255:red, 155; green, 155; blue, 155 }  ,draw opacity=1 ]   (386.63,88.14) -- (372.67,102.15) ;
    \draw [color={rgb, 255:red, 155; green, 155; blue, 155 }  ,draw opacity=1 ]   (245.95,153.59) -- (245.95,131.78) ;
    \draw [color={rgb, 255:red, 155; green, 155; blue, 155 }  ,draw opacity=1 ]   (273.88,125.57) -- (273.88,103.76) ;
    \draw [color={rgb, 255:red, 155; green, 155; blue, 155 }  ,draw opacity=1 ]   (292.84,153.59) -- (292.84,131.78) ;
    \draw [color={rgb, 255:red, 155; green, 155; blue, 155 }  ,draw opacity=1 ]   (320.78,125.57) -- (320.78,103.76) ;
    \draw [color={rgb, 255:red, 155; green, 155; blue, 155 }  ,draw opacity=1 ]   (339.74,153.59) -- (339.74,131.78) ;
    \draw [color={rgb, 255:red, 155; green, 155; blue, 155 }  ,draw opacity=1 ]   (386.63,153.59) -- (386.63,131.78) ;
    \draw [color={rgb, 255:red, 155; green, 155; blue, 155 }  ,draw opacity=1 ]   (414.57,125.57) -- (414.57,103.76) ;
    \draw [color={rgb, 255:red, 155; green, 155; blue, 155 }  ,draw opacity=1 ]   (367.67,125.57) -- (367.67,103.76) ;
    \draw  [draw opacity=0][fill={rgb, 255:red, 135; green, 206; blue, 250 }  ,fill opacity=0.5 ] (273.88,60.12) -- (273.88,103.76) -- (245.95,131.78) -- (245.95,88.14) -- cycle ;
    \draw [color={rgb, 255:red, 155; green, 155; blue, 155 }  ,draw opacity=1 ]   (273.88,60.12) -- (245.95,88.14) ;
    \draw [color={rgb, 255:red, 155; green, 155; blue, 155 }  ,draw opacity=1 ]   (273.88,103.76) -- (245.95,131.78) ;
    \draw [color={rgb, 255:red, 155; green, 155; blue, 155 }  ,draw opacity=1 ]   (273.88,103.76) -- (273.88,60.12) ;
    \draw [color={rgb, 255:red, 155; green, 155; blue, 155 }  ,draw opacity=1 ]   (245.95,131.78) -- (245.95,88.14) ;
    \draw  [draw opacity=0][fill={rgb, 255:red, 135; green, 206; blue, 250 }  ,fill opacity=0.5 ] (320.78,60.12) -- (320.78,103.76) -- (292.84,131.78) -- (292.84,88.14) -- cycle ;
    \draw [color={rgb, 255:red, 155; green, 155; blue, 155 }  ,draw opacity=1 ]   (292.84,131.78) -- (292.84,88.14) ;
    \draw [color={rgb, 255:red, 155; green, 155; blue, 155 }  ,draw opacity=1 ]   (320.78,103.76) -- (292.84,131.78) ;
    \draw [color={rgb, 255:red, 155; green, 155; blue, 155 }  ,draw opacity=1 ]   (320.78,103.76) -- (320.78,60.12) ;
    \draw  [draw opacity=0][fill={rgb, 255:red, 135; green, 206; blue, 250 }  ,fill opacity=0.5 ] (320.78,60.12) -- (367.67,60.12) -- (339.74,88.14) -- (292.84,88.14) -- cycle ;
    \draw [color={rgb, 255:red, 155; green, 155; blue, 155 }  ,draw opacity=1 ]   (339.74,88.14) -- (292.84,88.14) ;
    \draw [color={rgb, 255:red, 155; green, 155; blue, 155 }  ,draw opacity=1 ]   (320.78,60.12) -- (292.84,88.14) ;
    \draw [color={rgb, 255:red, 0; green, 0; blue, 0 }  ,draw opacity=1 , line width=1pt]   (339.74,44.51) -- (292.84,44.51) ;
    \draw [color={rgb, 255:red, 155; green, 155; blue, 155 }  ,draw opacity=1 ]   (320.78,60.12) -- (367.67,60.12) ;
    \draw  [draw opacity=0][fill={rgb, 255:red, 135; green, 206; blue, 250 } ,fill opacity=0.5 ] (367.67,16.49) -- (367.67,60.12) -- (339.74,88.14) -- (339.74,44.51) -- cycle ;
    \draw [color={rgb, 255:red, 155; green, 155; blue, 155 }  ,draw opacity=1 ]   (367.67,60.12) -- (339.74,88.14) ;
    \draw [color={rgb, 255:red, 155; green, 155; blue, 155 }  ,draw opacity=1 ]   (339.74,88.14) -- (339.74,44.51) ;
    \draw [color={rgb, 255:red, 155; green, 155; blue, 155 }  ,draw opacity=1 ][fill={rgb, 255:red, 155; green, 155; blue, 155 }  ,fill opacity=1 ]   (367.67,60.12) -- (367.67,16.49) ;
    \draw  [draw opacity=0][fill={rgb, 255:red, 135; green, 206; blue, 250 }  ,fill opacity=0.5 ] (366.66,16.49) -- (414.57,16.49) -- (386.64,44.51) -- (339.74,44.51) -- cycle ;
    \draw [color={rgb, 255:red, 0; green, 0; blue, 0 }  ,draw opacity=1, line width=1pt ]   (367.67,16.49) -- (339.74,44.51) ;
    \draw [color={rgb, 255:red, 0; green, 0; blue, 0 }  ,draw opacity=1, line width=1pt ]   (366.66,16.49) -- (413.56,16.49) ;
    \draw [color={rgb, 255:red, 0; green, 0; blue, 0 }  ,draw opacity=1, line width=1pt ]   (414.56,16.49) -- (386.63,44.51) ;
    \draw [color={rgb, 255:red, 0; green, 0; blue, 0 }  ,draw opacity=1, line width=1pt ]   (386.64,44.51) -- (339.74,44.51) ;
    \draw [color={rgb, 255:red, 155; green, 155; blue, 155 }  ,draw opacity=1 ]   (292.84,88.14) -- (245.95,88.14) ;
    \draw  [draw opacity=0][fill={rgb, 255:red, 255; green, 0; blue, 0 }  ,fill opacity=1 ] (383.63,88.14) .. controls (383.63,86.48) and (384.98,85.14) .. (386.63,85.14) .. controls (388.29,85.14) and (389.64,86.48) .. (389.64,88.14) .. controls (389.64,89.8) and (388.29,91.14) .. (386.63,91.14) .. controls (384.98,91.14) and (383.63,89.8) .. (383.63,88.14) -- cycle ;
    \draw [color={rgb, 255:red, 0; green, 0; blue, 255 }  ,draw opacity=1, line width=0.75pt ] [dash pattern={on 3.75pt off 1.5pt}]  (353.71,52.31) -- (376.14,52.31) ;
    \draw [color={rgb, 255:red, 0; green, 0; blue, 255 }  ,draw opacity=1, line width=0.75pt ] [dash pattern={on 3.75pt off 1.5pt}]  (331.27,74.13) -- (331.27,52.31) ;
    \draw [color={rgb, 255:red, 0; green, 0; blue, 255 }  ,draw opacity=1, line width=0.75pt ] [dash pattern={on 3.75pt off 1.5pt}]  (308.83,95.95) -- (331.27,95.95) ;
    \draw [color={rgb, 255:red, 0; green, 0; blue, 255 }  ,draw opacity=1, line width=0.75pt ] [dash pattern={on 3.75pt off 1.5pt}]  (263.96,95.95) -- (286.4,95.95) ;
    \draw [color={rgb, 255:red, 155; green, 155; blue, 155 }  ,draw opacity=1]   (292.84,88.14) -- (278.88,102.15) ;
    \draw [color={rgb, 255:red, 0; green, 0; blue, 0 }  ,draw opacity=1, line width=1pt]   (339.74,44.51) -- (325.77,58.52) ;

    \end{tikzpicture}
    \caption{Charge excitation (red dot) and string-like excitation (deep blue dashed line, consisting of a series of light blue plaquettes where the local flatness condition is violated) that terminat on the boundary in the three-dimensional GT model with gapped boundaries. }
    \label{fig:2}
\end{figure}

A gapped boundary condition is characterized by a subgroup $K\subseteq G$. The boundary Hamiltonian consists of respectively the sums of boundary vertex, plaquette, and edge operators: 
\begin{equation}\label{eq:2.6}
    \overline{H^\text{GT}_{K}}=-\sum_{v\in\partial\Gamma}\overline{A^\text{GT}_{v}}-\sum_{p\in\partial\Gamma}\overline{B^\text{GT}_p}-\sum_{e\in\partial\Gamma}\overline{C^\text{GT}_e},
\end{equation}
where the sums run over all the boundary vertices, plaquettes, and edges of $\Gamma$. The definition of the boundary vertex operators and plaquette operators are similar to the bulk operators: $\overline{A^\text{GT}_{v}}$ is again defined as a gauge transformation averaged instead in the subgroup $K$, and the definition of $\overline{B^\text{GT}_p}$ is just the same as $B^\text{GT}_p$, which imposes the local flatness condition in boundary plaquettes. The edge operator $\overline{C^\text{GT}_e}$ is a projector: 
\begin{equation}\label{eq:2.7}
    \overline{C_{e}^{\text{GT}}} \Bigg|\begin{tikzpicture}[x=0.75pt,y=0.75pt,yscale=-1,xscale=1, baseline=(XXXX.south) ]
\path (0,51);\path (94.99246215820312,0);\draw    ($(current bounding box.center)+(0,0.3em)$) node [anchor=south] (XXXX) {};
\draw [line width=1.5]    (23.87,12.44) -- (69.29,27.89) ;
\draw [shift={(46.58,20.16)}, rotate = 18.79] [fill={rgb, 255:red, 0; green, 0; blue, 0 }  ][line width=0.08]  [draw opacity=0] (6.97,-3.35) -- (0,0) -- (6.97,3.35) -- cycle    ;
\draw [line width=1.5]    (1.16,4.72) -- (23.87,12.44) ;
\draw [line width=1.5]    (69.29,27.89) -- (92,35.61) ;
\draw [line width=1.5]    (49.58,38.66) -- (69.29,27.89) ;
\draw [line width=1.5]    (4.16,23.22) -- (23.87,12.44) ;
\draw  [dash pattern={on 3.75pt off 2.25pt}]  (69.29,27.89) -- (69.29,48.36) ;
\draw [line width=1.5]    (69.29,27.89) -- (89,17.11) ;
\draw [line width=1.5]    (23.87,12.44) -- (43.58,1.66) ;
\draw  [dash pattern={on 3.75pt off 2.25pt}]  (23.87,12.44) -- (23.87,32.91) ;
\draw (44.46,7.4) node [anchor=north west][inner sep=0.75pt]    {\small $l$};
\draw (33.87,24.07) node [anchor=north west][inner sep=0.75pt]    {\small $e$};
\end{tikzpicture}
\Bigg\rangle =\delta _{l\in K} \Bigg|\begin{tikzpicture}[x=0.75pt,y=0.75pt,yscale=-1,xscale=1, baseline=(XXXX.south) ]
\path (0,51);\path (94.99246215820312,0);\draw    ($(current bounding box.center)+(0,0.3em)$) node [anchor=south] (XXXX) {};
\draw [line width=1.5]    (23.87,12.44) -- (69.29,27.89) ;
\draw [shift={(46.58,20.16)}, rotate = 18.79] [fill={rgb, 255:red, 0; green, 0; blue, 0 }  ][line width=0.08]  [draw opacity=0] (6.97,-3.35) -- (0,0) -- (6.97,3.35) -- cycle    ;
\draw [line width=1.5]    (1.16,4.72) -- (23.87,12.44) ;
\draw [line width=1.5]    (69.29,27.89) -- (92,35.61) ;
\draw [line width=1.5]    (49.58,38.66) -- (69.29,27.89) ;
\draw [line width=1.5]    (4.16,23.22) -- (23.87,12.44) ;
\draw  [dash pattern={on 3.75pt off 2.25pt}]  (69.29,27.89) -- (69.29,48.36) ;
\draw [line width=1.5]    (69.29,27.89) -- (89,17.11) ;
\draw [line width=1.5]    (23.87,12.44) -- (43.58,1.66) ;
\draw  [dash pattern={on 3.75pt off 2.25pt}]  (23.87,12.44) -- (23.87,32.91) ;
\draw (44.46,7.4) node [anchor=north west][inner sep=0.75pt]    {\small $l$};
\draw (33.87,24.07) node [anchor=north west][inner sep=0.75pt]    {\small $e$};
\end{tikzpicture}
\Bigg\rangle.
\end{equation}
The boundary vertex, plaquette, and edge operators are all commute with each other and with the bulk vertex and plaquette operators. Therefore, the total Hamiltonian \eqref{eq:2.2} is exactly solvable. The ground states are the common $+1$ eigenstates of all operators in the total Hamiltonian. The ground state degeneracy (GSD) can be computed by 
\begin{equation}
    \text{GSD}=\operatorname{Tr}\left(\prod_{v\in\Gamma\setminus\partial\Gamma}A^\text{GT}_{v}\prod_{p\in\Gamma\setminus\partial\Gamma}B^\text{GT}_p\prod_{v\in\partial\Gamma}\overline{A^\text{GT}_{v}}\prod_{p\in\partial\Gamma}\overline{B^\text{GT}_p}\prod_{e\in\partial\Gamma}\overline{C^\text{GT}_e}\right),
\end{equation}
where the trace is taken over the total Hilbert space \eqref{eq:2.1}. The elementary excitations in the model without boundary are charges on the vertices and loop-like excitations. A charge at vertex $v$ arises when the local Gauss constraint is violated; a loop-like excitation occurs when the local flatness condition is violated on a series of plaquettes which forms a loop\cite{Moradi2015,Kong2020a}. If the gapped boundary is included, there still exists one more type of elementary excitations, that is, the bulk string-like excitations that terminate at gapped boundaries (see \autoref{fig:2}). In three dimensions, the braiding between point-like charges is trivial, while the braiding with loop-like excitations or string-like excitations is highly non-trivial. Thus, despite some recent progress \cite{Lan2017a,Lan2018b,Kong2020a,Xi2021}, our understanding of the elementary excitations in three-dimensional topological orders is still incomplete. 

\section{Fourier transforming and rewriting the boundary of the 3-dimensional GT model}\label{sec:3}
In this section, we Fourier-transform the basis of the Hilbert space of the three-dimensional GT model with gapped boundaries from the group space to the representation space. This transformation then allows us to rewrite the three-dimensional GT model with gapped boundaries on a trivalent lattice. We then Fourier-transform the boundary Hamiltonian of the three-dimensional GT model. 

\subsection{A graphical tool for group representation theory}
To facilitate the derivations in this paper, we introduce the following graphical tool for group representation theory\cite{Hu2013a}. Let $L_G$ be the set of all unitary irreducible representations $V_\mu$ (up to equivalence) of a finite group $G$. We will use Greek indices $\mu$ to label irreducible representations, and use Latin indices $m_\mu$ to label the basis of the representation space $V_\mu$. 

\paragraph*{Duality map}
Every irreducible representation $\mu\in L_G$ has a dual $\mu^*\in L_G$, which is equivalent to (but not necessarily identical to) the complex conjugate representation of $\mu$. We can thus define an invertible duality map 
\begin{equation}\label{eq:3.1}
    \omega_\mu:\mathbb C\to V_\mu\otimes V_\mu;\;\;1\mapsto\sum_{m_\mu,n_{\mu^*}}\Omega^\mu_{m_\mu n_{\mu^*}}e_{m_\mu}\otimes e_{n_{\mu^*}},
\end{equation}
where $\Omega^\mu_{m_\mu n_{\mu^*}}$ is a complex matrix satisfies normalization $\Omega^\dagger\Omega=\mathbb 1$ and maps $\mu$ to $\mu^*$ by similarity transformation 
\begin{equation}\label{eq:3.2}
    (\Omega^\mu)^{-1}D^\mu(g)\Omega^\mu=(D^{\mu^*}(g))^*,
\end{equation}
where $D^\mu(g)$ is the representation matrix of group element $g$. 

The duality map $\Omega^\mu$ can only be determined up to a complex number. If $\mu$ is self-dual, the matrix $\Omega^\mu$ is either symmetric or antisymmetric depends on whether $\mu$ is pseudo-real or real. This is an intrinsic property characterized by a number $\beta_\mu$, called the Forbenius-Schur (FS) indicator. We have $(\Omega^{\mu^*})^\mathtt{T}=\beta_\mu\Omega^\mu$ with $\beta_\mu=\pm 1$ if $\mu$ is real or pseudo-real. If $\mu$ is not self-dual, we always set $\beta_\mu=1$. 

Graphically, the duality map and its inverse has presentations 
\begin{equation}

\equiv [( \Omega^{\mu })^{-1}]_{m_{\mu ^{*}} n_{\mu }}.
\end{equation}
Being an intertwiner, the duality map commutes with group actions, namely 
\begin{equation}
    %
,
\end{equation}
where we have introduced the graphical presentation for group actions: 
\begin{equation}
    %
=D_{m_{\mu } n_{\mu }}^{\mu }( g),
\end{equation}
and the contraction of Latin indices is presented by concatenating two lines. Note that the direction of all lines in our graphical presentation are upward by default. A line with label $\mu$ directed downward should always be regarded as the line with label $\mu^*$ directed upward. 

Since all irreducible representations $\mu\in L_G$ are unitary, 
\begin{equation}
    [D^\mu_{m_\mu n_\mu}(g)]^*=D^\mu_{n_\mu m_\mu}(\bar g)\;\;\Rightarrow\;\;\left[%
.
\end{equation}
Then, \eqref{eq:3.2} can be graphically presented as 
\begin{equation}
    %
.
\end{equation}

\paragraph*{$3j$-symbol}
Frequently in later derivations, we will need $3j$-symbols to deal with the coupling of three representations of $G$. A $3j$-symbol is a tensor $C_{\mu\nu\rho;m_\mu m_\nu m_\rho}$ that is defined as an intertwiner: 
\begin{equation}
    C_{\mu\nu\rho}:V_{\mu}\otimes V_{\nu}\otimes V_{\rho}\to\mathbb C,\;\;|\mu m_{\mu},\nu m_{\nu},\rho m_{\rho}\rangle\mapsto C_{\mu\nu\rho;m_{\mu}m_{\nu}m_{\rho}}.
\end{equation}
A $3j$-symbol is depicted as 
\begin{equation}
    %
\equiv ( C_{\mu \nu \rho ;m_{\mu} m_{\nu} m_{\rho}})^{*}.
\end{equation}
The normalization condition of $3j$-symbols can then be presented as 
\begin{equation}
    \sum_{m_\mu m_\nu m_\rho}C_{\mu\nu\rho;m_\mu m_\nu m_\rho}(C_{\mu\nu\rho;m_\mu m_\nu m_\rho})^*=%
=1.
\end{equation}
Being an intertwiner, a $3j$-symbol is invariant under group actions, namely 
\begin{equation}
    %
.
\end{equation}
In this paper, we will also use a lot of Clebsch-Gordan (CG) coefficients $C^\rho_{\mu\nu}:V_\mu\otimes V_\nu\to V_{\rho}$ and $C^{\mu\nu}_\rho:V_\rho\to V_{\mu}\otimes V_\nu$, which can be defined using $3j$-symbols and duality maps: 
\begin{equation}
    C_{\mu \nu ;m_{\mu} m_{\nu}}^{\rho^{*} m_{\rho^{*}}} \equiv %
.
\end{equation}
The CG coefficients enable the following basis transformation:
\begin{equation}\label{eq:3.14}
    |\mu m_\mu\rangle\otimes|\nu m_\nu\rangle=\sum_{\rho,m_\rho}C^{\mu\nu;m_\mu m_\nu}_{\rho m_\rho}|\rho m_\rho\rangle,
\end{equation}
which is the foundation of rewriting the basis of the Fourier-transformed Hilbert space on a trivalent lattice. 

For later convenience, we list a few properties of $3j$-symbols as follows. For a generic group $G$, we can always construct such $3j$-symbols satisfying the following properties\cite{Hu2013a}:
\begin{equation}\label{eq:3.15}
    \begin{tikzpicture}[x=0.75pt,y=0.75pt,yscale=-1,xscale=1, baseline=(XXXX.south) ]
\path (0,123);\path (64.99651336669922,0);\draw    ($(current bounding box.center)+(0,0.3em)$) node [anchor=south] (XXXX) {};
\draw    (13,27) -- (33.93,60.93) ;
\draw [shift={(23.46,43.96)}, rotate = 58.33] [fill={rgb, 255:red, 0; green, 0; blue, 0 }  ][line width=0.08]  [draw opacity=0] (5.36,-2.57) -- (0,0) -- (5.36,2.57) -- cycle    ;
\draw    (33.93,60.93) -- (33.93,26.93) ;
\draw [shift={(33.93,43.93)}, rotate = 90] [fill={rgb, 255:red, 0; green, 0; blue, 0 }  ][line width=0.08]  [draw opacity=0] (5.36,-2.57) -- (0,0) -- (5.36,2.57) -- cycle    ;
\draw    (33.93,60.93) -- (55.01,26.86) ;
\draw [shift={(44.47,43.89)}, rotate = 121.74] [fill={rgb, 255:red, 0; green, 0; blue, 0 }  ][line width=0.08]  [draw opacity=0] (5.36,-2.57) -- (0,0) -- (5.36,2.57) -- cycle    ;
\draw [shift={(33.93,60.93)}, rotate = 301.74] [color={rgb, 255:red, 0; green, 0; blue, 0 }  ][fill={rgb, 255:red, 0; green, 0; blue, 0 }  ][line width=0.75]      (0, 0) circle [x radius= 2.01, y radius= 2.01]   ;
\draw    (12.86,103.01) -- (33.93,68.93) ;
\draw [shift={(23.39,85.97)}, rotate = 121.74] [fill={rgb, 255:red, 0; green, 0; blue, 0 }  ][line width=0.08]  [draw opacity=0] (5.36,-2.57) -- (0,0) -- (5.36,2.57) -- cycle    ;
\draw    (33.93,68.93) -- (54.87,102.87) ;
\draw [shift={(44.4,85.9)}, rotate = 58.33] [fill={rgb, 255:red, 0; green, 0; blue, 0 }  ][line width=0.08]  [draw opacity=0] (5.36,-2.57) -- (0,0) -- (5.36,2.57) -- cycle    ;
\draw [shift={(33.93,68.93)}, rotate = 58.33] [color={rgb, 255:red, 0; green, 0; blue, 0 }  ][fill={rgb, 255:red, 0; green, 0; blue, 0 }  ][line width=0.75]      (0, 0) circle [x radius= 2.01, y radius= 2.01]   ;
\draw    (33.93,102.93) -- (33.93,68.93) ;
\draw [shift={(33.93,85.93)}, rotate = 90] [fill={rgb, 255:red, 0; green, 0; blue, 0 }  ][line width=0.08]  [draw opacity=0] (5.36,-2.57) -- (0,0) -- (5.36,2.57) -- cycle    ;
\draw    (10,2) -- (58,2) -- (58,27) -- (10,27) -- cycle  ;
\draw (13,6) node [anchor=north west][inner sep=0.75pt]    {$\Gamma _{\text{closed}}$};
\draw (12,41.4) node [anchor=north west][inner sep=0.75pt]    {$\mu $};
\draw (24,35) node [anchor=north west][inner sep=0.75pt]    {$\nu $};
\draw (44,41.4) node [anchor=north west][inner sep=0.75pt]    {$\rho $};
\draw (12,78) node [anchor=north west][inner sep=0.75pt]    {$\mu $};
\draw (23,86.4) node [anchor=north west][inner sep=0.75pt]    {$\nu $};
\draw (47,78) node [anchor=north west][inner sep=0.75pt]    {$\rho $};
\draw (2,105) node [anchor=north west][inner sep=0.75pt]    {$m_{\mu }$};
\draw (25,105) node [anchor=north west][inner sep=0.75pt]    {$m_{\nu }$};
\draw (46,105) node [anchor=north west][inner sep=0.75pt]    {$m_{\rho }$};
\end{tikzpicture}
=\frac{1}{|G|}\sum _{g\in G}\begin{tikzpicture}[x=0.75pt,y=0.75pt,yscale=-1,xscale=1, baseline=(XXXX.south) ]
\path (0,123);\path (72.99651336669922,0);\draw    ($(current bounding box.center)+(0,0.3em)$) node [anchor=south] (XXXX) {};
\draw    (18,27) -- (17.86,103.01) ;
\draw [shift={(17.93,65)}, rotate = 90.1] [fill={rgb, 255:red, 0; green, 0; blue, 0 }  ][line width=0.08]  [draw opacity=0] (5.36,-2.57) -- (0,0) -- (5.36,2.57) -- cycle    ;
\draw    (38.93,102.93) -- (38.93,26.93) ;
\draw [shift={(38.93,64.93)}, rotate = 90] [fill={rgb, 255:red, 0; green, 0; blue, 0 }  ][line width=0.08]  [draw opacity=0] (5.36,-2.57) -- (0,0) -- (5.36,2.57) -- cycle    ;
\draw    (59.87,102.87) -- (60.01,26.86) ;
\draw [shift={(59.94,64.86)}, rotate = 90.1] [fill={rgb, 255:red, 0; green, 0; blue, 0 }  ][line width=0.08]  [draw opacity=0] (5.36,-2.57) -- (0,0) -- (5.36,2.57) -- cycle    ;
\draw  [fill={rgb, 255:red, 255; green, 255; blue, 255 }  ,fill opacity=1 ] (8.58,44.66) .. controls (8.58,39.5) and (12.77,35.31) .. (17.93,35.31) .. controls (23.09,35.31) and (27.27,39.5) .. (27.27,44.66) .. controls (27.27,49.82) and (23.09,54) .. (17.93,54) .. controls (12.77,54) and (8.58,49.82) .. (8.58,44.66) -- cycle ;
\draw  [fill={rgb, 255:red, 255; green, 255; blue, 255 }  ,fill opacity=1 ] (29.59,44.59) .. controls (29.59,39.43) and (33.77,35.24) .. (38.93,35.24) .. controls (44.09,35.24) and (48.28,39.43) .. (48.28,44.59) .. controls (48.28,49.75) and (44.09,53.93) .. (38.93,53.93) .. controls (33.77,53.93) and (29.59,49.75) .. (29.59,44.59) -- cycle ;
\draw  [fill={rgb, 255:red, 255; green, 255; blue, 255 }  ,fill opacity=1 ] (50.59,44.52) .. controls (50.59,39.36) and (54.78,35.17) .. (59.94,35.17) .. controls (65.1,35.17) and (69.28,39.36) .. (69.28,44.52) .. controls (69.28,49.68) and (65.1,53.86) .. (59.94,53.86) .. controls (54.78,53.86) and (50.59,49.68) .. (50.59,44.52) -- cycle ;
\draw    (15,2) -- (63,2) -- (63,27) -- (15,27) -- cycle  ;
\draw (18,6) node [anchor=north west][inner sep=0.75pt]    {$\Gamma _{\text{closed}}$};
\draw (2,60.4) node [anchor=north west][inner sep=0.75pt]    {$\mu $};
\draw (24,60.4) node [anchor=north west][inner sep=0.75pt]    {$\nu $};
\draw (46,59.4) node [anchor=north west][inner sep=0.75pt]    {$\rho $};
\draw (7,105) node [anchor=north west][inner sep=0.75pt]    {$m_{\mu }$};
\draw (30,105) node [anchor=north west][inner sep=0.75pt]    {$m_{\nu }$};
\draw (51,105) node [anchor=north west][inner sep=0.75pt]    {$m_{\rho }$};
\draw (13,40) node [anchor=north west][inner sep=0.75pt]    {$g$};
\draw (34,40) node [anchor=north west][inner sep=0.75pt]    {$g$};
\draw (55,40) node [anchor=north west][inner sep=0.75pt]    {$g$};
\end{tikzpicture}
=\begin{tikzpicture}[x=0.75pt,y=0.75pt,yscale=-1,xscale=1, baseline=(XXXX.south) ]
\path (0,123);\path (72.99651336669922,0);\draw    ($(current bounding box.center)+(0,0.3em)$) node [anchor=south] (XXXX) {};
\draw    (18,27) -- (17.86,103.01) ;
\draw [shift={(17.93,65)}, rotate = 90.1] [fill={rgb, 255:red, 0; green, 0; blue, 0 }  ][line width=0.08]  [draw opacity=0] (5.36,-2.57) -- (0,0) -- (5.36,2.57) -- cycle    ;
\draw    (38.93,102.93) -- (38.93,26.93) ;
\draw [shift={(38.93,64.93)}, rotate = 90] [fill={rgb, 255:red, 0; green, 0; blue, 0 }  ][line width=0.08]  [draw opacity=0] (5.36,-2.57) -- (0,0) -- (5.36,2.57) -- cycle    ;
\draw    (59.87,102.87) -- (60.01,26.86) ;
\draw [shift={(59.94,64.86)}, rotate = 90.1] [fill={rgb, 255:red, 0; green, 0; blue, 0 }  ][line width=0.08]  [draw opacity=0] (5.36,-2.57) -- (0,0) -- (5.36,2.57) -- cycle    ;
\draw    (15,2) -- (63,2) -- (63,27) -- (15,27) -- cycle  ;
\draw (18,6) node [anchor=north west][inner sep=0.75pt]    {$\Gamma _{\text{closed}}$};
\draw (2,60.4) node [anchor=north west][inner sep=0.75pt]    {$\mu $};
\draw (24,60.4) node [anchor=north west][inner sep=0.75pt]    {$\nu $};
\draw (46,59.4) node [anchor=north west][inner sep=0.75pt]    {$\rho $};
\draw (7,105) node [anchor=north west][inner sep=0.75pt]    {$m_{\mu }$};
\draw (30,105) node [anchor=north west][inner sep=0.75pt]    {$m_{\nu }$};
\draw (51,105) node [anchor=north west][inner sep=0.75pt]    {$m_{\rho }$};
\end{tikzpicture},
\end{equation} 
where $\Gamma_\text{closed}$ means that the part of the graph does not have any open edges. Moreover, we have the following orthogonality conditions: 
\begin{equation}\label{eq:3.16}
    \sum _{\rho }\mathtt{d}_{\rho }\begin{tikzpicture}[x=0.75pt,y=0.75pt,yscale=-1,xscale=1, baseline=(XXXX.south) ]
\path (0,99);\path (79.99651336669922,0);\draw    ($(current bounding box.center)+(0,0.3em)$) node [anchor=south] (XXXX) {};
\draw    (10.38,11.27) -- (38.51,27.51) ;
\draw [shift={(24.44,19.39)}, rotate = 30] [fill={rgb, 255:red, 0; green, 0; blue, 0 }  ][line width=0.08]  [draw opacity=0] (5.36,-2.57) -- (0,0) -- (5.36,2.57) -- cycle    ;
\draw    (38.51,27.51) -- (66.64,11.27) ;
\draw [shift={(52.58,19.39)}, rotate = 150] [fill={rgb, 255:red, 0; green, 0; blue, 0 }  ][line width=0.08]  [draw opacity=0] (5.36,-2.57) -- (0,0) -- (5.36,2.57) -- cycle    ;
\draw    (38.51,27.51) -- (38.51,62.7) ;
\draw [shift={(38.51,62.7)}, rotate = 90] [color={rgb, 255:red, 0; green, 0; blue, 0 }  ][fill={rgb, 255:red, 0; green, 0; blue, 0 }  ][line width=0.75]      (0, 0) circle [x radius= 2.01, y radius= 2.01]   ;
\draw [shift={(38.51,45.11)}, rotate = 90] [fill={rgb, 255:red, 0; green, 0; blue, 0 }  ][line width=0.08]  [draw opacity=0] (5.36,-2.57) -- (0,0) -- (5.36,2.57) -- cycle    ;
\draw [shift={(38.51,27.51)}, rotate = 90] [color={rgb, 255:red, 0; green, 0; blue, 0 }  ][fill={rgb, 255:red, 0; green, 0; blue, 0 }  ][line width=0.75]      (0, 0) circle [x radius= 2.01, y radius= 2.01]   ;
\draw    (10.38,78.95) -- (38.51,62.7) ;
\draw [shift={(24.44,70.83)}, rotate = 150] [fill={rgb, 255:red, 0; green, 0; blue, 0 }  ][line width=0.08]  [draw opacity=0] (5.36,-2.57) -- (0,0) -- (5.36,2.57) -- cycle    ;
\draw    (38.51,62.7) -- (66.64,78.95) ;
\draw [shift={(52.58,70.83)}, rotate = 30] [fill={rgb, 255:red, 0; green, 0; blue, 0 }  ][line width=0.08]  [draw opacity=0] (5.36,-2.57) -- (0,0) -- (5.36,2.57) -- cycle    ;
\draw (15,21) node [anchor=north west][inner sep=0.75pt]    {$\mu $};
\draw (50,21) node [anchor=north west][inner sep=0.75pt]    {$\nu $};
\draw (1,-4) node [anchor=north west][inner sep=0.75pt]    {$m_{\mu } '$};
\draw (57,-4) node [anchor=north west][inner sep=0.75pt]    {$m_{\nu } '$};
\draw (26,40.4) node [anchor=north west][inner sep=0.75pt]    {$\rho $};
\draw (1,80) node [anchor=north west][inner sep=0.75pt]    {$m_{\mu }$};
\draw (57,80) node [anchor=north west][inner sep=0.75pt]    {$m_{\nu }$};
\draw (19,74) node [anchor=north west][inner sep=0.75pt]    {$\mu $};
\draw (45,74) node [anchor=north west][inner sep=0.75pt]    {$\nu $};
\end{tikzpicture}
=\delta _{m_{\mu } m_{\mu } '} \delta _{m_{\nu } m_{\nu } '} ,\; \;  \begin{tikzpicture}[x=0.75pt,y=0.75pt,yscale=-1,xscale=1, baseline=(XXXX.south) ]
\path (0,117);\path (68.99651336669922,0);\draw    ($(current bounding box.center)+(0,0.3em)$) node [anchor=south] (XXXX) {};
\draw    (33.61,72.7) .. controls (6.81,72.7) and (8.72,34.7) .. (33.61,34) ;
\draw [shift={(14.23,53.62)}, rotate = 83.18] [fill={rgb, 255:red, 0; green, 0; blue, 0 }  ][line width=0.08]  [draw opacity=0] (5.36,-2.57) -- (0,0) -- (5.36,2.57) -- cycle    ;
\draw    (33.61,9.7) -- (33.61,34) ;
\draw [shift={(33.61,34)}, rotate = 90] [color={rgb, 255:red, 0; green, 0; blue, 0 }  ][fill={rgb, 255:red, 0; green, 0; blue, 0 }  ][line width=0.75]      (0, 0) circle [x radius= 2.01, y radius= 2.01]   ;
\draw [shift={(33.61,21.85)}, rotate = 90] [fill={rgb, 255:red, 0; green, 0; blue, 0 }  ][line width=0.08]  [draw opacity=0] (5.36,-2.57) -- (0,0) -- (5.36,2.57) -- cycle    ;
\draw    (33.61,72.7) .. controls (58.6,72.7) and (60.52,34.7) .. (33.61,34) ;
\draw [shift={(53.08,53.04)}, rotate = 98.02] [fill={rgb, 255:red, 0; green, 0; blue, 0 }  ][line width=0.08]  [draw opacity=0] (5.36,-2.57) -- (0,0) -- (5.36,2.57) -- cycle    ;
\draw    (33.61,72.7) -- (33.61,97) ;
\draw [shift={(33.61,84.85)}, rotate = 90] [fill={rgb, 255:red, 0; green, 0; blue, 0 }  ][line width=0.08]  [draw opacity=0] (5.36,-2.57) -- (0,0) -- (5.36,2.57) -- cycle    ;
\draw [shift={(33.61,72.7)}, rotate = 90] [color={rgb, 255:red, 0; green, 0; blue, 0 }  ][fill={rgb, 255:red, 0; green, 0; blue, 0 }  ][line width=0.75]      (0, 0) circle [x radius= 2.01, y radius= 2.01]   ;
\draw (1,50) node [anchor=north west][inner sep=0.75pt]    {$\mu $};
\draw (54,50) node [anchor=north west][inner sep=0.75pt]    {$\nu $};
\draw (18.92,15.4) node [anchor=north west][inner sep=0.75pt]    {$\rho '$};
\draw (24,-5) node [anchor=north west][inner sep=0.75pt]    {$m_{\rho } '$};
\draw (18.98,79.4) node [anchor=north west][inner sep=0.75pt]    {$\rho $};
\draw (24,98) node [anchor=north west][inner sep=0.75pt]    {$m_{\rho }$};
\end{tikzpicture}
=\frac{1}{\mathtt{d}_{\rho }} \delta _{\rho '\rho } \delta _{m_{\rho } 'm_{\rho }},
\end{equation}
where $\mathtt{d}_\mu=\beta_\mu d_\mu$ is the quantum dimension of the representation $\mu$. 

Similar to the duality maps, a $3j$-symbol $C_{\mu\nu\rho}$ is also determined up to a complex number. In order to construct symmetrized $6j$-symbols via $3j$-symbols, we further require that 
\begin{equation}\label{eq:3.17}
    C_{\mu\nu\rho;m_{\mu} m_{\nu} m_{\rho}}=\beta_{\rho}C_{\rho\mu\nu;m_{\rho}m_{\mu}m_{\nu}}
\end{equation}
and that
\begin{equation}
    (C_{\mu\nu\rho;m_{\mu} m_{\nu} m_{\rho}})^*=\sum_{m_{\mu^*} m_{\nu^*} m_{\rho^*}} C_{\rho^*\nu^*\mu^*;m_{\rho^*}m_{\nu^*}m_{\mu^*}}\Omega^{\rho^*}_{m_{\rho^*}m_{\rho}}\Omega^{\nu^*}_{m_{\nu^*}m_{\nu}}\Omega^{\mu^*}_{m_{\mu^*}m_{\mu}}.
\end{equation}
Combining the two requirements above and the definition of $3j$-symbols, we can completely determine all the $3j$-symbols for a given finite group $G$. Note that \eqref{eq:3.17} yields that $\beta_\mu\beta_\nu\beta_\rho=1$ if $C_{\mu\nu\rho}$ does not vanish. Some examples are listed in \cite{Hu2013a}. 

\paragraph*{Symmetrized $6j$-symbol}
A $6j$-symbol $F:L_G^6\to\mathbb C$ is defined by 
\begin{equation}\label{eq:Fmove}     
\begin{tikzpicture}[x=0.75pt,y=0.75pt,yscale=-1,xscale=1, baseline=(XXXX.south) ]
    \path (0,94);\path (79.9923324584961,0);\draw    ($(current bounding box.center)+(0,0.3em)$) node [anchor=south] (XXXX) {};
    \draw    (13,12) -- (28.01,35.98) ;
    \draw [shift={(18.33,20.51)}, rotate = 57.96] [fill={rgb, 255:red, 0; green, 0; blue, 0 }  ][line width=0.08]  [draw opacity=0] (5.36,-2.57) -- (0,0) -- (5.36,2.57) -- cycle    ;
    \draw    (28.01,35.98) -- (43.02,59.96) ;
    \draw [shift={(43.02,59.96)}, rotate = 57.96] [color={rgb, 255:red, 0; green, 0; blue, 0 }  ][fill={rgb, 255:red, 0; green, 0; blue, 0 }  ][line width=0.75]      (0, 0) circle [x radius= 2.01, y radius= 2.01]   ;
    \draw [shift={(33.34,44.5)}, rotate = 57.96] [fill={rgb, 255:red, 0; green, 0; blue, 0 }  ][line width=0.08]  [draw opacity=0] (5.36,-2.57) -- (0,0) -- (5.36,2.57) -- cycle    ;
    \draw    (43.02,59.96) -- (58.03,83.95) ;
    \draw [shift={(48.35,68.48)}, rotate = 57.96] [fill={rgb, 255:red, 0; green, 0; blue, 0 }  ][line width=0.08]  [draw opacity=0] (5.36,-2.57) -- (0,0) -- (5.36,2.57) -- cycle    ;
    \draw    (28.01,35.98) -- (43.02,12) ;
    \draw [shift={(36.89,21.78)}, rotate = 122.04] [fill={rgb, 255:red, 0; green, 0; blue, 0 }  ][line width=0.08]  [draw opacity=0] (5.36,-2.57) -- (0,0) -- (5.36,2.57) -- cycle    ;
    \draw [shift={(28.01,35.98)}, rotate = 302.04] [color={rgb, 255:red, 0; green, 0; blue, 0 }  ][fill={rgb, 255:red, 0; green, 0; blue, 0 }  ][line width=0.75]      (0, 0) circle [x radius= 2.01, y radius= 2.01]   ;
    \draw    (43.02,59.96) -- (73.03,12) ;
    \draw [shift={(59.4,33.78)}, rotate = 122.04] [fill={rgb, 255:red, 0; green, 0; blue, 0 }  ][line width=0.08]  [draw opacity=0] (5.36,-2.57) -- (0,0) -- (5.36,2.57) -- cycle    ;
\draw (5,18.2) node [anchor=north west][inner sep=0.75pt]    {\small $\nu ^{*}$};
\draw (35,17.19) node [anchor=north west][inner sep=0.75pt]    {\small $\mu ^{*}$};
\draw (58.03,31.18) node [anchor=north west][inner sep=0.75pt]    {\small $\kappa ^{*}$};
\draw (25,45) node [anchor=north west][inner sep=0.75pt]    {\small $\lambda $};
\draw (40,69.2) node [anchor=north west][inner sep=0.75pt]    {\small $\eta $};
\end{tikzpicture}
=\sum _{\rho }F_{\eta\kappa\rho }^{\mu \nu \lambda }
\begin{tikzpicture}[x=0.75pt,y=0.75pt,yscale=-1,xscale=1, baseline=(XXXX.south) ]
    \path (0,94);\path (79.9923324584961,0);\draw    ($(current bounding box.center)+(0,0.3em)$) node [anchor=south] (XXXX) {};
    \draw    (8,12) -- (38.02,59.96) ;
    \draw [shift={(20.83,32.51)}, rotate = 57.96] [fill={rgb, 255:red, 0; green, 0; blue, 0 }  ][line width=0.08]  [draw opacity=0] (5.36,-2.57) -- (0,0) -- (5.36,2.57) -- cycle    ;
    \draw    (38.02,59.96) -- (53.03,83.95) ;
    \draw [shift={(43.35,68.48)}, rotate = 57.96] [fill={rgb, 255:red, 0; green, 0; blue, 0 }  ][line width=0.08]  [draw opacity=0] (5.36,-2.57) -- (0,0) -- (5.36,2.57) -- cycle    ;
    \draw    (53.01,35.98) -- (38,12) ;
    \draw [shift={(44.12,21.78)}, rotate = 57.96] [fill={rgb, 255:red, 0; green, 0; blue, 0 }  ][line width=0.08]  [draw opacity=0] (5.36,-2.57) -- (0,0) -- (5.36,2.57) -- cycle    ;
    \draw    (38.02,59.96) -- (53.03,35.98) ;
    \draw [shift={(53.03,35.98)}, rotate = 302.04] [color={rgb, 255:red, 0; green, 0; blue, 0 }  ][fill={rgb, 255:red, 0; green, 0; blue, 0 }  ][line width=0.75]      (0, 0) circle [x radius= 2.01, y radius= 2.01]   ;
    \draw [shift={(46.9,45.77)}, rotate = 122.04] [fill={rgb, 255:red, 0; green, 0; blue, 0 }  ][line width=0.08]  [draw opacity=0] (5.36,-2.57) -- (0,0) -- (5.36,2.57) -- cycle    ;
    \draw [shift={(38.02,59.96)}, rotate = 302.04] [color={rgb, 255:red, 0; green, 0; blue, 0 }  ][fill={rgb, 255:red, 0; green, 0; blue, 0 }  ][line width=0.75]      (0, 0) circle [x radius= 2.01, y radius= 2.01]   ;
    \draw    (53.03,35.98) -- (68.03,12) ;
    \draw [shift={(61.91,21.78)}, rotate = 122.04] [fill={rgb, 255:red, 0; green, 0; blue, 0 }  ][line width=0.08]  [draw opacity=0] (5.36,-2.57) -- (0,0) -- (5.36,2.57) -- cycle    ;
\draw (8,33) node [anchor=north west][inner sep=0.75pt]    {\small $\nu ^{*}$};
\draw (29,17.19) node [anchor=north west][inner sep=0.75pt]    {\small $\mu ^{*}$};
\draw (61.03,16.18) node [anchor=north west][inner sep=0.75pt]    {\small $\kappa ^{*}$};
\draw (46,42.2) node [anchor=north west][inner sep=0.75pt]    {\small $\rho $};
\draw (35,69.2) node [anchor=north west][inner sep=0.75pt]    {\small $\eta $};
\end{tikzpicture},
\end{equation}
where we have omited Latin indices $m_\mu,m_\nu,\cdots$. The above expression relates the two equivalent ways of decomposing the tensor product representation $V_{\nu^*}\otimes V_{\mu^*}\otimes V_{\kappa^*}$. Composing the both sides of the above equation by the duality maps and the $3j$-symbols in an appropriate way as: 
\begin{equation*}
    \begin{tikzpicture}[x=0.75pt,y=0.75pt,yscale=-1,xscale=1, baseline=(XXXX.south) ]
\path (0,116);\path (93.8043212890625,0);\draw    ($(current bounding box.center)+(0,0.3em)$) node [anchor=south] (XXXX) {};
\draw    (32.78,59) .. controls (32.54,67.83) and (34.46,75.15) .. (38.46,79.82) ;
\draw [shift={(33.07,65.92)}, rotate = 78.51] [fill={rgb, 255:red, 0; green, 0; blue, 0 }  ][line width=0.08]  [draw opacity=0] (5.36,-2.57) -- (0,0) -- (5.36,2.57) -- cycle    ;
\draw    (53.97,59) .. controls (53.73,67.33) and (50.79,67.82) .. (38.46,79.82) ;
\draw [shift={(51.02,67.71)}, rotate = 134.72] [fill={rgb, 255:red, 0; green, 0; blue, 0 }  ][line width=0.08]  [draw opacity=0] (5.36,-2.57) -- (0,0) -- (5.36,2.57) -- cycle    ;
\draw    (75.16,59) .. controls (75.12,73.82) and (64.79,82.82) .. (51.78,96.39) ;
\draw [shift={(69.55,76.42)}, rotate = 126.8] [fill={rgb, 255:red, 0; green, 0; blue, 0 }  ][line width=0.08]  [draw opacity=0] (5.36,-2.57) -- (0,0) -- (5.36,2.57) -- cycle    ;
\draw [color={rgb, 255:red, 208; green, 2; blue, 27 }  ,draw opacity=1 ]   (56.16,21.6) .. controls (42.79,34.15) and (33.46,40.54) .. (32.78,59) ;
\draw [shift={(42.49,34.71)}, rotate = 128.45] [fill={rgb, 255:red, 208; green, 2; blue, 27 }  ,fill opacity=1 ][line width=0.08]  [draw opacity=0] (5.36,-2.57) -- (0,0) -- (5.36,2.57) -- cycle    ;
\draw [shift={(56.16,21.6)}, rotate = 137.37] [color={rgb, 255:red, 208; green, 2; blue, 27 }  ,draw opacity=1 ][fill={rgb, 255:red, 208; green, 2; blue, 27 }  ,fill opacity=1 ][line width=0.75]      (0, 0) circle [x radius= 2.01, y radius= 2.01]   ;
\draw [color={rgb, 255:red, 208; green, 2; blue, 27 }  ,draw opacity=1 ]   (69.49,38.17) .. controls (74.12,44.49) and (75.12,51.49) .. (75.16,59) ;
\draw [shift={(72.83,44.24)}, rotate = 74.41] [fill={rgb, 255:red, 208; green, 2; blue, 27 }  ,fill opacity=1 ][line width=0.08]  [draw opacity=0] (5.36,-2.57) -- (0,0) -- (5.36,2.57) -- cycle    ;
\draw [shift={(69.49,38.17)}, rotate = 53.7] [color={rgb, 255:red, 208; green, 2; blue, 27 }  ,draw opacity=1 ][fill={rgb, 255:red, 208; green, 2; blue, 27 }  ,fill opacity=1 ][line width=0.75]      (0, 0) circle [x radius= 2.01, y radius= 2.01]   ;
\draw [color={rgb, 255:red, 208; green, 2; blue, 27 }  ,draw opacity=1 ]   (69.49,38.17) .. controls (59.12,47.15) and (53.79,50.49) .. (53.97,59) ;
\draw [shift={(62.32,44.32)}, rotate = 135.94] [fill={rgb, 255:red, 208; green, 2; blue, 27 }  ,fill opacity=1 ][line width=0.08]  [draw opacity=0] (5.36,-2.57) -- (0,0) -- (5.36,2.57) -- cycle    ;
\draw [color={rgb, 255:red, 208; green, 2; blue, 27 }  ,draw opacity=1 ]   (33.79,5.49) .. controls (41.12,7.15) and (48.79,14.82) .. (56.16,21.6) ;
\draw [shift={(42.6,9.85)}, rotate = 38.26] [fill={rgb, 255:red, 208; green, 2; blue, 27 }  ,fill opacity=1 ][line width=0.08]  [draw opacity=0] (5.36,-2.57) -- (0,0) -- (5.36,2.57) -- cycle    ;
\draw [shift={(33.79,5.49)}, rotate = 12.8] [color={rgb, 255:red, 208; green, 2; blue, 27 }  ,draw opacity=1 ][fill={rgb, 255:red, 208; green, 2; blue, 27 }  ,fill opacity=1 ][line width=0.75]      (0, 0) circle [x radius= 2.01, y radius= 2.01]   ;
\draw    (38.46,79.82) -- (51.78,96.39) ;
\draw [shift={(51.78,96.39)}, rotate = 51.2] [color={rgb, 255:red, 0; green, 0; blue, 0 }  ][fill={rgb, 255:red, 0; green, 0; blue, 0 }  ][line width=0.75]      (0, 0) circle [x radius= 2.01, y radius= 2.01]   ;
\draw [shift={(42.55,84.91)}, rotate = 51.2] [fill={rgb, 255:red, 0; green, 0; blue, 0 }  ][line width=0.08]  [draw opacity=0] (5.36,-2.57) -- (0,0) -- (5.36,2.57) -- cycle    ;
\draw [shift={(38.46,79.82)}, rotate = 51.2] [color={rgb, 255:red, 0; green, 0; blue, 0 }  ][fill={rgb, 255:red, 0; green, 0; blue, 0 }  ][line width=0.75]      (0, 0) circle [x radius= 2.01, y radius= 2.01]   ;
\draw [color={rgb, 255:red, 208; green, 2; blue, 27 }  ,draw opacity=1 ]   (56.16,21.6) -- (69.49,38.17) ;
\draw [shift={(60.25,26.69)}, rotate = 51.2] [fill={rgb, 255:red, 208; green, 2; blue, 27 }  ,fill opacity=1 ][line width=0.08]  [draw opacity=0] (5.36,-2.57) -- (0,0) -- (5.36,2.57) -- cycle    ;
\draw    (51.78,96.39) .. controls (42.79,105.82) and (33.79,108.82) .. (26.46,111.49) ;
\draw [shift={(43.47,103.49)}, rotate = 149.07] [fill={rgb, 255:red, 0; green, 0; blue, 0 }  ][line width=0.08]  [draw opacity=0] (5.36,-2.57) -- (0,0) -- (5.36,2.57) -- cycle    ;
\draw [color={rgb, 255:red, 208; green, 2; blue, 27 }  ,draw opacity=1 ]   (33.79,5.49) .. controls (1.46,19.82) and (-6.88,87.15) .. (26.46,111.49) ;
\draw [shift={(26.46,111.49)}, rotate = 36.13] [color={rgb, 255:red, 208; green, 2; blue, 27 }  ,draw opacity=1 ][fill={rgb, 255:red, 208; green, 2; blue, 27 }  ,fill opacity=1 ][line width=0.75]      (0, 0) circle [x radius= 2.01, y radius= 2.01]   ;
\draw [shift={(5.7,52.79)}, rotate = 94.51] [fill={rgb, 255:red, 208; green, 2; blue, 27 }  ,fill opacity=1 ][line width=0.08]  [draw opacity=0] (5.36,-2.57) -- (0,0) -- (5.36,2.57) -- cycle    ;
\draw (36.06,48.28) node [anchor=north west][inner sep=0.75pt]    {\small $\nu ^{*}$};
\draw (54.92,47.69) node [anchor=north west][inner sep=0.75pt]    {\small $\mu ^{*}$};
\draw (77.42,48.56) node [anchor=north west][inner sep=0.75pt]    {\small $\kappa ^{*}$};
\draw (35.97,85.71) node [anchor=north west][inner sep=0.75pt]    {\small $\lambda $};
\draw (42.71,100.42) node [anchor=north west][inner sep=0.75pt]    {\small $\eta $};
\draw (10.49,46.18) node [anchor=north west][inner sep=0.75pt]    {\small $\eta ^{*}$};
\draw (47.65,-0.15) node [anchor=north west][inner sep=0.75pt]    {\small $\eta $};
\draw (65.33,16.01) node [anchor=north west][inner sep=0.75pt]    {\small $\rho $};
\end{tikzpicture}
=\sum _{\rho'} F_{\eta \kappa \rho'}^{\mu \nu \lambda }
\begin{tikzpicture}[x=0.75pt,y=0.75pt,yscale=-1,xscale=1, baseline=(XXXX.south) ]
\path (0,116);\path (93.8043212890625,0);\draw    ($(current bounding box.center)+(0,0.3em)$) node [anchor=south] (XXXX) {};
\draw    (32.78,59) .. controls (33.29,73.98) and (39.96,80.32) .. (51.78,96.39) ;
\draw [shift={(37.12,75.65)}, rotate = 58.59] [fill={rgb, 255:red, 0; green, 0; blue, 0 }  ][line width=0.08]  [draw opacity=0] (5.36,-2.57) -- (0,0) -- (5.36,2.57) -- cycle    ;
\draw    (53.97,59) .. controls (53.73,67.33) and (58.56,72.82) .. (67.89,78.82) ;
\draw [shift={(55.81,67.35)}, rotate = 53.41] [fill={rgb, 255:red, 0; green, 0; blue, 0 }  ][line width=0.08]  [draw opacity=0] (5.36,-2.57) -- (0,0) -- (5.36,2.57) -- cycle    ;
\draw    (75.16,59) .. controls (75.29,65.98) and (72.56,72.15) .. (67.89,78.82) ;
\draw [shift={(74.41,65.59)}, rotate = 109.84] [fill={rgb, 255:red, 0; green, 0; blue, 0 }  ][line width=0.08]  [draw opacity=0] (5.36,-2.57) -- (0,0) -- (5.36,2.57) -- cycle    ;
\draw [color={rgb, 255:red, 208; green, 2; blue, 27 }  ,draw opacity=1 ]   (56.16,21.6) .. controls (42.79,34.15) and (33.46,40.54) .. (32.78,59) ;
\draw [shift={(42.49,34.71)}, rotate = 128.45] [fill={rgb, 255:red, 208; green, 2; blue, 27 }  ,fill opacity=1 ][line width=0.08]  [draw opacity=0] (5.36,-2.57) -- (0,0) -- (5.36,2.57) -- cycle    ;
\draw [shift={(56.16,21.6)}, rotate = 137.37] [color={rgb, 255:red, 208; green, 2; blue, 27 }  ,draw opacity=1 ][fill={rgb, 255:red, 208; green, 2; blue, 27 }  ,fill opacity=1 ][line width=0.75]      (0, 0) circle [x radius= 2.01, y radius= 2.01]   ;
\draw [color={rgb, 255:red, 208; green, 2; blue, 27 }  ,draw opacity=1 ]   (69.49,38.17) .. controls (74.12,44.49) and (75.12,51.49) .. (75.16,59) ;
\draw [shift={(72.83,44.24)}, rotate = 74.41] [fill={rgb, 255:red, 208; green, 2; blue, 27 }  ,fill opacity=1 ][line width=0.08]  [draw opacity=0] (5.36,-2.57) -- (0,0) -- (5.36,2.57) -- cycle    ;
\draw [shift={(69.49,38.17)}, rotate = 53.7] [color={rgb, 255:red, 208; green, 2; blue, 27 }  ,draw opacity=1 ][fill={rgb, 255:red, 208; green, 2; blue, 27 }  ,fill opacity=1 ][line width=0.75]      (0, 0) circle [x radius= 2.01, y radius= 2.01]   ;
\draw [color={rgb, 255:red, 208; green, 2; blue, 27 }  ,draw opacity=1 ]   (69.49,38.17) .. controls (59.12,47.15) and (53.79,50.49) .. (53.97,59) ;
\draw [shift={(62.32,44.32)}, rotate = 135.94] [fill={rgb, 255:red, 208; green, 2; blue, 27 }  ,fill opacity=1 ][line width=0.08]  [draw opacity=0] (5.36,-2.57) -- (0,0) -- (5.36,2.57) -- cycle    ;
\draw [color={rgb, 255:red, 208; green, 2; blue, 27 }  ,draw opacity=1 ]   (33.79,5.49) .. controls (41.12,7.15) and (48.79,14.82) .. (56.16,21.6) ;
\draw [shift={(42.6,9.85)}, rotate = 38.26] [fill={rgb, 255:red, 208; green, 2; blue, 27 }  ,fill opacity=1 ][line width=0.08]  [draw opacity=0] (5.36,-2.57) -- (0,0) -- (5.36,2.57) -- cycle    ;
\draw [shift={(33.79,5.49)}, rotate = 12.8] [color={rgb, 255:red, 208; green, 2; blue, 27 }  ,draw opacity=1 ][fill={rgb, 255:red, 208; green, 2; blue, 27 }  ,fill opacity=1 ][line width=0.75]      (0, 0) circle [x radius= 2.01, y radius= 2.01]   ;
\draw    (67.89,78.82) -- (51.78,96.39) ;
\draw [shift={(51.78,96.39)}, rotate = 132.52] [color={rgb, 255:red, 0; green, 0; blue, 0 }  ][fill={rgb, 255:red, 0; green, 0; blue, 0 }  ][line width=0.75]      (0, 0) circle [x radius= 2.01, y radius= 2.01]   ;
\draw [shift={(62.6,84.58)}, rotate = 132.52] [fill={rgb, 255:red, 0; green, 0; blue, 0 }  ][line width=0.08]  [draw opacity=0] (5.36,-2.57) -- (0,0) -- (5.36,2.57) -- cycle    ;
\draw [shift={(67.89,78.82)}, rotate = 132.52] [color={rgb, 255:red, 0; green, 0; blue, 0 }  ][fill={rgb, 255:red, 0; green, 0; blue, 0 }  ][line width=0.75]      (0, 0) circle [x radius= 2.01, y radius= 2.01]   ;
\draw [color={rgb, 255:red, 208; green, 2; blue, 27 }  ,draw opacity=1 ]   (56.16,21.6) -- (69.49,38.17) ;
\draw [shift={(60.25,26.69)}, rotate = 51.2] [fill={rgb, 255:red, 208; green, 2; blue, 27 }  ,fill opacity=1 ][line width=0.08]  [draw opacity=0] (5.36,-2.57) -- (0,0) -- (5.36,2.57) -- cycle    ;
\draw    (51.78,96.39) .. controls (42.79,105.82) and (33.79,108.82) .. (26.46,111.49) ;
\draw [shift={(43.47,103.49)}, rotate = 149.07] [fill={rgb, 255:red, 0; green, 0; blue, 0 }  ][line width=0.08]  [draw opacity=0] (5.36,-2.57) -- (0,0) -- (5.36,2.57) -- cycle    ;
\draw [color={rgb, 255:red, 208; green, 2; blue, 27 }  ,draw opacity=1 ]   (33.79,5.49) .. controls (1.46,19.82) and (-6.88,87.15) .. (26.46,111.49) ;
\draw [shift={(26.46,111.49)}, rotate = 36.13] [color={rgb, 255:red, 208; green, 2; blue, 27 }  ,draw opacity=1 ][fill={rgb, 255:red, 208; green, 2; blue, 27 }  ,fill opacity=1 ][line width=0.75]      (0, 0) circle [x radius= 2.01, y radius= 2.01]   ;
\draw [shift={(5.7,52.79)}, rotate = 94.51] [fill={rgb, 255:red, 208; green, 2; blue, 27 }  ,fill opacity=1 ][line width=0.08]  [draw opacity=0] (5.36,-2.57) -- (0,0) -- (5.36,2.57) -- cycle    ;
\draw (36.06,48.28) node [anchor=north west][inner sep=0.75pt]    {\small $\nu ^{*}$};
\draw (54.92,47.69) node [anchor=north west][inner sep=0.75pt]    {\small $\mu ^{*}$};
\draw (77.42,48.56) node [anchor=north west][inner sep=0.75pt]    {\small $\kappa ^{*}$};
\draw (62.31,82.38) node [anchor=north west][inner sep=0.75pt]    {\small $\rho '$};
\draw (42.71,100.42) node [anchor=north west][inner sep=0.75pt]    {\small $\eta $};
\draw (7.82,47.51) node [anchor=north west][inner sep=0.75pt]    {\small $\eta ^{*}$};
\draw (47.65,-0.15) node [anchor=north west][inner sep=0.75pt]    {\small $\eta $};
\draw (65.33,16.01) node [anchor=north west][inner sep=0.75pt]    {\small $\rho $};
\end{tikzpicture},
\end{equation*}
which by the second equation of \eqref{eq:3.16} becomes 
\begin{equation}\label{eq:3.20x}
    F^{\mu\nu\lambda}_{\eta\kappa\rho}=\mathtt{d}_\rho\begin{tikzpicture}[x=0.75pt,y=0.75pt,yscale=-1,xscale=1, baseline=(XXXX.south) ]
        \path (0,114);\path (96.99053955078125,0);\draw    ($(current bounding box.center)+(0,0.3em)$) node [anchor=south] (XXXX) {};
        \draw    (53.8,65) .. controls (49.8,56.13) and (47.8,34.13) .. (63.99,27) ;
        \draw [shift={(51.62,43.55)}, rotate = 95.83] [fill={rgb, 255:red, 0; green, 0; blue, 0 }  ][line width=0.08]  [draw opacity=0] (5.36,-2.57) -- (0,0) -- (5.36,2.57) -- cycle    ;
        \draw    (53.8,65) -- (81.8,37.13) ;
        \draw [shift={(81.8,37.13)}, rotate = 315.13] [color={rgb, 255:red, 0; green, 0; blue, 0 }  ][fill={rgb, 255:red, 0; green, 0; blue, 0 }  ][line width=0.75]      (0, 0) circle [x radius= 2.01, y radius= 2.01]   ;
        \draw [shift={(67.8,51.06)}, rotate = 135.13] [fill={rgb, 255:red, 0; green, 0; blue, 0 }  ][line width=0.08]  [draw opacity=0] (5.36,-2.57) -- (0,0) -- (5.36,2.57) -- cycle    ;
        \draw [shift={(53.8,65)}, rotate = 315.13] [color={rgb, 255:red, 0; green, 0; blue, 0 }  ][fill={rgb, 255:red, 0; green, 0; blue, 0 }  ][line width=0.75]      (0, 0) circle [x radius= 2.01, y radius= 2.01]   ;
        \draw    (63.8,74.63) .. controls (59.95,74.24) and (57.29,70.59) .. (55.35,67.51) ;
        \draw [shift={(53.8,65)}, rotate = 57.41] [fill={rgb, 255:red, 0; green, 0; blue, 0 }  ][line width=0.08]  [draw opacity=0] (5.36,-2.57) -- (0,0) -- (5.36,2.57) -- cycle    ;
        \draw    (81.8,37.13) .. controls (73.8,23.13) and (65.8,26.13) .. (63.99,27) ;
        \draw [shift={(74.55,28.44)}, rotate = 44.53] [fill={rgb, 255:red, 0; green, 0; blue, 0 }  ][line width=0.08]  [draw opacity=0] (5.36,-2.57) -- (0,0) -- (5.36,2.57) -- cycle    ;
        \draw    (63.8,74.63) .. controls (81.8,80.13) and (88.8,52.13) .. (81.8,37.13) ;
        \draw [shift={(82.7,62.06)}, rotate = 113.79] [fill={rgb, 255:red, 0; green, 0; blue, 0 }  ][line width=0.08]  [draw opacity=0] (5.36,-2.57) -- (0,0) -- (5.36,2.57) -- cycle    ;
        \draw    (42.4,97.13) .. controls (9.8,97.13) and (7.8,8.13) .. (42.4,5) ;
        \draw [shift={(17.2,50.88)}, rotate = 88.3] [fill={rgb, 255:red, 0; green, 0; blue, 0 }  ][line width=0.08]  [draw opacity=0] (5.36,-2.57) -- (0,0) -- (5.36,2.57) -- cycle    ;
        \draw    (63.99,27) .. controls (60.8,17.13) and (56.8,7.13) .. (42.4,5) ;
        \draw [shift={(42.4,5)}, rotate = 188.41] [color={rgb, 255:red, 0; green, 0; blue, 0 }  ][fill={rgb, 255:red, 0; green, 0; blue, 0 }  ][line width=0.75]      (0, 0) circle [x radius= 2.01, y radius= 2.01]   ;
        \draw [shift={(56.71,12.27)}, rotate = 55.63] [fill={rgb, 255:red, 0; green, 0; blue, 0 }  ][line width=0.08]  [draw opacity=0] (5.36,-2.57) -- (0,0) -- (5.36,2.57) -- cycle    ;
        \draw [shift={(63.99,27)}, rotate = 252.1] [color={rgb, 255:red, 0; green, 0; blue, 0 }  ][fill={rgb, 255:red, 0; green, 0; blue, 0 }  ][line width=0.75]      (0, 0) circle [x radius= 2.01, y radius= 2.01]   ;
        \draw    (42.4,97.13) .. controls (55.8,97.13) and (60.8,82.13) .. (63.8,74.63) ;
        \draw [shift={(63.8,74.63)}, rotate = 291.81] [color={rgb, 255:red, 0; green, 0; blue, 0 }  ][fill={rgb, 255:red, 0; green, 0; blue, 0 }  ][line width=0.75]      (0, 0) circle [x radius= 2.01, y radius= 2.01]   ;
        \draw [shift={(56.5,89.38)}, rotate = 133.86] [fill={rgb, 255:red, 0; green, 0; blue, 0 }  ][line width=0.08]  [draw opacity=0] (5.36,-2.57) -- (0,0) -- (5.36,2.57) -- cycle    ;
        \draw [shift={(42.4,97.13)}, rotate = 0] [color={rgb, 255:red, 0; green, 0; blue, 0 }  ][fill={rgb, 255:red, 0; green, 0; blue, 0 }  ][line width=0.75]      (0, 0) circle [x radius= 2.01, y radius= 2.01]   ;
        \draw (0.99,45.39) node [anchor=north west][inner sep=0.75pt]    {\small $\eta ^{*}$};
        \draw (54.99,1.39) node [anchor=north west][inner sep=0.75pt]    {\small $\eta $};
        \draw (51.99,95.39) node [anchor=north west][inner sep=0.75pt]    {\small $\eta $};
        \draw (57.99,36.39) node [anchor=north west][inner sep=0.75pt]    {\small $\mu ^{*}$};
        \draw (34.99,38.39) node [anchor=north west][inner sep=0.75pt]    {\small $\nu ^{*}$};
        \draw (45.99,70.39) node [anchor=north west][inner sep=0.75pt]    {\small $\lambda $};
        \draw (78.99,64.39) node [anchor=north west][inner sep=0.75pt]    {\small $\kappa ^{*}$};
        \draw (71.99,15.39) node [anchor=north west][inner sep=0.75pt]    {\small $\rho $};
        \end{tikzpicture}\equiv\mathtt{d}_\rho G^{\mu\nu\lambda}_{\eta\kappa\rho}.
\end{equation}
The symmetrized $6j$-symbols are thus given by $G^{\mu\nu\lambda}_{\eta\kappa\rho}=F^{\mu\nu\lambda}_{\eta\kappa\rho}/\mathtt{d}_\rho$. In terms of $3j$-symbols and duality maps, we have
\begin{equation}
    \begin{aligned}
        G^{\mu\nu\lambda}_{\eta\kappa\rho}=\sum_{\text{all $m$'s and $n$'s}}&\Omega^\mu_{m_\mu n_{\mu^*}}\Omega^\nu_{m_\nu n_{\nu^*}}\Omega^\lambda_{m_\lambda n_{\lambda^*}}\Omega^\eta_{m_\eta n_{\eta^*}}\Omega^\kappa_{m_\kappa n_{\kappa^*}}\Omega^\rho_{m_\rho n_{\rho^*}}\\
        &\times C_{m_{\kappa} n_{\lambda^{*}} m_{\eta}}^{\kappa \lambda^{*} \eta} C_{n_{\eta^{*}} n_{\nu^{*}} m_{\rho}}^{\eta^{*} \nu^{*} \rho} C_{n_{\rho^{*}} n_{\mu^{*}} n_{\kappa^{*}}}^{\rho^{*} \mu^{*} \kappa^{*}} C_{m_{\mu} m_{\nu} m_{\lambda}}^{\mu \nu \lambda}.
    \end{aligned}
\end{equation} 
Graphically, symmetrized $6j$-symbol can be presented either by a planar graph or a tetrahedron 
\begin{equation}\label{eq:3.20}
    G_{\eta \kappa \rho }^{\mu \nu \lambda } =
.
\end{equation}

\paragraph*{Other conventions}
The great orthogonality theorem of finite group representations will be often used in our paper; it can be presented graphically as 
\begin{equation}
    \frac{1}{|G|}\sum_{g\in G}%
.
\end{equation*}
Here and hereafter, the horizontal lines should be understood through the following convention:
\begin{equation*}
    %
.
\end{equation*} 

\subsection{Fourier transform on the Hilbert space}\label{subsec:3.2}
The Fourier transform of the operators in the Hamiltonian requires defining the Fourier transform of the Hilbert space of a three-dimensional GT model with gapped boundaries with $G$ as its input data. The total Hilbert space $\mathcal{H}$ of the model is the tensor product of all local Hilbert spaces $\mathcal{H}_e$ on the edges. The basis vector $|g\rangle$ of $\mathcal{H}_e$ Fourier-transforms as: 
\begin{equation}\label{eq:3.22}
    |\mu,m_\mu,n_\mu\rangle=\frac{v_\mu}{\sqrt{G}}\sum_{g\in G}D^\mu_{m_\mu n_\mu}(g)|g\rangle,
\end{equation}
where $v_\mu=d_\mu^{1/2}$. We dub the Fourier-transformed basis $\{|\mu,m_\mu,n_\mu\rangle\}$ the local rep-basis. The local rep-basis $\{|\mu,m_\mu,n_\mu\rangle\}$ and the group-basis $|{g}\rangle$ have the same dimension due to $\sum_{\mu\in L_G} d_\mu^2=|G|$. Additionaly, using the great orthogonality theorem, we can prove that the rep-basis is also orthonormal: $\langle\mu',m',n'|\mu,m,n\rangle=\delta_{\mu'\mu}\delta_{m'm}\delta_{n'n}$. We can Fourier-transform $\mathcal{H}_e$ on each individual edge independently, resulting in a linear superposition of the rep-basis states, shown in \autoref{fig:3}. Here, the orientations of the rep-basis states inherit from those of the group-basis states. The explicit transformation will be dealt with a little later. Our focus here lies in the logic and physics of the basis transformations.

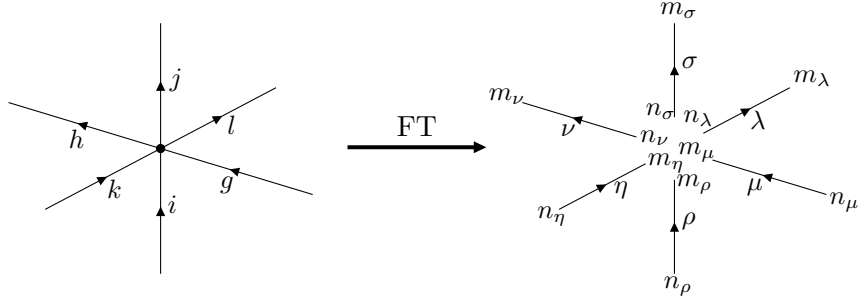
\begin{figure}[ht]
    \centering 
    \begin{tikzpicture}[x=0.75pt,y=0.75pt,yscale=-0.9,xscale=0.9]
    
    \draw    (93,57) -- (177.34,82.71) ;
    \draw [shift={(131.25,68.66)}, rotate = 16.95] [fill={rgb, 255:red, 0; green, 0; blue, 0 }  ][line width=0.08]  [draw opacity=0] (5.36,-2.57) -- (0,0) -- (5.36,2.57) -- cycle    ;
    \draw    (177.34,82.71) -- (261.67,108.41) ;
    \draw [shift={(215.58,94.36)}, rotate = 16.95] [fill={rgb, 255:red, 0; green, 0; blue, 0 }  ][line width=0.08]  [draw opacity=0] (5.36,-2.57) -- (0,0) -- (5.36,2.57) -- cycle    ;
    \draw    (177.34,82.71) -- (241.34,48.71) ;
    \draw [shift={(211.63,64.49)}, rotate = 152.02] [fill={rgb, 255:red, 0; green, 0; blue, 0 }  ][line width=0.08]  [draw opacity=0] (5.36,-2.57) -- (0,0) -- (5.36,2.57) -- cycle    ;
    \draw    (113.34,116.71) -- (177.34,82.71) ;
    \draw [shift={(147.63,98.49)}, rotate = 152.02] [fill={rgb, 255:red, 0; green, 0; blue, 0 }  ][line width=0.08]  [draw opacity=0] (5.36,-2.57) -- (0,0) -- (5.36,2.57) -- cycle    ;
    \draw    (177.34,82.71) -- (177.34,12.71) ;
    \draw [shift={(177.34,45.11)}, rotate = 90] [fill={rgb, 255:red, 0; green, 0; blue, 0 }  ][line width=0.08]  [draw opacity=0] (5.36,-2.57) -- (0,0) -- (5.36,2.57) -- cycle    ;
    \draw    (177.34,152.71) -- (177.34,82.71) ;
    \draw [shift={(177.34,82.71)}, rotate = 270] [color={rgb, 255:red, 0; green, 0; blue, 0 }  ][fill={rgb, 255:red, 0; green, 0; blue, 0 }  ][line width=0.75]      (0, 0) circle [x radius= 2.01, y radius= 2.01]   ;
    \draw [shift={(177.34,115.11)}, rotate = 90] [fill={rgb, 255:red, 0; green, 0; blue, 0 }  ][line width=0.08]  [draw opacity=0] (5.36,-2.57) -- (0,0) -- (5.36,2.57) -- cycle    ;
    \draw [line width=1.5]    (281,82) -- (354,82) ;
    \draw [shift={(358,82)}, rotate = 180] [fill={rgb, 255:red, 0; green, 0; blue, 0 }  ][line width=0.08]  [draw opacity=0] (6.97,-3.35) -- (0,0) -- (6.97,3.35) -- cycle    ;
    \draw    (378,57) -- (441.25,76.28) ;
    \draw [shift={(405.7,65.44)}, rotate = 16.95] [fill={rgb, 255:red, 0; green, 0; blue, 0 }  ][line width=0.08]  [draw opacity=0] (5.36,-2.57) -- (0,0) -- (5.36,2.57) -- cycle    ;
    \draw    (483.42,89.13) -- (546.67,108.41) ;
    \draw [shift={(511.13,97.58)}, rotate = 16.95] [fill={rgb, 255:red, 0; green, 0; blue, 0 }  ][line width=0.08]  [draw opacity=0] (5.36,-2.57) -- (0,0) -- (5.36,2.57) -- cycle    ;
    \draw    (478.34,74.21) -- (526.34,48.71) ;
    \draw [shift={(504.63,60.24)}, rotate = 152.02] [fill={rgb, 255:red, 0; green, 0; blue, 0 }  ][line width=0.08]  [draw opacity=0] (5.36,-2.57) -- (0,0) -- (5.36,2.57) -- cycle    ;
    \draw    (398.34,116.71) -- (446.34,91.21) ;
    \draw [shift={(424.63,102.74)}, rotate = 152.02] [fill={rgb, 255:red, 0; green, 0; blue, 0 }  ][line width=0.08]  [draw opacity=0] (5.36,-2.57) -- (0,0) -- (5.36,2.57) -- cycle    ;
    \draw    (462.34,65.21) -- (462.34,12.71) ;
    \draw [shift={(462.34,36.36)}, rotate = 90] [fill={rgb, 255:red, 0; green, 0; blue, 0 }  ][line width=0.08]  [draw opacity=0] (5.36,-2.57) -- (0,0) -- (5.36,2.57) -- cycle    ;
    \draw    (462.34,152.71) -- (462.34,100.21) ;
    \draw [shift={(462.34,123.86)}, rotate = 90] [fill={rgb, 255:red, 0; green, 0; blue, 0 }  ][line width=0.08]  [draw opacity=0] (5.36,-2.57) -- (0,0) -- (5.36,2.57) -- cycle    ;
    
    \draw (209,97) node [anchor=north west][inner sep=0.75pt]    {\small $g$};
    \draw (125,70.8) node [anchor=north west][inner sep=0.75pt]    {\small $h$};
    \draw (180,107.8) node [anchor=north west][inner sep=0.75pt]    {\small $i$};
    \draw (181,36.8) node [anchor=north west][inner sep=0.75pt]    {\small $j$};
    \draw (213,62.8) node [anchor=north west][inner sep=0.75pt]    {\small $l$};
    \draw (146.34,98.51) node [anchor=north west][inner sep=0.75pt]    {\small $k$};
    \draw (307,63) node [anchor=north west][inner sep=0.75pt]   [align=left] {FT};
    \draw (501,98) node [anchor=north west][inner sep=0.75pt]    {\small $\mu $};
    \draw (398,68) node [anchor=north west][inner sep=0.75pt]    {\small $\nu $};
    \draw (465,117) node [anchor=north west][inner sep=0.75pt]    {\small $\rho $};
    \draw (465,30) node [anchor=north west][inner sep=0.75pt]    {\small $\sigma $};
    \draw (503,60) node [anchor=north west][inner sep=0.75pt]    {\small $\lambda $};
    \draw (427.34,100) node [anchor=north west][inner sep=0.75pt]    {\small $\eta $};
    \draw (358,48) node [anchor=north west][inner sep=0.75pt]    {\small $m_{\nu }$};
    \draw (442,71) node [anchor=north west][inner sep=0.75pt]    {\small $n_{\nu }$};
    \draw (445,56) node [anchor=north west][inner sep=0.75pt]    {\small $n_{\sigma }$};
    \draw (446.25,84) node [anchor=north west][inner sep=0.75pt]    {\small $m_{\eta }$};
    \draw (455,153) node [anchor=north west][inner sep=0.75pt]    {\small $n_{\rho }$};
    \draw (385,114) node [anchor=north west][inner sep=0.75pt]    {\small $n_{\eta }$};
    \draw (462,96) node [anchor=north west][inner sep=0.75pt]    {\small $m_{\rho }$};
    \draw (463,78) node [anchor=north west][inner sep=0.75pt]    {\small $m_{\mu }$};
    \draw (466,60) node [anchor=north west][inner sep=0.75pt]    {\small $n_{\lambda }$};
    \draw (453,0) node [anchor=north west][inner sep=0.75pt]    {\small $m_{\sigma }$};
    \draw (527,38) node [anchor=north west][inner sep=0.75pt]    {\small $m_{\lambda }$};
    \draw (547,106) node [anchor=north west][inner sep=0.75pt]    {\small $n_{\mu }$};

    \end{tikzpicture}
\caption{The Fourier transform of a six-valent vertex in the bulk of the lattice $\Gamma$ on which the three-dimensional GT model is defined. }
\label{fig:3}
\end{figure}

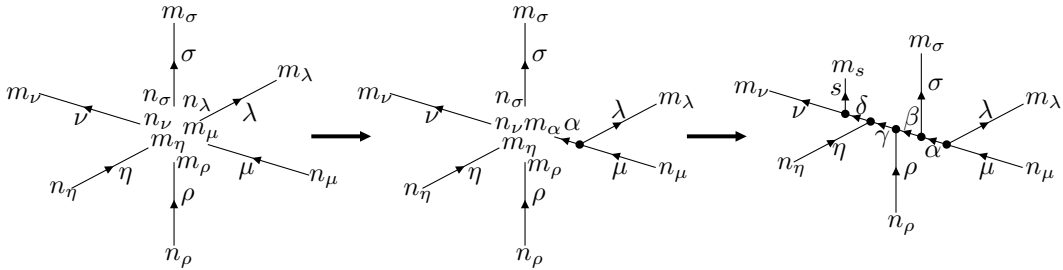
\begin{figure}[b]
    \centering 
\begin{tikzpicture}[x=0.75pt,y=0.75pt,yscale=-0.8,xscale=0.8]

\draw    (22,58) -- (85.25,77.28) ;
\draw [shift={(49.7,66.44)}, rotate = 16.95] [fill={rgb, 255:red, 0; green, 0; blue, 0 }  ][line width=0.08]  [draw opacity=0] (5.36,-2.57) -- (0,0) -- (5.36,2.57) -- cycle    ;
\draw    (127.42,90.13) -- (190.67,109.41) ;
\draw [shift={(155.13,98.58)}, rotate = 16.95] [fill={rgb, 255:red, 0; green, 0; blue, 0 }  ][line width=0.08]  [draw opacity=0] (5.36,-2.57) -- (0,0) -- (5.36,2.57) -- cycle    ;
\draw    (122.34,75.21) -- (170.34,49.71) ;
\draw [shift={(148.63,61.24)}, rotate = 152.02] [fill={rgb, 255:red, 0; green, 0; blue, 0 }  ][line width=0.08]  [draw opacity=0] (5.36,-2.57) -- (0,0) -- (5.36,2.57) -- cycle    ;
\draw    (42.34,117.71) -- (90.34,92.21) ;
\draw [shift={(68.63,103.74)}, rotate = 152.02] [fill={rgb, 255:red, 0; green, 0; blue, 0 }  ][line width=0.08]  [draw opacity=0] (5.36,-2.57) -- (0,0) -- (5.36,2.57) -- cycle    ;
\draw    (106.34,66.21) -- (106.34,13.71) ;
\draw [shift={(106.34,37.36)}, rotate = 90] [fill={rgb, 255:red, 0; green, 0; blue, 0 }  ][line width=0.08]  [draw opacity=0] (5.36,-2.57) -- (0,0) -- (5.36,2.57) -- cycle    ;
\draw    (106.34,153.71) -- (106.34,101.21) ;
\draw [shift={(106.34,124.86)}, rotate = 90] [fill={rgb, 255:red, 0; green, 0; blue, 0 }  ][line width=0.08]  [draw opacity=0] (5.36,-2.57) -- (0,0) -- (5.36,2.57) -- cycle    ;
\draw    (241,58) -- (304.25,77.28) ;
\draw [shift={(268.7,66.44)}, rotate = 16.95] [fill={rgb, 255:red, 0; green, 0; blue, 0 }  ][line width=0.08]  [draw opacity=0] (5.36,-2.57) -- (0,0) -- (5.36,2.57) -- cycle    ;
\draw    (359.42,90.13) -- (406.86,104.59) ;
\draw [shift={(379.22,96.17)}, rotate = 16.95] [fill={rgb, 255:red, 0; green, 0; blue, 0 }  ][line width=0.08]  [draw opacity=0] (5.36,-2.57) -- (0,0) -- (5.36,2.57) -- cycle    ;
\draw [shift={(359.42,90.13)}, rotate = 16.95] [color={rgb, 255:red, 0; green, 0; blue, 0 }  ][fill={rgb, 255:red, 0; green, 0; blue, 0 }  ][line width=0.75]      (0, 0) circle [x radius= 2.01, y radius= 2.01]   ;
\draw    (359.42,90.13) -- (407.42,64.63) ;
\draw [shift={(385.72,76.16)}, rotate = 152.02] [fill={rgb, 255:red, 0; green, 0; blue, 0 }  ][line width=0.08]  [draw opacity=0] (5.36,-2.57) -- (0,0) -- (5.36,2.57) -- cycle    ;
\draw    (261.34,117.71) -- (309.34,92.21) ;
\draw [shift={(287.63,103.74)}, rotate = 152.02] [fill={rgb, 255:red, 0; green, 0; blue, 0 }  ][line width=0.08]  [draw opacity=0] (5.36,-2.57) -- (0,0) -- (5.36,2.57) -- cycle    ;
\draw    (325.34,66.21) -- (325.34,13.71) ;
\draw [shift={(325.34,37.36)}, rotate = 90] [fill={rgb, 255:red, 0; green, 0; blue, 0 }  ][line width=0.08]  [draw opacity=0] (5.36,-2.57) -- (0,0) -- (5.36,2.57) -- cycle    ;
\draw    (325.34,153.71) -- (325.34,101.21) ;
\draw [shift={(325.34,124.86)}, rotate = 90] [fill={rgb, 255:red, 0; green, 0; blue, 0 }  ][line width=0.08]  [draw opacity=0] (5.36,-2.57) -- (0,0) -- (5.36,2.57) -- cycle    ;
\draw    (343.61,85.31) -- (359.42,90.13) ;
\draw [shift={(347.59,86.53)}, rotate = 16.95] [fill={rgb, 255:red, 0; green, 0; blue, 0 }  ][line width=0.08]  [draw opacity=0] (5.36,-2.57) -- (0,0) -- (5.36,2.57) -- cycle    ;
\draw [line width=1.5]    (192,85) -- (225.34,85) ;
\draw [shift={(229.34,85)}, rotate = 180] [fill={rgb, 255:red, 0; green, 0; blue, 0 }  ][line width=0.08]  [draw opacity=0] (6.97,-3.35) -- (0,0) -- (6.97,3.35) -- cycle    ;
\draw [line width=1.5]    (426.34,85) -- (459.67,85) ;
\draw [shift={(463.67,85)}, rotate = 180] [fill={rgb, 255:red, 0; green, 0; blue, 0 }  ][line width=0.08]  [draw opacity=0] (6.97,-3.35) -- (0,0) -- (6.97,3.35) -- cycle    ;
\draw    (477.73,56.39) -- (525.17,70.85) ;
\draw [shift={(525.17,70.85)}, rotate = 16.95] [color={rgb, 255:red, 0; green, 0; blue, 0 }  ][fill={rgb, 255:red, 0; green, 0; blue, 0 }  ][line width=0.75]      (0, 0) circle [x radius= 2.01, y radius= 2.01]   ;
\draw [shift={(497.53,62.43)}, rotate = 16.95] [fill={rgb, 255:red, 0; green, 0; blue, 0 }  ][line width=0.08]  [draw opacity=0] (5.36,-2.57) -- (0,0) -- (5.36,2.57) -- cycle    ;
\draw    (588.42,90.13) -- (635.86,104.59) ;
\draw [shift={(608.22,96.17)}, rotate = 16.95] [fill={rgb, 255:red, 0; green, 0; blue, 0 }  ][line width=0.08]  [draw opacity=0] (5.36,-2.57) -- (0,0) -- (5.36,2.57) -- cycle    ;
\draw [shift={(588.42,90.13)}, rotate = 16.95] [color={rgb, 255:red, 0; green, 0; blue, 0 }  ][fill={rgb, 255:red, 0; green, 0; blue, 0 }  ][line width=0.75]      (0, 0) circle [x radius= 2.01, y radius= 2.01]   ;
\draw    (588.42,90.13) -- (636.42,64.63) ;
\draw [shift={(614.72,76.16)}, rotate = 152.02] [fill={rgb, 255:red, 0; green, 0; blue, 0 }  ][line width=0.08]  [draw opacity=0] (5.36,-2.57) -- (0,0) -- (5.36,2.57) -- cycle    ;
\draw    (492.98,101.17) -- (540.98,75.67) ;
\draw [shift={(540.98,75.67)}, rotate = 332.02] [color={rgb, 255:red, 0; green, 0; blue, 0 }  ][fill={rgb, 255:red, 0; green, 0; blue, 0 }  ][line width=0.75]      (0, 0) circle [x radius= 2.01, y radius= 2.01]   ;
\draw [shift={(519.28,87.2)}, rotate = 152.02] [fill={rgb, 255:red, 0; green, 0; blue, 0 }  ][line width=0.08]  [draw opacity=0] (5.36,-2.57) -- (0,0) -- (5.36,2.57) -- cycle    ;
\draw    (572.61,85.31) -- (572.61,32.81) ;
\draw [shift={(572.61,56.46)}, rotate = 90] [fill={rgb, 255:red, 0; green, 0; blue, 0 }  ][line width=0.08]  [draw opacity=0] (5.36,-2.57) -- (0,0) -- (5.36,2.57) -- cycle    ;
\draw [shift={(572.61,85.31)}, rotate = 270] [color={rgb, 255:red, 0; green, 0; blue, 0 }  ][fill={rgb, 255:red, 0; green, 0; blue, 0 }  ][line width=0.75]      (0, 0) circle [x radius= 2.01, y radius= 2.01]   ;
\draw    (556.8,132.99) -- (556.8,80.49) ;
\draw [shift={(556.8,80.49)}, rotate = 270] [color={rgb, 255:red, 0; green, 0; blue, 0 }  ][fill={rgb, 255:red, 0; green, 0; blue, 0 }  ][line width=0.75]      (0, 0) circle [x radius= 2.01, y radius= 2.01]   ;
\draw [shift={(556.8,104.14)}, rotate = 90] [fill={rgb, 255:red, 0; green, 0; blue, 0 }  ][line width=0.08]  [draw opacity=0] (5.36,-2.57) -- (0,0) -- (5.36,2.57) -- cycle    ;
\draw    (572.61,85.31) -- (588.42,90.13) ;
\draw [shift={(576.59,86.53)}, rotate = 16.95] [fill={rgb, 255:red, 0; green, 0; blue, 0 }  ][line width=0.08]  [draw opacity=0] (5.36,-2.57) -- (0,0) -- (5.36,2.57) -- cycle    ;
\draw    (556.8,80.49) -- (572.61,85.31) ;
\draw [shift={(560.78,81.71)}, rotate = 16.95] [fill={rgb, 255:red, 0; green, 0; blue, 0 }  ][line width=0.08]  [draw opacity=0] (5.36,-2.57) -- (0,0) -- (5.36,2.57) -- cycle    ;
\draw    (540.98,75.67) -- (556.8,80.49) ;
\draw [shift={(544.97,76.89)}, rotate = 16.95] [fill={rgb, 255:red, 0; green, 0; blue, 0 }  ][line width=0.08]  [draw opacity=0] (5.36,-2.57) -- (0,0) -- (5.36,2.57) -- cycle    ;
\draw    (525.17,70.85) -- (540.98,75.67) ;
\draw [shift={(529.15,72.07)}, rotate = 16.95] [fill={rgb, 255:red, 0; green, 0; blue, 0 }  ][line width=0.08]  [draw opacity=0] (5.36,-2.57) -- (0,0) -- (5.36,2.57) -- cycle    ;
\draw    (525.17,70.85) -- (525.17,48.52) ;
\draw [shift={(525.17,57.09)}, rotate = 90] [fill={rgb, 255:red, 0; green, 0; blue, 0 }  ][line width=0.08]  [draw opacity=0] (5.36,-2.57) -- (0,0) -- (5.36,2.57) -- cycle    ;

\draw (145,100) node [anchor=north west][inner sep=0.75pt]    {\small $\mu $};
\draw (42,69) node [anchor=north west][inner sep=0.75pt]    {\small $\nu $};
\draw (109,116) node [anchor=north west][inner sep=0.75pt]    {\small $\rho $};
\draw (109,29) node [anchor=north west][inner sep=0.75pt]    {\small $\sigma $};
\draw (147,62) node [anchor=north west][inner sep=0.75pt]    {\small $\lambda $};
\draw (70,103) node [anchor=north west][inner sep=0.75pt]    {\small $\eta $};
\draw (0,53) node [anchor=north west][inner sep=0.75pt]    {\small $m_{\nu }$};
\draw (86,69) node [anchor=north west][inner sep=0.75pt]    {\small $n_{\nu }$};
\draw (85,54) node [anchor=north west][inner sep=0.75pt]    {\small $n_{\sigma }$};
\draw (90,82) node [anchor=north west][inner sep=0.75pt]    {\small $m_{\eta }$};
\draw (99,154) node [anchor=north west][inner sep=0.75pt]    {\small $n_{\rho }$};
\draw (27,112) node [anchor=north west][inner sep=0.75pt]    {\small $n_{\eta }$};
\draw (106,96) node [anchor=north west][inner sep=0.75pt]    {\small $m_{\rho }$};
\draw (110,75) node [anchor=north west][inner sep=0.75pt]    {\small $m_{\mu }$};
\draw (110,57.8) node [anchor=north west][inner sep=0.75pt]    {\small $n_{\lambda }$};
\draw (97,2) node [anchor=north west][inner sep=0.75pt]    {\small $m_{\sigma }$};
\draw (169,39) node [anchor=north west][inner sep=0.75pt]    {\small $m_{\lambda }$};
\draw (189,104) node [anchor=north west][inner sep=0.75pt]    {\small $n_{\mu }$};

\draw (378,100) node [anchor=north west][inner sep=0.75pt]    {\small $\mu $};
\draw (261,68) node [anchor=north west][inner sep=0.75pt]    {\small $\nu $};
\draw (328,116) node [anchor=north west][inner sep=0.75pt]    {\small $\rho $};
\draw (328,29) node [anchor=north west][inner sep=0.75pt]    {\small $\sigma $};
\draw (377.34,60.26) node [anchor=north west][inner sep=0.75pt]    {\small $\lambda $};
\draw (288,103) node [anchor=north west][inner sep=0.75pt]    {\small $\eta $};
\draw (219,53) node [anchor=north west][inner sep=0.75pt]    {\small $m_{\nu }$};
\draw (305,71.8) node [anchor=north west][inner sep=0.75pt]    {\small $n_{\nu }$};
\draw (305,55) node [anchor=north west][inner sep=0.75pt]    {\small $n_{\sigma }$};
\draw (309.25,84.08) node [anchor=north west][inner sep=0.75pt]    {\small $m_{\eta }$};
\draw (318,154) node [anchor=north west][inner sep=0.75pt]    {\small $n_{\rho }$};
\draw (248,113) node [anchor=north west][inner sep=0.75pt]    {\small $n_{\eta }$};
\draw (325,98) node [anchor=north west][inner sep=0.75pt]    {\small $m_{\rho }$};
\draw (323,72) node [anchor=north west][inner sep=0.75pt]    {\small $m_{\alpha }$};
\draw (316,2) node [anchor=north west][inner sep=0.75pt]    {\small $m_{\sigma }$};
\draw (407,56) node [anchor=north west][inner sep=0.75pt]    {\small $m_{\lambda }$};
\draw (407,100) node [anchor=north west][inner sep=0.75pt]    {\small $n_{\mu }$};
\draw (348,72) node [anchor=north west][inner sep=0.75pt]    {\small $\alpha $};

\draw (607,100) node [anchor=north west][inner sep=0.75pt]    {\small $\mu $};
\draw (490,65) node [anchor=north west][inner sep=0.75pt]    {\small $\nu $};
\draw (560,100) node [anchor=north west][inner sep=0.75pt]    {\small $\rho $};
\draw (575,47) node [anchor=north west][inner sep=0.75pt]    {\small $\sigma $};
\draw (606.34,60) node [anchor=north west][inner sep=0.75pt]    {\small $\lambda $};
\draw (515.34,88) node [anchor=north west][inner sep=0.75pt]    {\small $\eta $};
\draw (454,49) node [anchor=north west][inner sep=0.75pt]    {\small $m_{\nu }$};
\draw (550,133) node [anchor=north west][inner sep=0.75pt]    {\small $n_{\rho }$};
\draw (481,97) node [anchor=north west][inner sep=0.75pt]    {\small $n_{\eta }$};
\draw (563,20) node [anchor=north west][inner sep=0.75pt]    {\small $m_{\sigma }$};
\draw (636,56) node [anchor=north west][inner sep=0.75pt]    {\small $m_{\lambda }$};
\draw (636,100) node [anchor=north west][inner sep=0.75pt]    {\small $n_{\mu }$};
\draw (573,89) node [anchor=north west][inner sep=0.75pt]    {\small $\alpha $};
\draw (560,65) node [anchor=north west][inner sep=0.75pt]    {\small $\beta $};
\draw (542,80) node [anchor=north west][inner sep=0.75pt]    {\small $\gamma $};
\draw (530,57.8) node [anchor=north west][inner sep=0.75pt]    {\small $\delta $};
\draw (514,50) node [anchor=north west][inner sep=0.75pt]    {\small $s$};
\draw (515,36) node [anchor=north west][inner sep=0.75pt]    {\small $m_{s}$};
\end{tikzpicture}
\caption{Rewrite the rep-basis as defined on a trivalent lattice. In the first step, we fuse $\mu$ and $\lambda$ by contracting there indices $m_\mu$ and $n_\lambda$, which results in a linear combination of irreducible representations $\{\alpha\}$ with a free end and labeled by $m_\alpha$. Repeating this procedure and in the end, we fuse $\delta$ and $\nu$ and obtain a tail attached to the original vertex with an free end, labeled by $(s,m_s)$. }
\label{fig:4}
\end{figure}

Then we can rewrite the rep-basis state and identify it as a trivalent lattice. Recall that the definition of CG coefficients \eqref{eq:3.14}, which can be graphically presented as a basis rewriting:
\begin{equation}\label{eq:3.23}
    \Bigg|\begin{tikzpicture}[x=0.75pt,y=0.75pt,yscale=-1,xscale=1, baseline=(XXXX.south) ]
\path (0,84);\path (55.99651336669922,0);\draw    ($(current bounding box.center)+(0,0.3em)$) node [anchor=south] (XXXX) {};
\draw    (15,69) -- (15,10.64) ;
\draw [shift={(15,37.22)}, rotate = 90] [fill={rgb, 255:red, 0; green, 0; blue, 0 }  ][line width=0.08]  [draw opacity=0] (5.36,-2.57) -- (0,0) -- (5.36,2.57) -- cycle    ;
\draw    (43,69) -- (43,10.64) ;
\draw [shift={(43,37.22)}, rotate = 90] [fill={rgb, 255:red, 0; green, 0; blue, 0 }  ][line width=0.08]  [draw opacity=0] (5.36,-2.57) -- (0,0) -- (5.36,2.57) -- cycle    ;
\draw (1,36.4) node [anchor=north west][inner sep=0.75pt]    {\small $\mu $};
\draw (45,36.4) node [anchor=north west][inner sep=0.75pt]    {\small $\nu $};
\draw (7,69) node [anchor=north west][inner sep=0.75pt]    {\small $n_{\mu }$};
\draw (36,69) node [anchor=north west][inner sep=0.75pt]    {\small $n_{\nu }$};
\draw (5,0) node [anchor=north west][inner sep=0.75pt]    {\small $m_{\mu }$};
\draw (33,0) node [anchor=north west][inner sep=0.75pt]    {\small $m_{\nu }$};
\end{tikzpicture}
\Bigg\rangle \propto\sum _{\lambda,m_\lambda}\left(\begin{tikzpicture}[x=0.75pt,y=0.75pt,yscale=-1,xscale=1, baseline=(XXXX.south) ]
\path (0,88);\path (81.99651336669922,0);\draw    ($(current bounding box.center)+(0,0.3em)$) node [anchor=south] (XXXX) {};
\draw    (12.72,15) -- (40.72,43) ;
\draw [shift={(40.72,43)}, rotate = 45] [color={rgb, 255:red, 0; green, 0; blue, 0 }  ][fill={rgb, 255:red, 0; green, 0; blue, 0 }  ][line width=0.75]      (0, 0) circle [x radius= 2.01, y radius= 2.01]   ;
\draw [shift={(26.72,29)}, rotate = 45] [fill={rgb, 255:red, 0; green, 0; blue, 0 }  ][line width=0.08]  [draw opacity=0] (5.36,-2.57) -- (0,0) -- (5.36,2.57) -- cycle    ;
\draw    (68.72,15) -- (40.72,43) ;
\draw [shift={(54.72,29)}, rotate = 135] [fill={rgb, 255:red, 0; green, 0; blue, 0 }  ][line width=0.08]  [draw opacity=0] (5.36,-2.57) -- (0,0) -- (5.36,2.57) -- cycle    ;
\draw    (40.72,43) -- (40.72,75) ;
\draw [shift={(40.72,59)}, rotate = 90] [fill={rgb, 255:red, 0; green, 0; blue, 0 }  ][line width=0.08]  [draw opacity=0] (5.36,-2.57) -- (0,0) -- (5.36,2.57) -- cycle    ;
\draw (42,58) node [anchor=north west][inner sep=0.75pt]    {\small $\lambda $};
\draw (15,32) node [anchor=north west][inner sep=0.75pt]    {\small $\mu $};
\draw (53,32) node [anchor=north west][inner sep=0.75pt]    {\small $\nu $};
\draw (4,2) node [anchor=north west][inner sep=0.75pt]    {\small $m_{\mu }$};
\draw (60,2) node [anchor=north west][inner sep=0.75pt]    {\small $m_{\nu }$};
\draw (31,75.42) node [anchor=north west][inner sep=0.75pt]    {\small $m_{\lambda }$};
\end{tikzpicture}
\right) \Bigg|\begin{tikzpicture}[x=0.75pt,y=0.75pt,yscale=-1,xscale=1, baseline=(XXXX.south) ]
\path (0,88);\path (81.99651336669922,0);\draw    ($(current bounding box.center)+(0,0.3em)$) node [anchor=south] (XXXX) {};
\draw    (40.72,43) -- (68.72,71) ;
\draw [shift={(54.72,57)}, rotate = 45] [fill={rgb, 255:red, 0; green, 0; blue, 0 }  ][line width=0.08]  [draw opacity=0] (5.36,-2.57) -- (0,0) -- (5.36,2.57) -- cycle    ;
\draw [shift={(40.72,43)}, rotate = 45] [color={rgb, 255:red, 0; green, 0; blue, 0 }  ][fill={rgb, 255:red, 0; green, 0; blue, 0 }  ][line width=0.75]      (0, 0) circle [x radius= 2.01, y radius= 2.01]   ;
\draw    (40.72,43) -- (12.72,71) ;
\draw [shift={(26.72,57)}, rotate = 135] [fill={rgb, 255:red, 0; green, 0; blue, 0 }  ][line width=0.08]  [draw opacity=0] (5.36,-2.57) -- (0,0) -- (5.36,2.57) -- cycle    ;
\draw    (40.72,11) -- (40.72,43) ;
\draw [shift={(40.72,27)}, rotate = 90] [fill={rgb, 255:red, 0; green, 0; blue, 0 }  ][line width=0.08]  [draw opacity=0] (5.36,-2.57) -- (0,0) -- (5.36,2.57) -- cycle    ;
\draw (42,22.13) node [anchor=north west][inner sep=0.75pt]    {\small $\lambda $};
\draw (24,60) node [anchor=north west][inner sep=0.75pt]    {\small $\mu $};
\draw (45.72,60) node [anchor=north west][inner sep=0.75pt]    {\small $\nu $};
\draw (2,72) node [anchor=north west][inner sep=0.75pt]    {\small $n_{\mu }$};
\draw (61,72) node [anchor=north west][inner sep=0.75pt]    {\small $n_{\nu }$};
\draw (31,0) node [anchor=north west][inner sep=0.75pt]    {\small $m_{\lambda }$};
\end{tikzpicture}
\Bigg\rangle .
\end{equation}
The orthonormality of the rewritten basis still comes from the normalization condition of the $3j$-symbol. By the equation above, we can fuse the six representations and six Latin indices in a specific order at the vertex, as depicted in \autoref{fig:4}. Then, following the procedure shown in \autoref{fig:4}, we can rewrite the basis of the total Hilbert space as defined on an actual trivalent lattice $\tilde{\Gamma}$, with a tail attached to each of the original vertices to maintain the correct number of local degrees of freedom. If we restrict the total Hilbert space to a subspace wherein the degrees of freedom on the tails are all projected to the trivial representation, the newly obtained lattice $\tilde{\Gamma}$ can serve as a suitable lattice for defining a WW model. We will discuss the relationship between our Fourier-transformed GT model and the WW model in \autoref{sec:5} and revisit this point therein. 

To better understand the Fourier transform and basis rewriting illustrated in \autoref{fig:3} and \autoref{fig:4}, let us consider the local Hilbert space of a boundary 5-valent vertex in detail. By doing so, we aim to obtain a precise linear transformation of the basis with commensurate coefficients. In our graphical representation and by \eqref{eq:3.22}, the Fourier transform of the group-basis to the rep-basis of the local Hilbert space reads 
\begin{equation}\label{eq:3.24}
    \begin{aligned}
        &\Bigg|\begin{tikzpicture}[x=0.75pt,y=0.75pt,yscale=-1,xscale=1, baseline=(XXXX.south) ]
    \path (0,112);\path (166.95773315429688,0);\draw    ($(current bounding box.center)+(0,0.3em)$) node [anchor=south] (XXXX) {};
    \draw    (11.99,9.99) -- (57.41,25.44) ;
    \draw [shift={(34.7,17.72)}, rotate = 18.79] [fill={rgb, 255:red, 0; green, 0; blue, 0 }  ][line width=0.08]  [draw opacity=0] (5.36,-2.57) -- (0,0) -- (5.36,2.57) -- cycle    ;
    \draw    (98.41,47.44) -- (143.83,62.89) ;
    \draw [shift={(121.12,55.16)}, rotate = 18.79] [fill={rgb, 255:red, 0; green, 0; blue, 0 }  ][line width=0.08]  [draw opacity=0] (5.36,-2.57) -- (0,0) -- (5.36,2.57) -- cycle    ;
    \draw    (99.41,25.44) -- (138.83,3.89) ;
    \draw [shift={(119.12,14.66)}, rotate = 151.33] [fill={rgb, 255:red, 0; green, 0; blue, 0 }  ][line width=0.08]  [draw opacity=0] (5.36,-2.57) -- (0,0) -- (5.36,2.57) -- cycle    ;
    \draw    (12.99,69.99) -- (52.41,48.44) ;
    \draw [shift={(32.7,59.22)}, rotate = 151.33] [fill={rgb, 255:red, 0; green, 0; blue, 0 }  ][line width=0.08]  [draw opacity=0] (5.36,-2.57) -- (0,0) -- (5.36,2.57) -- cycle    ;
    \draw    (76.41,98.44) -- (76.41,52.44) ;
    \draw [shift={(76.41,75.44)}, rotate = 90] [fill={rgb, 255:red, 0; green, 0; blue, 0 }  ][line width=0.08]  [draw opacity=0] (5.36,-2.57) -- (0,0) -- (5.36,2.57) -- cycle    ;
    \draw (117.97,59.17) node [anchor=north west][inner sep=0.75pt]    {\small $\mu $};
    \draw (113,20) node [anchor=north west][inner sep=0.75pt]    {\small$\lambda $};
    \draw (26,63) node [anchor=north west][inner sep=0.75pt]    {\small$\eta $};
    \draw (33.7,23) node [anchor=north west][inner sep=0.75pt]    {\small$\nu $};
    \draw (76.41,78.64) node [anchor=north west][inner sep=0.75pt]    {\small$\rho $};
    \draw (67.97,99.16) node [anchor=north west][inner sep=0.75pt]    {\small$n_{\rho }$};
    \draw (55.96,22.15) node [anchor=north west][inner sep=0.75pt]    {\small$n_{\nu }$};
    \draw (-3,6) node [anchor=north west][inner sep=0.75pt]    {\small$m_{\nu }$};
    \draw (-1.04,68.15) node [anchor=north west][inner sep=0.75pt]    {\small$n_{\eta }$};
    \draw (48,40.15) node [anchor=north west][inner sep=0.75pt]    {\small$m_{\eta }$};
    \draw (66.96,43) node [anchor=north west][inner sep=0.75pt]    {\small$m_{\rho }$};
    \draw (143,60) node [anchor=north west][inner sep=0.75pt]    {\small$n_{\mu }$};
    \draw (85.95,23.15) node [anchor=north west][inner sep=0.75pt]    {\small$n_{\lambda }$};
    \draw (82,38) node [anchor=north west][inner sep=0.75pt]    {\small$m_{\mu }$};
    \draw (138,0) node [anchor=north west][inner sep=0.75pt]    {\small$m_{\lambda }$};
    \end{tikzpicture}
    \Bigg\rangle \\
    ={}&\frac{v_{\mu } v_{\nu } v_{\rho } v_{\eta } v_{\lambda }}{\sqrt{|G|^{5}}}\sum _{jikgh\in G}\left(\begin{tikzpicture}[x=0.75pt,y=0.75pt,yscale=-1,xscale=1, baseline=(XXXX.south) ]
    \path (0,112);\path (166.95773315429688,0);\draw    ($(current bounding box.center)+(0,0.3em)$) node [anchor=south] (XXXX) {};
    \draw    (11.99,9.99) -- (57.41,25.44) ;
    \draw [shift={(34.7,17.72)}, rotate = 18.79] [fill={rgb, 255:red, 0; green, 0; blue, 0 }  ][line width=0.08]  [draw opacity=0] (5.36,-2.57) -- (0,0) -- (5.36,2.57) -- cycle    ;
    \draw    (98.41,47.44) -- (143.83,62.89) ;
    \draw [shift={(121.12,55.16)}, rotate = 18.79] [fill={rgb, 255:red, 0; green, 0; blue, 0 }  ][line width=0.08]  [draw opacity=0] (5.36,-2.57) -- (0,0) -- (5.36,2.57) -- cycle    ;
    \draw    (99.41,25.44) -- (138.83,3.89) ;
    \draw [shift={(119.12,14.66)}, rotate = 151.33] [fill={rgb, 255:red, 0; green, 0; blue, 0 }  ][line width=0.08]  [draw opacity=0] (5.36,-2.57) -- (0,0) -- (5.36,2.57) -- cycle    ;
    \draw    (12.99,69.99) -- (52.41,48.44) ;
    \draw [shift={(32.7,59.22)}, rotate = 151.33] [fill={rgb, 255:red, 0; green, 0; blue, 0 }  ][line width=0.08]  [draw opacity=0] (5.36,-2.57) -- (0,0) -- (5.36,2.57) -- cycle    ;
    \draw    (76.41,98.44) -- (76.41,52.44) ;
    \draw [shift={(76.41,75.44)}, rotate = 90] [fill={rgb, 255:red, 0; green, 0; blue, 0 }  ][line width=0.08]  [draw opacity=0] (5.36,-2.57) -- (0,0) -- (5.36,2.57) -- cycle    ;
    \draw  [fill={rgb, 255:red, 255; green, 255; blue, 255 }  ,fill opacity=1 ] (102.93,52.16) .. controls (102.93,47.42) and (106.78,43.57) .. (111.53,43.57) .. controls (116.27,43.57) and (120.12,47.42) .. (120.12,52.16) .. controls (120.12,56.91) and (116.27,60.76) .. (111.53,60.76) .. controls (106.78,60.76) and (102.93,56.91) .. (102.93,52.16) -- cycle ;
    \draw  [fill={rgb, 255:red, 255; green, 255; blue, 255 }  ,fill opacity=1 ] (67.82,64.03) .. controls (67.82,59.29) and (71.66,55.44) .. (76.41,55.44) .. controls (81.15,55.44) and (85,59.29) .. (85,64.03) .. controls (85,68.78) and (81.15,72.62) .. (76.41,72.62) .. controls (71.66,72.62) and (67.82,68.78) .. (67.82,64.03) -- cycle ;
    \draw  [fill={rgb, 255:red, 255; green, 255; blue, 255 }  ,fill opacity=1 ] (119.93,10.16) .. controls (119.93,5.42) and (123.78,1.57) .. (128.53,1.57) .. controls (133.27,1.57) and (137.12,5.42) .. (137.12,10.16) .. controls (137.12,14.91) and (133.27,18.76) .. (128.53,18.76) .. controls (123.78,18.76) and (119.93,14.91) .. (119.93,10.16) -- cycle ;
    \draw  [fill={rgb, 255:red, 255; green, 255; blue, 255 }  ,fill opacity=1 ] (33.82,54.03) .. controls (33.82,49.29) and (37.66,45.44) .. (42.41,45.44) .. controls (47.15,45.44) and (51,49.29) .. (51,54.03) .. controls (51,58.78) and (47.15,62.62) .. (42.41,62.62) .. controls (37.66,62.62) and (33.82,58.78) .. (33.82,54.03) -- cycle ;
    \draw  [fill={rgb, 255:red, 255; green, 255; blue, 255 }  ,fill opacity=1 ] (16.52,14.72) .. controls (16.52,9.97) and (20.36,6.12) .. (25.11,6.12) .. controls (29.85,6.12) and (33.7,9.97) .. (33.7,14.72) .. controls (33.7,19.46) and (29.85,23.31) .. (25.11,23.31) .. controls (20.36,23.31) and (16.52,19.46) .. (16.52,14.72) -- cycle ;
    \draw (37,49) node [anchor=north west][inner sep=0.75pt]    {\small$g$};
    \draw (123,4) node [anchor=north west][inner sep=0.75pt]    {\small$h$};
    \draw (21,9) node [anchor=north west][inner sep=0.75pt]    {\small$i$};
    \draw (107,45) node [anchor=north west][inner sep=0.75pt]    {\small $j$};
    \draw (71,58) node [anchor=north west][inner sep=0.75pt]    {\small $k$};
    \draw (117.97,59.17) node [anchor=north west][inner sep=0.75pt]    {\small $\mu $};
    \draw (112.97,21.2) node [anchor=north west][inner sep=0.75pt]    {\small $\lambda $};
    \draw (22,66) node [anchor=north west][inner sep=0.75pt]    {\small $\eta $};
    \draw (33.7,23) node [anchor=north west][inner sep=0.75pt]    {\small $\nu $};
    \draw (76.41,78.64) node [anchor=north west][inner sep=0.75pt]    {\small $\rho $};
    \draw (69,99) node [anchor=north west][inner sep=0.75pt]    {\small $n_{\rho }$};
    \draw (55.96,22.15) node [anchor=north west][inner sep=0.75pt]    {\small $n_{\nu }$};
    \draw (-3.04,6) node [anchor=north west][inner sep=0.75pt]    {\small $m_{\nu }$};
    \draw (-1.04,68.15) node [anchor=north west][inner sep=0.75pt]    {\small $n_{\eta }$};
    \draw (49,40) node [anchor=north west][inner sep=0.75pt]    {\small $m_{\eta }$};
    \draw (67,43) node [anchor=north west][inner sep=0.75pt]    {\small $m_{\rho }$};
    \draw (142.96,62.15) node [anchor=north west][inner sep=0.75pt]    {\small $n_{\mu }$};
    \draw (85.95,23.15) node [anchor=north west][inner sep=0.75pt]    {\small $n_{\lambda }$};
    \draw (82,38) node [anchor=north west][inner sep=0.75pt]    {\small $m_{\mu }$};
    \draw (138,0) node [anchor=north west][inner sep=0.75pt]    {\small $m_{\lambda }$};
    \end{tikzpicture}
    \right) \Bigg|\begin{tikzpicture}[x=0.75pt,y=0.75pt,yscale=-1,xscale=1, baseline=(XXXX.south) ]
    \path (0,70);\path (95.99015045166016,0);\draw    ($(current bounding box.center)+(0,0.3em)$) node [anchor=south] (XXXX) {};
    \draw    (0.99,7.99) -- (46.41,23.44) ;
    \draw [shift={(23.7,15.72)}, rotate = 18.79] [fill={rgb, 255:red, 0; green, 0; blue, 0 }  ][line width=0.08]  [draw opacity=0] (5.36,-2.57) -- (0,0) -- (5.36,2.57) -- cycle    ;
    \draw    (46.41,23.44) -- (91.83,38.89) ;
    \draw [shift={(69.12,31.16)}, rotate = 18.79] [fill={rgb, 255:red, 0; green, 0; blue, 0 }  ][line width=0.08]  [draw opacity=0] (5.36,-2.57) -- (0,0) -- (5.36,2.57) -- cycle    ;
    \draw    (46.41,23.44) -- (85.83,1.89) ;
    \draw [shift={(66.12,12.66)}, rotate = 151.33] [fill={rgb, 255:red, 0; green, 0; blue, 0 }  ][line width=0.08]  [draw opacity=0] (5.36,-2.57) -- (0,0) -- (5.36,2.57) -- cycle    ;
    \draw    (6.99,44.99) -- (46.41,23.44) ;
    \draw [shift={(46.41,23.44)}, rotate = 331.33] [color={rgb, 255:red, 0; green, 0; blue, 0 }  ][fill={rgb, 255:red, 0; green, 0; blue, 0 }  ][line width=0.75]      (0, 0) circle [x radius= 2.01, y radius= 2.01]   ;
    \draw [shift={(26.7,34.22)}, rotate = 151.33] [fill={rgb, 255:red, 0; green, 0; blue, 0 }  ][line width=0.08]  [draw opacity=0] (5.36,-2.57) -- (0,0) -- (5.36,2.57) -- cycle    ;
    \draw    (46.41,66.44) -- (46.41,23.44) ;
    \draw [shift={(46.41,44.94)}, rotate = 90] [fill={rgb, 255:red, 0; green, 0; blue, 0 }  ][line width=0.08]  [draw opacity=0] (5.36,-2.57) -- (0,0) -- (5.36,2.57) -- cycle    ;
    \draw (20,38) node [anchor=north west][inner sep=0.75pt]    {\small $g$};
    \draw (59,0) node [anchor=north west][inner sep=0.75pt]    {\small $h$};
    \draw (16,16) node [anchor=north west][inner sep=0.75pt]    {\small $i$};
    \draw (64,17) node [anchor=north west][inner sep=0.75pt]    {\small $j$};
    \draw (48,44) node [anchor=north west][inner sep=0.75pt]    {\small $k$};
    \end{tikzpicture}
    \Bigg\rangle .
    \end{aligned}
\end{equation}
Here, the edges $i,j,g$, and $h$ are lying on the boundary, while the edge $k$ lies in the bulk. We denote the group-basis and rep-basis states in the above equation as $|jikgh\rangle$ and $|\mu\nu\rho\eta\lambda\rangle$. Then, we can rewrite the rep-basis by first fusing $\mu$ and $\rho$, resulting in a set of representations $\{\alpha\}$, and then fuse $\alpha$ and $\lambda$ to get $\beta$, and then fuse $\beta$ and $\eta$ to get $\gamma$, and finally we fuse $\gamma$ and $\nu$, resulting in a set of representations $\{s\}$. This procedure yields four $3j$-symbols with a pair of indices contracted, yielding the expansion coefficients: 
\begin{equation}\label{eq:3.25}
    |\mu\nu\rho\eta\lambda\rangle=\sum_{\alpha\beta\gamma\in L_G}\sum_{s,m_s}v_\alpha v_\beta v_\gamma v_s\left(\begin{tikzpicture}[x=0.75pt,y=0.75pt,yscale=-0.8,xscale=0.8, baseline=(XXXX.south) ]
        \path (0,177);\path (91.98590850830078,0);\draw    ($(current bounding box.center)+(0,0.3em)$) node [anchor=south] (XXXX) {};
        \draw    (47,131.99) -- (47,162.46) ;
        \draw [shift={(47,143.13)}, rotate = 90] [fill={rgb, 255:red, 0; green, 0; blue, 0 }  ][line width=0.08]  [draw opacity=0] (5.36,-2.57) -- (0,0) -- (5.36,2.57) -- cycle    ;
        \draw    (47,101.53) -- (47,131.99) ;
        \draw [shift={(47,131.99)}, rotate = 90] [color={rgb, 255:red, 0; green, 0; blue, 0 }  ][fill={rgb, 255:red, 0; green, 0; blue, 0 }  ][line width=0.75]      (0, 0) circle [x radius= 2.01, y radius= 2.01]   ;
        \draw [shift={(47,112.66)}, rotate = 90] [fill={rgb, 255:red, 0; green, 0; blue, 0 }  ][line width=0.08]  [draw opacity=0] (5.36,-2.57) -- (0,0) -- (5.36,2.57) -- cycle    ;
        \draw [shift={(47,101.53)}, rotate = 90] [color={rgb, 255:red, 0; green, 0; blue, 0 }  ][fill={rgb, 255:red, 0; green, 0; blue, 0 }  ][line width=0.75]      (0, 0) circle [x radius= 2.01, y radius= 2.01]   ;
        \draw    (47,71.06) -- (47,101.53) ;
        \draw [shift={(47,82.19)}, rotate = 90] [fill={rgb, 255:red, 0; green, 0; blue, 0 }  ][line width=0.08]  [draw opacity=0] (5.36,-2.57) -- (0,0) -- (5.36,2.57) -- cycle    ;
        \draw    (47,40.59) -- (47,71.06) ;
        \draw [shift={(47,71.06)}, rotate = 90] [color={rgb, 255:red, 0; green, 0; blue, 0 }  ][fill={rgb, 255:red, 0; green, 0; blue, 0 }  ][line width=0.75]      (0, 0) circle [x radius= 2.01, y radius= 2.01]   ;
        \draw [shift={(47,51.73)}, rotate = 90] [fill={rgb, 255:red, 0; green, 0; blue, 0 }  ][line width=0.08]  [draw opacity=0] (5.36,-2.57) -- (0,0) -- (5.36,2.57) -- cycle    ;
        \draw [shift={(47,40.59)}, rotate = 90] [color={rgb, 255:red, 0; green, 0; blue, 0 }  ][fill={rgb, 255:red, 0; green, 0; blue, 0 }  ][line width=0.75]      (0, 0) circle [x radius= 2.01, y radius= 2.01]   ;
        \draw    (47,10.13) -- (47,40.59) ;
        \draw [shift={(47,21.26)}, rotate = 90] [fill={rgb, 255:red, 0; green, 0; blue, 0 }  ][line width=0.08]  [draw opacity=0] (5.36,-2.57) -- (0,0) -- (5.36,2.57) -- cycle    ;
        \draw    (47,131.99) -- (77.99,131.99) ;
        \draw [shift={(58.39,131.99)}, rotate = 0] [fill={rgb, 255:red, 0; green, 0; blue, 0 }  ][line width=0.08]  [draw opacity=0] (5.36,-2.57) -- (0,0) -- (5.36,2.57) -- cycle    ;
        \draw    (16.01,101.53) -- (47,101.53) ;
        \draw [shift={(27.4,101.53)}, rotate = 0] [fill={rgb, 255:red, 0; green, 0; blue, 0 }  ][line width=0.08]  [draw opacity=0] (5.36,-2.57) -- (0,0) -- (5.36,2.57) -- cycle    ;
        \draw    (16.01,40.59) -- (47,40.59) ;
        \draw [shift={(27.4,40.59)}, rotate = 0] [fill={rgb, 255:red, 0; green, 0; blue, 0 }  ][line width=0.08]  [draw opacity=0] (5.36,-2.57) -- (0,0) -- (5.36,2.57) -- cycle    ;
        \draw    (47,71.06) -- (77.99,71.06) ;
        \draw [shift={(58.39,71.06)}, rotate = 0] [fill={rgb, 255:red, 0; green, 0; blue, 0 }  ][line width=0.08]  [draw opacity=0] (5.36,-2.57) -- (0,0) -- (5.36,2.57) -- cycle    ;
    \draw (59,133) node [anchor=north west][inner sep=0.75pt]    {\small $s$};
    \draw (77,127) node [anchor=north west][inner sep=0.75pt]    {\small $m_{s}$};
    \draw (48,147) node [anchor=north west][inner sep=0.75pt]    {\small $\nu $};
    \draw (40,164) node [anchor=north west][inner sep=0.75pt]    {\small $n_{\nu }$};
    \draw (48,114) node [anchor=north west][inner sep=0.75pt]    {\small $\gamma $};
    \draw (28,103) node [anchor=north west][inner sep=0.75pt]    {\small $\eta $};
    \draw (-1,97) node [anchor=north west][inner sep=0.75pt]    {\small $m_{\eta }$};
    \draw (48,84) node [anchor=north west][inner sep=0.75pt]    {\small $\beta $};
    \draw (48,54) node [anchor=north west][inner sep=0.75pt]    {\small $\alpha $};
    \draw (27,42) node [anchor=north west][inner sep=0.75pt]    {\small $\rho $};
    \draw (-1,36) node [anchor=north west][inner sep=0.75pt]    {\small $m_{\rho }$};
    \draw (60,73) node [anchor=north west][inner sep=0.75pt]    {\small $\lambda $};
    \draw (77,67) node [anchor=north west][inner sep=0.75pt]    {\small $n_{\lambda }$};
    \draw (48,22) node [anchor=north west][inner sep=0.75pt]    {\small $\mu $};
    \draw (38,-1) node [anchor=north west][inner sep=0.75pt]    {\small $m_{\mu }$};
    \end{tikzpicture}
    \right) \Bigg|\begin{tikzpicture}[x=0.75pt,y=0.75pt,yscale=-0.8,xscale=0.8, baseline=(XXXX.south) ]
    \path (0,108);\path (181.98782348632812,0);\draw    ($(current bounding box.center)+(0,0.3em)$) node [anchor=south] (XXXX) {};
    \draw    (21.73,17.39) -- (69.17,31.85) ;
    \draw [shift={(69.17,31.85)}, rotate = 16.95] [color={rgb, 255:red, 0; green, 0; blue, 0 }  ][fill={rgb, 255:red, 0; green, 0; blue, 0 }  ][line width=0.75]      (0, 0) circle [x radius= 2.01, y radius= 2.01]   ;
    \draw [shift={(41.53,23.43)}, rotate = 16.95] [fill={rgb, 255:red, 0; green, 0; blue, 0 }  ][line width=0.08]  [draw opacity=0] (5.36,-2.57) -- (0,0) -- (5.36,2.57) -- cycle    ;
    \draw    (116.61,46.31) -- (164.05,60.77) ;
    \draw [shift={(136.41,52.35)}, rotate = 16.95] [fill={rgb, 255:red, 0; green, 0; blue, 0 }  ][line width=0.08]  [draw opacity=0] (5.36,-2.57) -- (0,0) -- (5.36,2.57) -- cycle    ;
    \draw    (100.8,41.49) -- (148.8,15.99) ;
    \draw [shift={(127.09,27.52)}, rotate = 152.02] [fill={rgb, 255:red, 0; green, 0; blue, 0 }  ][line width=0.08]  [draw opacity=0] (5.36,-2.57) -- (0,0) -- (5.36,2.57) -- cycle    ;
    \draw [shift={(100.8,41.49)}, rotate = 332.02] [color={rgb, 255:red, 0; green, 0; blue, 0 }  ][fill={rgb, 255:red, 0; green, 0; blue, 0 }  ][line width=0.75]      (0, 0) circle [x radius= 2.01, y radius= 2.01]   ;
    \draw    (36.98,62.17) -- (84.98,36.67) ;
    \draw [shift={(84.98,36.67)}, rotate = 332.02] [color={rgb, 255:red, 0; green, 0; blue, 0 }  ][fill={rgb, 255:red, 0; green, 0; blue, 0 }  ][line width=0.75]      (0, 0) circle [x radius= 2.01, y radius= 2.01]   ;
    \draw [shift={(63.28,48.2)}, rotate = 152.02] [fill={rgb, 255:red, 0; green, 0; blue, 0 }  ][line width=0.08]  [draw opacity=0] (5.36,-2.57) -- (0,0) -- (5.36,2.57) -- cycle    ;
    \draw    (116.61,98.81) -- (116.61,46.31) ;
    \draw [shift={(116.61,46.31)}, rotate = 270] [color={rgb, 255:red, 0; green, 0; blue, 0 }  ][fill={rgb, 255:red, 0; green, 0; blue, 0 }  ][line width=0.75]      (0, 0) circle [x radius= 2.01, y radius= 2.01]   ;
    \draw [shift={(116.61,69.96)}, rotate = 90] [fill={rgb, 255:red, 0; green, 0; blue, 0 }  ][line width=0.08]  [draw opacity=0] (5.36,-2.57) -- (0,0) -- (5.36,2.57) -- cycle    ;
    \draw    (100.8,41.49) -- (116.61,46.31) ;
    \draw [shift={(104.78,42.71)}, rotate = 16.95] [fill={rgb, 255:red, 0; green, 0; blue, 0 }  ][line width=0.08]  [draw opacity=0] (5.36,-2.57) -- (0,0) -- (5.36,2.57) -- cycle    ;
    \draw    (84.98,36.67) -- (100.8,41.49) ;
    \draw [shift={(88.97,37.89)}, rotate = 16.95] [fill={rgb, 255:red, 0; green, 0; blue, 0 }  ][line width=0.08]  [draw opacity=0] (5.36,-2.57) -- (0,0) -- (5.36,2.57) -- cycle    ;
    \draw    (69.17,31.85) -- (84.98,36.67) ;
    \draw [shift={(73.15,33.07)}, rotate = 16.95] [fill={rgb, 255:red, 0; green, 0; blue, 0 }  ][line width=0.08]  [draw opacity=0] (5.36,-2.57) -- (0,0) -- (5.36,2.57) -- cycle    ;
    \draw    (69.17,31.85) -- (69.17,9.52) ;
    \draw [shift={(69.17,18.09)}, rotate = 90] [fill={rgb, 255:red, 0; green, 0; blue, 0 }  ][line width=0.08]  [draw opacity=0] (5.36,-2.57) -- (0,0) -- (5.36,2.57) -- cycle    ;
    \draw (129,56) node [anchor=north west][inner sep=0.75pt]    {\small $\mu $};
    \draw (34,27.2) node [anchor=north west][inner sep=0.75pt]    {\small $\nu $};
    \draw (103,67) node [anchor=north west][inner sep=0.75pt]    {\small $\rho $};
    \draw (116,14) node [anchor=north west][inner sep=0.75pt]    {\small $\lambda $};
    \draw (59.34,52) node [anchor=north west][inner sep=0.75pt]    {\small $\eta $};
    \draw (2,13) node [anchor=north west][inner sep=0.75pt]    {\small $m_{\nu }$};
    \draw (109,99) node [anchor=north west][inner sep=0.75pt]    {\small $n_{\rho }$};
    \draw (27,61.2) node [anchor=north west][inner sep=0.75pt]    {\small $n_{\eta }$};
    \draw (144,6.2) node [anchor=north west][inner sep=0.75pt]    {\small $m_{\lambda }$};
    \draw (163.05,58.97) node [anchor=north west][inner sep=0.75pt]    {\small $n_{\mu }$};
    \draw (100,46) node [anchor=north west][inner sep=0.75pt]    {\small $\alpha $};
    \draw (83,42) node [anchor=north west][inner sep=0.75pt]    {\small $\beta $};
    \draw (72,21) node [anchor=north west][inner sep=0.75pt]    {\small $\gamma $};
    \draw (56,18) node [anchor=north west][inner sep=0.75pt]    {\small $s$};
    \draw (60,-1.8) node [anchor=north west][inner sep=0.75pt]    {\small $m_{s}$};
    \end{tikzpicture}
    \Bigg\rangle,
\end{equation}
where the coefficients $v_\alpha v_\beta v_\gamma v_s$ are introduced such that the rewritten basis is still orthonormal. We denote the rewritten rep-basis state at the vertex by $|\Psi_{sm_s}\rangle$ and write down the inverse transformation as: 
\begin{equation}\label{eq:3.26}
    |\Psi_{sm_s}\rangle=\sum_{\substack{m_\mu m_\rho m_\eta\\n_\lambda n_\nu m_s}}v_\alpha v_\beta v_\gamma v_s\left(\begin{tikzpicture}[x=0.75pt,y=0.75pt,yscale=-0.8,xscale=0.8, baseline=(XXXX.south) ]
        \path (0,179);\path (93.99651336669922,0);\draw    ($(current bounding box.center)+(0,0.3em)$) node [anchor=south] (XXXX) {};
        \draw    (48,131.99) -- (48,162.46) ;
        \draw [shift={(48,143.13)}, rotate = 90] [fill={rgb, 255:red, 0; green, 0; blue, 0 }  ][line width=0.08]  [draw opacity=0] (5.36,-2.57) -- (0,0) -- (5.36,2.57) -- cycle    ;
        \draw    (48,101.53) -- (48,131.99) ;
        \draw [shift={(48,131.99)}, rotate = 90] [color={rgb, 255:red, 0; green, 0; blue, 0 }  ][fill={rgb, 255:red, 0; green, 0; blue, 0 }  ][line width=0.75]      (0, 0) circle [x radius= 2.01, y radius= 2.01]   ;
        \draw [shift={(48,112.66)}, rotate = 90] [fill={rgb, 255:red, 0; green, 0; blue, 0 }  ][line width=0.08]  [draw opacity=0] (5.36,-2.57) -- (0,0) -- (5.36,2.57) -- cycle    ;
        \draw [shift={(48,101.53)}, rotate = 90] [color={rgb, 255:red, 0; green, 0; blue, 0 }  ][fill={rgb, 255:red, 0; green, 0; blue, 0 }  ][line width=0.75]      (0, 0) circle [x radius= 2.01, y radius= 2.01]   ;
        \draw    (48,71.06) -- (48,101.53) ;
        \draw [shift={(48,82.19)}, rotate = 90] [fill={rgb, 255:red, 0; green, 0; blue, 0 }  ][line width=0.08]  [draw opacity=0] (5.36,-2.57) -- (0,0) -- (5.36,2.57) -- cycle    ;
        \draw    (48,40.59) -- (48,71.06) ;
        \draw [shift={(48,71.06)}, rotate = 90] [color={rgb, 255:red, 0; green, 0; blue, 0 }  ][fill={rgb, 255:red, 0; green, 0; blue, 0 }  ][line width=0.75]      (0, 0) circle [x radius= 2.01, y radius= 2.01]   ;
        \draw [shift={(48,51.73)}, rotate = 90] [fill={rgb, 255:red, 0; green, 0; blue, 0 }  ][line width=0.08]  [draw opacity=0] (5.36,-2.57) -- (0,0) -- (5.36,2.57) -- cycle    ;
        \draw [shift={(48,40.59)}, rotate = 90] [color={rgb, 255:red, 0; green, 0; blue, 0 }  ][fill={rgb, 255:red, 0; green, 0; blue, 0 }  ][line width=0.75]      (0, 0) circle [x radius= 2.01, y radius= 2.01]   ;
        \draw    (48,10.13) -- (48,40.59) ;
        \draw [shift={(48,21.26)}, rotate = 90] [fill={rgb, 255:red, 0; green, 0; blue, 0 }  ][line width=0.08]  [draw opacity=0] (5.36,-2.57) -- (0,0) -- (5.36,2.57) -- cycle    ;
            \draw    (48,101.53) -- (78.99,101.53) ; 
            \draw [shift={(66.09,101.53)}, rotate = 180] [fill={rgb, 255:red, 0; green, 0; blue, 0 }  ][line width=0.08]  [draw opacity=0] (5.36,-2.57) -- (0,0) -- (5.36,2.57) -- cycle    ;
            \draw    (17.01,131.99) -- (48,131.99) ; 
            \draw [shift={(35.1,131.99)}, rotate = 180] [fill={rgb, 255:red, 0; green, 0; blue, 0 }  ][line width=0.08]  [draw opacity=0] (5.36,-2.57) -- (0,0) -- (5.36,2.57) -- cycle    ;
        \draw    (17.01,71.06) -- (48,71.06) ;
        \draw [shift={(35.1,71.06)}, rotate = 180] [fill={rgb, 255:red, 0; green, 0; blue, 0 }  ][line width=0.08]  [draw opacity=0] (5.36,-2.57) -- (0,0) -- (5.36,2.57) -- cycle    ;
        \draw    (48,40.59) -- (78.99,40.59) ;
        \draw [shift={(66.09,40.59)}, rotate = 180] [fill={rgb, 255:red, 0; green, 0; blue, 0 }  ][line width=0.08]  [draw opacity=0] (5.36,-2.57) -- (0,0) -- (5.36,2.57) -- cycle    ;
        \draw (60,44.2) node [anchor=north west][inner sep=0.75pt]    {\small $s$};
        \draw (78,37) node [anchor=north west][inner sep=0.75pt]    {\small $m_{s}$};
        \draw (49,146) node [anchor=north west][inner sep=0.75pt]    {\small $\mu $};
        \draw (39,164) node [anchor=north west][inner sep=0.75pt]    {\small $m_{\mu }$};
        \draw (35,50) node [anchor=north west][inner sep=0.75pt]    {\small $\gamma $};
        \draw (28,75.2) node [anchor=north west][inner sep=0.75pt]    {\small $\eta $};
        \draw (0,67) node [anchor=north west][inner sep=0.75pt]    {\small $m_{\eta }$};
        \draw (49,80) node [anchor=north west][inner sep=0.75pt]    {\small $\beta $};
        \draw (34,113) node [anchor=north west][inner sep=0.75pt]    {\small $\alpha $};
        \draw (28,135) node [anchor=north west][inner sep=0.75pt]    {\small $\rho $};
        \draw (0,127) node [anchor=north west][inner sep=0.75pt]    {\small $m_{\rho }$};
        \draw (60,105) node [anchor=north west][inner sep=0.75pt]    {\small $\lambda $};
        \draw (78,97) node [anchor=north west][inner sep=0.75pt]    {\small $n_{\lambda }$};
        \draw (50,22) node [anchor=north west][inner sep=0.75pt]    {\small $\nu $};
        \draw (39,0) node [anchor=north west][inner sep=0.75pt]    {\small $n_{\nu }$};
        \end{tikzpicture}
        \right)|\mu\nu\rho\eta\lambda\rangle.
\end{equation} 
The linear transformation \eqref{eq:3.25} only rewrites a subspace of the Hilbert space spanned by the basis on the LHS of the equation. Note that the degrees of freedom $m_\mu$, $m_\rho$, $m_\eta$, $n_\nu$, and $n_\lambda$ in the left-hand-side state are transformed into $\alpha,\beta,\gamma,s$, and $m_s$, while other degrees of freedom remain unchanged. Additionally, $\alpha,\beta,\gamma$, and $s$ are not all independent. If we choose all the representations with an open end, i.e., $\mu,\nu,\rho,\eta,\lambda$, and $s$ independent, then $\alpha,\beta$, and $\gamma$ are determined. Therefore, we can denote the new basis after the rewriting by $|\Psi_{sm_s}\rangle$ for simplicity, while keeping all the other labels in the graph inexplicit. This simplification causes no confusion because in the actual calculation, e.g., in computing the inner product of two such local basis states, i.e., $\langle\Psi'_{s'm'_s}|\Psi_{sm_s}\rangle$, the prime in $\Psi'$ implies that the hidden labels in $\Psi$ should all be primed. In the next subsection, we will see another advantage of this simplified notation. 

Equations \eqref{eq:3.26} and \eqref{eq:3.24} lead to a direct inner product between the local basis states in the rewritten rep-basis and those in the group-basis at a boundary vertex: 
\begin{equation}\label{eq:3.27}
    \langle jikgh|\Psi_{sm_s} \rangle=\frac{v_\mu v_\nu v_\rho v_\eta v_\lambda v_\alpha v_\beta v_\gamma v_s}{\sqrt{|G|^5}}\left(\begin{tikzpicture}[x=0.75pt,y=0.75pt,yscale=-0.9,xscale=0.9, baseline=(XXXX.south) ]
        \path (0,172);\path (130.97605895996094,0);\draw    ($(current bounding box.center)+(0,0.3em)$) node [anchor=south] (XXXX) {};
        \draw    (62,9) -- (62,26.45) ;
        \draw    (13.31,74.32) -- (34.31,74.32) ;
        \draw    (11.4,114.04) -- (32.4,114.04) ;
        \draw    (62,141.91) -- (62,159.35) ;
        \draw    (91.59,93.76) -- (112.59,93.76) ;
        \draw    (62,113.76) -- (62,133.32) ;
        \draw [shift={(62,119.44)}, rotate = 90] [fill={rgb, 255:red, 0; green, 0; blue, 0 }  ][line width=0.08]  [draw opacity=0] (5.36,-2.57) -- (0,0) -- (5.36,2.57) -- cycle    ;
        \draw    (62,74.32) -- (62,94.04) ;
        \draw [shift={(62,94.04)}, rotate = 90] [color={rgb, 255:red, 0; green, 0; blue, 0 }  ][fill={rgb, 255:red, 0; green, 0; blue, 0 }  ][line width=0.75]      (0, 0) circle [x radius= 2.01, y radius= 2.01]   ;
        \draw [shift={(62,80.08)}, rotate = 90] [fill={rgb, 255:red, 0; green, 0; blue, 0 }  ][line width=0.08]  [draw opacity=0] (5.36,-2.57) -- (0,0) -- (5.36,2.57) -- cycle    ;
        \draw [shift={(62,74.32)}, rotate = 90] [color={rgb, 255:red, 0; green, 0; blue, 0 }  ][fill={rgb, 255:red, 0; green, 0; blue, 0 }  ][line width=0.75]      (0, 0) circle [x radius= 2.01, y radius= 2.01]   ;
        \draw    (62,54.59) -- (62,74.32) ;
        \draw [shift={(62,74.32)}, rotate = 90] [color={rgb, 255:red, 0; green, 0; blue, 0 }  ][fill={rgb, 255:red, 0; green, 0; blue, 0 }  ][line width=0.75]      (0, 0) circle [x radius= 2.01, y radius= 2.01]   ;
        \draw [shift={(62,60.35)}, rotate = 90] [fill={rgb, 255:red, 0; green, 0; blue, 0 }  ][line width=0.08]  [draw opacity=0] (5.36,-2.57) -- (0,0) -- (5.36,2.57) -- cycle    ;
        \draw [shift={(62,54.59)}, rotate = 90] [color={rgb, 255:red, 0; green, 0; blue, 0 }  ][fill={rgb, 255:red, 0; green, 0; blue, 0 }  ][line width=0.75]      (0, 0) circle [x radius= 2.01, y radius= 2.01]   ;
        \draw    (62,94.04) -- (83,94.04) ;
        \draw [shift={(75.1,94.04)}, rotate = 180] [fill={rgb, 255:red, 0; green, 0; blue, 0 }  ][line width=0.08]  [draw opacity=0] (5.36,-2.57) -- (0,0) -- (5.36,2.57) -- cycle    ;
        \draw    (62,54.59) -- (92.99,54.59) ;
        \draw [shift={(80.09,54.59)}, rotate = 180] [fill={rgb, 255:red, 0; green, 0; blue, 0 }  ][line width=0.08]  [draw opacity=0] (5.36,-2.57) -- (0,0) -- (5.36,2.57) -- cycle    ;
        \draw    (62,94.04) -- (62,113.76) ;
        \draw [shift={(62,113.76)}, rotate = 90] [color={rgb, 255:red, 0; green, 0; blue, 0 }  ][fill={rgb, 255:red, 0; green, 0; blue, 0 }  ][line width=0.75]      (0, 0) circle [x radius= 2.01, y radius= 2.01]   ;
        \draw [shift={(62,99.8)}, rotate = 90] [fill={rgb, 255:red, 0; green, 0; blue, 0 }  ][line width=0.08]  [draw opacity=0] (5.36,-2.57) -- (0,0) -- (5.36,2.57) -- cycle    ;
        \draw  [fill={rgb, 255:red, 255; green, 255; blue, 255 }  ,fill opacity=1 ] (83,94.04) .. controls (83,89.29) and (86.85,85.45) .. (91.59,85.45) .. controls (96.34,85.45) and (100.18,89.29) .. (100.18,94.04) .. controls (100.18,98.78) and (96.34,102.63) .. (91.59,102.63) .. controls (86.85,102.63) and (83,98.78) .. (83,94.04) -- cycle ;
        \draw  [fill={rgb, 255:red, 255; green, 255; blue, 255 }  ,fill opacity=1 ] (53.4,141.91) .. controls (53.4,137.16) and (57.25,133.32) .. (62,133.32) .. controls (66.74,133.32) and (70.59,137.16) .. (70.59,141.91) .. controls (70.59,146.65) and (66.74,150.5) .. (62,150.5) .. controls (57.25,150.5) and (53.4,146.65) .. (53.4,141.91) -- cycle ;
        \draw    (40.99,114.04) -- (62,114.04) ;
        \draw [shift={(54.1,114.04)}, rotate = 180] [fill={rgb, 255:red, 0; green, 0; blue, 0 }  ][line width=0.08]  [draw opacity=0] (5.36,-2.57) -- (0,0) -- (5.36,2.57) -- cycle    ;
        \draw    (40.99,74.32) -- (62,74.32) ;
        \draw [shift={(54.1,74.32)}, rotate = 180] [fill={rgb, 255:red, 0; green, 0; blue, 0 }  ][line width=0.08]  [draw opacity=0] (5.36,-2.57) -- (0,0) -- (5.36,2.57) -- cycle    ;
        \draw  [fill={rgb, 255:red, 255; green, 255; blue, 255 }  ,fill opacity=1 ] (23.81,114.04) .. controls (23.81,109.29) and (27.66,105.45) .. (32.4,105.45) .. controls (37.15,105.45) and (40.99,109.29) .. (40.99,114.04) .. controls (40.99,118.78) and (37.15,122.63) .. (32.4,122.63) .. controls (27.66,122.63) and (23.81,118.78) .. (23.81,114.04) -- cycle ;
        \draw  [fill={rgb, 255:red, 255; green, 255; blue, 255 }  ,fill opacity=1 ] (23.81,74.32) .. controls (23.81,69.57) and (27.66,65.72) .. (32.4,65.72) .. controls (37.15,65.72) and (40.99,69.57) .. (40.99,74.32) .. controls (40.99,79.06) and (37.15,82.91) .. (32.4,82.91) .. controls (27.66,82.91) and (23.81,79.06) .. (23.81,74.32) -- cycle ;
        \draw    (62,35.04) -- (62,54.59) ;
        \draw [shift={(62,40.72)}, rotate = 90] [fill={rgb, 255:red, 0; green, 0; blue, 0 }  ][line width=0.08]  [draw opacity=0] (5.36,-2.57) -- (0,0) -- (5.36,2.57) -- cycle    ;
        \draw  [fill={rgb, 255:red, 255; green, 255; blue, 255 }  ,fill opacity=1 ] (53.4,26.45) .. controls (53.4,21.7) and (57.25,17.85) .. (62,17.85) .. controls (66.74,17.85) and (70.59,21.7) .. (70.59,26.45) .. controls (70.59,31.19) and (66.74,35.04) .. (62,35.04) .. controls (57.25,35.04) and (53.4,31.19) .. (53.4,26.45) -- cycle ;
        \draw (74,58.2) node [anchor=north west][inner sep=0.75pt]    {\small $s$};
        \draw (50,59.2) node [anchor=north west][inner sep=0.75pt]    {\small $\gamma $};
        \draw (63,77.52) node [anchor=north west][inner sep=0.75pt]    {\small $\beta $};
        \draw (49,100) node [anchor=north west][inner sep=0.75pt]    {\small $\alpha $};
        \draw (64,38.33) node [anchor=north west][inner sep=0.75pt]    {\small $\nu $};
        \draw (86,88) node [anchor=north west][inner sep=0.75pt]    {\small $h$};
        \draw (70,96) node [anchor=north west][inner sep=0.75pt]    {\small $\lambda $};
        \draw (111,89) node [anchor=north west][inner sep=0.75pt]    {\small $m_{\lambda }$};
        \draw (57,135) node [anchor=north west][inner sep=0.75pt]    {\small $j$};
        \draw (61.98,121.2) node [anchor=north west][inner sep=0.75pt]    {\small $\mu $};
        \draw (55,159.55) node [anchor=north west][inner sep=0.75pt]    {\small $n_{\mu }$};
        \draw (46,77.52) node [anchor=north west][inner sep=0.75pt]    {\small $\eta $};
        \draw (46,116) node [anchor=north west][inner sep=0.75pt]    {\small $\rho $};
        \draw (27,69) node [anchor=north west][inner sep=0.75pt]    {\small $g$};
        \draw (27,108) node [anchor=north west][inner sep=0.75pt]    {\small $k$};
        \draw (57.5,21) node [anchor=north west][inner sep=0.75pt]    {\small $i$};
        \draw (91,50) node [anchor=north west][inner sep=0.75pt]    {\small $m_{s}$};
        \draw (0,70) node [anchor=north west][inner sep=0.75pt]    {\small $n_{\eta }$};
        \draw (0,110) node [anchor=north west][inner sep=0.75pt]    {\small $n_{\rho }$};
        \draw (53,-1.8) node [anchor=north west][inner sep=0.75pt]    {\small $m_{\nu }$};
        \end{tikzpicture}
        \right).
\end{equation} 
We verify in \autoref{appendix:A} that the local basis states $|\Psi_{sm_s}\rangle$ indeed form a well-defined local basis by showing that they are orthonormal and complete. 

Nevertheless, the total Hilbert space of the Fourier-transformed 3-dimensional GT model cannot be trivially viewed as the tensor product of the Fourier-transformed local Hilbert spaces of single vertices defined above. The key lies in braiding. Without loss of generality, let us consider the state in the Hilbert space with a whole plaquette in the bulk, as shown in \autoref{figx}. One can observe that when a larger Hilber space is considered, edges are unavoidably braided in the rewritten rep-basis. Therefore, defining the inner product between the states that include braided edges in the rewritten rep-basis and those in the group-basis becomes necessary. To achieve this, let us consider the following inner product. 
\begin{equation}\label{eq:3.30x}
    \bigg\langle \begin{tikzpicture}[x=0.75pt,y=0.75pt,yscale=-1,xscale=1, baseline=(XXXX.south) ]
\path (0,62);\path (56.99652862548828,0);\draw    ($(current bounding box.center)+(0,0.3em)$) node [anchor=south] (XXXX) {};
\draw    (26,30) -- (49.67,30) ;
\draw [shift={(33.73,30)}, rotate = 0] [fill={rgb, 255:red, 0; green, 0; blue, 0 }  ][line width=0.08]  [draw opacity=0] (5.36,-2.57) -- (0,0) -- (5.36,2.57) -- cycle    ;
\draw    (26,1.84) -- (26,26) ;
\draw [shift={(26,9.82)}, rotate = 90] [fill={rgb, 255:red, 0; green, 0; blue, 0 }  ][line width=0.08]  [draw opacity=0] (5.36,-2.57) -- (0,0) -- (5.36,2.57) -- cycle    ;
\draw    (26,34) -- (26,58.16) ;
\draw    (2.33,30) -- (26,30) ;
\draw (34,32) node [anchor=north west][inner sep=0.75pt]    {\small $g$};
\draw (28,6.04) node [anchor=north west][inner sep=0.75pt]    {\small $h$};
\end{tikzpicture}
\bigg|\begin{tikzpicture}[x=0.75pt,y=0.75pt,yscale=-1,xscale=1, baseline=(XXXX.south) ]
\path (0,62);\path (56.99652862548828,0);\draw    ($(current bounding box.center)+(0,0.3em)$) node [anchor=south] (XXXX) {};
\draw    (26,26) -- (26,49.16) ;
\draw [color={rgb, 255:red, 255; green, 255; blue, 255 }  ,draw opacity=1 ][line width=3]    (11.33,30) -- (41.67,30) ;
\draw    (26,30) -- (41.67,30) ;
\draw [shift={(29.73,30)}, rotate = 0] [fill={rgb, 255:red, 0; green, 0; blue, 0 }  ][line width=0.08]  [draw opacity=0] (5.36,-2.57) -- (0,0) -- (5.36,2.57) -- cycle    ;
\draw    (26,9.84) -- (26,26) ;
\draw [shift={(26,13.82)}, rotate = 90] [fill={rgb, 255:red, 0; green, 0; blue, 0 }  ][line width=0.08]  [draw opacity=0] (5.36,-2.57) -- (0,0) -- (5.36,2.57) -- cycle    ;
\draw    (11.33,30) -- (26,30) ;
\draw (34,32) node [anchor=north west][inner sep=0.75pt]    {\small $\mu $};
\draw (28,9) node [anchor=north west][inner sep=0.75pt]    {\small $\nu $};
\draw (19,-2) node [anchor=north west][inner sep=0.75pt]    {\small $m_{\nu }$};
\draw (21,49) node [anchor=north west][inner sep=0.75pt]    {\small $n_{\nu }$};
\draw (-2,26) node [anchor=north west][inner sep=0.75pt]    {\small $m_{\mu }$};
\draw (42,26) node [anchor=north west][inner sep=0.75pt]    {\small $n_{\mu }$};
\end{tikzpicture}
\bigg\rangle \equiv\frac{v_\mu v_\nu}{|G|}\begin{tikzpicture}[x=0.75pt,y=0.75pt,yscale=-1,xscale=1, baseline=(XXXX.south) ]
\path (0,87);\path (92.99652862548828,0);\draw    ($(current bounding box.center)+(0,0.3em)$) node [anchor=south] (XXXX) {};
\draw    (56,50) -- (56,73.16) ;
\draw [color={rgb, 255:red, 255; green, 255; blue, 255 }  ,draw opacity=1 ][line width=3]    (52,62) -- (75.67,62) ;
\draw    (52,62) -- (75.67,62) ;
\draw [shift={(59.73,62)}, rotate = 0] [fill={rgb, 255:red, 0; green, 0; blue, 0 }  ][line width=0.08]  [draw opacity=0] (5.36,-2.57) -- (0,0) -- (5.36,2.57) -- cycle    ;
\draw  [fill={rgb, 255:red, 255; green, 255; blue, 255 }  ,fill opacity=1 ] (34.81,62) .. controls (34.81,57.25) and (38.66,53.4) .. (43.4,53.4) .. controls (48.15,53.4) and (52,57.25) .. (52,62) .. controls (52,66.74) and (48.15,70.59) .. (43.4,70.59) .. controls (38.66,70.59) and (34.81,66.74) .. (34.81,62) -- cycle ;
\draw    (19.73,62) -- (34.81,62) ;
\draw  [fill={rgb, 255:red, 255; green, 255; blue, 255 }  ,fill opacity=1 ] (47.4,41.4) .. controls (47.4,36.66) and (51.25,32.81) .. (56,32.81) .. controls (60.74,32.81) and (64.59,36.66) .. (64.59,41.4) .. controls (64.59,46.15) and (60.74,50) .. (56,50) .. controls (51.25,50) and (47.4,46.15) .. (47.4,41.4) -- cycle ;
\draw    (56,16.65) -- (56,32.81) ;
\draw [shift={(56,20.63)}, rotate = 90] [fill={rgb, 255:red, 0; green, 0; blue, 0 }  ][line width=0.08]  [draw opacity=0] (5.36,-2.57) -- (0,0) -- (5.36,2.57) -- cycle    ;
\draw (39,56) node [anchor=north west][inner sep=0.75pt]    {\small $g$};
\draw (63,62) node [anchor=north west][inner sep=0.75pt]    {\small $\mu $};
\draw (1,57) node [anchor=north west][inner sep=0.75pt]    {\small $m_{\mu }$};
\draw (75,57) node [anchor=north west][inner sep=0.75pt]    {\small $n_{\mu }$};
\draw (51,35) node [anchor=north west][inner sep=0.75pt]    {\small $h$};
\draw (47,5) node [anchor=north west][inner sep=0.75pt]    {\small $m_{\nu }$};
\draw (51,73) node [anchor=north west][inner sep=0.75pt]    {\small $n_{\nu }$};
\draw (58,17) node [anchor=north west][inner sep=0.75pt]    {\small $\nu $};
\end{tikzpicture}
\overset{?}{=}\frac{v_\mu v_\nu}{|G|} D_{m_{\mu } n_{\mu }}^{\mu }( g) D_{m_{\nu } n_{\nu }}^{\nu }(h),
\end{equation}
where the first equivalence comes from the Fourier-transformation of the basis vector of $\mathcal{H}_e$ \eqref{eq:3.22}, and the equality with a question mark is the evaluation of the graphical presentation if the braiding is treated trivially. 
\begin{figure}[ht]
    \centering 
\begin{tikzpicture}[x=0.75pt,y=0.75pt,yscale=-1,xscale=1]

\draw    (326.01,70.16) -- (397.17,91.85) ;
\draw [shift={(397.17,91.85)}, rotate = 16.95] [color={rgb, 255:red, 0; green, 0; blue, 0 }  ][fill={rgb, 255:red, 0; green, 0; blue, 0 }  ][line width=0.75]      (0, 0) circle [x radius= 2.01, y radius= 2.01]   ;
\draw [shift={(357.67,79.81)}, rotate = 16.95] [fill={rgb, 255:red, 0; green, 0; blue, 0 }  ][line width=0.08]  [draw opacity=0] (5.36,-2.57) -- (0,0) -- (5.36,2.57) -- cycle    ;
\draw [color={rgb, 255:red, 255; green, 255; blue, 255 }  ,draw opacity=1 ][line width=3.75]    (334.2,52.59) -- (333.36,112.53) ;
\draw    (326.01,122.66) -- (326.01,70.16) ;
\draw [shift={(326.01,70.16)}, rotate = 270] [color={rgb, 255:red, 0; green, 0; blue, 0 }  ][fill={rgb, 255:red, 0; green, 0; blue, 0 }  ][line width=0.75]      (0, 0) circle [x radius= 2.01, y radius= 2.01]   ;
\draw [shift={(326.01,93.81)}, rotate = 90] [fill={rgb, 255:red, 0; green, 0; blue, 0 }  ][line width=0.08]  [draw opacity=0] (5.36,-2.57) -- (0,0) -- (5.36,2.57) -- cycle    ;
\draw [color={rgb, 255:red, 255; green, 255; blue, 255 }  ,draw opacity=1 ][line width=3.75]    (262.2,90.84) -- (333.36,112.53) ;
\draw    (444.61,106.31) -- (492.05,120.77) ;
\draw [shift={(464.41,112.35)}, rotate = 16.95] [fill={rgb, 255:red, 0; green, 0; blue, 0 }  ][line width=0.08]  [draw opacity=0] (5.36,-2.57) -- (0,0) -- (5.36,2.57) -- cycle    ;
\draw    (428.8,101.49) -- (476.8,75.99) ;
\draw [shift={(455.09,87.52)}, rotate = 152.02] [fill={rgb, 255:red, 0; green, 0; blue, 0 }  ][line width=0.08]  [draw opacity=0] (5.36,-2.57) -- (0,0) -- (5.36,2.57) -- cycle    ;
\draw [shift={(428.8,101.49)}, rotate = 332.02] [color={rgb, 255:red, 0; green, 0; blue, 0 }  ][fill={rgb, 255:red, 0; green, 0; blue, 0 }  ][line width=0.75]      (0, 0) circle [x radius= 2.01, y radius= 2.01]   ;
\draw    (444.61,158.81) -- (444.61,106.31) ;
\draw [shift={(444.61,106.31)}, rotate = 270] [color={rgb, 255:red, 0; green, 0; blue, 0 }  ][fill={rgb, 255:red, 0; green, 0; blue, 0 }  ][line width=0.75]      (0, 0) circle [x radius= 2.01, y radius= 2.01]   ;
\draw [shift={(444.61,129.96)}, rotate = 90] [fill={rgb, 255:red, 0; green, 0; blue, 0 }  ][line width=0.08]  [draw opacity=0] (5.36,-2.57) -- (0,0) -- (5.36,2.57) -- cycle    ;
\draw    (428.8,101.49) -- (444.61,106.31) ;
\draw [shift={(432.78,102.71)}, rotate = 16.95] [fill={rgb, 255:red, 0; green, 0; blue, 0 }  ][line width=0.08]  [draw opacity=0] (5.36,-2.57) -- (0,0) -- (5.36,2.57) -- cycle    ;
\draw    (412.98,96.67) -- (428.8,101.49) ;
\draw [shift={(416.97,97.89)}, rotate = 16.95] [fill={rgb, 255:red, 0; green, 0; blue, 0 }  ][line width=0.08]  [draw opacity=0] (5.36,-2.57) -- (0,0) -- (5.36,2.57) -- cycle    ;
\draw    (397.17,91.85) -- (412.98,96.67) ;
\draw [shift={(401.15,93.07)}, rotate = 16.95] [fill={rgb, 255:red, 0; green, 0; blue, 0 }  ][line width=0.08]  [draw opacity=0] (5.36,-2.57) -- (0,0) -- (5.36,2.57) -- cycle    ;
\draw    (397.17,91.85) -- (397.17,39.35) ;
\draw [shift={(397.17,63)}, rotate = 90] [fill={rgb, 255:red, 0; green, 0; blue, 0 }  ][line width=0.08]  [draw opacity=0] (5.36,-2.57) -- (0,0) -- (5.36,2.57) -- cycle    ;
\draw    (380.8,126.99) -- (428.23,141.45) ;
\draw [shift={(400.59,133.03)}, rotate = 16.95] [fill={rgb, 255:red, 0; green, 0; blue, 0 }  ][line width=0.08]  [draw opacity=0] (5.36,-2.57) -- (0,0) -- (5.36,2.57) -- cycle    ;
\draw    (364.98,122.17) -- (412.98,96.67) ;
\draw [shift={(412.98,96.67)}, rotate = 332.02] [color={rgb, 255:red, 0; green, 0; blue, 0 }  ][fill={rgb, 255:red, 0; green, 0; blue, 0 }  ][line width=0.75]      (0, 0) circle [x radius= 2.01, y radius= 2.01]   ;
\draw [shift={(391.28,108.2)}, rotate = 152.02] [fill={rgb, 255:red, 0; green, 0; blue, 0 }  ][line width=0.08]  [draw opacity=0] (5.36,-2.57) -- (0,0) -- (5.36,2.57) -- cycle    ;
\draw [shift={(364.98,122.17)}, rotate = 332.02] [color={rgb, 255:red, 0; green, 0; blue, 0 }  ][fill={rgb, 255:red, 0; green, 0; blue, 0 }  ][line width=0.75]      (0, 0) circle [x radius= 2.01, y radius= 2.01]   ;
\draw    (301.17,142.85) -- (349.17,117.35) ;
\draw [shift={(349.17,117.35)}, rotate = 332.02] [color={rgb, 255:red, 0; green, 0; blue, 0 }  ][fill={rgb, 255:red, 0; green, 0; blue, 0 }  ][line width=0.75]      (0, 0) circle [x radius= 2.01, y radius= 2.01]   ;
\draw [shift={(327.46,128.88)}, rotate = 152.02] [fill={rgb, 255:red, 0; green, 0; blue, 0 }  ][line width=0.08]  [draw opacity=0] (5.36,-2.57) -- (0,0) -- (5.36,2.57) -- cycle    ;
\draw    (380.8,179.49) -- (380.8,126.99) ;
\draw [shift={(380.8,126.99)}, rotate = 270] [color={rgb, 255:red, 0; green, 0; blue, 0 }  ][fill={rgb, 255:red, 0; green, 0; blue, 0 }  ][line width=0.75]      (0, 0) circle [x radius= 2.01, y radius= 2.01]   ;
\draw [shift={(380.8,150.64)}, rotate = 90] [fill={rgb, 255:red, 0; green, 0; blue, 0 }  ][line width=0.08]  [draw opacity=0] (5.36,-2.57) -- (0,0) -- (5.36,2.57) -- cycle    ;
\draw    (364.98,122.17) -- (380.8,126.99) ;
\draw [shift={(368.97,123.39)}, rotate = 16.95] [fill={rgb, 255:red, 0; green, 0; blue, 0 }  ][line width=0.08]  [draw opacity=0] (5.36,-2.57) -- (0,0) -- (5.36,2.57) -- cycle    ;
\draw    (349.17,117.35) -- (364.98,122.17) ;
\draw [shift={(353.15,118.57)}, rotate = 16.95] [fill={rgb, 255:red, 0; green, 0; blue, 0 }  ][line width=0.08]  [draw opacity=0] (5.36,-2.57) -- (0,0) -- (5.36,2.57) -- cycle    ;
\draw    (333.36,112.53) -- (349.17,117.35) ;
\draw [shift={(337.34,113.75)}, rotate = 16.95] [fill={rgb, 255:red, 0; green, 0; blue, 0 }  ][line width=0.08]  [draw opacity=0] (5.36,-2.57) -- (0,0) -- (5.36,2.57) -- cycle    ;
\draw    (333.36,112.53) -- (333.36,60.03) ;
\draw [shift={(333.36,83.68)}, rotate = 90] [fill={rgb, 255:red, 0; green, 0; blue, 0 }  ][line width=0.08]  [draw opacity=0] (5.36,-2.57) -- (0,0) -- (5.36,2.57) -- cycle    ;
\draw    (231.13,41.24) -- (278.57,55.7) ;
\draw [shift={(278.57,55.7)}, rotate = 16.95] [color={rgb, 255:red, 0; green, 0; blue, 0 }  ][fill={rgb, 255:red, 0; green, 0; blue, 0 }  ][line width=0.75]      (0, 0) circle [x radius= 2.01, y radius= 2.01]   ;
\draw [shift={(250.93,47.28)}, rotate = 16.95] [fill={rgb, 255:red, 0; green, 0; blue, 0 }  ][line width=0.08]  [draw opacity=0] (5.36,-2.57) -- (0,0) -- (5.36,2.57) -- cycle    ;
\draw    (310.2,65.34) -- (358.2,39.84) ;
\draw [shift={(336.49,51.37)}, rotate = 152.02] [fill={rgb, 255:red, 0; green, 0; blue, 0 }  ][line width=0.08]  [draw opacity=0] (5.36,-2.57) -- (0,0) -- (5.36,2.57) -- cycle    ;
\draw [shift={(310.2,65.34)}, rotate = 332.02] [color={rgb, 255:red, 0; green, 0; blue, 0 }  ][fill={rgb, 255:red, 0; green, 0; blue, 0 }  ][line width=0.75]      (0, 0) circle [x radius= 2.01, y radius= 2.01]   ;
\draw    (246.38,86.02) -- (294.38,60.52) ;
\draw [shift={(294.38,60.52)}, rotate = 332.02] [color={rgb, 255:red, 0; green, 0; blue, 0 }  ][fill={rgb, 255:red, 0; green, 0; blue, 0 }  ][line width=0.75]      (0, 0) circle [x radius= 2.01, y radius= 2.01]   ;
\draw [shift={(272.68,72.05)}, rotate = 152.02] [fill={rgb, 255:red, 0; green, 0; blue, 0 }  ][line width=0.08]  [draw opacity=0] (5.36,-2.57) -- (0,0) -- (5.36,2.57) -- cycle    ;
\draw [shift={(246.38,86.02)}, rotate = 332.02] [color={rgb, 255:red, 0; green, 0; blue, 0 }  ][fill={rgb, 255:red, 0; green, 0; blue, 0 }  ][line width=0.75]      (0, 0) circle [x radius= 2.01, y radius= 2.01]   ;
\draw    (310.2,65.34) -- (326.01,70.16) ;
\draw [shift={(314.18,66.56)}, rotate = 16.95] [fill={rgb, 255:red, 0; green, 0; blue, 0 }  ][line width=0.08]  [draw opacity=0] (5.36,-2.57) -- (0,0) -- (5.36,2.57) -- cycle    ;
\draw    (294.38,60.52) -- (310.2,65.34) ;
\draw [shift={(298.37,61.74)}, rotate = 16.95] [fill={rgb, 255:red, 0; green, 0; blue, 0 }  ][line width=0.08]  [draw opacity=0] (5.36,-2.57) -- (0,0) -- (5.36,2.57) -- cycle    ;
\draw    (278.57,55.7) -- (294.38,60.52) ;
\draw [shift={(282.55,56.92)}, rotate = 16.95] [fill={rgb, 255:red, 0; green, 0; blue, 0 }  ][line width=0.08]  [draw opacity=0] (5.36,-2.57) -- (0,0) -- (5.36,2.57) -- cycle    ;
\draw    (278.57,55.7) -- (278.57,3.2) ;
\draw [shift={(278.57,26.85)}, rotate = 90] [fill={rgb, 255:red, 0; green, 0; blue, 0 }  ][line width=0.08]  [draw opacity=0] (5.36,-2.57) -- (0,0) -- (5.36,2.57) -- cycle    ;
\draw    (167.32,61.92) -- (214.76,76.38) ;
\draw [shift={(214.76,76.38)}, rotate = 16.95] [color={rgb, 255:red, 0; green, 0; blue, 0 }  ][fill={rgb, 255:red, 0; green, 0; blue, 0 }  ][line width=0.75]      (0, 0) circle [x radius= 2.01, y radius= 2.01]   ;
\draw [shift={(187.11,67.96)}, rotate = 16.95] [fill={rgb, 255:red, 0; green, 0; blue, 0 }  ][line width=0.08]  [draw opacity=0] (5.36,-2.57) -- (0,0) -- (5.36,2.57) -- cycle    ;
\draw    (182.57,106.7) -- (230.57,81.2) ;
\draw [shift={(230.57,81.2)}, rotate = 332.02] [color={rgb, 255:red, 0; green, 0; blue, 0 }  ][fill={rgb, 255:red, 0; green, 0; blue, 0 }  ][line width=0.75]      (0, 0) circle [x radius= 2.01, y radius= 2.01]   ;
\draw [shift={(208.87,92.73)}, rotate = 152.02] [fill={rgb, 255:red, 0; green, 0; blue, 0 }  ][line width=0.08]  [draw opacity=0] (5.36,-2.57) -- (0,0) -- (5.36,2.57) -- cycle    ;
\draw    (262.2,143.34) -- (262.2,90.84) ;
\draw [shift={(262.2,90.84)}, rotate = 270] [color={rgb, 255:red, 0; green, 0; blue, 0 }  ][fill={rgb, 255:red, 0; green, 0; blue, 0 }  ][line width=0.75]      (0, 0) circle [x radius= 2.01, y radius= 2.01]   ;
\draw [shift={(262.2,114.49)}, rotate = 90] [fill={rgb, 255:red, 0; green, 0; blue, 0 }  ][line width=0.08]  [draw opacity=0] (5.36,-2.57) -- (0,0) -- (5.36,2.57) -- cycle    ;
\draw    (246.38,86.02) -- (262.2,90.84) ;
\draw [shift={(250.37,87.24)}, rotate = 16.95] [fill={rgb, 255:red, 0; green, 0; blue, 0 }  ][line width=0.08]  [draw opacity=0] (5.36,-2.57) -- (0,0) -- (5.36,2.57) -- cycle    ;
\draw    (230.57,81.2) -- (246.38,86.02) ;
\draw [shift={(234.55,82.42)}, rotate = 16.95] [fill={rgb, 255:red, 0; green, 0; blue, 0 }  ][line width=0.08]  [draw opacity=0] (5.36,-2.57) -- (0,0) -- (5.36,2.57) -- cycle    ;
\draw    (214.76,76.38) -- (230.57,81.2) ;
\draw [shift={(218.74,77.6)}, rotate = 16.95] [fill={rgb, 255:red, 0; green, 0; blue, 0 }  ][line width=0.08]  [draw opacity=0] (5.36,-2.57) -- (0,0) -- (5.36,2.57) -- cycle    ;
\draw    (214.76,76.38) -- (214.76,23.88) ;
\draw [shift={(214.76,47.53)}, rotate = 90] [fill={rgb, 255:red, 0; green, 0; blue, 0 }  ][line width=0.08]  [draw opacity=0] (5.36,-2.57) -- (0,0) -- (5.36,2.57) -- cycle    ;
\draw    (262.2,90.84) -- (333.36,112.53) ;
\draw [shift={(333.36,112.53)}, rotate = 16.95] [color={rgb, 255:red, 0; green, 0; blue, 0 }  ][fill={rgb, 255:red, 0; green, 0; blue, 0 }  ][line width=0.75]      (0, 0) circle [x radius= 2.01, y radius= 2.01]   ;
\draw [shift={(293.85,100.49)}, rotate = 16.95] [fill={rgb, 255:red, 0; green, 0; blue, 0 }  ][line width=0.08]  [draw opacity=0] (5.36,-2.57) -- (0,0) -- (5.36,2.57) -- cycle    ;

\draw (335,79.4) node [anchor=north west][inner sep=0.75pt]    {$\mu $};
\draw (358,67) node [anchor=north west][inner sep=0.75pt]    {$\nu $};
\draw (289,102) node [anchor=north west][inner sep=0.75pt]    {$\rho $};
\draw (315,86) node [anchor=north west][inner sep=0.75pt]    {$\sigma $};

\end{tikzpicture}
\caption{A whole plaquette in the bulk, where the edges labeled by $\mu$ and $\nu$ are braided, as are edges labeleds by $\rho$ and $\sigma$. }\label{figx}
\end{figure}
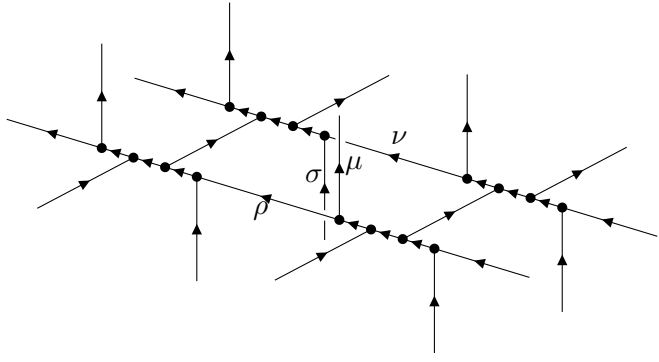

Clearly, the convention given by ``$\overset{?}{=}$'' is insufficient because it cannot distinguish the over- and under-crossing relation between the two edges. Nonetheless, we can indeed find a reasonable convention to evaluate the graphical presentation in \eqref{eq:3.30x} that encodes the braiding information of the edges in the inner product. Using the $F$-move given by the first equation in \eqref{eq:3.16}, formally we have: 
\begin{equation}\label{eq:3.31x}
    \begin{aligned}

    \\
    &\equiv \sum _{\rho ,m_{\rho } m'_{\mu } m'_{\nu }}\mathtt{d}_{\rho } D_{m_{\mu } m'_{\mu }}^{\mu }( g) D_{m_{\nu } m'_{\nu }}^{\nu }( h)\overline{R_{\rho }^{\mu \nu }} C_{\rho m_{\rho }}^{\mu \nu ;m'_{\mu } m'_{\nu }} C_{\nu \mu ;n_{\nu } n_{\mu }}^{\rho m_{\rho }},
    \end{aligned}
\end{equation}
where we have introduced new equivalences in the second row above. Namely 
\begin{equation}
    %
.
\end{equation}

Here, $R:L_G^3\to\mathbb C$ is called an $R$-matrix, which cannot be expressed explicitly in terms of the duality maps and $3j$-symbols of the representations of a finite group $G$. To determine the $R$-matrix, we impose the following symmetry constraint of the rewritten rep-basis (where Latin indices $m_\mu,m_\nu,\cdots$ are omitted): 
\begin{equation}\label{eq:3.33x}
    \bigg|%
\bigg\rangle.
\end{equation}
As a result, the $R$-matrix elements must be a complex phase and satisfy the Hexagon identity. Such an $R$-matrix satisfying \eqref{eq:3.33x} can always be found for a gauge theory\cite{Levin2004}. The data consisting a label set $L_G$, quantum dimension $\mathtt{d}:L_G\to\mathbb C$ (where $\mathtt d_\mu=\beta_\mu d_\mu$), $6j$-symbol $F:L_G^6\to\mathbb C$, and an $R$-matrix form a unitary braided fusion category (UBFC) $\mathcal{R}ep(G)$, which are the input data of the Fourier-transformed 3-dimensional GT model with input data $G$. 

Therefore, as the UBFC $\mathcal{R}ep(G)$ is given, we can define the Fourier transform of the total Hilbert space through \eqref{eq:3.31x} and the local Hilbert space Fourier transform \eqref{eq:3.27}. We verify that \eqref{eq:3.31x} preserves the orthonormality and completeness of the rewritten rep-basis in \autoref{appendix:A}.

\subsection{Fourier transform of the boundary vertex operators}
We are now ready to study how a boundary vertex operator $\overline{A_v^\text{GT}}$ acts on a local basis state $|\Psi_{sm_s}\rangle$. That is, we need to find the Fourier-transformed version $\overline{\tilde{A}_v^\text{GT}}$ of $\overline{A_v^\text{GT}}$. Since the group-basis and the rewritten rep-basis $\{|\Psi_{sm_s}\rangle\}$ are both orthonormal and complete in the local Hilbert space, we can compute the action of $\overline{\tilde{A}_v^\text{GT}}$ by inserting resolution of identity: 
\begin{equation*}
    \begin{aligned}
        &{\overline{\tilde A^\text{GT}_v}}|\Psi_{sm_s}\rangle\\
    ={} &\sum_{\begin{subarray}{c}
        \mu'\nu'\rho'\eta'\lambda'\\ \alpha'\beta'\gamma's'm_{s'}\\ n_{\mu'} n_{\rho'} n_{\eta'} m_{\nu'} m_{\lambda'}
    \end{subarray}}|\Psi'_{s'm_{s'}}\rangle\langle\Psi'_{s'm_{s'}}|\sum_{jikgh\in G}\overline{A_v^\text{QD}}|jikgh\rangle\langle jikgh|\Psi^{sm_s} \rangle\\
    ={}&\sum_{\begin{subarray}{c}
        \mu'\nu'\rho'\eta'\lambda'\\ \alpha'\beta'\gamma's'm_{s'}\\ n_{\mu'} n_{\rho'} n_{\eta'} m_{\nu'} m_{\lambda'}
    \end{subarray}}|\Psi'_{s'm_{s'}}\rangle\langle\Psi'_{s'm_{s'}}|\sum_{jikgh\in G}\frac{1}{|K|}\sum_{x\in K}|xj,i\bar x,xk,xg,h\bar x\rangle\langle jikgh|\Psi^{sm_s} \rangle\\
    ={}&\sum_{\begin{subarray}{c}
        \mu'\nu'\rho'\eta'\lambda'\\ \alpha'\beta'\gamma's'm_{s'}\\ n_{\mu'} n_{\rho'} n_{\eta'} m_{\nu'} m_{\lambda'}
    \end{subarray}}\sum_{jikgh\in G}\frac{1}{|K|}\sum_{x\in K}\bigl(\langle\Psi'_{s'm_{s'}}|xj,i\bar x,xk,xg,h\bar x\rangle\langle jikgh|\Psi_{sm_s} \rangle\bigr)|\Psi'_{s'm_{s'}}\rangle
    \end{aligned}.
\end{equation*}
We leave the simplification of the two inner products in the above expression in \autoref{appendix:A} but present the result here: 
\begin{equation}\label{eq:3.34o}
    {\overline{\tilde A^\text{GT}_v}}|\Psi_{sm_s}\rangle=\sum_{m_s'}\frac{1}{|K|}\sum_{x\in K}D^s_{m_sm_s'}(x)|\Psi_{sm_s'}\rangle\equiv\sum_{m_s'}(P_K^s)_{m_sm'_s}|\Psi_{sm_s'}\rangle.
\end{equation}
We observe that the Fourier-transformed vertex operator $\overline{\tilde{A}_v^\text{GT}}$ is automatically diagonalized in the entire local Hilbert space of the considered vertex spanned by $|\Psi_{sm_s}\rangle$ but the small subspace --- the representation space $V_s$ of $s$, which is spanned by $m_s$. Here, $P_K^s=(1/|K|)\sum_{x\in K}D^s(x)$ is a projector in $V_s$. As the eigenvalues of a projector must be 0 or 1, a linear transformation $V_s\to V_s,|m_s\rangle\mapsto|\tilde{m}_s\rangle$ can be applied to diagonalize the matrix $P^s_K$. This transformation will also transform the basis $|\Psi_{sm_s}\rangle$ to $|\Psi_{s\tilde{m}_s}\rangle$. In such a basis, \eqref{eq:3.34o} can be simplified as
\begin{equation}\label{eq:3.29}
    {\overline{\tilde A^\text{GT}_v}}|\Psi_{s\tilde m_s}\rangle=P^s_K|\Psi_{s\tilde m_s}\rangle=\delta_{(s,\tilde{m}_s)\in L_A}|\Psi_{s\tilde m_s}\rangle,
\end{equation}
where the set $L_A$ collects all the $+1$ eigenstates of $\overline{\tilde{A}_v^\text{GT}}$ or its representation $P^s_K$. More precisely, we can write
\begin{equation}\label{eq:3.30}
    L_A:=\{(s,\alpha_s)|P_K^s|\Psi_{s\alpha_s}\rangle=|\Psi_{s\alpha_s}\rangle, s\in L_G\}.
\end{equation}
The states $|\Psi_{s\tilde{m}_s}\rangle$, where $(s,\tilde{m}_s)\ne L_A$, are zero eigenstates of $\overline{\tilde{A}_v^\text{GT}}$. According to Hamiltonian \eqref{eq:2.6}, they have higher energy than the $+1$ eigenstates and are thus excited states. The excitations emerge at the end of the dangling edges with representations $s$ and are point-like pure charge excitations on the boundary. 

\subsection{Fourier transform of the boundary edge operators}
We can now proceed to investigate how a boundary edge operator $\overline{C_e^\text{GT}}$ \eqref{eq:2.7} of the GT model with gapped boundary should be Fourier-transformed to act on a local basis state in the rewritten rep-basis. According to \eqref{eq:2.7}, $\overline{C_e^\text{GT}}$ acts on the edge between two vertices. Consequently, in the rep-basis, a local basis state on which $\overline{\skew{-1}{\tilde}{C}_e^\text{GT}}$ acts would involve the two end vertices of edge $e$. In what follows, we denote the local basis states in the group-basis acted by a boundary edge operator as: 
\begin{equation}
    |\Psi _{l} \rangle :=\Bigg|\begin{tikzpicture}[x=0.75pt,y=0.75pt,yscale=-1,xscale=1, baseline=(XXXX.south) ]
\path (0,54);\path (93.97857666015625,0);\draw    ($(current bounding box.center)+(0,0.3em)$) node [anchor=south] (XXXX) {};
\draw    (69.29,30.89) -- (69.29,51.36) ;
\draw [shift={(69.29,37.02)}, rotate = 90] [fill={rgb, 255:red, 0; green, 0; blue, 0 }  ][line width=0.08]  [draw opacity=0] (5.36,-2.57) -- (0,0) -- (5.36,2.57) -- cycle    ;
\draw    (23.87,15.44) -- (23.87,35.91) ;
\draw [shift={(23.87,21.57)}, rotate = 90] [fill={rgb, 255:red, 0; green, 0; blue, 0 }  ][line width=0.08]  [draw opacity=0] (5.36,-2.57) -- (0,0) -- (5.36,2.57) -- cycle    ;
\draw    (89,20.11) -- (69.29,30.89) ;
\draw [shift={(82.74,23.53)}, rotate = 151.33] [fill={rgb, 255:red, 0; green, 0; blue, 0 }  ][line width=0.08]  [draw opacity=0] (5.36,-2.57) -- (0,0) -- (5.36,2.57) -- cycle    ;
\draw    (69.29,30.89) -- (49.58,41.66) ;
\draw [shift={(63.03,34.31)}, rotate = 151.33] [fill={rgb, 255:red, 0; green, 0; blue, 0 }  ][line width=0.08]  [draw opacity=0] (5.36,-2.57) -- (0,0) -- (5.36,2.57) -- cycle    ;
\draw    (23.87,15.44) -- (69.29,30.89) ;
\draw [shift={(42.7,21.84)}, rotate = 18.79] [fill={rgb, 255:red, 0; green, 0; blue, 0 }  ][line width=0.08]  [draw opacity=0] (5.36,-2.57) -- (0,0) -- (5.36,2.57) -- cycle    ;
\draw    (69.29,30.89) -- (92,38.61) ;
\draw [shift={(76.76,33.43)}, rotate = 18.79] [fill={rgb, 255:red, 0; green, 0; blue, 0 }  ][line width=0.08]  [draw opacity=0] (5.36,-2.57) -- (0,0) -- (5.36,2.57) -- cycle    ;
\draw    (1.16,7.72) -- (23.87,15.44) ;
\draw [shift={(8.64,10.26)}, rotate = 18.79] [fill={rgb, 255:red, 0; green, 0; blue, 0 }  ][line width=0.08]  [draw opacity=0] (5.36,-2.57) -- (0,0) -- (5.36,2.57) -- cycle    ;
\draw    (43.58,4.66) -- (23.87,15.44) ;
\draw [shift={(37.32,8.09)}, rotate = 151.33] [fill={rgb, 255:red, 0; green, 0; blue, 0 }  ][line width=0.08]  [draw opacity=0] (5.36,-2.57) -- (0,0) -- (5.36,2.57) -- cycle    ;
\draw    (23.87,15.44) -- (4.16,26.22) ;
\draw [shift={(17.61,18.86)}, rotate = 151.33] [fill={rgb, 255:red, 0; green, 0; blue, 0 }  ][line width=0.08]  [draw opacity=0] (5.36,-2.57) -- (0,0) -- (5.36,2.57) -- cycle    ;
\draw (44.46,10.2) node [anchor=north west][inner sep=0.75pt]    {\small $l$};
\draw (9,24.2) node [anchor=north west][inner sep=0.75pt]    {\small $g$};
\draw (28,-3) node [anchor=north west][inner sep=0.75pt]    {\small $h$};
\draw (8,-0.8) node [anchor=north west][inner sep=0.75pt]    {\small $i$};
\draw (25,25) node [anchor=north west][inner sep=0.75pt]    {\small $j$};
\draw (55.58,39.36) node [anchor=north west][inner sep=0.75pt]    {\small $x$};
\draw (73,13) node [anchor=north west][inner sep=0.75pt]    {\small $y$};
\draw (71.29,42.32) node [anchor=north west][inner sep=0.75pt]    {\small $z$};
\draw (79.5,26.7) node [anchor=north west][inner sep=0.75pt]    {\small $w$};
\end{tikzpicture}
\Bigg\rangle,
\end{equation}
and denote the local basis states in the rewritten rep-basis as:
\begin{equation}\label{eq:3.32}
    |\Psi ^{\nu \pi \phi \lambda }_{s\tilde{m}_{s} ,r\tilde{m}_{r}} \rangle =\Bigg|\begin{tikzpicture}[x=0.75pt,y=0.75pt,yscale=-0.9,xscale=0.9, baseline=(XXXX.south) ]
        \path (0,143);\path (198.99090576171875,0);\draw    ($(current bounding box.center)+(0,0.3em)$) node [anchor=south] (XXXX) {};
        \draw    (165.81,91.49) -- (195.66,103.53) ;
        \draw [shift={(176.93,95.98)}, rotate = 21.98] [fill={rgb, 255:red, 0; green, 0; blue, 0 }  ][line width=0.08]  [draw opacity=0] (5.36,-2.57) -- (0,0) -- (5.36,2.57) -- cycle    ;
        \draw    (150.89,85.47) -- (165.81,91.49) ;
        \draw [shift={(165.81,91.49)}, rotate = 21.98] [color={rgb, 255:red, 0; green, 0; blue, 0 }  ][fill={rgb, 255:red, 0; green, 0; blue, 0 }  ][line width=0.75]      (0, 0) circle [x radius= 2.01, y radius= 2.01]   ;
        \draw [shift={(154.55,86.94)}, rotate = 21.98] [fill={rgb, 255:red, 0; green, 0; blue, 0 }  ][line width=0.08]  [draw opacity=0] (5.36,-2.57) -- (0,0) -- (5.36,2.57) -- cycle    ;
        \draw [shift={(150.89,85.47)}, rotate = 21.98] [color={rgb, 255:red, 0; green, 0; blue, 0 }  ][fill={rgb, 255:red, 0; green, 0; blue, 0 }  ][line width=0.75]      (0, 0) circle [x radius= 2.01, y radius= 2.01]   ;
        \draw    (135.97,79.45) -- (150.89,85.47) ;
        \draw [shift={(139.63,80.92)}, rotate = 21.98] [fill={rgb, 255:red, 0; green, 0; blue, 0 }  ][line width=0.08]  [draw opacity=0] (5.36,-2.57) -- (0,0) -- (5.36,2.57) -- cycle    ;
        \draw    (180.74,65.6) -- (150.89,85.47) ;
        \draw [shift={(169.22,73.26)}, rotate = 146.34] [fill={rgb, 255:red, 0; green, 0; blue, 0 }  ][line width=0.08]  [draw opacity=0] (5.36,-2.57) -- (0,0) -- (5.36,2.57) -- cycle    ;
        \draw    (165.81,91.49) -- (165.81,135.25) ;
        \draw [shift={(165.81,109.27)}, rotate = 90] [fill={rgb, 255:red, 0; green, 0; blue, 0 }  ][line width=0.08]  [draw opacity=0] (5.36,-2.57) -- (0,0) -- (5.36,2.57) -- cycle    ;
        \draw    (135.97,79.45) -- (106.12,99.32) ;
        \draw [shift={(124.46,87.11)}, rotate = 146.34] [fill={rgb, 255:red, 0; green, 0; blue, 0 }  ][line width=0.08]  [draw opacity=0] (5.36,-2.57) -- (0,0) -- (5.36,2.57) -- cycle    ;
        \draw    (121.04,73.42) -- (135.97,79.45) ;
        \draw [shift={(135.97,79.45)}, rotate = 21.98] [color={rgb, 255:red, 0; green, 0; blue, 0 }  ][fill={rgb, 255:red, 0; green, 0; blue, 0 }  ][line width=0.75]      (0, 0) circle [x radius= 2.01, y radius= 2.01]   ;
        \draw [shift={(124.7,74.9)}, rotate = 21.98] [fill={rgb, 255:red, 0; green, 0; blue, 0 }  ][line width=0.08]  [draw opacity=0] (5.36,-2.57) -- (0,0) -- (5.36,2.57) -- cycle    ;
        \draw [shift={(121.04,73.42)}, rotate = 21.98] [color={rgb, 255:red, 0; green, 0; blue, 0 }  ][fill={rgb, 255:red, 0; green, 0; blue, 0 }  ][line width=0.75]      (0, 0) circle [x radius= 2.01, y radius= 2.01]   ;
        \draw    (121.04,48.36) -- (121.04,73.42) ;
        \draw [shift={(121.04,56.79)}, rotate = 90] [fill={rgb, 255:red, 0; green, 0; blue, 0 }  ][line width=0.08]  [draw opacity=0] (5.36,-2.57) -- (0,0) -- (5.36,2.57) -- cycle    ;
        \draw    (76.27,55.36) -- (121.04,73.42) ;
        \draw [shift={(94.86,62.86)}, rotate = 21.98] [fill={rgb, 255:red, 0; green, 0; blue, 0 }  ][line width=0.08]  [draw opacity=0] (5.36,-2.57) -- (0,0) -- (5.36,2.57) -- cycle    ;
        \draw    (91.2,29.46) -- (61.35,49.33) ;
        \draw [shift={(79.69,37.13)}, rotate = 146.34] [fill={rgb, 255:red, 0; green, 0; blue, 0 }  ][line width=0.08]  [draw opacity=0] (5.36,-2.57) -- (0,0) -- (5.36,2.57) -- cycle    ;
        \draw    (61.35,49.33) -- (76.27,55.36) ;
        \draw [shift={(76.27,55.36)}, rotate = 21.98] [color={rgb, 255:red, 0; green, 0; blue, 0 }  ][fill={rgb, 255:red, 0; green, 0; blue, 0 }  ][line width=0.75]      (0, 0) circle [x radius= 2.01, y radius= 2.01]   ;
        \draw [shift={(65.01,50.81)}, rotate = 21.98] [fill={rgb, 255:red, 0; green, 0; blue, 0 }  ][line width=0.08]  [draw opacity=0] (5.36,-2.57) -- (0,0) -- (5.36,2.57) -- cycle    ;
        \draw [shift={(61.35,49.33)}, rotate = 21.98] [color={rgb, 255:red, 0; green, 0; blue, 0 }  ][fill={rgb, 255:red, 0; green, 0; blue, 0 }  ][line width=0.75]      (0, 0) circle [x radius= 2.01, y radius= 2.01]   ;
        \draw    (46.43,43.31) -- (61.35,49.33) ;
        \draw [shift={(50.09,44.79)}, rotate = 21.98] [fill={rgb, 255:red, 0; green, 0; blue, 0 }  ][line width=0.08]  [draw opacity=0] (5.36,-2.57) -- (0,0) -- (5.36,2.57) -- cycle    ;
        \draw    (31.5,37.29) -- (46.43,43.31) ;
        \draw [shift={(46.43,43.31)}, rotate = 21.98] [color={rgb, 255:red, 0; green, 0; blue, 0 }  ][fill={rgb, 255:red, 0; green, 0; blue, 0 }  ][line width=0.75]      (0, 0) circle [x radius= 2.01, y radius= 2.01]   ;
        \draw [shift={(35.16,38.77)}, rotate = 21.98] [fill={rgb, 255:red, 0; green, 0; blue, 0 }  ][line width=0.08]  [draw opacity=0] (5.36,-2.57) -- (0,0) -- (5.36,2.57) -- cycle    ;
        \draw [shift={(31.5,37.29)}, rotate = 21.98] [color={rgb, 255:red, 0; green, 0; blue, 0 }  ][fill={rgb, 255:red, 0; green, 0; blue, 0 }  ][line width=0.75]      (0, 0) circle [x radius= 2.01, y radius= 2.01]   ;
        \draw    (76.27,55.36) -- (76.27,99.12) ;
        \draw [shift={(76.27,73.14)}, rotate = 90] [fill={rgb, 255:red, 0; green, 0; blue, 0 }  ][line width=0.08]  [draw opacity=0] (5.36,-2.57) -- (0,0) -- (5.36,2.57) -- cycle    ;
        \draw    (46.43,43.31) -- (16.58,63.18) ;
        \draw [shift={(34.92,50.98)}, rotate = 146.34] [fill={rgb, 255:red, 0; green, 0; blue, 0 }  ][line width=0.08]  [draw opacity=0] (5.36,-2.57) -- (0,0) -- (5.36,2.57) -- cycle    ;
        \draw    (31.5,12.23) -- (31.5,37.29) ;
        \draw [shift={(31.5,20.66)}, rotate = 90] [fill={rgb, 255:red, 0; green, 0; blue, 0 }  ][line width=0.08]  [draw opacity=0] (5.36,-2.57) -- (0,0) -- (5.36,2.57) -- cycle    ;
        \draw    (1.66,25.25) -- (31.5,37.29) ;
        \draw [shift={(12.78,29.73)}, rotate = 21.98] [fill={rgb, 255:red, 0; green, 0; blue, 0 }  ][line width=0.08]  [draw opacity=0] (5.36,-2.57) -- (0,0) -- (5.36,2.57) -- cycle    ;
\draw (7,34) node [anchor=north west][inner sep=0.75pt]    {\small $\mu $};
\draw (35.5,30) node [anchor=north west][inner sep=0.75pt]    {\small $\nu $};
\draw (20,20) node [anchor=north west][inner sep=0.75pt]    {\small $s$};
\draw (22,0) node [anchor=north west][inner sep=0.75pt]    {\small $\tilde{m}_{s}$};
\draw (29.5,55) node [anchor=north west][inner sep=0.75pt]    {\small $\rho $};
\draw (71,28) node [anchor=north west][inner sep=0.75pt]    {\small $\sigma $};
\draw (49,35) node [anchor=north west][inner sep=0.75pt]    {\small $\pi $};
\draw (60,55) node [anchor=north west][inner sep=0.75pt]    {\small $\phi $};
\draw (77,75) node [anchor=north west][inner sep=0.75pt]    {\small $\eta $};
\draw (89,66) node [anchor=north west][inner sep=0.75pt]    {\small $\lambda $};
\draw (110,56.18) node [anchor=north west][inner sep=0.75pt]    {\small $r$};
\draw (112,36.18) node [anchor=north west][inner sep=0.75pt]    {\small $\tilde{m}_{r}$};
\draw (124.04,62) node [anchor=north west][inner sep=0.75pt]    {\small $\delta $};
\draw (139,69) node [anchor=north west][inner sep=0.75pt]    {\small $\gamma $};
\draw (150,92) node [anchor=north west][inner sep=0.75pt]    {\small $\beta $};
\draw (170.74,100) node [anchor=north west][inner sep=0.75pt]    {\small $\alpha $};
\draw (117,91) node [anchor=north west][inner sep=0.75pt]    {\small $\psi $};
\draw (167,112) node [anchor=north west][inner sep=0.75pt]    {\small $\kappa $};
\draw (161,64) node [anchor=north west][inner sep=0.75pt]    {\small $\epsilon $};
\end{tikzpicture}
\Bigg\rangle,
\end{equation}
where we have omitted some Latin indices, and the indices $\tilde{m}_r$ and $\tilde{m}_s$ diagonalize the boundary vertex operators acting on the two vertices in \eqref{eq:3.32}. In the rewritten rep-basis, the boundary edge operator acts on an open plaquette outlined by the edges with degrees of freedom $\tilde{m}_s, s,\nu,\pi,\phi,\lambda,r$, and $\tilde{m}_r$, on which $\overline{\skew{-1}{\tilde}{C}_e^\text{GT}}$ is not diagonalized. 

Similar to the Fourier transform of vertex operators, we can compute the matrix elements of $\overline{\skew{-1}{\tilde}{C}_e^\text{GT}}$ in the local basis by inserting resolution of identity: 
\begin{equation}\label{eq:3.33}
    \begin{aligned}
        &\langle \Psi ^{\nu' \pi' \phi' \lambda' }_{s'\tilde{m}'_{s} ,r'\tilde{m}'_{r}}|\overline{\skew{-1}{\tilde}{C}_e^\text{GT}}|\Psi^{\nu \pi \phi \lambda }_{s\tilde{m}_{s} ,r\tilde{m}_{r}} \rangle\\
        ={}&\langle \Psi^{\nu' \pi' \phi' \lambda' }_{s'\tilde{m}'_{s} ,r'\tilde{m}'_{r}}|\sum_{ghijlxyzw\in G}\overline{C_e^\text{GT}}|\Psi_l\rangle\langle\Psi_l|\Psi^{\nu \pi \phi \lambda }_{s\tilde{m}_{s} ,r\tilde{m}_{r}} \rangle\\
        ={}&\sum_{ghijlxyzw\in G}\delta_{l\in K}\langle \Psi^{\nu' \pi' \phi' \lambda' }_{s'\tilde{m}'_{s} ,r'\tilde{m}'_{r}}|\Psi_l\rangle\langle\Psi_l|\Psi^{\nu \pi \phi \lambda }_{s\tilde{m}_{s} ,r\tilde{m}_{r}} \rangle\\
        ={}&\sum_{ghijlxyzw\in G}\sum_{(t,\alpha_t)\in L_A}\frac{|K|}{|G|}d_tD^t_{\alpha_t\alpha_t}(l)\langle \Psi^{\nu' \pi' \phi' \lambda' }_{s'\tilde{m}'_{s} ,r'\tilde{m}'_{r}}|\Psi_l\rangle\langle\Psi_l|\Psi^{\nu \pi \phi \lambda }_{s\tilde{m}_{s} ,r\tilde{m}_{r}} \rangle,
    \end{aligned}
\end{equation}
where use is made of that 
\begin{equation}\label{eq:3.34}
    \delta_{l\in K}=\sum_{(t,\alpha_t)\in L_A}\frac{|K|}{|G|}d_tD^t_{\alpha_t\alpha_t}(l).
\end{equation}
The proof of the the above equation and the simplification of the two inner products in \eqref{eq:3.33} are left in \autoref{appendix:A}. Here, we just write down the result: 
\begin{equation}\label{eq:3.35}
    \begin{aligned}
        &\langle \Psi^{\nu' \pi' \phi' \lambda' }_{s'\tilde{m}'_{s} ,r'\tilde{m}'_{r}}|\overline{\skew{-1}{\tilde}{C}_e^\text{GT}}|\Psi^{\nu \pi \phi \lambda }_{s\tilde{m}_{s} ,r\tilde{m}_{r}} \rangle\\
        ={}&\sum_{(t,\alpha_t)\in L_A}\frac{|K|}{|G|}\mathtt d_t\mathtt{v}_\nu\mathtt{v}_\pi\mathtt{v}_\phi\mathtt{v}_\lambda\mathtt{v}_s\mathtt{v}_r[\mathtt{v}']R^{\pi\sigma}_\phi\overline{R^{\pi'\sigma}_{\phi'}}C^{st;\tilde{m}_s\alpha_t}_{s'\tilde{m}_{s'}}C^{r\tilde{m}_r}_{tr';\alpha_t\tilde{m}_{r'}}\\
        &\;\;\;\;\;\;\;\;\;\;\;\;\;\;\times G^{\mu^*\nu s^*}_{ts'^*\nu'}G^{\rho\pi\nu^*}_{t\nu'^*\pi'}G^{\sigma^*\phi\pi^*}_{t\pi'^*\phi'}G^{\eta\lambda\phi^*}_{t\phi'^*\lambda'}G^{t\lambda'^*\lambda}_{\delta r^*r'}\\
        ={}&\sum_{(t,\alpha_t)\in L_A}\frac{|K|}{|G|}\mathtt{d}_t\mathtt{v}_\nu\mathtt{v}_\pi\mathtt{v}_\phi\mathtt{v}_\lambda\mathtt{v}_s\mathtt{v}_r[\mathtt{v}']R^{\pi\sigma}_\phi\overline{R^{\pi'\sigma}_{\phi'}}(C_{r'^*t^*r,\tilde{m}_{r'^*}\tilde{m}_{t^*}\tilde{m}_r})^*(\Omega^{r'^*})^{-1}_{\tilde{m}_{r'}\tilde{m}_{r'^*}}(\Omega^{t^*})^{-1}_{\alpha_t\tilde{m}_{t^*}}\\
        &\;\;\;\;\;\;\;\;\;\;\;\;\;\;\times (C_{s'^*st,\tilde{m}_{s'^*}\tilde{m}_s\alpha_t})^*(\Omega^{s'^*})^{-1}_{\tilde{m}_s'\tilde{m}_{s'^*}}G^{\mu^*\nu s^*}_{ts'^*\nu'}G^{\rho\pi\nu^*}_{t\nu'^*\pi'}G^{\sigma^*\phi\pi^*}_{t\pi'^*\phi'}G^{\eta\lambda\phi^*}_{t\phi'^*\lambda'}G^{t\lambda'^*\lambda}_{\delta r^*r'},
\end{aligned}
\end{equation}
where $\mathtt{v}_\mu=\sqrt{\mathtt{d}_\mu}$, and $\mathtt{v}_\mu\cdots \mathtt{v}_{\nu}[\mathtt{v}']:=\mathtt{v}_\mu\cdots \mathtt{v}_{\nu}\mathtt{v}_{\mu'}\cdots \mathtt{v}_{\nu'}$ is introduced as a shorthand notation. The local state $|\Psi^{\nu \pi \phi \lambda }_{s\tilde{m}_{s} ,r\tilde{m}_{r}} \rangle$ may be a $+1$ or zero eigenstate of the boundary vertex operators acting on the relevant vertices. In general, we would like to study the matrix elements of $\overline{\skew{-1}{\tilde}{C}_e^\text{GT}}$ in the local basis states free of any charge excitations. This is accomplished by acting the boundary vertex operators on the states $|\Psi^{\nu \pi \phi \lambda }_{s\tilde{m}_{s} ,r\tilde{m}_{r}} \rangle$ to project out all the charge excitations. Equivalently, we can simply replace the indices $\tilde{m}_s$ and $\tilde{m}_r$ in \eqref{eq:3.35} by $\alpha_s$ and $\alpha_r$ in \eqref{eq:3.30} and obtain 
\begin{equation}\label{eq:3.42}
    \begin{aligned}
        &\langle \Psi^{\nu' \pi' \phi' \lambda' }_{s'\alpha_{s'} ,r'\alpha_{r'}}|\overline{\skew{-1}{\tilde}{C}_e^\text{GT}}|\Psi^{\nu \pi \phi \lambda }_{s\alpha_{s} ,r\alpha_{r}} \rangle\\
        ={}&\sum_{(t,\alpha_t)\in L_A}\frac{|K|}{|G|}\mathtt{d}_t\mathtt{v}_\nu\mathtt{v}_\pi\mathtt{v}_\phi\mathtt{v}_\lambda\mathtt{v}_s\mathtt{v}_r[\mathtt{v}']R^{\pi\sigma}_\phi\overline{R^{\pi'\sigma}_{\phi'}}(C_{r'^*t^*r,\tilde{m}_{r'^*}\tilde{m}_{t^*}\alpha_r})^*(\Omega^{r'^*})^{-1}_{\alpha_{r'}\tilde{m}_{r'^*}}(\Omega^{t^*})^{-1}_{\alpha_t\tilde{m}_{t^*}}\\
        &\;\;\;\;\;\;\;\;\;\;\;\;\;\;\times (C_{s'^*st,\tilde{m}_{s'^*}\alpha_s\alpha_t})^*(\Omega^{s'^*})^{-1}_{\alpha_s'\tilde{m}_{s'^*}}G^{\mu^*\nu s^*}_{ts'^*\nu'}G^{\rho\pi\nu^*}_{t\nu'^*\pi'}G^{\sigma^*\phi\pi^*}_{t\pi'^*\phi'}G^{\eta\lambda\phi^*}_{t\phi'^*\lambda'}G^{t\lambda'^*\lambda}_{\delta r^*r'}\\
        ={}&\sum_{(t,\alpha_t)\in L_A}\frac{|K|}{|G|}\mathtt{v}_t(\mathtt{v}_\nu\mathtt{v}_\pi\mathtt{v}_\phi\mathtt{v}_\lambda[\mathtt{v}'])\mathtt{u}_s\mathtt{u}_r[\mathtt{u}']R^{\pi\sigma}_\phi\overline{R^{\pi'\sigma}_{\phi'}}G^{\mu^*\nu s^*}_{ts'^*\nu'}G^{\rho\pi\nu^*}_{t\nu'^*\pi'}G^{\sigma^*\phi\pi^*}_{t\pi'^*\phi'}G^{\eta\lambda\phi^*}_{t\phi'^*\lambda'}G^{t\lambda'^*\lambda}_{\delta r^*r'}\\
        &\;\;\;\;\;\;\;\;\;\;\;\;\;\;\times f_{r'^*\alpha_{r'}t^*\alpha_t r\alpha_r}f_{s'^*\alpha_{s'} s\alpha_s t\alpha_t},
    \end{aligned}
\end{equation}
where in the last equality we define a map $f:L_A\times L_A\times L_A\to\mathbb C$ as
\begin{equation}\label{eq:3.37}
    f_{c^*\alpha_c a\alpha_a b\alpha_b}=\sum_{\tilde{m}_{c^*}}\mathtt{u}_a\mathtt{u}_b\mathtt{u}_c(C_{c^*ab;\tilde{m}_{c^*}\alpha_a\alpha_b})^*(\Omega^{c^*})^{-1}_{\alpha_c\tilde{m}_{c^*}},\;\;\text{with}\;(a,\alpha_a),(b,\alpha_b),(c,\alpha_c)\in L_A,
\end{equation}
where $\mathtt{u}_a=\sqrt{\mathtt{v}_a}$. 

\subsection{Fourier transform of the boundary plaquette operators}
Finally, we discuss the Fourier transform of the boundary plaquette operators $\overline{\tilde{B}^\text{GT}_p}$. The local Hilbert space where $\overline{\tilde{B}^\text{GT}_p}$ acts on is shown in \autoref{fig:5}. 

\begin{figure}[ht]
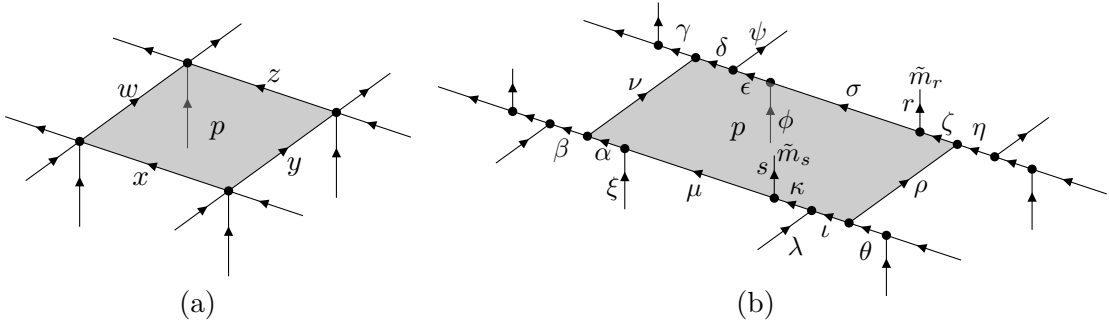

    \centering

\caption{The local Hilbert space where the boundary plaquette operators $\overline{B_p^\text{GT}}$ and $\overline{\tilde{B}_p^\text{GT}}$  act on. (a) is the group basis, denoted as $|wxyz\rangle$, and (b) is the rewritten rep-basis, denoted as $|\Psi^{\mu\alpha\nu\delta\epsilon\sigma\zeta\rho\iota\kappa}_{r\tilde{m}_r,s\tilde{m}_s}\rangle$. Here we have ignored all the group element labels and irreducible representation labels which will not appear in the matrix elements of the boundary plaquette operator. }
\label{fig:5}
\end{figure}

We can calculate the group elements of $\overline{\tilde{B}^\text{GT}_p}$ through a similar procedure of the calculation in the previous section, resulting in 
\begin{equation}\label{eq:3.38}
    \begin{aligned}
        &\langle \Psi^{\mu'\alpha'\nu'\delta'\epsilon'\sigma'\zeta'\rho'\iota'\kappa'}_{r'\tilde{m}_r',s'\tilde{m}_s'}|\overline{\tilde{B}_p^\text{GT}}|\Psi^{\mu\alpha\nu\delta\epsilon\sigma\zeta\rho\iota\kappa}_{r\tilde{m}_r,s\tilde{m}_s}\rangle\\
        ={}&\sum\delta_{xw\bar z\bar y,e}\langle \Psi^{\mu'\alpha'\nu'\delta'\epsilon'\sigma'\zeta'\rho'\iota'\kappa'}_{r'\tilde{m}_r',s'\tilde{m}_s'}|xyzw\rangle\langle xyzw|\Psi^{\mu\alpha\nu\delta\epsilon\sigma\zeta\rho\iota\kappa}_{r\tilde{m}_r,s\tilde{m}_s}\rangle\\
        ={}&\sum\sum_{t\in L_G,m_t}\frac{1}{|G|}d_t D^t_{m_t m_t}(xw\bar z\bar y)\langle \Psi^{\mu'\alpha'\nu'\delta'\epsilon'\sigma'\zeta'\rho'\iota'\kappa'}_{r'\tilde{m}_r',s'\tilde{m}_s'}|xyzw\rangle\langle xyzw|\Psi^{\mu\alpha\nu\delta\epsilon\sigma\zeta\rho\iota\kappa}_{r\tilde{m}_r,s\tilde{m}_s}\rangle,
    \end{aligned}
\end{equation}
where the first summation is over all the relevant group element labels in the group basis state $|xyzw\rangle$, including those that we haven't explicitly specified in \autoref{fig:5}. Here, we shall focus on the boundary degrees of freedom $(r,\tilde{m}_r)$ and $(s,\tilde{m}_s)$, which will be projected into the set $L_A$ by boundary vertex operators. We want to find out whether the Fourier-transformed boundary plaquette operators would introduce new structures on the set $L_A$, like what we have found in the Fourier transformation of the boundary edge operator where a map $f:L_A\times L_A\times L_A\to\mathbb R$ emerges. The answer is however negative. 

Since evaluating \eqref{eq:3.38} is very tedious, and since a similar procedure will be shown in \autoref{sec:5}, we simply present the result of \eqref{eq:3.38}:
\begin{equation}\label{eq:3.45}
        \langle \Psi'|\overline{\tilde{B}_p^\text{GT}}|\Psi\rangle=\sum_{t\in L_G}\mathtt d_t \mathtt{v}_\mu\mathtt{v}_\alpha\cdots\mathtt{v}_\kappa[\mathtt v'] R^{\mu s}_\kappa\overline{R^{\phi\epsilon}_\sigma}\overline{R^{\mu's}_{\kappa}}R^{\phi\epsilon'}_{\sigma'}G^{\lambda\iota\kappa}_{t\kappa'\iota'}G^{\theta\rho^*\iota}_{t\iota'\rho'^*}\cdots G^{s\kappa\mu}_{t\mu'\kappa'},
\end{equation}
where $|\Psi\rangle=|\Psi^{\mu\alpha\nu\delta\epsilon\sigma\zeta\rho\iota\kappa}_{r\tilde{m}_r,s\tilde{m}_s}\rangle$, $\langle\Psi'|=\langle\Psi^{\mu'\alpha'\nu'\delta'\epsilon'\sigma'\zeta'\rho'\iota'\kappa'}_{r'\tilde{m}_r',s'\tilde{m}_s'}|$, and the ellipsis between those $6j$-symbols represent other $6j$-symbols corresponding to the vertices along the boundary of plaquette $p$. 

While factors $R^{\mu s}_\kappa$ and $\overline{R^{\mu's}_{\kappa'}}$ in \eqref{eq:3.45} depend on $s$, which labels the tail, these $R$-matrices do not introduce any new structures on $L_A$ because $s$ should be viewed as an element of $L_G$ here. In fact, each Fourier-transformed boundary plaquette operators $\overline{\tilde{B}^\text{GT}_p}$ is an identity operator within the subspace spanned by $(s,\tilde{m}_s)$ and $(r,\tilde{m}_r)$. Hence, unlike $\overline{\tilde{A}_v^\text{GT}}$ and $\overline{\skew{-1}{\tilde}{C}_e^\text{GT}}$, the boundary plaquette operators $\overline{\tilde{B}^\text{GT}_p}$ do not project out any degrees of freedom on the tails attached to the relevant boundary vertices, and thus do not provide any new gapped boundary condition for the Fourier-transformed 3-dimensional GT model. 

So far we have Fourier-transformed and rewritten the three-dimensional GT model with gapped boundaries on a trivalent lattice $\tilde{\Gamma}$. In the sequel sections, we shall study the physics revealed by the Fourier transform. 

\section{Emergence of Frobenius algebras and boundary charge condensation}\label{sec:4}
In this section, we first examine the gapped boundary condition of the Fourier-transformed GT model, then characterize the gapped boundaries by charge condensation, and finally explain why the Fourier transform makes the physics at the gapped boundaries of the GT model more explicit than it was before the transformation.

\subsection{The gapped boundary condition of the Fourier-transformed model}
In the previous section, we have found that each Fourier-transformed boundary vertex operator $\overline{\tilde{A}_v^\text{GT}}$ projects the degrees of freedom $(s,\tilde{m}_s)$ on the tail attached to the vertex $v$ into a set $L_A$. Along with the Fourier-transformed boundary edge operators, an emergent map $f:L_A\times L_A\times L_A\to\mathbb R$ appears. Nevertheless, within the subspace spanned by $(s,\tilde{m}_s)$, the relevant boundary plaquette operators behave as identity operators. Hence, the gapped boundary condition of the Fourier-transformed three-dimensional GT model can be specified by a pair $(L_A,f)$, which is determined by the input data $G$ and the boundary condition $K$ of the original GT model. 

In the two-dimensional case\cite{Wang2020b}, the gapped boundary condition $(L_A,f)$ indeed forms a Frobenius algebra $A$, which is an object in the UBFC $\mathcal{R}ep(G)$. Generally, an element of $L_A$ is denoted as a pair $(s,\alpha_s)$ (or simply $s\alpha_s$), where $s$ is a simple object of a UBFC $\mathfrak{F}$. The multiplicity of $s$ in $A$ is denoted by $|s|$ and refers to the number of different pairs $(s,\alpha_s)$ with the same $s$. The multiplication is a map $f:L_A\times L_A\times L_A\to\mathbb C$ that satisfies the following associativity and non-degeneracy: 
\begin{equation}\label{eq:4.1}
    \begin{aligned}
\sum_{c\alpha_c}f_{a\alpha_ab\alpha_bc^*\alpha_c}f_{c\alpha_cr\alpha_rs^*\alpha_s}G^{abc^*}_{rs^*t}\mathtt{v}_c\mathtt{v}_t&=\sum_{\alpha_t}f_{a\alpha_at\alpha_ts^*\alpha_s}f_{b\alpha_br\alpha_rt^*\alpha_t}\\
        f_{b\alpha_bb'^*\alpha_{b'}0}&=\delta_{bb'}\delta_{\alpha_b\alpha_b'}\beta_b,
    \end{aligned}
\end{equation}
where 0 is the unit element of $A$ and has multiplicity 1, i.e. $0\equiv(0,1)$. Here, $\mathtt v_c=\sqrt{\mathtt{d}_c}$ with $\mathtt{d}_c$ the quantum dimension of element $c$. That the Frobenius algebra $A$ defined above is an object of the corresponding UBFC $\mathfrak{F}$ is understood by writing $A$ as $A=\bigoplus_{s|_{(s,\alpha_s)\in L_A}}s^{\oplus|s|}$, which is generally a non-simple object in $\mathfrak{F}$. For computational convenience, one may also write $A=\bigoplus_{(s,\alpha_s)\in L_A}s_{\alpha_s}$, explicitly treating different appearances of $s$ as distinct elements of $A$. 
 
Therefore, given any finite group $G$ and a subgroup $K\subseteq G$, the set $L_A$ defined in \eqref{eq:3.30} equipped with the multiplication $f$ defined by \eqref{eq:3.37} forms a Frobenius algebra $A_{G,K}=(L_A,f)_{G,K}$, specifying a gapped boundary of the Fourier-transformed three-dimensional GT model.

\subsection{Charge condensation at the gapped boundary}\label{sec:4.2}
We will subsequently see that the boundary input data $A_{G,K}$ of the Fourier-transformed model precisely mirrors the charge condensation at the boundary. Recall that according to \eqref{eq:3.29}, each pair $(s,\alpha_s)$ labels a local $+1$ eigenstate $|\Psi_{s\alpha_s}\rangle$ at vertex $v$ of $\overline{\tilde{A}_v^\text{GT}}$, which is a projector. All such eigenstates sharing the same hidden labels span a subspace in the $d_s$-dimensional representation space $V_s$. The dimension $|s|$ of this subspace is the number of pairs $(s,\alpha_s)$ with the same $s$. The emergence of the Frobenius algebra then identifies this dimension $|s|$ as the multiplicity $|s|$ of $s$ appearing in $A_{G,K}$. This identification is closely related to the mechanism of charge condensation in three-dimensional topological orders. We briefly describe this relation as follows. 

In two-dimensional topological orders, there exist only point-like excitations called anyons including charges, fluxes, and dyons. The anyon condensation in two-dimensional topological orders has been extensively studied recently\cite{Bais2009,HungWan2014,Kong2013,Gu2014a,Hung2015,Wan2017,Burnell2018,Hu2022}. On the other hand, point-like excitations in three-dimensional topological orders are pure charges, with additional loop-like and string-like excitations present. Nevertheless, the investigation of charge condensation in three-dimensional topological orders remains limited to specific cases\cite{Zhao2022a}. 

Generally, in a topological order $\mathcal{C}$, certain types of elementary excitations may condense and cause a phase transition that takes the topological order to a simpler child topological order $\mathcal{U}$. In an extreme case, $\mathcal{U}$ could be merely a vacuum (a symmetry-protected topological phase precisely speaking)\cite{Hung2013,Gu2014a,Hu2022}, rendering the original topological order entirely broken. This process can also be viewed from the perspective of creating a gapped domain wall that separates $\mathcal{C}$ and $\mathcal{U}$\cite{Hung2015}. When $\mathcal{U}$ is a vacuum, we say certain types of elementary excitations of $\mathcal{C}$ can move to and condense at the gapped boundary between $\mathcal{C}$ and the vacuum. 

Layer construction offers another interpretation for condensation of charge excitations at the boundary of a three-dimensional topological order\cite{jian2014,Gaiotto2019}. Here, three-dimensional topological orders are achieved by sequentially stacking two-dimensional topological orders. For instance, a three-dimensional GT model with input data $G$ can be built by stacking the QD model with input data $G$ as well. The layers are then glued together by condensing specific types of quasiparticle pairs between them. Nevertheless, at the final layer of the two-dimensional topological order, which is the boundary of the three-dimensional topological order, excitations at the boundary are allowed to condense separately. Different boundary conditions correspond to different condensates at the boundary.

While the general classification of loop-like excitations in three-dimensional topological orders remains unclear, the irreducible representations of group $G$ still classify pure charge excitations in the three-dimensional GT model. Therefore, we can investigate charge condensation in three-dimensional topological orders. For the three-dimensional GT model with gapped boundaries with input data $G$, the gapped boundary condition is specified by the subgroup $K\subseteq G$. Consider the case where $K=\{e\}$ called rough boundary condition. Then by Eqs.\eqref{eq:3.29} and \eqref{eq:3.30}, all irreducible representations must appear in $L_A$. Namely, $L_A=\{(s,\alpha_s)|s\in L_G,\alpha_s=1,2,\cdots,d_s\}$. Since $s$ labels a type of pure charge excitation in the bulk, and since each pair $(s,\alpha_s)$ is an independent element of $A_{G,K}$, the pure charge $s$ splits into $d_s$ pieces, each of which condenses at the boundary. Thus, the multiplicity of the charge $s$ in the boundary condensate is $d_s=|s|$, the multiplicity of $s$ in the Frobenius algebra $A_{G,K}$. In the case where $K$ is a nontrivial subgroup, we may have for some $s$, only a subset $\{(s,\alpha_s)|\alpha=1,2,\cdots,|s|<d_s\}\subset L_A$. That is, although the pure charge $s$ splits into $d_s$ pieces, only $|s|$ pieces of them contribute to the boundary condensate. 

\begin{figure}[ht]
    \centering 
\begin{tikzpicture}[x=0.75pt,y=0.75pt,yscale=-0.7,xscale=0.7]

\draw    (254.53,60) -- (557.47,60) ;
\draw    (214.53,100) -- (517.47,100) ;
\draw    (174.53,140) -- (477.47,140) ;
\draw    (134.53,180) -- (437.47,180) ;
\draw    (361.41,45.6) -- (211.47,195.55) ;
\draw    (168.53,180) -- (168.53,205.4) ;
\draw    (228.68,180) -- (228.68,205.4) ;
\draw    (420.41,46.6) -- (270.47,196.55) ;
\draw    (479.41,46.6) -- (329.47,196.55) ;
\draw    (539.41,45.6) -- (389.47,195.55) ;
\draw    (302.41,45.6) -- (152.47,195.55) ;
\draw    (286,180) -- (286,205.4) ;
\draw    (346,180) -- (346,205.4) ;
\draw    (406,180) -- (406,205.4) ;
\draw    (208.53,140) -- (208.53,165.4) ;
\draw    (268.68,140) -- (268.68,165.4) ;
\draw    (327,140) -- (327,165.4) ;
\draw    (386,140) -- (386,165.4) ;
\draw    (446,140) -- (446,165.4) ;
\draw    (248.53,100) -- (248.53,125.4) ;
\draw    (307.68,100) -- (307.68,125.4) ;
\draw    (367,100) -- (367,125.4) ;
\draw    (426,100) -- (426,125.4) ;
\draw    (486,100) -- (486,125.4) ;
\draw    (288.53,60) -- (288.53,85.4) ;
\draw    (348.68,60) -- (348.68,85.4) ;
\draw    (407,60) -- (407,85.4) ;
\draw    (467,60) -- (467,85.4) ;
\draw    (526,60) -- (526,85.4) ;
\draw    (254.53,291.5) -- (557.47,291.5) ;
\draw    (201.53,331.5) -- (504.47,331.5) ;
\draw    (149.53,371.5) -- (452.47,371.5) ;
\draw    (96.53,412.5) -- (399.47,412.5) ;
\draw    (130.53,412.5) -- (130.53,437.9) ;
\draw    (190.68,412.5) -- (190.68,437.9) ;
\draw    (248,412.5) -- (248,437.9) ;
\draw    (308,412.5) -- (308,437.9) ;
\draw    (370,412.5) -- (370,437.9) ;
\draw    (183.53,371.5) -- (183.53,396.9) ;
\draw    (243.68,371.5) -- (243.68,396.9) ;
\draw    (302,371.5) -- (302,396.9) ;
\draw    (361,371.5) -- (361,396.9) ;
\draw    (423,371.5) -- (423,396.9) ;
\draw    (235.53,331.5) -- (235.53,356.9) ;
\draw    (294.68,331.5) -- (294.68,356.9) ;
\draw    (354,331.5) -- (354,356.9) ;
\draw    (413,331.5) -- (413,356.9) ;
\draw    (475,331.5) -- (475,356.9) ;
\draw    (288.53,291.5) -- (288.53,316.9) ;
\draw    (349.68,291.5) -- (349.68,316.9) ;
\draw    (407,291.5) -- (407,316.9) ;
\draw    (467,291.5) -- (467,316.9) ;
\draw    (528,291.5) -- (528,316.9) ;
\draw    (343.47,291.08) -- (303,331.55) ;
\draw    (401.47,291.08) -- (361,331.55) ;
\draw    (461.47,291.08) -- (421,331.55) ;
\draw    (282.47,291.08) -- (242,331.55) ;
\draw    (522.47,291.08) -- (482,331.55) ;
\draw    (289.47,331.08) -- (249,371.55) ;
\draw    (349.47,331.08) -- (309,371.55) ;
\draw    (408.47,331.08) -- (368,371.55) ;
\draw    (230.47,331.08) -- (190,371.55) ;
\draw    (468.47,332.08) -- (429,371.55) ;
\draw    (236.47,372.08) -- (196,412.55) ;
\draw    (296.47,372.08) -- (256,412.55) ;
\draw    (355.47,372.08) -- (315,412.55) ;
\draw    (177.47,372.08) -- (137,412.55) ;
\draw    (416.47,372.08) -- (375.81,412.74) ;
\draw [color={rgb, 255:red, 208; green, 2; blue, 27 }  ,draw opacity=1 ]   (274,277.94) .. controls (275.67,279.61) and (275.67,281.27) .. (274,282.94) .. controls (272.33,284.61) and (272.33,286.27) .. (274,287.94) -- (274,291.45) -- (274,291.45) ;
\draw [color={rgb, 255:red, 208; green, 2; blue, 27 }  ,draw opacity=1 ]   (335,277.94) .. controls (336.67,279.61) and (336.67,281.27) .. (335,282.94) .. controls (333.33,284.61) and (333.33,286.27) .. (335,287.94) -- (335,291.45) -- (335,291.45) ;
\draw [color={rgb, 255:red, 208; green, 2; blue, 27 }  ,draw opacity=1 ]   (394,277.94) .. controls (395.67,279.61) and (395.67,281.27) .. (394,282.94) .. controls (392.33,284.61) and (392.33,286.27) .. (394,287.94) -- (394,291.45) -- (394,291.45) ;
\draw [color={rgb, 255:red, 208; green, 2; blue, 27 }  ,draw opacity=1 ]   (452,277.94) .. controls (453.67,279.61) and (453.67,281.27) .. (452,282.94) .. controls (450.33,284.61) and (450.33,286.27) .. (452,287.94) -- (452,291.45) -- (452,291.45) ;
\draw [color={rgb, 255:red, 208; green, 2; blue, 27 }  ,draw opacity=1 ]   (514,277.94) .. controls (515.67,279.61) and (515.67,281.27) .. (514,282.94) .. controls (512.33,284.61) and (512.33,286.27) .. (514,287.94) -- (514,291.45) -- (514,291.45) ;
\draw [color={rgb, 255:red, 208; green, 2; blue, 27 }  ,draw opacity=1 ]   (221,317.94) .. controls (222.67,319.61) and (222.67,321.27) .. (221,322.94) .. controls (219.33,324.61) and (219.33,326.27) .. (221,327.94) -- (221,331.45) -- (221,331.45) ;
\draw [color={rgb, 255:red, 208; green, 2; blue, 27 }  ,draw opacity=1 ]   (282,317.94) .. controls (283.67,319.61) and (283.67,321.27) .. (282,322.94) .. controls (280.33,324.61) and (280.33,326.27) .. (282,327.94) -- (282,331.45) -- (282,331.45) ;
\draw [color={rgb, 255:red, 208; green, 2; blue, 27 }  ,draw opacity=1 ]   (341,317.94) .. controls (342.67,319.61) and (342.67,321.27) .. (341,322.94) .. controls (339.33,324.61) and (339.33,326.27) .. (341,327.94) -- (341,331.45) -- (341,331.45) ;
\draw [color={rgb, 255:red, 208; green, 2; blue, 27 }  ,draw opacity=1 ]   (399,317.94) .. controls (400.67,319.61) and (400.67,321.27) .. (399,322.94) .. controls (397.33,324.61) and (397.33,326.27) .. (399,327.94) -- (399,331.45) -- (399,331.45) ;
\draw [color={rgb, 255:red, 208; green, 2; blue, 27 }  ,draw opacity=1 ]   (461,317.94) .. controls (462.67,319.61) and (462.67,321.27) .. (461,322.94) .. controls (459.33,324.61) and (459.33,326.27) .. (461,327.94) -- (461,331.45) -- (461,331.45) ;
\draw [color={rgb, 255:red, 208; green, 2; blue, 27 }  ,draw opacity=1 ]   (170,357.94) .. controls (171.67,359.61) and (171.67,361.27) .. (170,362.94) .. controls (168.33,364.61) and (168.33,366.27) .. (170,367.94) -- (170,371.45) -- (170,371.45) ;
\draw [color={rgb, 255:red, 208; green, 2; blue, 27 }  ,draw opacity=1 ]   (231,357.94) .. controls (232.67,359.61) and (232.67,361.27) .. (231,362.94) .. controls (229.33,364.61) and (229.33,366.27) .. (231,367.94) -- (231,371.45) -- (231,371.45) ;
\draw [color={rgb, 255:red, 208; green, 2; blue, 27 }  ,draw opacity=1 ]   (290,357.94) .. controls (291.67,359.61) and (291.67,361.27) .. (290,362.94) .. controls (288.33,364.61) and (288.33,366.27) .. (290,367.94) -- (290,371.45) -- (290,371.45) ;
\draw [color={rgb, 255:red, 208; green, 2; blue, 27 }  ,draw opacity=1 ]   (348,357.94) .. controls (349.67,359.61) and (349.67,361.27) .. (348,362.94) .. controls (346.33,364.61) and (346.33,366.27) .. (348,367.94) -- (348,371.45) -- (348,371.45) ;
\draw [color={rgb, 255:red, 208; green, 2; blue, 27 }  ,draw opacity=1 ]   (410,357.94) .. controls (411.67,359.61) and (411.67,361.27) .. (410,362.94) .. controls (408.33,364.61) and (408.33,366.27) .. (410,367.94) -- (410,371.45) -- (410,371.45) ;
\draw [color={rgb, 255:red, 208; green, 2; blue, 27 }  ,draw opacity=1 ]   (123,398.94) .. controls (124.67,400.61) and (124.67,402.27) .. (123,403.94) .. controls (121.33,405.61) and (121.33,407.27) .. (123,408.94) -- (123,412.45) -- (123,412.45) ;
\draw [color={rgb, 255:red, 208; green, 2; blue, 27 }  ,draw opacity=1 ]   (184,398.94) .. controls (185.67,400.61) and (185.67,402.27) .. (184,403.94) .. controls (182.33,405.61) and (182.33,407.27) .. (184,408.94) -- (184,412.45) -- (184,412.45) ;
\draw [color={rgb, 255:red, 208; green, 2; blue, 27 }  ,draw opacity=1 ]   (243,398.94) .. controls (244.67,400.61) and (244.67,402.27) .. (243,403.94) .. controls (241.33,405.61) and (241.33,407.27) .. (243,408.94) -- (243,412.45) -- (243,412.45) ;
\draw [color={rgb, 255:red, 208; green, 2; blue, 27 }  ,draw opacity=1 ]   (301,398.94) .. controls (302.67,400.61) and (302.67,402.27) .. (301,403.94) .. controls (299.33,405.61) and (299.33,407.27) .. (301,408.94) -- (301,412.45) -- (301,412.45) ;
\draw [color={rgb, 255:red, 208; green, 2; blue, 27 }  ,draw opacity=1 ]   (363,398.94) .. controls (364.67,400.61) and (364.67,402.27) .. (363,403.94) .. controls (361.33,405.61) and (361.33,407.27) .. (363,408.94) -- (363,412.45) -- (363,412.45) ;

\draw   (313,250) -- (316.85,250) -- (316.85,221.63) -- (324.55,221.63) -- (324.55,250) -- (328.39,250) -- (320.7,261.63) -- cycle ;

\end{tikzpicture}
\caption{Upper: The original boundary lattice. Lower: The Fourier-transformed boundary lattice. Red wiggly lines are tails attached to original vertices, with degrees of freedom taking value in Frobenius algebra $A_{G,K}$. The degrees of freedom on the black straight lines still take value in $\mathcal{R}ep(G)$. }\label{figq}
\end{figure}
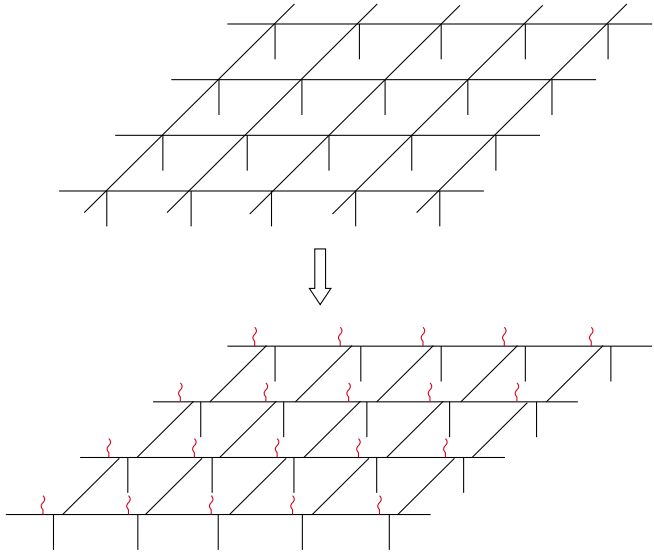

\subsection{The Boundary of the Fourier-transformed three-dimensional GT model}
In the three-dimensional GT model with a bulk gauge group $G$, the boundary is specified by a subgroup $K\subseteq G$. In view of input data and the boundary Hamiltonian, the boundary appears to be a standalone $2$-dimensional QD model with gauge group $K$. Consequently, the effects at the gapped boundary of the GT model that originate from the bulk, like charge condensation at the boundary, are not clearly reflected from the boundary gauge group and Hamiltonian. Nevertheless, unlike the Fourier-transformed $2$-dimensional QD model\cite{Wang2020b}, the Fourier-transformed boundary of the GT model cannot be mistaken as a standalone $2$-dimensional LW model. Moreover, Fourier transform enables us to accurately account for the charge condensation at the boundary within the Fourier-transformed input data of the boundary, explained as follows.

As depicted in \autoref{figq}, upon the Fourier transform, the original cubic lattice becomes trivalent with an additional tail (dangling edge) attached to each original vertex. In a ground state, the tails in the bulk are projected by the vertex operators into the trivial representation of $G$; however, tails on the boundary may survive such projection, depending on the boundary conditions specified by the boundary Frobenius algebra $A_{G,K}$. As discussed in \autoref{sec:4.2}, this boundary Frobenius algebra $A_{G,K}$ determines the charge condensation at the boundary. Since $A_{G,K}$ depends on both the boundary group $K$ and the bulk group $G$ of the original GT model, it is now clear that the gapped boundary is not a standalone two-dimensional model. In fact, as shown in \autoref{sec:5}, the Fourier transformed GT model with a gapped boundary can be mapped to the WW model with a gapped boundary LW model.  

\section{Mapping the three-dimensional GT model to WW model}\label{sec:5}
The 2-dimensional QD model with group $G$ as its input data has already been proven to be identified with a LW model with UFC $\mathcal{R}ep(G)$ as its input data, via Fourier transform\cite{Wang2020b}. A natural question arises: will the Fourier transform map the GT model, i.e., the 3-dimensional version of the QD model, to the WW model, i.e., the three-dimensional version of the LW model? The answer is yes. In this section, we will show that the Fourier transform defined in \autoref{subsec:3.2} indeed maps the three-dimensional GT model with input data $G$ to the WW model with UBFC $\mathcal{R}ep(G)$ as its input data. Since the original WW model doesn't have a boundary, and since there has not been any systematic construction of the gapped boundary of the WW model, we shall consider the bulk first and then the boundary. We also suppose that all the representations of $G$ are self-dual for simplicity, as was done in \cite{Walker2011}. It is straightforward to generalize to the case where non-self-dual representations exist. Details of the definition of the WW model can be found in \autoref{appendix:A0}. 

\subsection{Mapping the bulk Hilbert space}
By \eqref{eq:3.23} and the basis rewriting in \autoref{fig:4}, one can always rewrite a Fourier-transformed six-valent vertex as a trivalent lattice as shown in \autoref{fig:6}, with an extra tail attached to the vertex. The edges are labeled by irreducible representations $\mu\in L_G$ of group $G$. As the representations of a finite group $G$ do form a UBFC $\mathcal{R}ep(G)$, the sub Hilbert space of the three-dimensional GT model where all the degrees of freedom on the tails are restricted to the trivial representation, denoted as $\tilde{\mathcal{H}}^\text{GT}_0$, is the same as the Hilbert space of the WW model. Moreover, since \eqref{eq:3.23} requires that each vertex in the rewritten rep-basis of the Fourier-transformed three-dimensional GT model satisfies the fusion rules, all the states in $\tilde{\mathcal{H}}^\text{GT}_0$ are already the $+1$ eigenstates of the vertex operators in the WW model. Although the zero eigenstates of vertex operators do not appear in $\tilde{\mathcal{H}}^\text{GT}_0$, no loss of information is caused because the WW model does not contain charge and dyon excitations, just like the LW model. 

\subsection{Mapping the bulk Hamiltonian}
In order to Fourier-transform the bulk plaquette operator of the GT model, we consider two plaquette states, $|\Psi_{abcdpqruvw}\rangle$ defined in \eqref{eq:5.4} and state $|g_1g_2g_3g_4\rangle$ defined by 
\begin{equation}
    |g_1g_2g_3g_4\rangle=\begin{tikzpicture}[x=0.75pt,y=0.75pt,yscale=-0.7,xscale=0.7, baseline=(XXXX.south) ]
    \path (0,167);\path (227.99087524414062,0);\draw    ($(current bounding box.center)+(0,0.3em)$) node [anchor=south] (XXXX) {};
    \draw    (102.12,47.74) -- (102.12,95.44) ;
    \draw [shift={(102.12,67.49)}, rotate = 90] [fill={rgb, 255:red, 0; green, 0; blue, 0 }  ][line width=0.08]  [draw opacity=0] (5.36,-2.57) -- (0,0) -- (5.36,2.57) -- cycle    ;
    \draw  [draw opacity=0][fill={rgb, 255:red, 155; green, 155; blue, 155 }  ,fill opacity=0.5 ] (102.12,47.74) -- (184.71,75.42) -- (124.88,119.3) -- (42.29,91.62) -- cycle ;
    \draw    (102.12,47.74) -- (184.71,75.42) ;
    \draw [shift={(139.53,60.27)}, rotate = 18.53] [fill={rgb, 255:red, 0; green, 0; blue, 0 }  ][line width=0.08]  [draw opacity=0] (5.36,-2.57) -- (0,0) -- (5.36,2.57) -- cycle    ;
    \draw    (102.12,47.74) -- (42.29,91.62) ;
    \draw [shift={(42.29,91.62)}, rotate = 143.74] [color={rgb, 255:red, 0; green, 0; blue, 0 }  ][fill={rgb, 255:red, 0; green, 0; blue, 0 }  ][line width=0.75]      (0, 0) circle [x radius= 2.01, y radius= 2.01]   ;
    \draw [shift={(75.51,67.25)}, rotate = 143.74] [fill={rgb, 255:red, 0; green, 0; blue, 0 }  ][line width=0.08]  [draw opacity=0] (5.36,-2.57) -- (0,0) -- (5.36,2.57) -- cycle    ;
    \draw [shift={(102.12,47.74)}, rotate = 143.74] [color={rgb, 255:red, 0; green, 0; blue, 0 }  ][fill={rgb, 255:red, 0; green, 0; blue, 0 }  ][line width=0.75]      (0, 0) circle [x radius= 2.01, y radius= 2.01]   ;
    \draw    (184.71,75.42) -- (124.88,119.3) ;
    \draw [shift={(124.88,119.3)}, rotate = 143.74] [color={rgb, 255:red, 0; green, 0; blue, 0 }  ][fill={rgb, 255:red, 0; green, 0; blue, 0 }  ][line width=0.75]      (0, 0) circle [x radius= 2.01, y radius= 2.01]   ;
    \draw [shift={(158.1,94.93)}, rotate = 143.74] [fill={rgb, 255:red, 0; green, 0; blue, 0 }  ][line width=0.08]  [draw opacity=0] (5.36,-2.57) -- (0,0) -- (5.36,2.57) -- cycle    ;
    \draw [shift={(184.71,75.42)}, rotate = 143.74] [color={rgb, 255:red, 0; green, 0; blue, 0 }  ][fill={rgb, 255:red, 0; green, 0; blue, 0 }  ][line width=0.75]      (0, 0) circle [x radius= 2.01, y radius= 2.01]   ;
    \draw    (42.29,91.62) -- (124.88,119.3) ;
    \draw [shift={(79.7,104.15)}, rotate = 18.53] [fill={rgb, 255:red, 0; green, 0; blue, 0 }  ][line width=0.08]  [draw opacity=0] (5.36,-2.57) -- (0,0) -- (5.36,2.57) -- cycle    ;
    \draw    (124.88,119.3) -- (166.17,133.14) ;
    \draw [shift={(141.64,124.91)}, rotate = 18.53] [fill={rgb, 255:red, 0; green, 0; blue, 0 }  ][line width=0.08]  [draw opacity=0] (5.36,-2.57) -- (0,0) -- (5.36,2.57) -- cycle    ;
    \draw    (184.71,75.42) -- (226,89.26) ;
    \draw [shift={(201.47,81.03)}, rotate = 18.53] [fill={rgb, 255:red, 0; green, 0; blue, 0 }  ][line width=0.08]  [draw opacity=0] (5.36,-2.57) -- (0,0) -- (5.36,2.57) -- cycle    ;
    \draw    (60.83,33.9) -- (102.12,47.74) ;
    \draw [shift={(77.59,39.52)}, rotate = 18.53] [fill={rgb, 255:red, 0; green, 0; blue, 0 }  ][line width=0.08]  [draw opacity=0] (5.36,-2.57) -- (0,0) -- (5.36,2.57) -- cycle    ;
    \draw    (1,77.78) -- (42.29,91.62) ;
    \draw [shift={(17.76,83.4)}, rotate = 18.53] [fill={rgb, 255:red, 0; green, 0; blue, 0 }  ][line width=0.08]  [draw opacity=0] (5.36,-2.57) -- (0,0) -- (5.36,2.57) -- cycle    ;
    \draw    (42.29,91.62) -- (42.29,139.32) ;
    \draw [shift={(42.29,111.37)}, rotate = 90] [fill={rgb, 255:red, 0; green, 0; blue, 0 }  ][line width=0.08]  [draw opacity=0] (5.36,-2.57) -- (0,0) -- (5.36,2.57) -- cycle    ;
    \draw    (124.88,119.3) -- (124.88,167) ;
    \draw [shift={(124.88,139.05)}, rotate = 90] [fill={rgb, 255:red, 0; green, 0; blue, 0 }  ][line width=0.08]  [draw opacity=0] (5.36,-2.57) -- (0,0) -- (5.36,2.57) -- cycle    ;
    \draw    (184.71,75.42) -- (184.71,123.12) ;
    \draw [shift={(184.71,95.17)}, rotate = 90] [fill={rgb, 255:red, 0; green, 0; blue, 0 }  ][line width=0.08]  [draw opacity=0] (5.36,-2.57) -- (0,0) -- (5.36,2.57) -- cycle    ;
    \draw    (214.62,53.48) -- (184.71,75.42) ;
    \draw [shift={(202.97,62.02)}, rotate = 143.74] [fill={rgb, 255:red, 0; green, 0; blue, 0 }  ][line width=0.08]  [draw opacity=0] (5.36,-2.57) -- (0,0) -- (5.36,2.57) -- cycle    ;
    \draw    (132.03,25.8) -- (102.12,47.74) ;
    \draw [shift={(120.38,34.34)}, rotate = 143.74] [fill={rgb, 255:red, 0; green, 0; blue, 0 }  ][line width=0.08]  [draw opacity=0] (5.36,-2.57) -- (0,0) -- (5.36,2.57) -- cycle    ;
    \draw    (124.88,119.3) -- (94.96,141.24) ;
    \draw [shift={(113.23,127.84)}, rotate = 143.74] [fill={rgb, 255:red, 0; green, 0; blue, 0 }  ][line width=0.08]  [draw opacity=0] (5.36,-2.57) -- (0,0) -- (5.36,2.57) -- cycle    ;
    \draw    (42.29,91.62) -- (12.38,113.56) ;
    \draw [shift={(30.64,100.16)}, rotate = 143.74] [fill={rgb, 255:red, 0; green, 0; blue, 0 }  ][line width=0.08]  [draw opacity=0] (5.36,-2.57) -- (0,0) -- (5.36,2.57) -- cycle    ;
    \draw    (184.71,27.71) -- (184.71,75.42) ;
    \draw [shift={(184.71,47.46)}, rotate = 90] [fill={rgb, 255:red, 0; green, 0; blue, 0 }  ][line width=0.08]  [draw opacity=0] (5.36,-2.57) -- (0,0) -- (5.36,2.57) -- cycle    ;
    \draw    (102.12,0.03) -- (102.12,47.74) ;
    \draw [shift={(102.12,19.79)}, rotate = 90] [fill={rgb, 255:red, 0; green, 0; blue, 0 }  ][line width=0.08]  [draw opacity=0] (5.36,-2.57) -- (0,0) -- (5.36,2.57) -- cycle    ;
    \draw    (124.88,71.59) -- (124.88,119.3) ;
    \draw [shift={(124.88,91.34)}, rotate = 90] [fill={rgb, 255:red, 0; green, 0; blue, 0 }  ][line width=0.08]  [draw opacity=0] (5.36,-2.57) -- (0,0) -- (5.36,2.57) -- cycle    ;
    \draw    (42.29,43.91) -- (42.29,91.62) ;
    \draw [shift={(42.29,63.67)}, rotate = 90] [fill={rgb, 255:red, 0; green, 0; blue, 0 }  ][line width=0.08]  [draw opacity=0] (5.36,-2.57) -- (0,0) -- (5.36,2.57) -- cycle    ;
    \draw (70.23,105) node [anchor=north west][inner sep=0.75pt]    {\small $g_{1}$};
    \draw (157.85,91.76) node [anchor=north west][inner sep=0.75pt]    {\small $g_{4}$};
    \draw (145.47,42.04) node [anchor=north west][inner sep=0.75pt]    {\small $g_{3}$};
    \draw (57.65,52.14) node [anchor=north west][inner sep=0.75pt]    {\small $g_{2}$};
    \draw (110,77.84) node [anchor=north west][inner sep=0.75pt]    {\small $p$};
    \end{tikzpicture}.
\end{equation}

As discussed earlier, state \eqref{eq:5.4} can also be viewed as a state in the Hilbert space of the Fourier-transformed GT model. Therefore, by \eqref{eq:3.27}, we can construct the inner product between the two states above as 
\begin{equation}
    \langle g_1g_2g_3g_4|\Psi_{a\cdots w}\rangle=\mathcal{N}\times v_a\cdots v_w
    \begin{tikzpicture}[x=0.75pt,y=0.75pt,yscale=-0.75,xscale=0.75, baseline=(XXXX.south) ]
    \path (0,211);\path (415.99176025390625,0);\draw    ($(current bounding box.center)+(0,0.3em)$) node [anchor=south] (XXXX) {};
    \draw    (212.72,57) -- (332.71,57) ;
    \draw    (0.59,145.11) -- (26.59,145.11) ;
    \draw    (26.59,145.11) -- (52.59,145.11) ;
    \draw    (52.59,145.11) -- (78.6,145.11) ;
    \draw    (9.31,162.39) -- (26.59,145.11) ;
    \draw    (91.6,126.11) -- (52.59,145.11) ;
    \draw    (78.6,145.11) -- (78.6,209) ;
    \draw    (78.6,145.11) -- (198.59,145.11) ;
    \draw    (91.6,126.11) -- (160.71,57) ;
    \draw    (91.6,89) -- (91.6,126.11) ;
    \draw    (181.31,162.39) -- (198.59,145.11) ;
    \draw    (198.59,145.11) -- (224.59,145.11) ;
    \draw    (224.59,145.11) -- (250.6,145.11) ;
    \draw    (263.6,126.11) -- (224.59,145.11) ;
    \draw    (250.6,145.11) -- (250.6,209) ;
    \draw    (250.6,145.11) -- (276.6,145.11) ;
    \draw    (263.6,126.11) -- (332.71,57) ;
    \draw    (134.71,57) -- (160.71,57) ;
    \draw    (160.71,57) -- (186.72,57) ;
    \draw    (225.72,38) -- (186.72,57) ;
    \draw    (225.72,38) -- (243,20.72) ;
    \draw    (225.72,0.89) -- (225.72,38) ;
    \draw    (186.72,57) -- (212.72,57) ;
    \draw    (212.72,57) -- (212.72,120.89) ;
    \draw    (332.71,57) -- (358.72,57) ;
    \draw    (358.72,57) -- (384.72,57) ;
    \draw    (397.72,38) -- (358.72,57) ;
    \draw    (384.72,57) -- (384.72,120.89) ;
    \draw    (384.72,57) -- (410.72,57) ;
    \draw    (397.72,38) -- (415,20.72) ;
    \draw    (397.72,0.89) -- (397.72,38) ;
    \draw  [fill={rgb, 255:red, 255; green, 255; blue, 255 }  ,fill opacity=1 ] (286.77,91.56) .. controls (286.77,85.27) and (291.87,80.17) .. (298.16,80.17) .. controls (304.44,80.17) and (309.54,85.27) .. (309.54,91.56) .. controls (309.54,97.84) and (304.44,102.94) .. (298.16,102.94) .. controls (291.87,102.94) and (286.77,97.84) .. (286.77,91.56) -- cycle ;
    \draw  [fill={rgb, 255:red, 255; green, 255; blue, 255 }  ,fill opacity=1 ] (283.33,57) .. controls (283.33,50.71) and (288.43,45.62) .. (294.71,45.62) .. controls (301,45.62) and (306.09,50.71) .. (306.09,57) .. controls (306.09,63.29) and (301,68.38) .. (294.71,68.38) .. controls (288.43,68.38) and (283.33,63.29) .. (283.33,57) -- cycle ;
    \draw    (263.6,89) -- (263.6,126.11) ;
    \draw  [fill={rgb, 255:red, 255; green, 255; blue, 255 }  ,fill opacity=1 ] (127.21,145.11) .. controls (127.21,138.83) and (132.31,133.73) .. (138.59,133.73) .. controls (144.88,133.73) and (149.98,138.83) .. (149.98,145.11) .. controls (149.98,151.4) and (144.88,156.5) .. (138.59,156.5) .. controls (132.31,156.5) and (127.21,151.4) .. (127.21,145.11) -- cycle ;
    \draw  [fill={rgb, 255:red, 255; green, 255; blue, 255 }  ,fill opacity=1 ] (114.77,91.56) .. controls (114.77,85.27) and (119.87,80.17) .. (126.16,80.17) .. controls (132.44,80.17) and (137.54,85.27) .. (137.54,91.56) .. controls (137.54,97.84) and (132.44,102.94) .. (126.16,102.94) .. controls (119.87,102.94) and (114.77,97.84) .. (114.77,91.56) -- cycle ;
    \draw (110.59,145) node [anchor=north west][inner sep=0.75pt]    {\small $a$};
    \draw (274,94) node [anchor=north west][inner sep=0.75pt]    {\small $b$};
    \draw (269.59,57) node [anchor=north west][inner sep=0.75pt]    {\small $c$};
    \draw (141.59,71.31) node [anchor=north west][inner sep=0.75pt]    {\small $d$};
    \draw (76,96.31) node [anchor=north west][inner sep=0.75pt]    {\small $d'$};
    \draw (62.59,122) node [anchor=north west][inner sep=0.75pt]    {\small $v$};
    \draw (35.59,128) node [anchor=north west][inner sep=0.75pt]    {\small $v'$};
    \draw (60.59,145.31) node [anchor=north west][inner sep=0.75pt]    {\small $w$};
    \draw (59,174.31) node [anchor=north west][inner sep=0.75pt]    {\small $w'$};
    \draw (187,150) node [anchor=north west][inner sep=0.75pt]    {\small $a'$};
    \draw (206.59,145.31) node [anchor=north west][inner sep=0.75pt]    {\small $p$};
    \draw (233.59,145.31) node [anchor=north west][inner sep=0.75pt]    {\small $p'$};
    \draw (235.59,121) node [anchor=north west][inner sep=0.75pt]    {\small $q$};
    \draw (249,96.31) node [anchor=north west][inner sep=0.75pt]    {\small $q'$};
    \draw (341.59,40) node [anchor=north west][inner sep=0.75pt]    {\small $b'$};
    \draw (199,79.31) node [anchor=north west][inner sep=0.75pt]    {\small $c'$};
    \draw (197.59,57.31) node [anchor=north west][inner sep=0.75pt]    {\small $r$};
    \draw (201.59,29.31) node [anchor=north west][inner sep=0.75pt]    {\small $r'$};
    \draw (170.59,44) node [anchor=north west][inner sep=0.75pt]    {\small $u$};
    \draw (142.59,40) node [anchor=north west][inner sep=0.75pt]    {\small $u'$};
    \draw (118,83) node [anchor=north west][inner sep=0.75pt]    {\small $g_{2}$};
    \draw (287,49) node [anchor=north west][inner sep=0.75pt]    {\small $g_{3}$};
    \draw (130,137) node [anchor=north west][inner sep=0.75pt]    {\small $g_{1}$};
    \draw (290,83) node [anchor=north west][inner sep=0.75pt]    {\small $g_{4}$};
    \end{tikzpicture},
\end{equation}
where we have collected all unimportant coefficients into $\mathcal{N}$, and we also neglect all the Latin indices $m_{q'},\cdots$. Note that 
\begin{equation*}
    \mathcal{N}=\frac{v_{p'}v_{q'}v_{b'}v_{r'}v_{u'}v_{d'}v_{v'}v_{w'}v_{a'}}{|G|^2}
\end{equation*}
will not be affected by the action of the plaquette operator. 

Inserting resolution of identity, we can write the action of the plaquette operator $\tilde{B}_p^\text{GT}$ of the Fourier-transformed GT model as 
\begin{equation}\label{eq:5.8}
    \begin{aligned}
        &\phantom{=}\tilde{B}_p^\text{GT}|\Psi_{abcdpqruvw}\rangle\\
        &=\sum_{g_1g_2g_3g_4\in G}B_p^\text{GT}|g_1g_2g_3g_4\rangle\langle g_1g_2g_3g_4|\Psi_{abcdpqruvw}\rangle\\
        &=\sum_{g_1g_2g_3g_4\in G}\delta_{g_1g_2\bar g_3\bar g_4,e}|g_1g_2g_3g_4\rangle\langle g_1g_2g_3g_4|\Psi_{abcdpqruvw}\rangle\\
        &=\sum_{g_1g_2g_3g_4\in G}\sum_{s\in L_G,m_t}\frac{1}{|G|}d_sD^s_{m_sm_s}(g_1g_2\bar g_3\bar g_4)|g_1g_2g_3g_4\rangle\langle g_1g_2g_3g_4|\Psi_{abcdpqruvw}\rangle
    \end{aligned}
\end{equation}
In order to compare $\tilde{B}_p^\text{GT}$ to $B_p^\text{WW}$, by means of our graphical tool, we re-express \eqref{eq:5.8} as 
\begin{equation}\label{eq:5.9}
    \begin{aligned}
        &\phantom{=}\tilde{B}_p^\text{GT}|\Psi_{abcdpqruvw}\rangle\\
        &=\sum_{g_1g_2g_3g_4\in G}\sum_{s\in L_G}\frac{\mathtt d_s}{|G|}\mathcal{N}v_a\cdots v_w\begin{tikzpicture}[x=0.75pt,y=0.75pt,yscale=-0.8,xscale=0.8, baseline=(XXXX.south) ]
        \path (0,212);\path (373.9913024902344,0);\draw    ($(current bounding box.center)+(0,0.3em)$) node [anchor=south] (XXXX) {};
        \draw    (219.8,126.01) -- (253.01,90.07) ;
        \draw [color={rgb, 255:red, 255; green, 255; blue, 255 }  ,draw opacity=1 ][line width=3.75]    (236.06,93.62) -- (236.06,131.87) ;
        \draw    (190.2,60.65) -- (190.2,107.48) ;
        \draw [color={rgb, 255:red, 255; green, 255; blue, 255 }  ,draw opacity=1 ][fill={rgb, 255:red, 255; green, 255; blue, 255 }  ,fill opacity=1 ][line width=3.75]    (162.05,71.33) -- (243.83,71.33) ;
        \draw    (-1,151.44) -- (22.43,151.44) ;
        \draw    (22.43,151.44) -- (45.87,151.44) ;
        \draw    (45.87,151.44) -- (69.31,151.44) ;
        \draw    (6.86,169.25) -- (22.43,151.44) ;
        \draw    (81.03,131.87) -- (45.87,151.44) ;
        \draw    (69.31,151.44) -- (69.31,198.28) ;
        \draw    (69.31,151.44) -- (177.47,151.44) ;
        \draw    (81.03,131.87) -- (143.32,60.65) ;
        \draw    (81.03,93.62) -- (81.03,131.87) ;
        \draw    (161.89,169.25) -- (177.47,151.44) ;
        \draw    (177.47,151.44) -- (200.9,151.44) ;
        \draw    (200.9,151.44) -- (224.34,151.44) ;
        \draw    (236.06,131.87) -- (200.9,151.44) ;
        \draw    (224.34,151.44) -- (247.78,151.44) ;
        \draw    (236.06,131.87) -- (298.36,60.65) ;
        \draw    (236.06,93.62) -- (236.06,131.87) ;
        \draw    (119.89,60.65) -- (143.32,60.65) ;
        \draw    (143.32,60.65) -- (166.76,60.65) ;
        \draw    (201.92,41.07) -- (166.76,60.65) ;
        \draw    (201.92,41.07) -- (217.49,23.26) ;
        \draw    (201.92,2.82) -- (201.92,41.07) ;
        \draw    (166.76,60.65) -- (190.2,60.65) ;
        \draw    (190.2,60.65) -- (298.36,60.65) ;
        \draw    (298.36,60.65) -- (321.79,60.65) ;
        \draw    (321.79,60.65) -- (345.23,60.65) ;
        \draw    (356.95,41.07) -- (321.79,60.65) ;
        \draw    (345.23,60.65) -- (345.23,107.48) ;
        \draw    (345.23,60.65) -- (368.67,60.65) ;
        \draw    (356.95,41.07) -- (372.53,23.26) ;
        \draw    (356.95,2.82) -- (356.95,41.07) ;
        \draw    (162.05,71.33) -- (243.83,71.33) ;
        \draw [shift={(162.05,71.33)}, rotate = 0] [color={rgb, 255:red, 0; green, 0; blue, 0 }  ][fill={rgb, 255:red, 0; green, 0; blue, 0 }  ][line width=0.75]      (0, 0) circle [x radius= 2.01, y radius= 2.01]   ;
        \draw    (113.16,141.42) -- (194.95,141.42) ;
        \draw    (243.83,71.33) .. controls (263.04,71.02) and (258.03,85.03) .. (253.01,90.07) ;
        \draw    (113.16,141.42) .. controls (93.96,141.73) and (98.97,127.71) .. (103.98,122.68) ;
        \draw    (162.05,71.33) .. controls (151.04,71.64) and (144.46,78.47) .. (137.2,86.74) ;
        \draw    (103.98,122.68) -- (137.2,86.74) ;
        \draw    (194.95,141.42) .. controls (205.96,141.11) and (212.54,134.28) .. (219.8,126.01) ;
        \draw [shift={(194.95,141.42)}, rotate = 358.38] [color={rgb, 255:red, 0; green, 0; blue, 0 }  ][fill={rgb, 255:red, 0; green, 0; blue, 0 }  ][line width=0.75]      (0, 0) circle [x radius= 2.01, y radius= 2.01]   ;
        \draw  [fill={rgb, 255:red, 255; green, 255; blue, 255 }  ,fill opacity=1 ] (106.2,151.44) .. controls (106.2,146.7) and (110.05,142.85) .. (114.8,142.85) .. controls (119.54,142.85) and (123.39,146.7) .. (123.39,151.44) .. controls (123.39,156.19) and (119.54,160.04) .. (114.8,160.04) .. controls (110.05,160.04) and (106.2,156.19) .. (106.2,151.44) -- cycle ;
        \draw  [fill={rgb, 255:red, 255; green, 255; blue, 255 }  ,fill opacity=1 ] (136.87,141.42) .. controls (136.87,136.67) and (140.72,132.83) .. (145.46,132.83) .. controls (150.21,132.83) and (154.06,136.67) .. (154.06,141.42) .. controls (154.06,146.16) and (150.21,150.01) .. (145.46,150.01) .. controls (140.72,150.01) and (136.87,146.16) .. (136.87,141.42) -- cycle ;
        \draw  [fill={rgb, 255:red, 255; green, 255; blue, 255 }  ,fill opacity=1 ] (128.61,86.74) .. controls (128.61,81.99) and (132.45,78.14) .. (137.2,78.14) .. controls (141.94,78.14) and (145.79,81.99) .. (145.79,86.74) .. controls (145.79,91.48) and (141.94,95.33) .. (137.2,95.33) .. controls (132.45,95.33) and (128.61,91.48) .. (128.61,86.74) -- cycle ;
        \draw  [fill={rgb, 255:red, 255; green, 255; blue, 255 }  ,fill opacity=1 ] (267.62,86.26) .. controls (267.62,81.51) and (271.46,77.66) .. (276.21,77.66) .. controls (280.95,77.66) and (284.8,81.51) .. (284.8,86.26) .. controls (284.8,91) and (280.95,94.85) .. (276.21,94.85) .. controls (271.46,94.85) and (267.62,91) .. (267.62,86.26) -- cycle ;
        \draw  [fill={rgb, 255:red, 255; green, 255; blue, 255 }  ,fill opacity=1 ] (240.43,94.26) .. controls (240.43,89.51) and (244.28,85.66) .. (249.02,85.66) .. controls (253.77,85.66) and (257.62,89.51) .. (257.62,94.26) .. controls (257.62,99) and (253.77,102.85) .. (249.02,102.85) .. controls (244.28,102.85) and (240.43,99) .. (240.43,94.26) -- cycle ;
        \draw  [fill={rgb, 255:red, 255; green, 255; blue, 255 }  ,fill opacity=1 ] (227.09,60.65) .. controls (227.09,55.9) and (230.94,52.05) .. (235.69,52.05) .. controls (240.43,52.05) and (244.28,55.9) .. (244.28,60.65) .. controls (244.28,65.39) and (240.43,69.24) .. (235.69,69.24) .. controls (230.94,69.24) and (227.09,65.39) .. (227.09,60.65) -- cycle ;
        \draw  [fill={rgb, 255:red, 255; green, 255; blue, 255 }  ,fill opacity=1 ] (202.94,71.33) .. controls (202.94,66.58) and (206.79,62.74) .. (211.53,62.74) .. controls (216.28,62.74) and (220.12,66.58) .. (220.12,71.33) .. controls (220.12,76.07) and (216.28,79.92) .. (211.53,79.92) .. controls (206.79,79.92) and (202.94,76.07) .. (202.94,71.33) -- cycle ;
        \draw  [fill={rgb, 255:red, 255; green, 255; blue, 255 }  ,fill opacity=1 ] (103.59,96.26) .. controls (103.59,91.51) and (107.43,87.66) .. (112.18,87.66) .. controls (116.92,87.66) and (120.77,91.51) .. (120.77,96.26) .. controls (120.77,101) and (116.92,104.85) .. (112.18,104.85) .. controls (107.43,104.85) and (103.59,101) .. (103.59,96.26) -- cycle ;
        \draw    (224.34,151.44) -- (224.34,198.28) ;
        \draw (133.07,152) node [anchor=north west][inner sep=0.75pt]    {\footnotesize $a$};
        \draw (255.62,105.7) node [anchor=north west][inner sep=0.75pt]    {\footnotesize $b$};
        \draw (256.16,49) node [anchor=north west][inner sep=0.75pt]    {\footnotesize $c$};
        \draw (88.16,97.69) node [anchor=north west][inner sep=0.75pt]    {\footnotesize $d$};
        \draw (67,101.36) node [anchor=north west][inner sep=0.75pt]    {\footnotesize $d'$};
        \draw (54.48,130) node [anchor=north west][inner sep=0.75pt]    {\footnotesize $v$};
        \draw (30,135.37) node [anchor=north west][inner sep=0.75pt]    {\footnotesize $v'$};
        \draw (52.53,151.86) node [anchor=north west][inner sep=0.75pt]    {\footnotesize $w$};
        \draw (50,173.74) node [anchor=north west][inner sep=0.75pt]    {\footnotesize $w'$};
        \draw (165,155) node [anchor=north west][inner sep=0.75pt]    {\footnotesize $a'$};
        \draw (184.28,151.86) node [anchor=north west][inner sep=0.75pt]    {\footnotesize $p$};
        \draw (208.47,151.86) node [anchor=north west][inner sep=0.75pt]    {\footnotesize $p'$};
        \draw (224,138) node [anchor=north west][inner sep=0.75pt]    {\footnotesize $q$};
        \draw (236,103.88) node [anchor=north west][inner sep=0.75pt]    {\footnotesize $q'$};
        \draw (305.68,45) node [anchor=north west][inner sep=0.75pt]    {\footnotesize $b'$};
        \draw (177.82,83.85) node [anchor=north west][inner sep=0.75pt]    {\footnotesize $c'$};
        \draw (183,49) node [anchor=north west][inner sep=0.75pt]    {\footnotesize $r$};
        \draw (179.62,32.32) node [anchor=north west][inner sep=0.75pt]    {\footnotesize $r'$};
        \draw (151.78,49) node [anchor=north west][inner sep=0.75pt]    {\footnotesize $u$};
        \draw (126.39,45) node [anchor=north west][inner sep=0.75pt]    {\footnotesize $u'$};
        \draw (211.04,112.33) node [anchor=north west][inner sep=0.75pt]    {\footnotesize $s$};
        \draw (108,145) node [anchor=north west][inner sep=0.75pt]    {\footnotesize $g_{1}$};
        \draw (139,135) node [anchor=north west][inner sep=0.75pt]    {\footnotesize $g_{1}$};
        \draw (269,80) node [anchor=north west][inner sep=0.75pt]    {\footnotesize $g_{4}$};
        \draw (242,88) node [anchor=north west][inner sep=0.75pt]    {\footnotesize $g_{4}$};
        \draw (229,55) node [anchor=north west][inner sep=0.75pt]    {\footnotesize  $g_{3}$};
        \draw (204,65) node [anchor=north west][inner sep=0.75pt]    {\footnotesize $g_{3}$};
        \draw (105,90) node [anchor=north west][inner sep=0.75pt]    {\footnotesize  $g_{2}$};
        \draw (130,80) node [anchor=north west][inner sep=0.75pt]    {\footnotesize $g_{2}$};
        \end{tikzpicture}|g_1g_2g_3g_4\rangle
    \end{aligned},
\end{equation}
where use is made of 
\begin{equation}
    \begin{tikzpicture}[x=0.75pt,y=0.75pt,yscale=-1,xscale=1, baseline=(XXXX.south) ]
\path (0,86);\path (97.96504974365234,0);\draw    ($(current bounding box.center)+(0,0.3em)$) node [anchor=south] (XXXX) {};
\draw    (49.12,81.47) .. controls (33.58,80.6) and (12.23,70.21) .. (9.84,44.71) ;
\draw [shift={(49.12,81.47)}, rotate = 183.21] [color={rgb, 255:red, 0; green, 0; blue, 0 }  ][fill={rgb, 255:red, 0; green, 0; blue, 0 }  ][line width=0.75]      (0, 0) circle [x radius= 2.01, y radius= 2.01]   ;
\draw    (9.84,44.71) .. controls (8.33,23.47) and (27.66,2.71) .. (48.52,3.52) ;
\draw [shift={(48.52,3.52)}, rotate = 2.22] [color={rgb, 255:red, 0; green, 0; blue, 0 }  ][fill={rgb, 255:red, 0; green, 0; blue, 0 }  ][line width=0.75]      (0, 0) circle [x radius= 2.01, y radius= 2.01]   ;
\draw [shift={(22.58,14.04)}, rotate = 132.48] [fill={rgb, 255:red, 0; green, 0; blue, 0 }  ][line width=0.08]  [draw opacity=0] (5.36,-2.57) -- (0,0) -- (5.36,2.57) -- cycle    ;
\draw  [fill={rgb, 255:red, 255; green, 255; blue, 255 }  ,fill opacity=1 ] (1.25,36.12) .. controls (1.25,31.37) and (5.1,27.53) .. (9.84,27.53) .. controls (14.59,27.53) and (18.43,31.37) .. (18.43,36.12) .. controls (18.43,40.86) and (14.59,44.71) .. (9.84,44.71) .. controls (5.1,44.71) and (1.25,40.86) .. (1.25,36.12) -- cycle ;
\draw    (48.52,3.52) .. controls (66.67,4.03) and (86.25,14.44) .. (87.82,42.73) ;
\draw    (87.82,42.73) .. controls (86.65,65.81) and (74.34,78.36) .. (49.12,81.47) ;
\draw [shift={(79.48,67.38)}, rotate = 134.85] [fill={rgb, 255:red, 0; green, 0; blue, 0 }  ][line width=0.08]  [draw opacity=0] (5.36,-2.57) -- (0,0) -- (5.36,2.57) -- cycle    ;
\draw  [fill={rgb, 255:red, 255; green, 255; blue, 255 }  ,fill opacity=1 ] (79.23,42.73) .. controls (79.23,37.99) and (83.07,34.14) .. (87.82,34.14) .. controls (92.56,34.14) and (96.41,37.99) .. (96.41,42.73) .. controls (96.41,47.48) and (92.56,51.32) .. (87.82,51.32) .. controls (83.07,51.32) and (79.23,47.48) .. (79.23,42.73) -- cycle ;
\draw  [fill={rgb, 255:red, 255; green, 255; blue, 255 }  ,fill opacity=1 ] (10.25,66.12) .. controls (10.25,61.37) and (14.1,57.53) .. (18.84,57.53) .. controls (23.59,57.53) and (27.43,61.37) .. (27.43,66.12) .. controls (27.43,70.86) and (23.59,74.71) .. (18.84,74.71) .. controls (14.1,74.71) and (10.25,70.86) .. (10.25,66.12) -- cycle ;
\draw  [fill={rgb, 255:red, 255; green, 255; blue, 255 }  ,fill opacity=1 ] (62.23,10.73) .. controls (62.23,5.99) and (66.07,2.14) .. (70.82,2.14) .. controls (75.56,2.14) and (79.41,5.99) .. (79.41,10.73) .. controls (79.41,15.48) and (75.56,19.32) .. (70.82,19.32) .. controls (66.07,19.32) and (62.23,15.48) .. (62.23,10.73) -- cycle ;
\draw (13,61) node [anchor=north west][inner sep=0.75pt]    {\small $g_1$};
\draw (3,31) node [anchor=north west][inner sep=0.75pt]    {\small $g_2$};
\draw (9,5) node [anchor=north west][inner sep=0.75pt]    {\small $s$};
\draw (79,65.2) node [anchor=north west][inner sep=0.75pt]    {\small $s$};
\draw (82,37) node [anchor=north west][inner sep=0.75pt]    {\small $g_4$};
\draw (65,5) node [anchor=north west][inner sep=0.75pt]    {\small $g_3$};
\end{tikzpicture}
=\beta _{s} D_{m_{s} m_{s}}^{s}(g_1g_2\bar g_3\bar g_4).
\end{equation}
Note that here the braidings between loop $s$ and edges $c'$ and $q'$ are inevitable if we want to place loop $s$ along this plaquette correctly. 

To evaluate the graph in the equation above, we introduce the convention 
\begin{equation}\label{eq:5.10}
	C^{\mu0,n_\mu}_{\mu,m_\mu}C^{\nu,n_\nu}_{0\nu,m_\nu}=\frac{1}{\mathtt v_\mu \mathtt v_\nu}\delta_{n_\mu m_\mu}\delta_{n_\nu m_\nu},
\end{equation}
which can be expressed graphically as
\begin{equation*}
\begin{tikzpicture}[x=0.75pt,y=0.75pt,yscale=-1,xscale=1, baseline=(XXXX.south) ]
\path (0,100);\path (72.99652862548828,0);\draw    ($(current bounding box.center)+(0,0.3em)$) node [anchor=south] (XXXX) {};
\draw    (17,8) -- (17,91.22) ;
\draw [shift={(17,45.51)}, rotate = 90] [fill={rgb, 255:red, 0; green, 0; blue, 0 }  ][line width=0.08]  [draw opacity=0] (5.36,-2.57) -- (0,0) -- (5.36,2.57) -- cycle    ;
\draw    (53,8) -- (53,91.22) ;
\draw [shift={(53,45.51)}, rotate = 90] [fill={rgb, 255:red, 0; green, 0; blue, 0 }  ][line width=0.08]  [draw opacity=0] (5.36,-2.57) -- (0,0) -- (5.36,2.57) -- cycle    ;
\draw (21,40.2) node [anchor=north west][inner sep=0.75pt]    {\small $\mu $};
\draw (58,40.2) node [anchor=north west][inner sep=0.75pt]    {\small $\nu $};
\end{tikzpicture}
=\mathtt{v}_{\mu }\mathtt{v}_{\nu }
\begin{tikzpicture}[x=0.75pt,y=0.75pt,yscale=-1,xscale=1, baseline=(XXXX.south) ]
\path (0,100);\path (72.99652862548828,0);\draw    ($(current bounding box.center)+(0,0.3em)$) node [anchor=south] (XXXX) {};
\draw    (17,8) -- (17,91.22) ;
\draw [shift={(17,45.51)}, rotate = 90] [fill={rgb, 255:red, 0; green, 0; blue, 0 }  ][line width=0.08]  [draw opacity=0] (5.36,-2.57) -- (0,0) -- (5.36,2.57) -- cycle    ;
\draw    (53,8) -- (53,91.22) ;
\draw [shift={(53,45.51)}, rotate = 90] [fill={rgb, 255:red, 0; green, 0; blue, 0 }  ][line width=0.08]  [draw opacity=0] (5.36,-2.57) -- (0,0) -- (5.36,2.57) -- cycle    ;
\draw  [dash pattern={on 0.75pt off 1.5pt}]  (17,65.22) -- (53,40) ;
\draw (21,40.2) node [anchor=north west][inner sep=0.75pt]    {\small $\mu $};
\draw (58,40.2) node [anchor=north west][inner sep=0.75pt]    {\small $\nu $};
\end{tikzpicture},
\end{equation*}
where the dotted line is graced by the trivial representation. It is easy to prove that convention \eqref{eq:5.10} is consistent with the first equation in \eqref{eq:3.16} and the $F$-move \eqref{eq:Fmove}. We can then apply $F$-moves in the graph in \eqref{eq:5.9} and obtain
\begin{align}
        \tilde{B}_p^\text{GT}|\Psi_{abcdpqruvw}\rangle=&\sum_{g_1g_2g_3g_4\in G}\sum_{s\in L_G}\frac{\mathtt d_s}{|G|}R^{q'b}_q\overline{R^{c'r}_c}\overline{R^{q'b''}_{q''}}R^{c'r''}_{c''}F_{sa''p''}^{a'pa}F_{sp''q''}^{p'qp}\cdots F_{sw''a''}^{w'aw}\notag\\
        &\times\mathtt{d}_a\mathtt{d}_b\mathtt{d}_c\mathtt{d}_d\mathcal Nv_a\cdots v_w\notag\\
        &\times\begin{tikzpicture}[x=0.75pt,y=0.75pt,yscale=-0.8,xscale=0.8, baseline=(XXXX.south) ]
        \path (0,212);\path (373.9913024902344,0);\draw    ($(current bounding box.center)+(0,0.3em)$) node [anchor=south] (XXXX) {};
        \draw    (120.59,104.71) -- (137.2,86.74) ;
        \draw    (135,70.28) .. controls (143,79.28) and (141,83.28) .. (137.2,86.74) ;
        \draw [shift={(135,70.28)}, rotate = 48.37] [color={rgb, 255:red, 0; green, 0; blue, 0 }  ][fill={rgb, 255:red, 0; green, 0; blue, 0 }  ][line width=0.75]      (0, 0) circle [x radius= 2.01, y radius= 2.01]   ;
        \draw    (196.2,60.65) .. controls (203,72.28) and (208,71.28) .. (211.53,71.33) ;
        \draw [shift={(196.2,60.65)}, rotate = 59.7] [color={rgb, 255:red, 0; green, 0; blue, 0 }  ][fill={rgb, 255:red, 0; green, 0; blue, 0 }  ][line width=0.75]      (0, 0) circle [x radius= 2.01, y radius= 2.01]   ;
        \draw    (289,71.28) .. controls (273,72.28) and (262,77.28) .. (249.02,94.26) ;
        \draw [shift={(289,71.28)}, rotate = 176.42] [color={rgb, 255:red, 0; green, 0; blue, 0 }  ][fill={rgb, 255:red, 0; green, 0; blue, 0 }  ][line width=0.75]      (0, 0) circle [x radius= 2.01, y radius= 2.01]   ;
        \draw [color={rgb, 255:red, 255; green, 255; blue, 255 }  ,draw opacity=1 ][line width=3.75]    (236.06,93.62) -- (236.06,131.87) ;
        \draw    (190.2,60.65) -- (190.2,107.48) ;
        \draw    (-1,151.44) -- (22.43,151.44) ;
        \draw    (22.43,151.44) -- (45.87,151.44) ;
        \draw    (45.87,151.44) -- (69.31,151.44) ;
        \draw    (6.86,169.25) -- (22.43,151.44) ;
        \draw    (81.03,131.87) -- (45.87,151.44) ;
        \draw    (69.31,151.44) -- (69.31,198.28) ;
        \draw    (69.31,151.44) -- (177.47,151.44) ;
        \draw    (81.03,131.87) -- (143.32,60.65) ;
        \draw    (81.03,93.62) -- (81.03,131.87) ;
        \draw    (161.89,169.25) -- (177.47,151.44) ;
        \draw    (177.47,151.44) -- (200.9,151.44) ;
        \draw    (200.9,151.44) -- (224.34,151.44) ;
        \draw    (236.06,131.87) -- (200.9,151.44) ;
        \draw    (224.34,151.44) -- (247.78,151.44) ;
        \draw    (236.06,131.87) -- (298.36,60.65) ;
        \draw    (236.06,93.62) -- (236.06,131.87) ;
        \draw    (119.89,60.65) -- (143.32,60.65) ;
        \draw    (143.32,60.65) -- (166.76,60.65) ;
        \draw    (201.92,41.07) -- (166.76,60.65) ;
        \draw    (201.92,41.07) -- (217.49,23.26) ;
        \draw    (201.92,2.82) -- (201.92,41.07) ;
        \draw    (166.76,60.65) -- (190.2,60.65) ;
        \draw    (190.2,60.65) -- (298.36,60.65) ;
        \draw    (298.36,60.65) -- (321.79,60.65) ;
        \draw    (321.79,60.65) -- (345.23,60.65) ;
        \draw    (356.95,41.07) -- (321.79,60.65) ;
        \draw    (345.23,60.65) -- (345.23,107.48) ;
        \draw    (345.23,60.65) -- (368.67,60.65) ;
        \draw    (356.95,41.07) -- (372.53,23.26) ;
        \draw    (356.95,2.82) -- (356.95,41.07) ;
        \draw    (211.53,71.33) -- (243.83,71.33) ;
        \draw [shift={(211.53,71.33)}, rotate = 0] [color={rgb, 255:red, 0; green, 0; blue, 0 }  ][fill={rgb, 255:red, 0; green, 0; blue, 0 }  ][line width=0.75]      (0, 0) circle [x radius= 2.01, y radius= 2.01]   ;
        \draw    (113.16,141.42) -- (159.95,141.42) ;
        \draw    (242,125.28) .. controls (240.93,111.69) and (244,102.28) .. (249.02,95.26) ;
        \draw [shift={(242,125.28)}, rotate = 265.52] [color={rgb, 255:red, 0; green, 0; blue, 0 }  ][fill={rgb, 255:red, 0; green, 0; blue, 0 }  ][line width=0.75]      (0, 0) circle [x radius= 2.01, y radius= 2.01]   ;
        \draw  [fill={rgb, 255:red, 255; green, 255; blue, 255 }  ,fill opacity=1 ] (106.2,151.44) .. controls (106.2,146.7) and (110.05,142.85) .. (114.8,142.85) .. controls (119.54,142.85) and (123.39,146.7) .. (123.39,151.44) .. controls (123.39,156.19) and (119.54,160.04) .. (114.8,160.04) .. controls (110.05,160.04) and (106.2,156.19) .. (106.2,151.44) -- cycle ;
        \draw  [fill={rgb, 255:red, 255; green, 255; blue, 255 }  ,fill opacity=1 ] (136.87,141.42) .. controls (136.87,136.67) and (140.72,132.83) .. (145.46,132.83) .. controls (150.21,132.83) and (154.06,136.67) .. (154.06,141.42) .. controls (154.06,146.16) and (150.21,150.01) .. (145.46,150.01) .. controls (140.72,150.01) and (136.87,146.16) .. (136.87,141.42) -- cycle ;
        \draw  [fill={rgb, 255:red, 255; green, 255; blue, 255 }  ,fill opacity=1 ] (128.61,86.74) .. controls (128.61,81.99) and (132.45,78.14) .. (137.2,78.14) .. controls (141.94,78.14) and (145.79,81.99) .. (145.79,86.74) .. controls (145.79,91.48) and (141.94,95.33) .. (137.2,95.33) .. controls (132.45,95.33) and (128.61,91.48) .. (128.61,86.74) -- cycle ;
        \draw  [fill={rgb, 255:red, 255; green, 255; blue, 255 }  ,fill opacity=1 ] (267.62,86.26) .. controls (267.62,81.51) and (271.46,77.66) .. (276.21,77.66) .. controls (280.95,77.66) and (284.8,81.51) .. (284.8,86.26) .. controls (284.8,91) and (280.95,94.85) .. (276.21,94.85) .. controls (271.46,94.85) and (267.62,91) .. (267.62,86.26) -- cycle ;
        \draw  [fill={rgb, 255:red, 255; green, 255; blue, 255 }  ,fill opacity=1 ] (240.43,94.26) .. controls (240.43,89.51) and (244.28,85.66) .. (249.02,85.66) .. controls (253.77,85.66) and (257.62,89.51) .. (257.62,94.26) .. controls (257.62,99) and (253.77,102.85) .. (249.02,102.85) .. controls (244.28,102.85) and (240.43,99) .. (240.43,94.26) -- cycle ;
        \draw  [fill={rgb, 255:red, 255; green, 255; blue, 255 }  ,fill opacity=1 ] (227.09,60.65) .. controls (227.09,55.9) and (230.94,52.05) .. (235.69,52.05) .. controls (240.43,52.05) and (244.28,55.9) .. (244.28,60.65) .. controls (244.28,65.39) and (240.43,69.24) .. (235.69,69.24) .. controls (230.94,69.24) and (227.09,65.39) .. (227.09,60.65) -- cycle ;
        \draw  [fill={rgb, 255:red, 255; green, 255; blue, 255 }  ,fill opacity=1 ] (202.94,71.33) .. controls (202.94,66.58) and (206.79,62.74) .. (211.53,62.74) .. controls (216.28,62.74) and (220.12,66.58) .. (220.12,71.33) .. controls (220.12,76.07) and (216.28,79.92) .. (211.53,79.92) .. controls (206.79,79.92) and (202.94,76.07) .. (202.94,71.33) -- cycle ;
        \draw  [fill={rgb, 255:red, 255; green, 255; blue, 255 }  ,fill opacity=1 ] (103.59,96.26) .. controls (103.59,91.51) and (107.43,87.66) .. (112.18,87.66) .. controls (116.92,87.66) and (120.77,91.51) .. (120.77,96.26) .. controls (120.77,101) and (116.92,104.85) .. (112.18,104.85) .. controls (107.43,104.85) and (103.59,101) .. (103.59,96.26) -- cycle ;
        \draw    (224.34,151.44) -- (224.34,198.28) ;
        \draw    (113.16,141.42) .. controls (102.16,141.73) and (98.57,143.18) .. (91.31,151.44) ;
        \draw [shift={(91.31,151.44)}, rotate = 131.3] [color={rgb, 255:red, 0; green, 0; blue, 0 }  ][fill={rgb, 255:red, 0; green, 0; blue, 0 }  ][line width=0.75]      (0, 0) circle [x radius= 2.01, y radius= 2.01]   ;
        \draw    (159.95,141.42) .. controls (167,141.28) and (169,146.28) .. (169.47,151.44) ;
        \draw [shift={(169.47,151.44)}, rotate = 84.82] [color={rgb, 255:red, 0; green, 0; blue, 0 }  ][fill={rgb, 255:red, 0; green, 0; blue, 0 }  ][line width=0.75]      (0, 0) circle [x radius= 2.01, y radius= 2.01]   ;
        \draw    (267.36,60.65) .. controls (257,70.28) and (257,71.28) .. (243.83,71.33) ;
        \draw [shift={(267.36,60.65)}, rotate = 137.08] [color={rgb, 255:red, 0; green, 0; blue, 0 }  ][fill={rgb, 255:red, 0; green, 0; blue, 0 }  ][line width=0.75]      (0, 0) circle [x radius= 2.01, y radius= 2.01]   ;
        \draw    (120.59,104.71) .. controls (114,112.28) and (105,114.28) .. (96,114.28) ;
        \draw [shift={(96,114.28)}, rotate = 180] [color={rgb, 255:red, 0; green, 0; blue, 0 }  ][fill={rgb, 255:red, 0; green, 0; blue, 0 }  ][line width=0.75]      (0, 0) circle [x radius= 2.01, y radius= 2.01]   ;
        \draw (133.07,152) node [anchor=north west][inner sep=0.75pt]    {\footnotesize $a$};
        \draw (255.62,105.7) node [anchor=north west][inner sep=0.75pt]    {\footnotesize $b$};
        \draw (256.16,49) node [anchor=north west][inner sep=0.75pt]    {\footnotesize $c$};
        \draw (88.16,97.69) node [anchor=north west][inner sep=0.75pt]    {\footnotesize $d$};
        \draw (67,101.36) node [anchor=north west][inner sep=0.75pt]    {\footnotesize $d'$};
        \draw (54.48,127) node [anchor=north west][inner sep=0.75pt]    {\footnotesize $v''$};
        \draw (30,135.37) node [anchor=north west][inner sep=0.75pt]    {\footnotesize $v'$};
        \draw (50,151.86) node [anchor=north west][inner sep=0.75pt]    {\footnotesize $w''$};
        \draw (50,173.74) node [anchor=north west][inner sep=0.75pt]    {\footnotesize $w'$};
        \draw (165,155) node [anchor=north west][inner sep=0.75pt]    {\footnotesize $a'$};
        \draw (184.28,151.86) node [anchor=north west][inner sep=0.75pt]    {\footnotesize $p''$};
        \draw (208.47,151.86) node [anchor=north west][inner sep=0.75pt]    {\footnotesize $p'$};
        \draw (223,133) node [anchor=north west][inner sep=0.75pt]    {\footnotesize $q''$};
        \draw (222,103.88) node [anchor=north west][inner sep=0.75pt]    {\footnotesize $q'$};
        \draw (305.68,45) node [anchor=north west][inner sep=0.75pt]    {\footnotesize $b'$};
        \draw (177.82,83.85) node [anchor=north west][inner sep=0.75pt]    {\footnotesize $c'$};
        \draw (183,45) node [anchor=north west][inner sep=0.75pt]    {\footnotesize $r''$};
        \draw (179.62,32.32) node [anchor=north west][inner sep=0.75pt]    {\footnotesize $r'$};
        \draw (151.78,45) node [anchor=north west][inner sep=0.75pt]    {\footnotesize $u''$};
        \draw (126.39,45) node [anchor=north west][inner sep=0.75pt]    {\footnotesize $u'$};
        \draw (122,129) node [anchor=north west][inner sep=0.75pt]    {\footnotesize $s$};
        \draw (108,145) node [anchor=north west][inner sep=0.75pt]    {\footnotesize $g_{1}$};
        \draw (139,135) node [anchor=north west][inner sep=0.75pt]    {\footnotesize $g_{1}$};
        \draw (269,80) node [anchor=north west][inner sep=0.75pt]    {\footnotesize $g_{4}$};
        \draw (242,88) node [anchor=north west][inner sep=0.75pt]    {\footnotesize $g_{4}$};
        \draw (229,55) node [anchor=north west][inner sep=0.75pt]    {\footnotesize  $g_{3}$};
        \draw (204,65) node [anchor=north west][inner sep=0.75pt]    {\footnotesize $g_{3}$};
        \draw (105,90) node [anchor=north west][inner sep=0.75pt]    {\footnotesize  $g_{2}$};
        \draw (130,80) node [anchor=north west][inner sep=0.75pt]    {\footnotesize $g_{2}$};
        \draw (73.07,150.86) node [anchor=north west][inner sep=0.75pt]    {\footnotesize $a''$};
        \draw (170.07,135.86) node [anchor=north west][inner sep=0.75pt]    {\footnotesize $a''$};
        \draw (238.07,125.86) node [anchor=north west][inner sep=0.75pt]    {\footnotesize $b''$};
        \draw (295.07,61.86) node [anchor=north west][inner sep=0.75pt]    {\footnotesize $b''$};
        \draw (281.16,44.69) node [anchor=north west][inner sep=0.75pt]    {\footnotesize $c''$};
        \draw (256.04,80) node [anchor=north west][inner sep=0.75pt]    {\footnotesize $s$};
        \draw (223.68,71) node [anchor=north west][inner sep=0.75pt]    {\footnotesize $s$};
        \draw (190.2,59.85) node [anchor=north west][inner sep=0.75pt]    {\footnotesize $c''$};
        \draw (86.16,118.69) node [anchor=north west][inner sep=0.75pt]    {\footnotesize $d''$};
        \draw (139.16,60.69) node [anchor=north west][inner sep=0.75pt]    {\footnotesize $d''$};
        \draw (122.77,97) node [anchor=north west][inner sep=0.75pt]    {\footnotesize $s$};
        \end{tikzpicture}|g_1g_2g_3g_4\rangle.
\end{align}
Then, applying the second equation in \eqref{eq:3.16} and using the intertwiner property of $3j$-symbols, we find that the graph in the above equation is equal to
\begin{equation*}
    \frac{\langle g_1g_2g_3g_4|\Psi_{a''b''\cdots w''}\rangle}{\mathtt{d}_a\mathtt{d}_b\mathtt{d}_c\mathtt{d}_d\times\mathcal N v_{a''}\cdots v_{w''}}.
\end{equation*}
The results above enable us to write the matrix elements of the Fourier-transformed plaquette operators of the three-dimensional GT model explicitly as 
\begin{equation}
    \begin{aligned}
        \langle\Psi_{a''\cdots w''}|\tilde{B}^\text{GT}_p|\Psi_{a\cdots w}\rangle&=\sum_{s\in L_G}\frac{\mathtt d_s}{|G|}\frac{v_a\cdots v_w}{v_{a''}\cdots v_{w''}}R^{q'b}_q\overline{R^{c'r}_c}\overline{R^{q'b''}_{q''}}R^{c'r''}_{c''}F_{sa''p''}^{a'pa}F_{sp''q''}^{p'qp}\cdots F_{sw''a''}^{w'aw}\\
        &=\sum_{s\in L_G}\frac{\mathtt d_s}{|G|}\mathtt v_a\cdots\mathtt v_w[\mathtt v'']R^{q'b}_q\overline{R^{c'r}_c}\overline{R^{q'b''}_{q''}}R^{c'r''}_{c''}G_{sa''p''}^{a'pa}G_{sp''q''}^{p'qp}\cdots G_{sw''a''}^{w'aw}
    \end{aligned},
\end{equation} 
where the second equality is due to that $F^{\mu\nu\lambda}_{\eta\kappa\rho}=\mathtt d_\rho G^{\mu\nu\lambda}_{\eta\kappa\rho}$ \eqref{eq:3.20x} is used. 

Bearing in mind that $D^2=\sum_{s\in L_G}\mathtt{d}_s^2=\sum_{s\in L_G}d_s^2=|G|$, we find that $\tilde{B}_p^\text{GT}$ is exactly the same as $B_p^\text{WW}$. Therefore, as far as bulk is concerned, after the Fourier transform and basis rewriting, and finally projecting all the degrees of freedom on the tails to the trivial representation, the three-dimensional GT model with input data $G$ has the same Hilbert space and Hamiltonian term by term as the WW model with input data $\mathcal{R}ep(G)$. 

\subsection{Mapping the boundary}
Recall that as aforementioned, the gapped boundary theory of the WW model has not been systematically constructed before. Now that we have already mapped the bulk Hilbert space and Hamiltonian to the WW model via Fourier transform, the Fourier-transformed boundary of the three-dimensional GT model also offers a systematic construction of the gapped boundary theory of the WW model. The construction is understood as follows. The boundary lattice is shown in \autoref{figq}. The boundary Hilbert space is the tensor product of the local Hilbert spaces on those edges and tails. The black edges are labeled by the objects of the UBFC $\mathcal{R}ep(G)$, while the tails are labeled by elements in the Frobenius algebra $A_{G,K}\in\mathcal{R}ep(G)$. The boundary Hilbert space can thus be expressed as
\begin{equation}
    \overline{\mathcal{H}^\text{WW}}=\left(\bigotimes_{e\in\partial\tilde{\Gamma}}\text{span}_{j_e\in\mathcal Rep(G)}\{|j_e\rangle\}\right)\otimes\left(\bigotimes_{t\in\text{boundary tails}}\text{span}_{j_t\in A_{G,K}}\{|j_t\rangle\}\right).
\end{equation}
The boundary Hamiltonian consists of respectively the sums of boundary vertex, plaquette, and edge operators:
\begin{equation}
    \overline{H^\text{WW}}=-\sum_{v\in\partial\tilde{\Gamma}}\overline{A^\text{WW}_v}-\sum_{p\in\partial\tilde{\Gamma}}\overline{B^\text{WW}_p}-\sum_{e\in\partial\Gamma}\overline{C^\text{WW}_e}.
\end{equation}
The boundary vertex operator $\overline{A^\text{WW}_v}$ is the identity operator if the labels of the three edges (or tails) around $v$ obey the fusion rules; otherwise it is $0$. The boundary plaquette operator $\overline{B^\text{WW}_p}$ is given by \eqref{eq:3.45}. The boundary edge operator $\overline{C^\text{WW}_e}$ acts on the local Hilbert space corresponds to edge $e$ of the original cubic lattice, as shown in \eqref{eq:3.32}, with its matrix elements given by \eqref{eq:3.42}. Since these local operators are obtained from the boundary Hamiltonian of the three-dimensional GT model via Fourier transform, they commute with each other. Therefore, the total Hamiltonian of the WW model is still exactly solvable. 

\appendix

\section{The Walker-Wang model}\label{appendix:A0}
The Walker-Wang model is defined on a three-dimensional trivalent vertex $\Gamma$, which is deformed from a three-dimensional cubic lattice. At each six-valent vertex of the cubic lattice, the deformation is depicted in \autoref{fig:6}.  
\begin{figure}[ht]
    \centering
\begin{tikzpicture}[x=0.75pt,y=0.75pt,yscale=-0.75,xscale=0.7]

\draw    (150,61) -- (214,61) ;
\draw    (214,61) -- (278.01,61) ;
\draw    (214,61) -- (280,23.79) ;
\draw    (148,98.21) -- (214,61) ;
\draw    (214,120.79) -- (214,61) ;
\draw    (214,61) -- (214,1.21) ;
\draw    (330,62) -- (394.01,62) ;
\draw    (328.01,99.21) -- (394.01,62) ;
\draw    (436,121.79) -- (436,62) ;
\draw    (415.01,62) -- (436,62) ;
\draw    (436,62) -- (500.01,62) ;
\draw    (415.01,62) -- (447,48.79) ;
\draw    (447,48.79) -- (513,11.58) ;
\draw    (447,48.79) -- (447,1.79) ;
\draw    (394.01,62) -- (415.01,62) ;

\draw (292,54.4) node [anchor=north west][inner sep=0.75pt]    {$\rightsquigarrow $};
\end{tikzpicture}
\caption{Deform a vertex in a three-dimensional cubic lattice to a trivalent lattice in the WW model.}
\label{fig:6}
\end{figure}
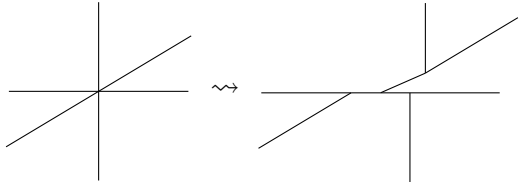

The input data of the model is a UBFC, in which the string types will be labeled by Latin letters $a,b,c,\cdots\in L$, are assigned to the edges of $\Gamma$. The Hilbert space $\mathcal{H}^\text{WW}$ of the model is spanned by all labels of the edges in the lattice $\Gamma$. The definition of UBFC also includes a set of symmetrized $6j$-symbols $G:L^6\to\mathbb C$ and a set of $R$-symbols $R:L^3\to\mathbb C$. The $6j$-symbols give the following basis transformation: 
\begin{equation}\label{eq:5.1}
   \begin{tikzpicture}[x=0.75pt,y=0.75pt,yscale=-1,xscale=1, baseline=(XXXX.south) ]
\path (0,76);\path (66.99651336669922,0);\draw    ($(current bounding box.center)+(0,0.3em)$) node [anchor=south] (XXXX) {};
\draw    (4,2) -- (19.01,25.98) ;
\draw    (19.01,25.98) -- (34.02,49.96) ;
\draw    (34.02,49.96) -- (49.03,73.95) ;
\draw    (19.01,25.98) -- (34.02,2) ;
\draw    (34.02,49.96) -- (49.03,25.98) ;
\draw    (49.03,25.98) -- (64.03,2) ;
\draw (2,15) node [anchor=north west][inner sep=0.75pt]    {$a$};
\draw (27,13) node [anchor=north west][inner sep=0.75pt]    {$b$};
\draw (50,26) node [anchor=north west][inner sep=0.75pt]    {$c$};
\draw (14,39.2) node [anchor=north west][inner sep=0.75pt]    {$m$};
\draw (31,64.2) node [anchor=north west][inner sep=0.75pt]    {$d$};
\end{tikzpicture}
=\sum _{n}\mathtt{v}_{m}\mathtt{v}_{n} G_{dcn}^{bam}\begin{tikzpicture}[x=0.75pt,y=0.75pt,yscale=-1,xscale=1, baseline=(XXXX.south) ]
\path (0,76);\path (66.99651336669922,0);\draw    ($(current bounding box.center)+(0,0.3em)$) node [anchor=south] (XXXX) {};
\draw    (4,2) -- (19.01,25.98) ;
\draw    (19.01,25.98) -- (34.02,49.96) ;
\draw    (34.02,49.96) -- (49.03,73.95) ;
\draw    (49.03,25.98) -- (34.02,2) ;
\draw    (34.02,49.96) -- (49.03,25.98) ;
\draw    (49.03,25.98) -- (64.03,2) ;
\draw (9,26) node [anchor=north west][inner sep=0.75pt]    {$a$};
\draw (32,13) node [anchor=north west][inner sep=0.75pt]    {$b$};
\draw (55,15.2) node [anchor=north west][inner sep=0.75pt]    {$c$};
\draw (40,39.2) node [anchor=north west][inner sep=0.75pt]    {$n$};
\draw (31,64.2) node [anchor=north west][inner sep=0.75pt]    {$d$};
\end{tikzpicture},
\end{equation}
and the $R$-symbols encode the braidings: 
\begin{equation}\label{eq:5.2}
    \begin{tikzpicture}[x=0.75pt,y=0.75pt,yscale=-1,xscale=1, baseline=(XXXX.south) ]
\path (0,88);\path (48.99651336669922,0);\draw    ($(current bounding box.center)+(0,0.3em)$) node [anchor=south] (XXXX) {};
\draw    (24,49) .. controls (45.01,44.13) and (24.99,16.13) .. (0.99,1.13) ;
\draw [color={rgb, 255:red, 255; green, 255; blue, 255 }  ,draw opacity=1 ][line width=4.5]    (24,49) .. controls (6.01,46.13) and (18.01,18.13) .. (47.01,1.13) ;
\draw    (24,49) -- (24,85.13) ;
\draw    (24,49) .. controls (6.01,46.13) and (18.01,18.13) .. (47.01,1.13) ;
\draw (2.99,12) node [anchor=north west][inner sep=0.75pt]    {$a$};
\draw (34,9.2) node [anchor=north west][inner sep=0.75pt]    {$b$};
\draw (12,61.2) node [anchor=north west][inner sep=0.75pt]    {$c$};
\end{tikzpicture}
=R_{c}^{ab}\begin{tikzpicture}[x=0.75pt,y=0.75pt,yscale=-1,xscale=1, baseline=(XXXX.south) ]
\path (0,88);\path (48.99651336669922,0);\draw    ($(current bounding box.center)+(0,0.3em)$) node [anchor=south] (XXXX) {};
\draw    (24,49) -- (24,85.13) ;
\draw    (0.99,1.13) -- (24,49) ;
\draw    (47.01,1.13) -- (24,49) ;
\draw (2.99,25) node [anchor=north west][inner sep=0.75pt]    {$a$};
\draw (36,22) node [anchor=north west][inner sep=0.75pt]    {$b$};
\draw (12,61.2) node [anchor=north west][inner sep=0.75pt]    {$c$};
\end{tikzpicture}.
\end{equation}
In the WW model, we also assume multiplicity-free in the fusion rules and self-duality of all labels, so edges in our lattices are not oriented. The Hamiltonian of the model is 
\begin{equation}
    H=-\sum_{v\in\Gamma}A^\text{WW}_v-\sum_{p\in\Gamma}B^\text{WW}_p,
\end{equation}
where $v$ ranges over all vertices in the trivalent lattice $\Gamma$, and $p$ ranges over all plaquettes in $\Gamma$ which correspond to the original squares of the cubic lattice. For $|\Psi\rangle\in\mathcal{H}^\text{WW}$, we have $A_v|\Psi\rangle=|\Psi\rangle$ if the three labels around vertex $v$ obey the fusion rules; otherwise $A_v|\Psi\rangle=0$. In order to derive the plaquette operator $B^\text{WW}_p$, we consider the following plaquette with the relevant labels: 
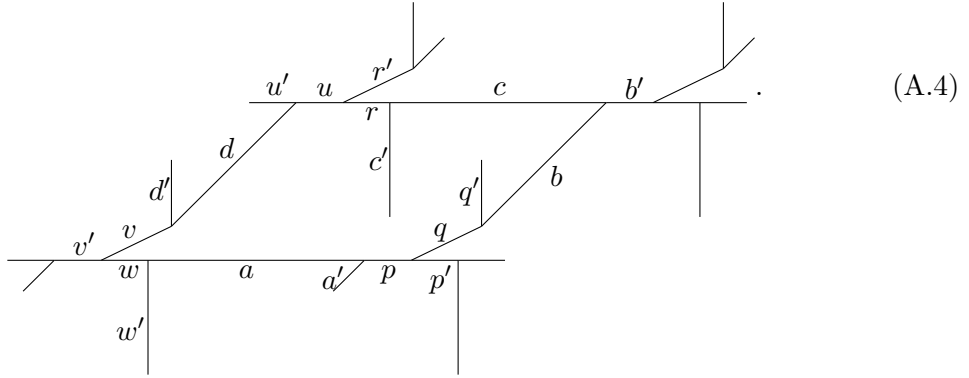
\begin{equation}\label{eq:5.4}
   \begin{tikzpicture}[x=0.75pt,y=0.75pt,yscale=-0.9,xscale=0.9,baseline=(XXXX.south)]

\draw    (123,146) -- (149,146) ;
\draw    (149,146) -- (175.01,146) ;
\draw    (175.01,146) -- (201.01,146) ;
\draw    (131.73,163.28) -- (149,146) ;
\draw    (214.01,127) -- (175.01,146) ;
\draw    (201.01,146) -- (201.01,209.89) ;
\draw    (201.01,146) -- (321,146) ;
\draw    (214.01,127) -- (283.13,57.89) ;
\draw    (214.01,89.89) -- (214.01,127) ;
\draw    (303.73,163.28) -- (321,146) ;
\draw    (321,146) -- (347.01,146) ;
\draw    (347.01,146) -- (373.01,146) ;
\draw    (386.01,127) -- (347.01,146) ;
\draw    (373.01,146) -- (373.01,209.89) ;
\draw    (373.01,146) -- (399.02,146) ;
\draw    (386.01,127) -- (455.13,57.89) ;
\draw    (386.01,89.89) -- (386.01,127) ;
\draw    (257.12,57.89) -- (283.13,57.89) ;
\draw    (283.13,57.89) -- (309.13,57.89) ;
\draw    (348.14,38.89) -- (309.13,57.89) ;
\draw    (348.14,38.89) -- (365.41,21.61) ;
\draw    (348.14,1.77) -- (348.14,38.89) ;
\draw    (309.13,57.89) -- (335.13,57.89) ;
\draw    (335.13,57.89) -- (335.13,121.77) ;
\draw    (335.13,57.89) -- (455.13,57.89) ;
\draw    (455.13,57.89) -- (481.13,57.89) ;
\draw    (481.13,57.89) -- (507.13,57.89) ;
\draw    (520.14,38.89) -- (481.13,57.89) ;
\draw    (507.13,57.89) -- (507.13,121.77) ;
\draw    (507.13,57.89) -- (533.14,57.89) ;
\draw    (520.14,38.89) -- (537.41,21.61) ;
\draw    (520.14,1.77) -- (520.14,38.89) ;

\draw (250,147.4) node [anchor=north west][inner sep=0.75pt]    {$a$};
\draw (422.57,92) node [anchor=north west][inner sep=0.75pt]    {$b$};
\draw (391,45) node [anchor=north west][inner sep=0.75pt]    {$c$};
\draw (239,76.4) node [anchor=north west][inner sep=0.75pt]    {$d$};
\draw (200,98.4) node [anchor=north west][inner sep=0.75pt]    {$d'$};
\draw (185,125) node [anchor=north west][inner sep=0.75pt]    {$v$};
\draw (158,129) node [anchor=north west][inner sep=0.75pt]    {$v'$};
\draw (183,147.4) node [anchor=north west][inner sep=0.75pt]    {$w$};
\draw (182,176.4) node [anchor=north west][inner sep=0.75pt]    {$w'$};
\draw (296,148) node [anchor=north west][inner sep=0.75pt]    {$a'$};
\draw (329,147.4) node [anchor=north west][inner sep=0.75pt]    {$p$};
\draw (356,147.4) node [anchor=north west][inner sep=0.75pt]    {$p'$};
\draw (358,124) node [anchor=north west][inner sep=0.75pt]    {$q$};
\draw (372,98.4) node [anchor=north west][inner sep=0.75pt]    {$q'$};
\draw (464,42) node [anchor=north west][inner sep=0.75pt]    {$b'$};
\draw (322,81.4) node [anchor=north west][inner sep=0.75pt]    {$c'$};
\draw (320,59.4) node [anchor=north west][inner sep=0.75pt]    {$r$};
\draw (324,31.4) node [anchor=north west][inner sep=0.75pt]    {$r'$};
\draw (293,46) node [anchor=north west][inner sep=0.75pt]    {$u$};
\draw (265,40) node [anchor=north west][inner sep=0.75pt]    {$u'$};
\end{tikzpicture}.
\end{equation}
Then, analogous to the LW model, we define $B_p^\text{WW}=\sum_s (\mathtt d_s/D^2)B_p^s$, where $D^2=\sum_{s\in L}\mathtt{d}_s^2$, and the action of $B_p^s$ on the state \eqref{eq:5.4} is to fuse a simple loop labeled by $s$ with the edges around the plaquette, through the basis transformation \eqref{eq:5.1}, as shown in \autoref{fig:7}(a). However, things get tricky when we encounter the vertex where $q,q'$, and $b$ meet. As shown in \autoref{fig:7}(b), explicitly applying the basis transformation \eqref{eq:5.1} results in a state where string $s$ fuses with $q'$, which is not what we want. Thus, we have to twist the vertex using \eqref{eq:5.2} before applying \eqref{eq:5.1}, and twist the vertex back finally to recover the original lattice. The same procedure is also applied when dealing with the vertex where $c,c'$, and $r$ meet. Therefore, we get four extra $R$-symbols in the expression of the matrix elements of $B_p^s$ comparing with the LW model. 

\begin{figure}[ht]
    \centering
\begin{tikzpicture}[x=0.75pt,y=0.75pt,yscale=-0.8,xscale=0.8]

\draw    (205.51,269.14) .. controls (197.67,250.01) and (270.67,217.01) .. (298.67,180.51) ;
\draw    (244.51,250.14) -- (313.63,181.03) ;
\draw [color={rgb, 255:red, 255; green, 255; blue, 255 }  ,draw opacity=1 ][line width=3.75]    (265.01,203.39) -- (265.01,240.5) ;
\draw    (47.01,269) .. controls (39.17,249.86) and (112.17,216.86) .. (140.17,180.36) ;
\draw [color={rgb, 255:red, 255; green, 255; blue, 255 }  ,draw opacity=1 ][line width=3.75]    (86.01,212.89) -- (86.01,250) ;
\draw    (181.16,92.46) -- (207.66,67.46) ;
\draw [color={rgb, 255:red, 255; green, 255; blue, 255 }  ,draw opacity=1 ][line width=3.75]    (194.13,69.93) -- (194.13,96.53) ;
\draw    (157.54,46.99) -- (157.54,92.78) ;
\draw [color={rgb, 255:red, 255; green, 255; blue, 255 }  ,draw opacity=1 ][fill={rgb, 255:red, 255; green, 255; blue, 255 }  ,fill opacity=1 ][line width=3.75]    (135.08,54.42) -- (200.33,54.42) ;
\draw    (5,110.15) -- (23.7,110.15) ;
\draw    (23.7,110.15) -- (42.4,110.15) ;
\draw    (42.4,110.15) -- (61.1,110.15) ;
\draw    (11.27,122.53) -- (23.7,110.15) ;
\draw    (70.45,96.53) -- (42.4,110.15) ;
\draw    (61.1,110.15) -- (61.1,155.94) ;
\draw    (61.1,110.15) -- (147.38,110.15) ;
\draw    (70.45,96.53) -- (120.15,46.99) ;
\draw    (70.45,69.93) -- (70.45,96.53) ;
\draw    (134.96,122.53) -- (147.38,110.15) ;
\draw    (147.38,110.15) -- (166.08,110.15) ;
\draw    (166.08,110.15) -- (184.78,110.15) ;
\draw    (194.13,96.53) -- (166.08,110.15) ;
\draw    (184.78,110.15) -- (184.78,155.94) ;
\draw    (184.78,110.15) -- (203.48,110.15) ;
\draw    (194.13,96.53) -- (243.83,46.99) ;
\draw    (194.13,69.93) -- (194.13,96.53) ;
\draw    (101.45,46.99) -- (120.15,46.99) ;
\draw    (120.15,46.99) -- (138.85,46.99) ;
\draw    (166.89,33.37) -- (138.85,46.99) ;
\draw    (166.89,33.37) -- (179.32,20.99) ;
\draw    (166.89,6.77) -- (166.89,33.37) ;
\draw    (138.85,46.99) -- (157.54,46.99) ;
\draw    (157.54,46.99) -- (243.83,46.99) ;
\draw    (243.83,46.99) -- (262.53,46.99) ;
\draw    (262.53,46.99) -- (281.23,46.99) ;
\draw    (290.58,33.37) -- (262.53,46.99) ;
\draw    (281.23,46.99) -- (281.23,92.78) ;
\draw    (281.23,46.99) -- (299.93,46.99) ;
\draw    (290.58,33.37) -- (303,20.99) ;
\draw    (290.58,6.77) -- (290.58,33.37) ;
\draw    (135.08,54.42) -- (200.33,54.42) ;
\draw    (96.08,103.17) -- (161.33,103.17) ;
\draw    (200.33,54.42) .. controls (215.66,54.21) and (211.66,63.96) .. (207.66,67.46) ;
\draw    (96.08,103.17) .. controls (80.76,103.39) and (84.76,93.64) .. (88.76,90.14) ;
\draw    (135.08,54.42) .. controls (126.3,54.64) and (121.05,59.39) .. (115.26,65.14) ;
\draw    (88.76,90.14) -- (115.26,65.14) ;
\draw    (161.33,103.17) .. controls (170.11,102.96) and (175.36,98.21) .. (181.16,92.46) ;
\draw    (532.16,92.46) -- (558.66,67.46) ;
\draw [color={rgb, 255:red, 255; green, 255; blue, 255 }  ,draw opacity=1 ][line width=3.75]    (545.13,69.93) -- (545.13,96.53) ;
\draw    (508.54,46.99) -- (508.54,92.78) ;
\draw [color={rgb, 255:red, 255; green, 255; blue, 255 }  ,draw opacity=1 ][fill={rgb, 255:red, 255; green, 255; blue, 255 }  ,fill opacity=1 ][line width=3.75]    (486.08,54.42) -- (551.33,54.42) ;
\draw    (374.7,110.15) -- (393.4,110.15) ;
\draw    (393.4,110.15) -- (412.1,110.15) ;
\draw    (421.45,96.53) -- (393.4,110.15) ;
\draw    (412.1,110.15) -- (412.1,155.94) ;
\draw    (412.1,110.15) -- (498.38,110.15) ;
\draw    (421.45,96.53) -- (471.15,46.99) ;
\draw    (421.45,69.93) -- (421.45,96.53) ;
\draw    (485.96,122.53) -- (498.38,110.15) ;
\draw    (498.38,110.15) -- (517.08,110.15) ;
\draw    (517.08,110.15) -- (535.78,110.15) ;
\draw    (545.13,96.53) -- (517.08,110.15) ;
\draw    (535.78,110.15) -- (535.78,155.94) ;
\draw    (535.78,110.15) -- (554.48,110.15) ;
\draw    (545.13,96.53) -- (594.83,46.99) ;
\draw    (545.13,69.93) -- (545.13,96.53) ;
\draw    (452.45,46.99) -- (471.15,46.99) ;
\draw    (471.15,46.99) -- (489.85,46.99) ;
\draw    (517.89,33.37) -- (489.85,46.99) ;
\draw    (517.89,33.37) -- (530.32,20.99) ;
\draw    (517.89,6.77) -- (517.89,33.37) ;
\draw    (489.85,46.99) -- (508.54,46.99) ;
\draw    (508.54,46.99) -- (594.83,46.99) ;
\draw    (594.83,46.99) -- (613.53,46.99) ;
\draw    (613.53,46.99) -- (632.23,46.99) ;
\draw    (641.58,33.37) -- (613.53,46.99) ;
\draw    (632.23,46.99) -- (632.23,92.78) ;
\draw    (632.23,46.99) -- (650.93,46.99) ;
\draw    (641.58,33.37) -- (654,20.99) ;
\draw    (641.58,6.77) -- (641.58,33.37) ;
\draw    (486.08,54.42) -- (551.33,54.42) ;
\draw    (491.98,103.17) -- (512.33,103.17) ;
\draw    (551.33,54.42) .. controls (566.66,54.21) and (562.66,63.96) .. (558.66,67.46) ;
\draw    (447.08,103.17) .. controls (431.76,103.39) and (435.76,93.64) .. (439.76,90.14) ;
\draw    (486.08,54.42) .. controls (477.3,54.64) and (472.05,59.39) .. (466.26,65.14) ;
\draw    (439.76,90.14) -- (466.26,65.14) ;
\draw    (512.33,103.17) .. controls (521.11,102.96) and (526.36,98.21) .. (532.16,92.46) ;
\draw    (356,110.15) -- (374.7,110.15) ;
\draw    (362.27,122.53) -- (374.7,110.15) ;
\draw    (447.08,103.17) .. controls (453.62,103.21) and (454.87,105.46) .. (451.12,109.96) ;
\draw    (491.98,103.17) .. controls (483.19,103.39) and (479.48,105.71) .. (477.23,110.21) ;
\draw [line width=1.5]    (311,87) -- (336.34,87) ;
\draw [shift={(340.34,87)}, rotate = 180] [fill={rgb, 255:red, 0; green, 0; blue, 0 }  ][line width=0.08]  [draw opacity=0] (8.75,-4.2) -- (0,0) -- (8.75,4.2) -- (5.81,0) -- cycle    ;
\draw    (86.01,250) -- (47.01,269) ;
\draw    (86.01,250) -- (155.13,180.89) ;
\draw    (86.01,212.89) -- (86.01,250) ;
\draw    (47.01,269) -- (27.5,278.5) ;
\draw    (244.51,250.14) -- (205.51,269.14) ;
\draw    (205.51,269.14) -- (186,278.64) ;
\draw [line width=1.5]    (155,230.5) -- (180.34,230.5) ;
\draw [shift={(184.34,230.5)}, rotate = 180] [fill={rgb, 255:red, 0; green, 0; blue, 0 }  ][line width=0.08]  [draw opacity=0] (8.75,-4.2) -- (0,0) -- (8.75,4.2) -- (5.81,0) -- cycle    ;
\draw    (265.01,203.39) -- (265.01,240.5) ;
\draw    (244.51,250.14) .. controls (253.32,266.76) and (265,274.84) .. (265.01,240.5) ;
\draw    (415.65,237.34) .. controls (402.15,223.34) and (429.67,217.01) .. (457.67,180.51) ;
\draw    (403.51,250.14) -- (472.63,181.03) ;
\draw [color={rgb, 255:red, 255; green, 255; blue, 255 }  ,draw opacity=1 ][line width=3.75]    (424.01,203.39) -- (424.01,240.5) ;
\draw    (403.51,250.14) -- (345,278.64) ;
\draw    (424.01,203.39) -- (424.01,240.5) ;
\draw    (403.51,250.14) .. controls (412.32,266.76) and (424,274.84) .. (424.01,240.5) ;
\draw    (575.65,237.2) .. controls (562.15,223.2) and (589.67,216.86) .. (617.67,180.36) ;
\draw    (563.51,250) -- (632.63,180.89) ;
\draw    (563.51,250) -- (505,278.5) ;
\draw    (563.51,212.89) -- (563.51,250) ;
\draw [line width=1.5]    (315,230.5) -- (340.34,230.5) ;
\draw [shift={(344.34,230.5)}, rotate = 180] [fill={rgb, 255:red, 0; green, 0; blue, 0 }  ][line width=0.08]  [draw opacity=0] (8.75,-4.2) -- (0,0) -- (8.75,4.2) -- (5.81,0) -- cycle    ;
\draw [line width=1.5]    (475.34,230.5) -- (500.67,230.5) ;
\draw [shift={(504.67,230.5)}, rotate = 180] [fill={rgb, 255:red, 0; green, 0; blue, 0 }  ][line width=0.08]  [draw opacity=0] (8.75,-4.2) -- (0,0) -- (8.75,4.2) -- (5.81,0) -- cycle    ;

\draw (95.2,111) node [anchor=north west][inner sep=0.75pt]    {\small $a$};
\draw (219.3,70) node [anchor=north west][inner sep=0.75pt]    {\small $b$};
\draw (196.59,35.02) node [anchor=north west][inner sep=0.75pt]    {\small $c$};
\draw (87.29,58.68) node [anchor=north west][inner sep=0.75pt]    {\small $d$};
\draw (56,74.44) node [anchor=north west][inner sep=0.75pt]    {\small $d'$};
\draw (48.5,92.36) node [anchor=north west][inner sep=0.75pt]    {\small $v$};
\draw (28.5,94) node [anchor=north west][inner sep=0.75pt]    {\small $v'$};
\draw (45,111) node [anchor=north west][inner sep=0.75pt]    {\small $w$};
\draw (43,130.35) node [anchor=north west][inner sep=0.75pt]    {\small $w'$};
\draw (130,120) node [anchor=north west][inner sep=0.75pt]    {\small $a'$};
\draw (152.01,111) node [anchor=north west][inner sep=0.75pt]    {\small $p$};
\draw (169,109.56) node [anchor=north west][inner sep=0.75pt]    {\small $p'$};
\draw (186.86,96.5) node [anchor=north west][inner sep=0.75pt]    {\small $q$};
\draw (194,74) node [anchor=north west][inner sep=0.75pt]    {\small $q'$};
\draw (248.67,31) node [anchor=north west][inner sep=0.75pt]    {\small $b'$};
\draw (144,62.26) node [anchor=north west][inner sep=0.75pt]    {\small $c'$};
\draw (153.79,36.8) node [anchor=north west][inner sep=0.75pt]    {\small $r$};
\draw (147.99,25) node [anchor=north west][inner sep=0.75pt]    {\small $r'$};
\draw (125.98,35.74) node [anchor=north west][inner sep=0.75pt]    {\small $u$};
\draw (105.43,30) node [anchor=north west][inner sep=0.75pt]    {\small $u'$};
\draw (12,8) node [anchor=north west][inner sep=0.75pt]   [align=left] {(a)};
\draw (173.36,82.07) node [anchor=north west][inner sep=0.75pt]    {\small $s$};

\draw (432.45,111) node [anchor=north west][inner sep=0.75pt]    {\small $a$};
\draw (570.3,70) node [anchor=north west][inner sep=0.75pt]    {\small $b$};
\draw (547.59,35.02) node [anchor=north west][inner sep=0.75pt]    {\small $c$};
\draw (438.29,58.68) node [anchor=north west][inner sep=0.75pt]    {\small $d$};
\draw (406,74.44) node [anchor=north west][inner sep=0.75pt]    {\small $d'$};
\draw (399.5,92.36) node [anchor=north west][inner sep=0.75pt]    {\small $v$};
\draw (379.5,94) node [anchor=north west][inner sep=0.75pt]    {\small $v'$};
\draw (396,111) node [anchor=north west][inner sep=0.75pt]    {\small $w$};
\draw (394,130.35) node [anchor=north west][inner sep=0.75pt]    {\small $w'$};
\draw (483,120) node [anchor=north west][inner sep=0.75pt]    {\small $a'$};
\draw (501,111) node [anchor=north west][inner sep=0.75pt]    {\small $p$};
\draw (520,109.56) node [anchor=north west][inner sep=0.75pt]    {\small $p'$};
\draw (537.86,96.5) node [anchor=north west][inner sep=0.75pt]    {\small $q$};
\draw (544,74) node [anchor=north west][inner sep=0.75pt]    {\small $q'$};
\draw (599.67,31) node [anchor=north west][inner sep=0.75pt]    {\small $b'$};
\draw (495,62.26) node [anchor=north west][inner sep=0.75pt]    {\small $c'$};
\draw (504.79,36.8) node [anchor=north west][inner sep=0.75pt]    {\small $r$};
\draw (498.99,25) node [anchor=north west][inner sep=0.75pt]    {\small $r'$};
\draw (476.98,35.74) node [anchor=north west][inner sep=0.75pt]    {\small $u$};
\draw (456.43,30) node [anchor=north west][inner sep=0.75pt]    {\small $u'$};
\draw (524.36,82.07) node [anchor=north west][inner sep=0.75pt]    {\small $s$};
\draw (455,110.57) node [anchor=north west][inner sep=0.75pt]    {\small $a''$};
\draw (479,111) node [anchor=north west][inner sep=0.75pt]    {\small $a$};

\draw (122.57,217.84) node [anchor=north west][inner sep=0.75pt]    {\small $b$};
\draw (65.5,258.4) node [anchor=north west][inner sep=0.75pt]    {\small $q$};
\draw (86,223) node [anchor=north west][inner sep=0.75pt]    {\small $q'$};
\draw (34,274.26) node [anchor=north west][inner sep=0.75pt]    {\small $q''$};
\draw (109,190.26) node [anchor=north west][inner sep=0.75pt]    {\small $s$};
\draw (281.07,217.99) node [anchor=north west][inner sep=0.75pt]    {\small $b$};
\draw (224,258.54) node [anchor=north west][inner sep=0.75pt]    {\small $q$};
\draw (267.01,242.9) node [anchor=north west][inner sep=0.75pt]    {\small $q'$};
\draw (192.5,274.4) node [anchor=north west][inner sep=0.75pt]    {\small $q''$};
\draw (233.5,216.9) node [anchor=north west][inner sep=0.75pt]    {\small $s$};
\draw (440.07,217.99) node [anchor=north west][inner sep=0.75pt]    {\small $b$};
\draw (426.01,242.9) node [anchor=north west][inner sep=0.75pt]    {\small $q'$};
\draw (375.51,262.54) node [anchor=north west][inner sep=0.75pt]    {\small $q''$};
\draw (436.5,183.4) node [anchor=north west][inner sep=0.75pt]    {\small $s$};
\draw (396,233.73) node [anchor=north west][inner sep=0.75pt]    {\small $b''$};
\draw (600.07,217.84) node [anchor=north west][inner sep=0.75pt]    {\small $b$};
\draw (549,226.26) node [anchor=north west][inner sep=0.75pt]    {\small $q'$};
\draw (535.51,262.4) node [anchor=north west][inner sep=0.75pt]    {\small $q''$};
\draw (596.5,183.26) node [anchor=north west][inner sep=0.75pt]    {\small $s$};
\draw (570.65,245.59) node [anchor=north west][inner sep=0.75pt]    {\small $b''$};
\draw (12,168) node [anchor=north west][inner sep=0.75pt]   [align=left] {(b)};

\end{tikzpicture}
\caption{(a) Fusing the simple loop labeled by $s$ with the edge labeled by $a$. (b) In order to fuse the string labeled by $s$ with edge labeled by $b$, first we have to act a $R$-symbol at the vertex and then apply the basis transformation \eqref{eq:5.1}; otherwise the string $s$ will fuse with edge labeled by $q'$, which is not in the boundary of the plaquette. }
\label{fig:7}
\end{figure}
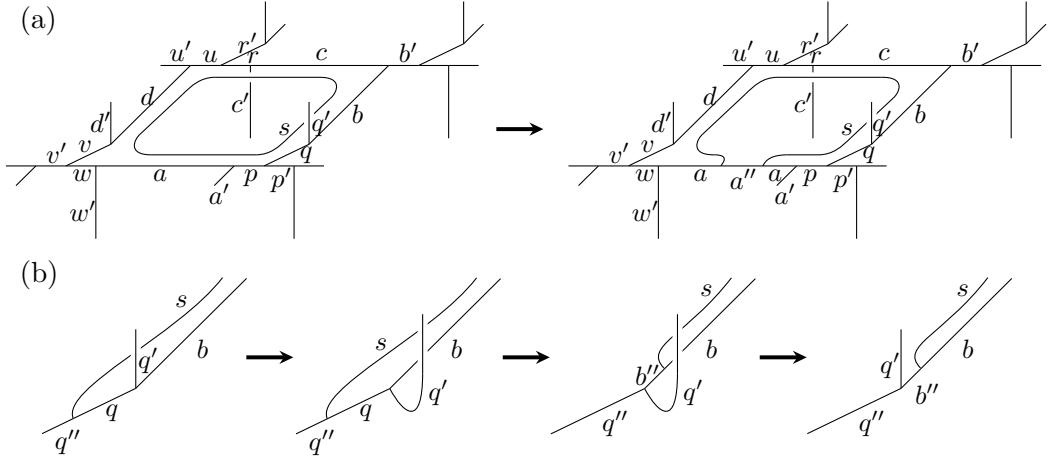

Denote the state \eqref{eq:5.4} as $|\Psi_{abcdpqruvw}\rangle$, then the explicit expression of the matrix elements of $B_p^s$ is given by 
\begin{equation}
    \begin{aligned}
        &\langle\Psi_{a''\cdots w''}|B_p^s|\Psi_{a\cdots w}\rangle\\
        ={}&\mathtt{v}_a\cdots\mathtt v_{w}[\mathtt v''] R^{q'b}_q\overline{R^{c'r}_c}\overline{R^{q'b''}_{q''}}R^{c'r''}_{c''}\\
        &\times G_{sa''p''}^{a'pa}G_{sp''q''}^{p'qp}G^{q'bq}_{sq''b''}G^{b'cb}_{sb''c''}G^{c'rc}_{sc''r''}G^{r'ur}_{sr''u''}G^{u'du}_{su''d''}G^{d'vd}_{sd''v''}G^{v'wv}_{sv''w''} G_{sw''a''}^{w'aw}.
    \end{aligned}
\end{equation}
Note that in order to evaluate the above expression, here we use the following convention to remove the bubble: 
\begin{equation}\label{eq:2.19}
    \bigg|

\bigg\rangle.
\end{equation}
which is different from our graphical presentation \eqref{eq:3.16}.

\section{Some proofs}\label{appendix:A}
Here we prove that the rewritten rep-basis defined in \eqref{eq:3.26} is orthonormal and complete. 
\begin{align}\label{eq:A.1}
    &\;\;\;\;\langle\Psi'_{s'm_s'}|\Psi_{sm_s}\rangle=\sum_{jikgh\in G}\langle\Psi'_{s'm_s'}|jikgh\rangle\langle jikgh|\Psi_{sm_s}\rangle\notag\\
    &=\frac{v_\mu v_\nu v_\rho v_\eta v_\lambda v_\alpha v_\beta v_\gamma v_s[v']}{|G|^5}\sum_{jikgh\in G}\left(%
    \right),
\end{align}
where \eqref{eq:3.16} is used, and the $\beta_\alpha$ comes from $v_{\alpha}v_{\alpha'}\delta_{\alpha'\alpha}/\mathtt{d}_\alpha$. Although the ends labeled by $m$ and $n$ in the graphical presentation in the last line of the above expression are summed, they cannot be connected explicitly. In order to evaluate the above graphical presentation, we use the following trick: 
\begin{align}\label{eq:A.2}
        \sum _{n}%
,
\end{align}
where the normalization condition of duality map and \eqref{eq:3.17} are used. Noting that 
\begin{equation}\label{eq:A.3}
    %
,
\end{equation}
we can simplify the graphical presentation in \eqref{eq:A.1} as 
\begin{equation}
    %
        =\frac{\beta _{\nu } \beta _{\lambda } \beta _{\beta } \beta _{\gamma } \beta _{s}}{d_{\beta } d_{\gamma } d_{s}} \delta _{\beta '\beta } \delta _{\gamma '\gamma } \delta _{s's} \delta _{m'_{s} m_{s}}.
\end{equation}
Inserting the above expression into \eqref{eq:A.1} yields
\begin{equation}
    \begin{aligned}
        \langle\Psi'_{s'm_s'}|\Psi_{sm_s}\rangle={}&\beta_\nu \beta_\lambda \beta_\alpha \beta_\beta \beta_\gamma \beta_s \delta_{\mu'\mu}\delta_{\nu'\nu}\delta_{\eta'\eta}\delta_{\lambda'\lambda}\delta_{\rho'\rho}\delta_{\alpha'\alpha}\delta _{\beta '\beta } \delta _{\gamma '\gamma } \delta _{s's}\\
    &\times\delta_{n'_\mu n_\mu}\delta_{m_\nu'm_\nu}\delta_{n'_\eta n_\eta}\delta_{m_\lambda' m_\lambda}\delta_{n'_\rho n_\rho}\delta _{m'_{s} m_{s}}.
    \end{aligned}
\end{equation}
Finally, using the fact that $\beta_\mu\beta_\nu\beta_\rho=1$ when $C_{\mu\nu\rho}\ne 0$, we have $\beta_\nu \beta_\lambda \beta_\alpha \beta_\beta \beta_\gamma \beta_s=1$, and thus it is proved that the basis $|\Psi_{sm_s}\rangle$ is orthonormal. 

The proof of completeness is as follows: 
\begin{align}\label{eq:A.6}
    &\;\;\;\;\sum_{\mu\cdots s}\sum_{m_\nu \cdots m_s}
    \langle jikgh|\Psi_{sm_s}\rangle\langle\Psi_{sm_s}|j'i'k'g'h'\rangle\notag\\
    &=\sum_{\mu\cdots s}\sum_{m_\nu \cdots m_s}\frac{d_\mu d_\nu d_\rho d_\eta d_\lambda d_\alpha d_\beta d_\gamma d_s}{|G|^5}\left(
,
\end{align}
where $\tilde{g}=g\bar g'$, $\tilde{k}=k\bar k'$, and $\tilde{j}=j\bar j'$. Then, note that 
\begin{equation}
    %
,
\end{equation}
where \eqref{eq:3.15} is used. Using the similar method, we can simplify \eqref{eq:A.6} as 
\begin{equation}\label{eq:A.8}
    \begin{aligned}
        &\sum_{\mu\cdots s}\sum_{m_\nu \cdots m_s}
        \langle jikgh|\Psi_{sm_s}\rangle\langle\Psi_{sm_s}|j'i'k'g'h'\rangle\\
        ={}&\sum_{\mu\cdots s}\sum_{m_\nu m_\lambda m_s}\frac{d_\mu d_\nu d_\rho d_\eta d_\lambda d_\alpha d_\beta d_\gamma d_s}{|G|^5}\sum_{l_1l_2l_3l_4\in G}\frac{1}{|G|^4}\\
        &\times %
.
    \end{aligned}
\end{equation}
Let us consider the first term in the graphical representation above, which reads 
\begin{equation}\label{eq:A.9}
    \sum_{m_\nu,k,l}D^\nu_{m_\nu k}(i)D^\nu_{kl}(l_4)D^\nu_{lm_\nu}(\bar{i}')=\operatorname{Tr}D^\nu(il_4\bar i')\equiv\operatorname{Tr}_\nu(\tilde{i}'l_4),
\end{equation}
where $\tilde{i}'=\bar i'i$. Recalling that 
\begin{equation}\label{eq:A.10}
    \delta_{l,e}=\sum_{\mu\in L_G}\frac{1}{|G|}d_\mu \operatorname{Tr}D^\mu(l),
\end{equation}
we find that the term \eqref{eq:A.9} gives a factor of $\delta_{\tilde{i}'l_4,e}$. Similarly, the second term of \eqref{eq:A.8} gives $\delta_{l_4,e}$ and the sixth term gives $\delta_{\tilde{h}'l_2,e}$ with $\tilde{h}'=\bar h'h$. The third term of \eqref{eq:A.8} reads 
\begin{align}
    \sum_{mnkl} D^{\gamma^*}_{mn}(l_4)(\Omega^\gamma)^{-1}_{mk}D^\gamma_{kl}(l_3)\Omega^{\gamma^*}_{nl}&=\sum_{mnkl}\beta_\gamma (\Omega^{\gamma^*})^{-1}_{km}D^{\gamma^*}_{mn}(l_4)\Omega^{\gamma^*}_{nl}D^\gamma_{kl}(l_3)\\
    &=\sum_{kl}\beta_\gamma [D^\gamma_{kl}(l_4)]^*D^\gamma_{kl}(l_3)=\beta_\gamma\operatorname{Tr}_\gamma(\bar l_4l_3),
\end{align}
where \eqref{eq:3.2}, the unitarity of $\Omega^\mu$, and the definition of $\beta_\mu$, $(\Omega^{\mu^*})^\mathtt{T}=\beta_\mu\Omega^\mu$ are used. Therefore, the third term gives $\beta_\gamma\delta_{\bar l_4l_3,e}$ in \eqref{eq:A.8}. \eqref{eq:A.8} thus can be further simplified as 
\begin{equation}
    \begin{aligned}
        &\sum_{\mu\cdots s}\sum_{m_\nu \cdots m_s}\langle jikgh|\Psi_{sm_s}\rangle\langle\Psi_{sm_s}|j'i'k'g'h'\rangle\\
        ={}&\sum_{l_1l_2l_3l_4\in G}\beta_{\alpha}\beta_\beta\beta_\gamma\beta_{\mu}\beta_\rho\beta_\eta\times\delta_{\tilde{i}'l_4,e}\delta_{l_4,e}\delta_{\bar l_4l_3,e}\delta_{\bar l_3\tilde{g},e}\delta_{\bar l_3l_2,e}\delta_{\tilde{h}'l_2,e}\delta_{\bar l_2l_1,e}\delta_{\bar l_1\tilde{k},e}\delta_{\bar l_1\tilde{j},e}.
    \end{aligned}
\end{equation}
Noting that $\beta_\alpha\beta_\rho\beta_\mu=1$ and $\beta_\beta\beta_\gamma\beta_\eta=1$, we finally get 
\begin{equation}
    \sum_{\mu\cdots s}\sum_{m_\nu \cdots m_s}\langle jikgh|\Psi_{sm_s}\rangle\langle\Psi_{sm_s}|j'i'k'g'h'\rangle=\delta_{\bar i'i,e}\delta_{g\bar g',e}\delta_{\bar h'h,e}\delta_{k\bar k',e}\delta_{j\bar j',e}=\langle jikgh|j'i'k'g'h'\rangle,
\end{equation}
which implies that
\begin{equation}
    \sum_{\mu\cdots s}\sum_{m_\nu \cdots m_s}|\Psi_{sm_s}\rangle\langle\Psi_{sm_s}|=\mathbb 1,
\end{equation}
i.e., the basis $|\Psi_{sm_s}\rangle$ is complete. 

We then prove that the rewritten rep-basis with braided edges defined by \eqref{eq:3.30x} and \eqref{eq:3.31x} is orthonormal and complete. We define 
\begin{equation}
|\mu \nu \rangle:=  \bigg|
\notag\\
&=\delta_{\rho'\rho}\delta_{\mu'\mu}\delta_{\nu'\nu}\delta_{m_{\mu'}m_\mu}\delta_{m_{\nu'}m_\nu}\delta_{n_{\mu'}n_\mu}\delta_{n_{\nu'}n_\nu},
\end{align}
which proves the orthonormality. 

To prove the completeness, we need to compute 
\begin{align*}
&\sum_{\mu,\nu}\sum_{m_\mu m_\nu n_\nu n_\mu}\langle gh|\mu\nu\rangle\langle\mu\nu|g'h'\rangle\\
    ={}&\sum_{\mu,\nu}\sum_{m_\mu m_\nu n_\nu n_\mu}\frac{d_\mu d_\rho}{|G|^2}\sum_{\rho,\rho'}\mathtt{d}_{\rho}\mathtt{d}_{\rho'}%
\\
={}&\sum_{\mu,\nu}\sum_{m_\mu m_\nu}\frac{d_\mu d_\rho}{|G|^2}D^\mu_{m_\mu m_\mu}(g\bar g')D^\nu_{m_\nu m_\nu}(h\bar h')\\
={}&\delta_{g\bar g',e}\delta_{h\bar h',e}=\delta_{gg'}\delta_{hh'}=\langle gh|g'h'\rangle.
\end{align*}
Thus 
\begin{equation}
    \sum_{\mu,\nu}\sum_{m_\mu m_\nu n_\mu n_\nu}|\mu\nu\rangle\langle\mu\nu|=\mathbb{1},
\end{equation}
i.e., the basis $|\mu\nu\rangle$ is complete. 

Details of the action of the operators $\overline{\tilde{A}_v^\text{GT}}$ are as follows. 
\begin{align}
        &{\overline{\tilde A^\text{GT}_v}}|\Psi_{sm_s}\rangle\notag\\
    ={}&\sum_{\begin{subarray}{c}
        \mu'\cdots s'\\n'_{\mu}\cdots m_s'
    \end{subarray}}\sum_{jikgh\in G}\frac{1}{|K|}\sum_{x\in K}\bigl(\langle\Psi'_{s'm_{s'}}|xj,i\bar x,xk,xg,h\bar x\rangle\langle jikgh|\Psi_{sm_s} \rangle\bigr)|\Psi'_{s'm_{s'}}\rangle\notag\\
    ={}&\sum_{\begin{subarray}{c}
        \mu'\cdots s'\\n'_{\mu}\cdots m_s'
    \end{subarray}}\sum_{jikgh\in G}\frac{1}{|K|}\sum_{x\in K}\frac{v_\mu v_\nu v_\rho v_\eta v_\lambda v_\alpha v_\beta v_\gamma v_s [v']}{|G|^5}\notag\\
    &\times\left( 

        \right)|\Psi'_{s'm_{s'}}\rangle\notag.
\end{align}
In the third equality the fact that $3j$-symbols are intertwiners is used. Applying the great orthogonality theorem and the trick introduced in \eqref{eq:A.2}, we have 
\begin{align}
    {\overline{\tilde A^\text{GT}_v}}|\Psi_{sm_s}\rangle ={}&\sum_{\begin{subarray}{c}
        \mu'\cdots s'\\n'_{\mu}\cdots m_s'
    \end{subarray}}\delta_{\mu'\mu}\cdots\delta_{\lambda'\lambda}\delta_{n_\mu'n_\mu}\cdots\delta_{m_\nu'm_\nu}\frac{1}{|K|}\sum_{x\in K}\beta_\lambda\beta_\nu%
|\Psi'_{s'm_s'}\rangle\notag\\
    &=\sum_{m_s'}\frac{1}{|K|}\sum_{x\in K}\beta_\lambda\beta_\nu\beta_\alpha\beta_\beta\beta_\gamma\beta_s D^s_{m_sm'_s}(\bar x)|\Psi_{sm_s'}\rangle\notag\\
    &=\sum_{m_s'}\frac{1}{|K|}\sum_{x\in K}D^s_{m_sm'_s}(x)|\Psi_{sm_s'}\rangle.
\end{align}

Details of the action of the operators $\overline{\skew{-1}{\tilde}{C}_e^\text{GT}}$ are as follows. Recall that 
\begin{align}\label{eq:A.17}
&\;\;\;\;\langle\Psi^{\nu'\pi'\phi'\lambda'}_{s'\tilde{m}_s',r'\tilde{m}'_r}|\overline{\skew{-1}{\tilde}{C}_e^\text{GT}}|\Psi^{\nu\pi\phi\lambda}_{s\tilde{m}_s,r\tilde{m}_r}\rangle\notag\\
    &=\sum_{g\cdots w\in G}\sum_{(t,\alpha_t)\in L_A}\frac{|K|}{|G|}d_tD^t_{\alpha_t\alpha_t}(l)\langle\Psi^{\nu'\pi'\phi'\lambda'}_{s'\tilde{m}_s',r'\tilde{m}'_r}|\Psi_l\rangle\langle\Psi_l|\Psi^{\nu\pi\phi\lambda}_{s\tilde{m}_a,r\tilde{m}_r}\rangle.
\end{align}
Similar to the calculation in \autoref{sec:5}, we can firstly evaluate $D^t_{\alpha_t\alpha_t}(l)\langle\Psi_l|\Psi^{\nu\pi\phi\lambda}_{s\tilde{m}_a,r\tilde{m}_r}\rangle$, which reads 
\begin{align}
    D^t_{\alpha_t\alpha_t}(l)\langle\Psi_l|\Psi^{\nu\pi\phi\lambda}_{s\tilde{m}_s,r\tilde{m}_r}\rangle={}&\frac{v_\nu v_\pi v_\phi v_\lambda v_s v_r v_\mu v_\rho v_\sigma v_\eta v_\delta }{\sqrt{|G|}}\notag\\
    &\times 
,
\end{align}
where some edges and Latin indices irrelevant to the calculation are omitted. Note that the braiding between the lines $t$ and $\sigma$ is unavoidable if we want to evaluate the expression correctly. Then we can use the convention \eqref{eq:5.10} and apply $F$-moves to the graph in the equation above and obtain 
\begin{align}
    &%
\notag\\
={}&\sum_{s'\nu'\pi'\phi'\lambda'r'}\mathtt d_{s'}F^{\mu^*\nu s^*}_{ts'^*\nu'}F^{\rho\pi\nu^*}_{t\nu'^*\pi'}R^{\pi\sigma}_\phi F^{\sigma^*\phi\pi^*}_{t\pi'^*\phi'}\overline{R^{\pi'\sigma}_{\phi'}}F^{\eta\lambda\phi^*}_{t\phi'^*\lambda'}F^{t\lambda'^*\lambda}_{\delta r^*r'}C^{st;\tilde{m}_s\alpha_t}_{s'\tilde{m}_{s'}}C^{t^*r;\tilde {m}_{t^*}\tilde{m}_r}_{r'\tilde{m}_{r'}}(\Omega^t)^{-1}_{\tilde{m}_{t^*}\alpha_t}\notag\\
&\times\frac{\sqrt{|G|}}{v_{\nu'}v_{\pi'}v_{\phi'}v_{\lambda'}v_{s'}v_{r'}v_\mu v_\rho v_\sigma v_\eta v_\delta}\langle\Psi_l|\Psi^{\nu'\pi'\phi'\lambda'}_{s'\tilde{m}_{s'},r'\tilde{m}_{r'}}\rangle
\end{align}
Substituting the result above back to \eqref{eq:A.17} and using the completeness and orthonormality of our basis, we get 
\begin{align}
    &\langle\Psi^{\nu'\pi'\phi'\lambda'}_{s'\tilde{m}_s',r'\tilde{m}'_r}|\overline{\skew{-1}{\tilde}{C}_e^\text{GT}}|\Psi^{\nu\pi\phi\lambda}_{s\tilde{m}_s,r\tilde{m}_r}\rangle\notag\\
    ={}&\sum_{(t,\alpha_t)\in L_A}\frac{|K|}{|G|}d_t v_\nu v_\pi v_\phi v_\lambda v_s v_r[v']^{-1}\mathtt{d}_{s'}\mathtt{d}_{\nu'}\mathtt{d}_{\pi'}\mathtt{d}_{\phi'}\mathtt{d}_{\lambda'}\mathtt{d}_{r'}\notag\\
    &\times R^{\pi\sigma}_\phi\overline{R^{\pi'\sigma}_{\phi'}}G^{\mu^*\nu s^*}_{ts'^*\nu'}G^{\rho\pi\nu^*}_{t\nu'^*\pi'}G^{\sigma^*\phi\pi^*}_{t\pi'^*\phi'}G^{\eta\lambda\phi^*}_{t\phi'^*\lambda'}G^{t\lambda'^*\lambda}_{\delta r^*r'}\notag\\
    &\times(C_{sts'^*;\tilde{m}_s\alpha_t\tilde{m}_{s'^*}})^*(\Omega^{s'})^{-1}_{\tilde{m}_{s'^*}\tilde{m}_{s'}}(C_{t^*r r'^*;\tilde{m}_{t^*}\tilde{m}_r\tilde{m}_{r'^*}})^*(\Omega^t)^{-1}_{\tilde{m}_{t^*}\alpha_t}(\Omega^{r'})^{-1}_{\tilde{m}_{r'^*}\tilde{m}_{r'}}\notag\\
    ={}&\sum_{(t,\alpha_t)\in L_A}\frac{|K|}{|G|}\mathtt d_t v_\nu v_\pi v_\phi v_\lambda v_s v_r[v']^{-1}\mathtt{d}_{s'}\mathtt{d}_{\nu'}\mathtt{d}_{\pi'}\mathtt{d}_{\phi'}\mathtt{d}_{\lambda'}\mathtt{d}_{r'}\notag\\
    &\times R^{\pi\sigma}_\phi\overline{R^{\pi'\sigma}_{\phi'}}G^{\mu^*\nu s^*}_{ts'^*\nu'}G^{\rho\pi\nu^*}_{t\nu'^*\pi'}G^{\sigma^*\phi\pi^*}_{t\pi'^*\phi'}G^{\eta\lambda\phi^*}_{t\phi'^*\lambda'}G^{t\lambda'^*\lambda}_{\delta r^*r'}\notag\\
    &\times(C_{s'^*st;\tilde{m}_{s'^*}\tilde{m}_s\alpha_t})^*(\Omega^{s'^*})^{-1}_{\tilde{m}_{s'}\tilde{m}_{s'^*}}(C_{r'^*t^*r ;\tilde{m}_{r'^*}\tilde{m}_{t^*}\tilde{m}_r})^*(\Omega^{t^*})^{-1}_{\alpha_t\tilde{m}_{t^*}}(\Omega^{r'^*})^{-1}_{\tilde{m}_{r'}\tilde{m}_{r'^*}}.
\end{align}
Finally, let us check:
\begin{equation}
    v_\nu v_\pi v_\phi v_\lambda v_s v_r[v']^{-1}\mathtt{d}_{s'}\mathtt{d}_{\nu'}\mathtt{d}_{\pi'}\mathtt{d}_{\phi'}\mathtt{d}_{\lambda'}\mathtt{d}_{r'}=v_\nu v_\pi v_\phi v_\lambda v_s v_r[v']\beta_{\nu'}\beta_{\pi'}\beta_{\phi'}\beta_{\lambda'}\beta_{r'}\beta_{s'}.
\end{equation}
Keeping mind mind that 
\begin{equation}
    \begin{aligned}
        \beta_\mu&=\beta_{s}\beta_\nu=\beta_{s'}\beta_{\nu'}=\sqrt{\beta_{s}\beta_\nu\beta_{s'}\beta_{\nu'}},\\
        \beta_\delta&=\beta_\lambda\beta_r=\beta_{\lambda'}\beta_{r'}=\sqrt{\beta_\lambda\beta_r\beta_{\lambda'}\beta_{r'}},\\
        \beta_\sigma&=\beta_\pi\beta_\phi=\beta_{\pi'}\beta_{\phi'}=\sqrt{\beta_\pi\beta_\phi\beta_{\pi'}\beta_{\phi'}},
    \end{aligned}
\end{equation}
we have 
\begin{equation}
    v_\nu v_\pi v_\phi v_\lambda v_s v_r[v']\beta_{\nu'}\beta_{\pi'}\beta_{\phi'}\beta_{\lambda'}\beta_{r'}\beta_{s'}=\mathtt v_{\nu}\mathtt v_\pi\mathtt v_\phi\mathtt v_\lambda\mathtt v_s\mathtt v_r[\mathtt v'],
\end{equation}
which completes our proof of \eqref{eq:3.35}. 

The proof of \eqref{eq:3.34} is as follows: multiplying the both sides of \eqref{eq:3.34} by $D^t_{\tilde{m}_t\tilde{m}_t'}(\bar l)$ and then summing over $l\in G$ will give 
\begin{align*}
    \frac{1}{|K|}\sum_{l\in G}\delta_{l\in K}D^t_{\tilde m_t\tilde m_t'}(\bar l)&=\sum_{(t,\alpha_t)\in L_A}\frac{d_t}{|G|}\sum_{l\in G}D^t_{\alpha_t\alpha_t}(l)D^t_{\tilde m_t\tilde m_t'}(\bar l)\\
    &=\sum_{(t,\alpha_t)\in L_A}\delta_{\alpha_t\tilde{m}_t}\delta_{\alpha_t\tilde{m}_t'}\equiv \delta_{(t,\tilde{m}_t)\in L_A}\delta_{\tilde{m}_t\tilde{m}_t'}.
\end{align*}
Recalling the definition of $L_A$, we find that the LHS of the equation above is 
\begin{equation*}
    \frac{1}{|K|}\sum_{l\in G}\delta_{l\in K}D^t_{\tilde m_t\tilde m_t'}(\bar l)=\frac{1}{|K|}\sum_{l\in K}D^t_{\tilde m_t\tilde m_t'}(l)=\delta_{(t,\tilde{m}_t)\in L_A}\delta_{\tilde{m}_t\tilde{m}_t'}=\text{RHS}.
\end{equation*}
Thus \eqref{eq:3.34} is proved. Requiring $K=\{e\}$ in \eqref{eq:3.34} gives \eqref{eq:A.10}.

\bibliographystyle{apsrev}
\bibliography{StringNet.bib}






\end{document}